\documentclass[adp,a4paper,fleqn]{w-art}

\pdfoutput=1

\usepackage{times,cite,w-thm}
\usepackage{amsmath}
\usepackage{graphicx}
\usepackage{graphics}
\usepackage{amssymb}
\usepackage{color}
\usepackage{subfigure}
\usepackage{bbm}

\usepackage{epsfig}

\theoremstyle{plain}

\theoremstyle{definition}

\usepackage[]{graphicx}
\def\CZ{{\mathcal Z}}

\newcommand{\Tr}{\mathrm{Tr}}
\newcommand{\STr}{\mathrm{STr}}
\newcommand{\tr}{\mathrm{tr}}
\newcommand{\trf}{\mathrm{tr}_{\rm F}}
\newcommand{\E}{\mathrm{e}}
\newcommand{\I}{\mathrm{i}}
\newcommand{\be}{\begin{eqnarray}}
\newcommand{\ee}{\end{eqnarray}}

\newcommand{\nn}{\nonumber }
\newcommand{\fslash}{\hspace{-0.1ex} \slash }
\def\slash#1{\setbox0=\hbox{$#1$}              
   \dimen0=\wd0                                
   \setbox1=\hbox{/} \dimen1=\wd1              
   \ifdim\dimen0>\dimen1                      
      \rlap{\hbox to \dimen0{\hfil/\hfil}}  
      #1                                       
   \else                            
           
      \rlap{\hbox to \dimen1{\hfil$#1$\hfil}}  
      /                                      
   \fi}

\newcommand{\yb}{\bar\psi}
\newcommand{\xsb}{$\chi$SB}
\newcommand{\Nf}{N_{\text{f}}}
\newcommand{\Nc}{N_{\text{c}}}

\newcommand{\pat}{\partial_t}

\newcommand{\bfe}{\langle A_0\rangle}
\newcommand{\Td}{T_{\rm d}}
\newcommand{\Tc}{T_{\chi}}
\newcommand{\luv}{\lambda_{\psi}^{\rm UV}}

\newcommand{\fderL}[1]{\frac{\overrightarrow{\delta}}{\delta #1}}
\newcommand{\fderR}[1]{\frac{\overleftarrow{\delta}}{\delta #1}}

\newcommand{\fss}[1]{#1\!\!\!/}

\newcommand{\LQCD}{\Lambda_{\text{QCD}}}
\newcommand{\MSbar}{\overline{\text{MS}}} 
\newcommand{\hath}[1]{#1}
\newcommand{\lp}{\hath{\lambda}_{+}}
\newcommand{\lm}{\hath{\lambda}_{-}}

\newcommand{\lsf}{\hath{\lambda}_{\sigma}}
\newcommand{\lva}{\hath{\lambda}_{\text{VA}}}
\newcommand{\lF}{l_1^{\text{(F)},(4)}(0;0)}
\newcommand{\lFna}{l_1^{\text{(F)},(4)}}
\newcommand{\lFB}{l^{\textrm{(FB)},(4)}_{1,2}}
\newcommand{\lFBwa}{l^{\textrm{(FB)},(4)}_{1,2}(0,0;0,\eta_{\rm A})}
\newcommand{\lFBo}{l^{\textrm{(FB)},(4)}_{1,1}}
\newcommand{\lFBowa}{l^{\textrm{(FB)},(4)}_{1,1}(0,0;0,\eta_{\rm A})}

\newcommand{\Cas}{C_2(\Nc)}

\newcommand{\Nfcr}{N_{\text{f,cr}}}
\newcommand{\kcr}{k_{\text{cr}}}
\newcommand{\ksb}{k_{\text{SB}}}

\newcommand{\GIES}[3]{\mbox{\raisebox{#3} 
{\epsfig{file=#1,scale=#2,clip=true
}}~}}

\newcommand{\FGRAPH}[1]{\GIES{#1}{0.5}{-7.4mm}}

\begin{document}
\pagespan{1}{}

\title[Fermion Interactions and Universal Behavior in Strongly Interacting Theories] 
{Fermion Interactions and Universal Behavior\\ in Strongly Interacting Theories} 

\author[Jens Braun]{Jens Braun 
  \footnote{Jens Braun\quad E-mail:~\textsf{j.braun@uni-jena.de}}
\address{Theoretisch-Physikalisches Institut,
  Friedrich-Schiller-Universit\"at Jena,\\ 
  Max-Wien-Platz~1, D-07743~Jena, Germany}}

\begin{abstract}
The theory of the strong interaction, Quantum Chromodynamics (QCD), describes the generation of hadronic masses 
and the state of hadronic matter during the early stages of the evolution of the universe. 
As a complement, experiments with ultracold fermionic atoms provide a clean environment to benchmark our understanding 
of dynamical formation of condensates and the generation of bound states in strongly interacting many-body systems. 

Renormalization group (RG) techniques offer great potential for theoretical advances in both hot and dense QCD 
as well as many-body physics, but their connections have not yet been investigated in great detail. 
We aim to take a further step to bridge this gap. 
A cross-fertilization is indeed promising since it may eventually provide us with an ab-initio description of 
hadronization, condensation, and bound-state formation in strongly interacting theories. 
After giving a thorough introduction to the derivation and analysis of fermionic RG flows, we give an 
introductory review of our present understanding of universal long-range behavior in various 
different theories, ranging from non-relativistic many-body problems to relativistic gauge theories, 
with an emphasis on scaling behavior of physical observables close to 
quantum phase transitions (i.~e. phase transitions at zero temperature) as well as thermal phase transitions.
\end{abstract}

\maketitle                   

\tableofcontents  


%
\section{Introduction}\label{sec:intro}

Strongly interacting fermions play a very prominent role in nature. The dynamics of a large variety of theories close
to the boundary between a phase of gapped and ungapped fermions is determined by strong fermion interactions.
For instance, the chiral finite-temperature phase boundary in quantum chromodynamics (QCD), the theory of the strong interaction, is 
governed by strong fermionic self-interactions. In the low-temperature phase the quark sector is driven to criticality 
due to strong quark-gluon interactions. These strong gluon-induced quark self-interactions eventually lead to a breaking of the chiral symmetry
and the quarks acquire a dynamically generated mass. The chirally symmetric high-temperature phase, on the other hand, is 
characterized by massless quarks. 
The investigation of the QCD phase boundary represents one of the major research fields in physics, both experimentally and theoretically. Since 
the dynamics of the quarks close to the chiral phase boundary affect the equation of state of the theory, a comprehensive
understanding of the quark dynamics is of great importance for the analysis of present and
future heavy-ion collision experiments at BNL, CERN and the FAIR facility~\cite{Braun-Munzinger:2001ip}.

While heavy-ion collision experiments provide us with information on hot and dense QCD, experiments with ultracold trapped atoms 
provide an accessible and controllable system where strongly-interacting quantum many-body phenomena can be investigated precisely. 
In contrast to the theory of strong interactions, the interaction strength can be considered a free parameter in these systems which can 
be tuned by hand. In fact, the interaction strength is directly proportional to the $s$-wave scattering length and can therefore 
be modulated via an external magnetic field using Feshbach resonances~\cite{Feshbach}. It is therefore possible to study quantum phenomena 
such as superfluidity and Bose-Einstein condensation in these systems. From a theorist's point of view, this strong degree of experimental 
control opens up the possibility to test non-perturbative methods for the description of strongly interacting systems.

Phases of ultracold Fermi gases at zero and finite temperature 
have been studied experimentally, see e.~g. Refs.~\cite{Zwierlein,Partridge,Partridge:2006zz,Regal,Nascimbene} as well as 
theoretically, see e.~g.~Refs.~\cite{Carlson:2003zz,Astrakharchik,Bulgac:2005pj,Bulgac:2007ah, Burovski,Ceperley,Bulgac:2008zz,CombMora,Chevy2,Chevy,Stoof,Duan,Forbes,Diehl:2007ri,Pilati,Stoof2,Recati,Bloch,Haussmann,Diehl:2009ma}, 
over the past few years. In particular, studies with renormalization group (RG) methods 
exhibit many technical similarities to studies 
of QCD at finite temperature and density, see e.~g. Refs.~\cite{Gies:2002hq,Gies:2005as,Braun:2005uj,Braun:2006jd,Braun:2008pi,Braun:2009si}. 
Physically, in both cases the phase boundary is determined by strong interactions of the fermions. 
While the asymptotic limits of the phase diagram of ultracold atoms
for small positive and small negative (s-wave) scattering length associated with Bose-Einstein condensation and
Bardeen-Cooper-Schrieffer (BCS) superfluidity~\cite{Bardeen:1957mv}, respectively, are under control 
theoretically~\cite{Melik,Heiselberg:2000ya,Arnold:2001mu,Kashurnikov:2001zz,Blaizot:2004qa}, our understanding
of the finite-temperature phase diagram in the limit of large scattering length (strong-coupling limit) is still 
incomplete~\cite{Carlson:2003zz,Bulgac:2005pj,Bulgac:2007ah,Bulgac:2008zz,Bloch,Diehl:2009ma}.

Aside from phase transitions at finite temperature, experiments with ultracold fermionic atoms provide a very clean environment
for studies of {\it quantum phase transitions}. Experiments with a dilute gas of atoms in two different hyperfine spin states
have been carried out in a harmonic trap at a finite spin-polarization~\cite{Zwierlein,Partridge}.
Since there is effectively no spin relaxation in these experiments, in contrast to most other condensed matter systems, 
the polarization remains constant for long times. Deforming the system by varying the polarization allows us 
to gain a deep insight into BCS superfluidity and its underlying mechanisms~\cite{Bardeen:1957mv}.  
Originally, BCS theory has been worked out for systems in which the Fermi surfaces of the 
two spin states are identical, i. e. the polarization of the system is zero.  
As a function of the polarization, a {\it quantum phase transition}
occurs at which the (fully polarized) normal phase becomes 
energetically more favorable than the superconducting phase~\cite{Chevy,Forbes,Pilati}. 
After giving a thorough introduction 
on the level of (advanced) graduate students to the derivation 
and analysis of fermionic RG flow equations\footnote{Our introduction is kept on a basic level. 
In order to explain the derivation of fermionic RG flows,  we employ  a simple four-fermion theory to explain the derivation of fermionic RG flows.
Our toy model already shares many aspects with more complicated theories, such as QCD. In addition to 
gaining first insights into symmetry breaking patterns encoded in the fixed-point structure of strongly-interacting systems, 
a simple four-fermion theory allows us to develop a simple terminology for the discussion of various other theories in the subsequent sections. 
In particular, the analysis of such a theory allows us to address many technically relevant questions such as Fierz ambiguities, the role of momentum
dependences of couplings, and Hubbard-Stratonovich transformations.
However, more advanced readers may readily skip Sects.~\ref{sec:rg} and~\ref{sec:example}.} 
in Sects.~\ref{sec:rg} and~\ref{sec:example}, 
we shall discuss aspects of symmetry breaking and condensate-formation in non-relativistic theories from a RG point of view in Sect.~\ref{sec:coldgases}.
For simplicity, we restrict ourselves to systems with a vanishing polarization. The generalization of our RG approach to
spin-polarized gases is straightforward and has been discussed in Refs.~\cite{Stoof2,Schmidt:2011zu}.

Phase separation between a superfluid core and a surrounding normal phase has been indeed observed in experiments
with an imbalanced population of trapped spin-polarized ${}^{6}\text{Li}$ atoms at unitarity
at MIT and Rice University~\cite{Zwierlein,Partridge}. The density profiles measured in these experiments prove the existence 
of a skin of the majority atoms. 
A critical polarization $P_{\rm c}$ associated with a quantum phase transition has been found in both the
MIT and Rice experiment. Aside from studies at zero temperature, finite-temperature studies of a spin-polarized gas have been performed at 
Rice University~\cite{Partridge}. In these experiments the critical polarization, above which the superfluid core disappears, 
has been measured as a function of the temperature. 
In accordance with theoretical studies~\cite{CombMora,Stoof,Stoof2}, 
the results from the Rice group suggest that a tricritical
point exists in the phase diagram spanned by temperature and polarization, at which
the superfluid-normal phase transition changes from second to first order as the 
temperature is lowered. Depending on the physical observable, it is in principle possible that finite-size 
and particle-number effects are visible in \mbox{the experimental data}. Concerning the critical polarization,
such effects have been studied in~Ref.~\cite{Ku:2008vk}.

There is indeed direct evidence that finite-size effects can alter the phase structure of
a given theory. For example, Monte-Carlo studies of the $1+1d$ Gross-Neveu model show that the 
finite-temperature phase diagram of the uniform system is modified significantly due to the non-commensurability of the spatial lattice size 
with the intrinsic length scale of the inhomogeneous condensate~\cite{Thies:2003kk,deForcrand:2006zz}. In particular, 
the phase with an inhomogeneous ground state shrinks. Such commensurability effects may be present in trapped ultracold
Fermi gases as well. Since the Gross-Neveu model in $1+1d$ is reminiscent of QCD in many ways, the existence 
of a stable ground state governed by an inhomogeneous condensate is subject of an ongoing debate, see e.~g. Refs.~\cite{Nickel:2009wj,Kojo:2009ha}.
In any case, it is well-known that the mass spectrum and the thermodynamics of QCD has an intriguing dependence 
on the volume size and the boundary conditions of the fields, see e.~g.
Refs.~\cite{Colangelo:2002hy,AliKhan:2003cu,Colangelo:2003hf,Guagnelli:2004ww,Colangelo:2004xr,Braun:2004yk,%
Colangelo:2005gd,Braun:2005fj,Braun:2005gy,Braun:2010vd,Klein:2010tk,Braun:2010pp}.

Our theoretical understanding of the phase structure of trapped fermions is currently 
mostly based on Density Functional Theory (DFT)~\cite{Kohn:1965zzb} 
in a local density approximation (LDA) in which, for example, derivatives of densities are omitted in the
ansatz for the action, see e.~g. Refs.~\cite{Stoof,Chevy2,Duan,Recati,Haussmann}. 
From a field-theoretical point of view, DFT corresponds to a mapping of the (effective) action of a fermionic theory onto
an action which depends solely on the density. The latter then plays the role of a composite degree of freedom of
fermions. Thus, the underlying idea is reminiscent of the {\it Hubbard-Stratonovich} transformation~\cite{Hubbard:1959ub,Stratonovich}
widely used in low-energy QCD models and spin systems. In any case, the introduction of an effective degree of freedom, such as the density,
turns out to be advantageous for a description of theories with an inhomogeneous ground-state.
Again, experiments with ultracold atoms allow us to test different approaches and approximation schemes.
In Refs.~\cite{Chevy2,Recati}, the equation of states of the superfluid and the normal phase 
of a uniform system have been employed to construct a density functional which allows
to study the ground-state properties of trapped Fermi gases. Such a procedure corresponds to an LDA.
While there is some evidence that Fermi gases in isotropic traps can be quantitatively understood within DFT in LDA~\cite{Recati},
the description of atoms in a highly-elongated trap in LDA seem to fail and derivatives of the density need to be taken
into account~\cite{Mueller1,Sensarma}. In the spirit of these studies, 
we shall discuss a functional RG approach to DFT in Sect.~\ref{sec:DFTRG}
which relies on an expansion of the energy density functional in terms of correlation functions and 
allows to include effects beyond LDA in a systematic fashion. 

Heavy nuclei combine aspects of dense and hot QCD and systems of ultracold atoms. 
We again need to describe strong interactions between fermions, the nucleons, which form a stable bound
state depending on, e.~g., the number of protons and neutrons. These interactions 
are repulsive at short range and attractive at long range as in the case of ultracold atomic gases.
Loosely speaking, heavy nuclei can be viewed as spin-polarized systems of two fermion species 
comparable to those systems studied in experiments with trapped spin-polarized atoms at MIT and Rice University~\cite{Zwierlein,Partridge}. 
In fact, almost all nuclei have more neutrons than protons.\footnote{Protons and neutrons correspond to two different isospin states of the nucleon.}
Therefore density profiles of protons and neutrons in heavy nuclei 
are evocative of the profiles associated with the two fermion states in experiments with ultracold atoms. 
For heavy nuclei, DFT remains currently to be the only feasible approach for a calculation of ground-state properties associated
with inhomogeneous densities. State-of-the-art density functional approaches are essentially based on fitting the parameters of a 
given density functional such that one reproduces the experimentally determined values of the ground-state properties of various heavy 
nuclei~\cite{Dobaczewski:2001ed,Stoitsov:2003pd,Bender:2003jk}. These density functionals are then employed to describe ground-state properties of other heavy nuclei. 
As mentioned above, we shall briefly discuss an RG approach to DFT in Sect.~\ref{sec:DFTRG}, which opens up the
possibility  to study ground-state properties of heavy nuclei from the underlying nucleon-nucleon interactions.
Such an ab-initio DFT approach might prove to be useful for future studies of ground-state properties of (heavy) self-bound systems.

Hot and dense QCD, ultracold atoms and nuclear physics represent just three examples for systems in which the dynamics
are governed by strong fermion interactions. Of course, the list can be extended almost arbitrarily. In the context of
condensed-matter theory, we encounter systems such as so-called high-$T_{\rm c}$ superconductors.
In this case the challenge is to describe reliably the dynamics of electrons
at finite temperature in an ambient solid-state system. 
The so-called Hubbard model provides a theory to describe these
superconductors~\cite{Anderson,FuldeBook} and
has been extensively studied with renormalization-group techniques, see e.~g. 
Refs.~\cite{Honerkamp1,Honerkamp2,Honerkamp3,Honerkamp4}. 
It is worth noting that both the mechanisms as well as the techniques are remarkably similar to the ones in
renormalization-group studies of gauged fermionic systems interacting strongly via competing channels~\cite{Braun:2005uj,Braun:2006jd,Braun:2008pi},
such as QCD, and of imbalanced Fermi gases in free space-time~\cite{Stoof2}. In Sect.~\ref{sec:njlgn}, we discuss more general aspects of
(non-gauged) Gross-Neveu- and Nambu-Jona-Lasinio-type models which also exhibit technical similarities to studies 
of condensed-matter systems. Nambu-Jona-Lasinio-type models are widely used as effective QCD low-energy models.
On the other hand, Gross-Neveu-type models have been employed as toy models to study certain aspects of the QCD phase diagram but
they are also related to models in condensed-matter theory, e.~g. to 
models of ferromagnetic (relativistic) superconductors~\cite{PSSB:PSSB2221030242,Schnetz:2004vr}.
In this review, we shall use Gross-Neveu- and Nambu-Jona-Lasinio-type models to discuss dynamical 
chiral symmetry breaking (via competing channels) and the role of momentum dependences of fermionic interactions.

In addition to fermion dynamics at finite temperature, quantum phase transitions play a prominent role in condensed-matter 
theory, e.~g., in the context of graphene. Effective theories of graphene, such as QED${}_3$ and the Thirring model,
are expected to approach a quantum critical point when the number of fermion species, namely the number of electron species, is 
varied~\cite{Drut:2007zx,Gies:2010st}. 
RG studies of these effective theories, see e.~g. Refs.~\cite{Drut:2007zx,Gies:2009da,Gies:2010st}, are closely related 
to studies of quantum phase transitions in QCD~\cite{Jaeckel:2003uz,Gies:2005as,Braun:2005uj,Braun:2006jd,Braun:2009ns,Braun:2010qs}.
Similar to the situation in QED${}_3$, a quantum phase transition from a chirally broken to a conformal phase 
is expected in QCD when the number of (massless) quark flavors is increased.
Studies of the dependence on the number of fermion species seem to be a purely academic question. Depending on the theory under
consideration, however, such a deformation of the theory may allow us to gain insights into the dynamics of fermions close to a 
phase boundary in a controlled fashion. For example, the gauge coupling in QCD becomes
small when the number of quark flavors is increased and therefore perturbative approaches in the gauge sector become meaningful. Moreover, 
an understanding of strongly-flavored QCD-like gauge theories is crucial for applications beyond the standard-model, namely for so-called
walking technicolor scenarios for the Higgs sector~\cite{Weinberg:1979bn,Holdom:1981rm,Hong:2004td,Sannino:2004qp,Dietrich:2005jn,
Dietrich:2006cm,Ryttov:2007sr,Antipin:2009wr,Sannino:2009za}.
In Sect.~\ref{sec:gaugetheories}, we shall discuss chiral symmetry breaking in gauge theories with~$\Nf$ fermion flavors. 
In particular, we shall present a detailed discussion of the scaling
behavior of physical observables close to the quantum phase transition which occurs for large~$\Nf$.

Our discussion shows that systems of strongly interacting fermions play indeed are very prominent role in nature and that their dynamics
determine the behavior of a wide class of physical systems with seemingly substantial differences. 
However, our discussion also shows that the underlying 
mechanisms of symmetry breaking and the applied techniques are very similar in these different fields. Therefore a phenomenological
and technical cross-fertilization offers great potential to gain a better understanding of the associated physical processes.
As outlined, examples include an understanding of the dynamical generation of hadron
masses as well as of the dynamical formation of condensates and bound-states in ultracold gases 
from first principles. The main intent of the present review is to 
give a general introduction to the underlying mechanisms of symmetry breaking and bound-state formation 
in strongly-interacting fermionic theories. In particular, we aim to give a thorough introduction into the scaling behavior of 
physical observables close to critical points, ranging from power-law scaling behavior to essential scaling. 
As a universal tool for studies of quantum field theories we employ mainly Wilsonian-type renormalization-group 
techniques~\cite{Wilson:1971bg,Wilson:1971dh,Wilson:1973jj,Wegner:1972ih,Nicoll:1977hi,Polchinski:1983gv,Wetterich:1992yh}. 
For concrete calculations we shall use the so-called Wetterich equation~\cite{Wetterich:1992yh} which we briefly introduce in the 
next section. Reviews focussing on various different aspects of renormalization-group approaches can be found 
in Refs.~\cite{Shankar:1993pf,Litim:1998nf,SalmhoferBook,Bagnuls:2000ae,Berges:2000ew,Polonyi:2001se,Bogner:2003wn,Delamotte:2003dw,%
Pawlowski:2005xe,Gies:2006wv,Delamotte:2007pf,Bogner:2009bt,Rosten:2010vm,Kopietz:2010zz,Honerkamp5}.

%



%
\section{Renormalization Group - Basic Ideas}\label{sec:rg}
We begin with a brief introduction of the basic ideas of RG approaches including a discussion of the Wetterich equation.
The latter describes the scale dependence of the quantum effective action which underlies our studies in this and the following sections.

In perturbation theory, the correlation functions of a given quantum field theory contain divergences which can be removed by a 
renormalization prescription. The choice of such a prescription defines a renormalization scheme and renders all 
(coupling) constants of a given theory scheme-dependent. Since the renormalized (coupling) constants are nothing but 
mathematical parameters, their values can be arbitrarily changed by changing the renormalization prescription. We stress that
these renormalized constants should not be confused with physical observables such as, for example, the phase transition
temperature or the {\em physical} mass of a particle. Physical observables are, of course, 
invariant under a variation of the renormalization prescription, provided we have not truncated the 
perturbation series. If we consider a truncated perturbation series, we find that there is a residual dependence on the 
renormalization scheme which can be controlled to some extent by the 
so-called "Principle of Minimum Sensitivity"~\cite{Stevenson:1981vj}, see also discussion below.

At this point we are then still free to perform additional finite renormalizations. 
This results in different {\it effective} renormalization prescriptions. A given renormalization prescription can then be considered 
as a particular reordering of the perturbative expansion which expresses it in terms of new renormalized constants~\cite{Pokorski:1987ed}.
Let us assume that the transformations between the finite renormalizations can be parametrized by introducing 
an auxiliary single mass scale~$\mu$. This scale corresponds to a UV (cutoff) scale at which the parameters of the theory are fixed.
A set of RG equations for a given theory then describes the changes of the renormalized parameters of this theory (e. g. 
the coupling constant) induced by a variation of the auxiliary mass scale~$\mu$. The set of renormalization {\em transformations} is called 
the renormalization group.

Let us now consider a (renormalized) microscopic theory at some large momentum scale $\Lambda$ defined by a (classical) action~$S$.
Wilson's basic idea of the renormalization group is to start with such a classical action $S$ and then to integrate out successively 
all fluctuations from high to low momentum scales~\cite{Wilson:1971bg,Wilson:1971dh,Wilson:1973jj}. 
This procedure results in an action which depends on an IR regulator scale, 
say~$k$, which plays the role of a reference scale. The values of the (scale-dependent) couplings defining 
this action on the different scales are related by continuous RG transformations. We shall refer to the change of a coupling under a variation 
of the scale~$k$ as the RG flow of the coupling. In this picture, 
{\it universality} means that the RG flow of the couplings is governed by a
{\it fixed point}. The possibility of identifying fixed points of a theory makes the RG such a powerful tool for 
studying statistical field theories as well as quantum field theories. As we shall discuss below, critical behavior near phase transitions 
is intimately linked to the fixed-point structure of the theory under consideration.

In this review we employ a non-perturbative RG flow equation, the Wetterich equation~\cite{Wetterich:1992yh}, 
for the so-called effective average action in order to analyze critical behavior in physical systems. The effective average 
action~$\Gamma _k$  depends on an intrinsic momentum scale $k$ which parameterizes the Wilsonian RG transformations.
We note that such an approach is based on the fact that an infinitesimal RG transformation (i.~e. an RG step), 
performed by an integration over a single momentum shell of width~$\Delta k$, is finite. For this reason we are able to 
integrate out all quantum fluctuations through an infinite sequence of such RG steps. The flow equation 
for $\Gamma_k$ then describes the continuous trajectory from the microscopic theory $S$ at large 
momentum scales to the full quantum effective action (macroscopic theory) at small momentum scales.  Thus, it 
allows us to cover physics over a wide range of scales.

Here, we only discuss briefly the derivation and the properties of the RG flow equation for the effective average
action~$\Gamma_k$; for details we refer to the original work by Wetterich~\cite{Wetterich:1992yh}. 
The scale-dependent effective action $\Gamma_k$ is a generalization of the (quantum) effective action $\Gamma$ but only includes
the effects of fluctuations with momenta $p^2 \gtrsim k^2$.  Therefore $\Gamma _k$ is sometimes called a coarse-grained effective action 
since quantum fluctuations on length scales smaller than $1/k$ are integrated out. 
The underlying idea is to calculate the generating functional $\Gamma$ 
of one-particle irreducible (1PI) graphs of a given theory by starting at an ultraviolet~(UV) scale~$\Lambda$ with the microscopic (classical) action $S$ and 
then successively integrating out quantum fluctuations by lowering the scale $k$. 
The quantum effective action $\Gamma$ is then obtained in the limit~$k\rightarrow 0$.
In other words, the coarse-grained effective action~$\Gamma_k$ interpolates between the classical action $S$ at the UV scale~$\Lambda$ and 
the 1PI generating functional $\Gamma$ in the infrared limit (IR)~$k\rightarrow 0$. 
The starting point for the derivation of the flow equation of $\Gamma$ is a UV- and IR-regularized 
generating functional~$Z_k$ for the Greens functions:\footnote{Throughout this review we work in 
Euclidean space-time. We refer the reader to App.~\ref{sec:conv} for details on our conventions.}
\be
Z_{k} [J]
=\int _{\Lambda} \mathcal{D}\phi (\{p_i\})\E ^{-S[\phi]-\Delta S_k [\phi]+ J^{\rm T}\cdot \phi}  
\equiv \E ^{W_k [J]}\,,\label{eq:ZkDef}
\ee
where  $\{p_i\}\equiv \{p_0,\dots,p_d\}$ and $W_k$ is the scale-dependent generating functional for the connected Greens functions.
The field variable $\phi$ as well as the source $J$ are considered as generalized vectors in field space and are defined as
\be
\phi=\left(\begin{array}{c}
\psi\\
\bar{\psi}^{\rm T}\\
\varphi\\
\vdots
\end{array}\right)\quad\text{and}\quad J^{\rm T}=\left(\bar{\eta},\,\eta ^{\rm T},\,j,\,\dots\right)\,.
\ee
Moreover, we have introduced a generalized scalar product in field space:~$J^{\rm T}\cdot \phi \equiv \int d^d x \{ \bar{\eta}\psi + \dots \}$.
Here, the field~$\psi$ represents a Dirac spinor, and $\varphi$ denotes a real-valued scalar field. The dots indicate that other 
types of fields, e. g. gauge-fields, are allowed as well. 
For non-relativistic theories of fermions, the generating functional
can be defined accordingly. We assume that the theory is well-defined by a UV-regularized 
generating functional: The index $\Lambda$ indicates that we only integrate over fields $\phi(\{p_i\})$ 
with momenta $|p| \lesssim \Lambda$, i. e. we implicitly take $\phi(\{p_i\})=0$ for $|p|>\Lambda$. 
To regularize the infrared modes a cutoff term has been inserted into the path integral. It is defined as
\be
\Delta S_k [\phi] &=& \frac{1}{2}\sum_{a,b}\int \frac{d^d p}{(2\pi)^d}\,\phi _a(\{-p_i\})R_k ^{ab} (\{p_i\})\phi _b (\{p_i\})
\equiv\frac{1}{2}\,\phi ^{\rm T}\cdot R_k  \cdot\phi\,,
\ee
where $R_k$ is a matrix-valued regulator function. 
Through the insertion of the cutoff term, we have defined a generating functional which now 
depends on the scale~$k$.

The cutoff function $R_k$ has to fulfill three conditions. Since $R_k$ has been introduced to
regularize the IR, it must fulfill
\be
\lim _{\frac{p^2}{k^2}\rightarrow 0} R_k (\{p_i\}) > 0\,,\label{eq:regprop1}
\ee
where $p^2=p_0^2 + \dots + p_{d-1}^2$. Second, the function $R_k$ must vanish in the IR-limit, 
i.~e. for $k\rightarrow 0$:
\be
\lim _{\frac{k^2}{p^2}\rightarrow 0} R_k  (\{p_i\}) =0\,.\label{eq:regprop2}
\ee
This condition ensures that we obtain the 1PI generating functional for $k\to0$.
Third, the cutoff function should obey
\be
\lim _{k\rightarrow \Lambda} R_k (\{p_i\}) \rightarrow \infty\label{eq:regprop3}
\ee
for fixed $p^2$. This property guarantees that $\Gamma _{k\rightarrow\Lambda}\to S$ for $k\to \Lambda$.

In this review, we shall always use cutoff functions which can be written in terms of 
a dimensionless regulator shape function $r(p^2/k^2)$. For simple relativistic scalar theories, we may choose
\be
R_k (p^2) \propto  
p^2 r\left(\frac{p^2}{k^2}\right)\,.\label{eq:bosregexample}
\ee
For studies of theories with chiral fermions, it is convenient to 
employ a cutoff function which {\it preserves} chiral symmetry. An appropriate choice is~\cite{Jungnickel:1995fp}
\be
R_k ^{\psi}(\{p_i\}) \propto 
\fslash{p}\, r_{\psi}\left(\frac{p^2}{k^2}\right)\,.
\ee
On the other hand, for non-relativistic fermionic many-body problems 
the choice of the cutoff function should respect the presence of a Fermi surface. An appropriate choice for such 
a cutoff function is given by~\cite{Diehl:2009ma} 
\be
R_k ^{\psi}(\vec{p}^{\,2})=k^2 \, r_{\psi}({\mathcal Z})\,\quad\text{with}\quad {\mathcal Z}=(\vec{p}^{\,2}-\mu)/k^2\,,\label{eq:nr_cutoffexample}
\ee
where, for instance,
\be
r_{\psi}({\mathcal Z})=(\text{sign}({\mathcal Z})-{\mathcal Z})\theta(1-|{\mathcal Z}|)\,.
\ee
The chemical potential of the fermions is given by~$\mu$ and defines the associated Fermi surface. This choice for the regulator function
arranges the momentum-shell integrations around the Fermi surface, i.~e. modes with momenta $(\vec{p}^{\,2}-\mu) \geq k^2$
remain unchanged while the momenta of modes with $(\vec{p}^{\,2} -\mu) < k^2$ are cut off.

For scalar field theories, the presence of a cutoff function of the form $\sim R_k(p^2)$ is in general not problematic. 
For gauge theories, however, it causes difficulties due to condition~\eqref{eq:regprop1} which
essentially requires that the cutoff function acts like a mass term for small momenta. Therefore the cutoff function 
necessarily breaks gauge symmetry. We stress that this observation does by no means imply that such an approach cannot be 
applied to gauge theories. In fact, it is always necessary to fix the gauge in order to treat gauge theories perturbatively 
within a path-integral approach. This gauge-fixing procedure also breaks gauge invariance. 
Gauge-invariant results are then obtained by resolving Ward-Takahashi identities. Consequently, we can think of the cutoff 
function as an additional source of gauge-symmetry breaking. In analogy to perturbation theory, one then needs 
to deal with {\em modified} Ward-Takahashi identities in order to recover 
gauge invariance~\cite{Bonini:1994kp,Ellwanger:1995qf,Ellwanger:1996wy,Reuter:1996ub,Reuter:1997gx,Freire:2000bq}. 
In addition, there are essentially two alternatives: first, one can construct manifestly 
gauge-invariant flows as proposed in~\cite{Morris:2000fs,Arnone:2005fb,Rosten:2010vm}.
Second, we can apply special (useful) gauges, such as the background-field gauge~\cite{Abbott:1980hw,Abbott:1981ke}. 
We refer the reader to Ref.~\cite{Gies:2006wv} for a detailed introduction to RG flows in gauge theories.

The coarse-grained effective action~$\Gamma_k$ can in principle be obtained from the IR-regularized functional~$W _k [J]$ in 
a standard fashion, see, e.~g., the standard textbook derivation of 
the quantum effective action~$\Gamma$ in Refs.~\cite{Abbott:1981ke,Pokorski:1987ed}. 
However, we employ here a {\it modified} Legendre 
transformation to calculate the coarse-grained effective action:\footnote{Note that a functional 
obtained by an ordinary Legendre transformation, e.~g.~$\Gamma[\Phi]=\sup_J\{-W[J]+J^{\rm T}\cdot\Phi\}$, 
is convex. However, the coarse-grained effective action $\Gamma _k$ is not necessarily convex 
for finite $k$ due to the insertion of the cutoff term. Since $R_k \to 0$ for $k\to 0$, 
convexity is recovered in the limit $k\rightarrow 0$.} 
\be
\Gamma _k [\Phi]=\sup_J\left\{-W_k [J] + J^{\rm T}\cdot\Phi\right\} - \Delta S_k [\Phi]\,.\label{eq:GkDef}
\ee
The so-called classical field~$\Phi$ is implicitly defined by the supremum prescription.
The modification of the Legendre transformation is necessary for the connection of $\Gamma _k$ with the classical action $S$ in the limit $k\to\Lambda$. From this 
definition of $\Gamma_k$ we find the RG flow equation of the coarse-grained effective action, the so-called Wetterich equation~\cite{Wetterich:1992yh}, 
by taking the derivative with respect to the scale~$k$:
\be
\partial _t\,\Gamma _k [\Phi] = \frac{1}{2}\,\STr \, 
\left[\Gamma ^{(2)} _k [\Phi] + R_k \right]^{-1}\cdot (\partial _t R_k) 
=\phantom{-}\frac{1}{2}\;\;\FGRAPH{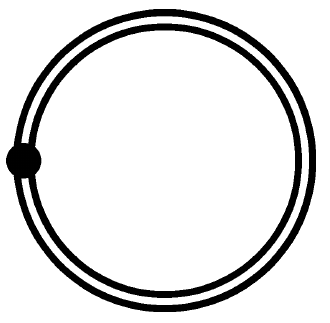}\,,\label{eq:FLOWEQ}
\ee
with $t=\ln (k/\Lambda)$ being the RG ``time" and~$\Gamma^{(2)}\equiv \Gamma^{(1,1)}$.
The $(n+m)$-point functions are defined as follows:
\be
\Gamma ^{(n,m)}_k [\Phi]
=\stackrel{n-\text{times}}{\overbrace{\fderL{\Phi ^{T}}\cdots\fderL{\Phi ^{T}}}}\Gamma _k [\Phi]
\stackrel{m-\text{times}}{\overbrace{\fderR{\Phi }\cdots\fderR{\Phi}}}\,.\label{eq:funcder}
\ee
Thus, $\Gamma ^{(1,1)} _k$ is matrix-valued in field space. 
The super-trace arises since $\Phi$ contains both fermionic as well as bosonic degrees of 
freedom and it provides a minus sign in the fermionic subspace of the matrix. 
The double-line in Eq.~\eqref{eq:FLOWEQ} represents the {\it full} propagator of the theory which includes the {\it complete} 
field dependence. The solid black dot in the loop stands for the insertion of $\partial _t R_k$. The structure of the
flow equation reveals that the regulator function $R_k$ specifies the Wilsonian momentum-shell integrations, such
that the RG flow of $\Gamma_k$ is dominated by fluctuations with momenta $|p|\simeq k$.

The flow equation~\eqref{eq:FLOWEQ} has been obtained by taking the derivative of $\Gamma _k$ with respect to the scale $k$.
However, we have not taken into account a possible scale dependence of the classical field $\Phi$ yielding a 
term~$\sim\partial_t \Phi_k$ on the right-hand side of Eq.~\eqref{eq:FLOWEQ}. 
We stress that the inclusion of this term is a powerful extension of the flow equation discussed here, since 
it allows to bridge the gap between microscopic and macroscopic degrees of freedom in the RG flow, e.~g. between quarks and gluons
and hadronic degrees of freedom, without any fine-tuning~\cite{Gies:2001nw,Gies:2002hq,Gies:2002kd}. 
More technically speaking, this extension makes it possible to perform continuous Hubbard-Stratonovich transformations in the RG flow. 
We shall not employ these techniques here since they do not provide us with additional insights into the 
fermionic fixed-point structure to which the scope of the present review is limited.
For details concerning such an extension of the
flow equation~\eqref{eq:FLOWEQ}, we refer the reader to 
Refs.~\cite{Gies:2001nw,Gies:2002hq,Gies:2002kd,Pawlowski:2005xe,Gies:2006wv,Floerchinger:2009uf,Floerchinger:2010da}. 
In Ref.~\cite{Braun:2008pi} these so-called re-bosonization techniques\footnote{In the context of QCD these techniques are sometimes referred to
as "dynamical hadronization".} 
have been employed for a first-principles study of the QCD phase boundary.

As should be the case for an {\it exact} one loop flow~\cite{Litim:2002xm}, the Wetterich equation~\eqref{eq:FLOWEQ} is linear in the inverse of 
the full propagator. Moreover, it is a nonlinear functional differential equation, since it involves the inverse of the second functional derivative of the 
effective action. 
We stress, however, that the loop in Eq.~\eqref{eq:FLOWEQ} is not a simple perturbative loop 
since it depends on the {\it full} propagator. In fact, it can be shown that arbitrarily high loop-orders are 
summed up by integrating this flow equation \cite{Litim:2002xm}. Nonetheless it is possible and sometimes even 
technically convenient to rewrite~\eqref{eq:FLOWEQ} in a form which is reminiscent of the textbook form of the one-loop
contribution to the effective action: 
\be
\partial _t\,\Gamma _k [\Phi] = \frac{1}{2}\,\STr \, \tilde{\partial}_t \ln \left( \Gamma ^{(1,1)} _k [\Phi] + R_k \right)\,.\label{eq:trlogrep}
\ee
Here, $\tilde{\partial}_{t}$ denotes a formal derivative acting only on the $k$-dependence of the regulator function $R_k$. Replacing 
$\Gamma ^{(1,1)} _k$ by the (scale-independent) second functional derivative of the classical action, $S^{(1,1)} $, we can perform 
the integration over the RG scale $k$ analytically and obtain the standard one-loop expression for the effective action:
\be
\Gamma _{\rm 1-loop} [\Phi] =S_{\rm UV}[\Phi] +\frac{1}{2}\,\STr \,  \ln  S ^{(1,1)} [\Phi] \,,\label{eq:effactoneloop}
\ee
where
\be
S_{\rm UV} [\Phi] = S[\Phi] - \frac{1}{2}\,\STr \,  \ln  \left( S ^{(1,1)} [\Phi] + R_{\Lambda} \right)\,.
\ee
Here, the second term on the right-hand side corresponds to the boundary condition for the RG flow at the UV scale~$\Lambda$, which
renders $\Gamma _{\rm 1-loop} $ finite.
 
From a technical point of view, the representation~\eqref{eq:trlogrep} turns out to be a convenient starting point
for our studies of the fixed-point structure of four-fermion interactions.
In order to calculate flow equations of four-fermion interactions, 
we decompose the inverse regularized propagator $\Gamma ^{(1,1)} _k [\Phi]$ on the right-hand side of the flow equation into 
a field-independent (${\mathcal P}_k$) and a field-dependent (${\mathcal F}_k$) part,
\be
\Gamma _{k} ^{(1,1)}[\Phi]+R_k ={\mathcal P}_k + {\mathcal F}_k\,.\label{eq:PF}
\ee
We can then expand the flow equation in powers of the fields according to
\be
\label{eq:flowexp}
\partial_{t}\Gamma_{k}&=&\frac{1}{2}\STr\bigg\{\tilde{\partial}_{t}\ln(\mathcal{P}_k+\mathcal{F}_k)\bigg\}\\
&=&\frac{1}{2}\STr\bigg\{\tilde{\partial}_{t}\left(\frac{1}{\mathcal{P}}_{k}\mathcal{F}_{k}\right)\bigg\}
\!-\!\frac{1}{4}\STr\bigg\{\tilde{\partial}_{t}\left(\frac{1}{\mathcal{P}}_{k}\mathcal{F}_{k}\right)^{2}\bigg\}
\!+\!\frac{1}{6}\STr\bigg\{\tilde{\partial}_{t}\left(\frac{1}{\mathcal{P}}_{k}\mathcal{F}_{k}\right)^{3}\bigg\}+\dots.
\nn
\ee
The powers of $\frac{1}{\mathcal{P}}_{k}\mathcal{F}_{k}$ can be calculated by simple matrix multiplications.
The flow equations for the various couplings can now be obtained by comparing the coefficients of the four-fermion operators on the
right-hand side of Eq.~\eqref{eq:flowexp} with the couplings specified in the definition of the effective action. In other words,
the flow equation of higher $n$-point functions are obtained straightforwardly from the flow equation \eqref{eq:FLOWEQ} 
(or, equivalently, from Eq.~\eqref{eq:flowexp}) by taking the appropriate number of functional derivatives. 
From this, we observe that the RG flow of the $n$-point function depends in general on the flow of the $(n+1)$- and $(n+2)$-point function.
This means that we obtain an infinite tower of coupled flow equations by taking functional derivatives of the flow equation \eqref{eq:FLOWEQ}. 
In most cases we are not able to solve this infinite tower of flow equations. Thus, we need to truncate the effective action and 
restrict it to include only correlation functions with $N_{\text{max}}$ external fields. However, such a {\em truncation} 
poses severe problems: first, the system of flow equations is no longer closed and, second, neglecting higher $n$-point 
functions may render the flow unstable in the IR region of strongly coupled theories.
For example in QCD, 
one would naively expect that contributions from higher $n$-point functions are important. 

Finding reliable truncations of the effective action is the most difficult step and requires a lot of physical insight. We stress that
an expansion in terms of $n$-point functions must not be confused with an expansion in some small parameter as in perturbation theory.
The assumption here is that the influence of neglected operators on the already included operators is small.
Once we have chosen a truncation for studying a given theory, we need to check its reliability. One possibility for such a check
is to extend the truncation by including additional operators and then check if the results obtained from this new truncation 
are in agreement with the earlier results. If this is not the case, one must rethink the chosen truncation. However, 
even if the results are not sensitive to the specific set of additional operators added to the truncation, this 
does not necessarily mean that one has included all relevant operators in the calculation. 
A second possibility to assess the reliability of a given truncation is to exploit the fact that physical observables should not depend on the regularization 
scheme. Since the scheme is specified by the cutoff function, the physical observables should be independent of this choice. 
In the present approach the scheme is defined by our choice for the regulator function $R_k$. Thus, we can vary $R_k$ and then 
check if the results depend on the choice of the cutoff function. If this is the case, an extension of the truncation might be required.
In addition to a simple variation of regulator functions, we may actually exploit the dependence on $R_k$ 
to {\it optimize} the truncated RG flow of a given theory. For example, an optimization criterion can be based on the size of the gap induced in the
effective propagator $(\Gamma _{k} ^{(1,1)}[\Phi]+R_k)^{-1}$, see Refs.~\cite{Litim:2001fd,Litim:2000ci,Litim:2001up}. 
We then denote those regulators to be optimized for which
the gap is maximized with respect to the cutoff scheme. In addition to such an optimization of RG flows within a given regulator class,
a more general criterion has been put forward in Ref.~\cite{Pawlowski:2005xe}. The latter defines the optimized regulator to be the one for which
the regularized theory is already closest to the full theory at~$k=0$, for a given gap induced in the effective propagator $(\Gamma _{k} ^{(1,1)}[\Phi]+R_k)^{-1}$. 
This optimization criterion yields an RG trajectory which defines the shortest path in theory space between the UV theory at $k=\Lambda$ and the
full theory at~$k=0$.
Both optimization criteria naturally encompass the so-called ``Principle of Minimum Sensitivity". 
However, in contradistinction to the ``Principle of Minimum Sensitivity", the optimization of (truncated) RG flows does not
rely on the existence of extremal values of physical observables which may arise from a variation of the regularization scheme.
For a detailed discussion of optimization 
criteria and properties of optimized RG flows, we refer the reader to Refs.~\cite{Litim:2000ci,Litim:2001fd,Litim:2001up,Pawlowski:2005xe}. 

Nonetheless, even an approximate solution of the flow equation~\eqref{eq:FLOWEQ} can describe non-perturbative physics 
reliably, provided the relevant degrees of freedom in the form of RG relevant operators are kept in the ansatz for the effective action. 



%
\section{RG Flow of Four-Fermion Interactions - A Simple Example}\label{sec:example}
In this section we discuss a simple four-fermion theory which already allows us to  
gain some important insight into the mechanisms of symmetry breaking in strongly-interacting theories.
A study of a simple four-fermion theory is useful for many reasons. First, it allows us to 
highlight various methods and technical aspects such as Fierz ambiguities, (partial) bosonization 
and the role of explicit symmetry breaking. Second, a confrontation of this model
study with our analysis of symmetry breaking in gauge theories is instructive:
To be specific, we will consider the mechanisms of chiral symmetry breaking
to point out the substantial differences between these theories.

\subsection{A Simple Example and the Fierz Ambiguity}\label{subsec:simpleex}
In this section we discuss the basic concepts and problems in describing strongly-interacting fermionic theories, 
with a particular emphasis on the application of RG approaches. To this end, we employ a Nambu--Jona-Lasinio-type model. Such models play
a very prominent role in theoretical physics. Originally, the Nambu--Jona-Lasinio (NJL) model has been used as an effective theory to describe spontaneous
symmetry breaking in particle physics based on an analogy with superconducting materials~\cite{Nambu:1961tp,Nambu:1961fr}, 
see Ref.~\cite{Klevansky:1992qe} for a review. RG methods have been extensively employed to study critical behavior in QCD with the aid of NJL-type models,
see e.~g.~Refs.~\cite{Jungnickel:1995fp,Berges:1997eu,Schaefer:1999em,Berges:2000ew,Braun:2003ii,%
Schaefer:2004en,Braun:2005fj,Nakano:2009ps,Braun:2009si,Braun:2010vd}.
Usually these model studies rely on a (partially) bosonized version of the action. 
We shall discuss aspects of bosonization in Sect.~\ref{subsec:bos}. 
For the sake of simplicity we start with a purely fermionic formulation of the NJL model with only one fermion species. This 
model has been extensively studied at zero temperature with the functional RG in Refs.~\cite{Jaeckel:2002rm,Jaeckel:2003uz}.
In particular, the ambiguities arising from Fierz transformations have been explicitly worked out and discussed. We shall follow
the discussion in Refs.~\cite{Jaeckel:2002rm,Jaeckel:2003uz} but extend it with respect to issues arising at finite temperature and 
for a finite (explicit) fermion mass. In addition, we exploit this model to discuss general aspects of theories with many fermion flavors as well 
as quantum critical behavior.

In the following we consider a simple ansatz for the effective action in $d=4$ Euclidean space-time dimensions:
\be
\label{eq:NJLtruncBasic}
\Gamma_{\text{NJL}}\left[\bar{\psi},\psi\right]&=&\int d^4 x
\left\{Z_{\psi}\bar{\psi} \I\fslash{\partial}\psi
+\frac{1}{2}\bar{\lambda}_{\sigma}[(\bar{\psi}\psi)^{2}-(\bar{\psi}\gamma_{5}\psi)^2]\right\}\,,
\ee
where $\bar{\lambda}_{\sigma}$ is the bare four-fermion coupling and $Z_{\psi}$ is the so-called
fermionic wave-function renormalization. The coupling $\bar{\lambda}_{\sigma}$ is considered
to be RG-scale dependent. Here, we consider four-fermion couplings as fundamental parameters. However, in other theories
 fermionic self-interactions might be fluctuation-induced. In QCD, for example, 
they are induced by two-gluon exchange 
and are therefore not fundamental as we shall discuss in Sect.~\ref{sec:gaugetheories}, 
see also Refs.~\cite{Gies:2002hq,Gies:2005as,Braun:2005uj,Braun:2006jd,Braun:2008pi,Braun:2009ns}.  
We would like to add that the NJL model in~$d=4$ is perturbatively non-renormalizable. In the following we define it with
a fixed UV cutoff~$\Lambda$. Also the regularization scheme therefore belongs to the definition of the model. We shall
come back to this issue in Sects.~\ref{subsec:bos} and~\ref{sec:GNmodel}. 

Our ansatz~\eqref{eq:NJLtruncBasic} for the effective action can be considered as the leading order
approximation in a systematic expansion in derivatives. The associated small parameter of such an expansion is the so-called 
anomalous dimension $\eta_{\psi}=-\partial_t \ln Z_{\psi}$ of the fermion fields. If this parameter is small, then such
a derivative expansion is indeed justified. We shall come back to this issue below. In any case, we will 
drop terms in our studies which are of higher order in derivatives, such as terms~$\sim(\bar{\psi} \I\fslash{\partial}\psi)^2$.

The action~\eqref{eq:NJLtruncBasic} is clearly invariant under simple phase transformations,
\be
\psi (x) \longmapsto \E^{\I\alpha}\psi(x)\,, \label{eq:U1reltrans}
\ee
but also under chiral~U($1$) transformations (axial phase transformations),
\be
\psi (x) \longmapsto \E^{\I \gamma_5 \alpha}\psi(x)\,,\quad 
\bar{\psi} (x) \longmapsto \bar{\psi}(x) \,\E^{\I \gamma_5 \alpha}\,,
\label{eq:chiraltrafo}
\ee
where $\alpha$ is an arbitrary ``rotation" angle. A necessary condition for the chiral symmetry of the NJL model is the
absence of explicit mass terms for the fermion fields in the action, such as $\sim\bar{\psi}\bar{m}\psi$.
As we shall discuss in more detail below, the chiral symmetry can be still broken spontaneously, if a finite 
vacuum expectation value $\langle\bar{\psi}\psi\rangle$ is generated by loop corrections associated with (strong)
fermionic self-interactions. 
Breaking of chiral symmetry in the ground state of the theory is then indicated by a dynamically generated 
mass term for the fermions. This mass term is associated, e.~g., with 
a constituent quark mass in low-energy models of QCD
and similar to the gap in condensed-matter theory. The relation between the strength of the four-fermion interactions
and the symmetry properties of the ground-state are discussed in detail in Sects.~\ref{subsec:bos} and~\ref{subsec:fourfermionSSB}. 

We may now ask whether the action~\eqref{eq:NJLtruncBasic} is complete or whether other four-fermion couplings, such 
as a vector interaction $\sim(\bar{\psi}\gamma_{\mu}\psi)^2$, can be generated dynamically due to quantum fluctuations. 
We first realize that the four-fermion interaction in our ansatz~\eqref{eq:NJLtruncBasic} can be expressed in terms of a vector and axial-vector 
interaction term with the aid of so-called Fierz transformations, see App.~\ref{sec:dirac} for details:
\be
\left[(\bar{\psi}\psi)^{2}-(\bar{\psi}\gamma_{5}\psi)^{2}\right ]
=\frac{1}{2}\left[(\bar{\psi}\gamma_{\mu}\gamma _{5}\psi)^{2}-(\bar{\psi}\gamma _{\mu}\psi)^{2}\right]\,.\label{eq:expFierz}
\ee
This ambiguity in the representation of a four-fermion interaction term arises due to the fact that an arbitrary $d\times d$-matrix $M$ can
be expanded in terms of a complete and orthonormalized set $\{O^{(1)},\dots,O^{(n)}\}$ of $d\times d$-matrices as follows:
\be
M_{ab} = \sum_{j=1}^{n} O^{(j)}_{ab}\, \tr (O_j M ) \equiv\sum_{j=1}^{n} O^{(j)}_{ab}\, \sum_{c,d}\left(O^{(j)}_{cd} M_{dc}\right) \;\,\text{with}\;\, \tr (O^{(j)}O^{(k)})
=\mathbbm{1} _d\delta_{jk}\,.
\ee
The expansion of a combination of two matrices $M^{(1)}$ and $M^{(2)}$ then reads (say for fixed~$b$ and~$c$)
\be
(M_{bc})_{ad}\equiv M_{ad}:=M^{(1)}_{ab} M^{(2)}_{cd} = \sum_{j=1}^{n} O^{(j)}_{ad}\, \sum_{e,f} (M^{(2)}_{ce} O^{(j)}_{ef} M^{(1)}_{fb} )\,. 
\label{eq:genFierz}
\ee
In the case of four-fermion interactions we may classify the basis elements $O_i$ according to the transformation properties 
of the corresponding interaction terms $(\bar{\psi}O_i\psi)^2$, i.~e. scalar channel, vector channel, tensor channel, 
axial-vector channel and pseudo-scalar channel. 
To be specific, we choose $O_{\text{S}}=\mathbbm{1} _d$, $O_{\text{V}}=\gamma _{\mu}$,
$O_{\text{T}}=\frac{1}{\sqrt{2}}\sigma _{\mu\nu}=\frac{\I}{2\sqrt{2}}[\gamma_{\mu},\gamma_{\nu}]$, 
$O_{\text{A}}=\gamma _{\mu}\gamma _{5}$ and $O_{\text{P}}=\gamma _5$ as 
basis elements of the Clifford algebra defined by the $\gamma$ matrices, see App.~\ref{sec:dirac} for details. 
To obtain Eq.~\eqref{eq:expFierz} we then simply apply Eq.~\eqref{eq:genFierz} to the matrix products $(\mathbbm{1})_{ab}(\mathbbm{1})_{cd}$
and $(\gamma_5)_{ab}(\gamma_5)_{cd}$, respectively. Thus, a Fierz transformation can be considered as a reordering of the fermion fields. We
stress that this is by no means related to quantum effects but a simple algebraic operation. Nonetheless it suggests that other four-fermion
couplings compatible with the underlying symmetries of our model exist and are potentially generated by quantum effects.

With our choice for the set of basis elements $\{O_{\rm S},\dots,O_{\rm P}\}$ it is straightforward to write down the most general ansatz
for the effective action~$\Gamma_{\text{NJL}}$ which is compatible with the underlying symmetries of the model, i.~e. 
the symmetries with respect to U($1$) phase transformations, U($1$) chiral transformations and Lorentz 
transformations:\footnote{Note that $(\bar{\psi}O_{\rm T}\psi)^2$ is not invariant under chiral U($1$) transformations.}
\be
\label{eq:NJLtruncBasic2}
\Gamma_{\text{NJL}}\left[\bar{\psi},\psi\right]&=&\int d^4 x
\left\{Z_{\psi}\bar{\psi} \I\fslash{\partial}\psi
+\frac{1}{2}\bar{\lambda}_{\sigma}[(\bar{\psi}\psi)^{2}-(\bar{\psi}\gamma_{5}\psi)^2]\right.\nn\\
& &\qquad\qquad\qquad\qquad\quad -\frac{1}{2}\bar{\lambda} _{\rm V}[(\bar{\psi}\gamma_{\mu}\psi)^2]
-\frac{1}{2}\bar{\lambda} _{\rm A}[(\bar{\psi}\gamma_ {\mu}\gamma_{5}\psi)^{2}]\bigg\}\,.
\ee
Because of Eq.~\eqref{eq:expFierz} only two of the three couplings $\bar{\lambda}_{\sigma}$, 
$\bar{\lambda} _{\rm V}$ and $\bar{\lambda} _{\rm A}$ are independent. Thus, it suffices to consider the
following action with implicitly redefined four-fermion couplings:
\be
\label{eq:NJLtruncBasic3}
\Gamma_{\text{NJL}}\left[\bar{\psi},\psi\right]&=&\int d^4 x
\left\{Z_{\psi}\bar{\psi} \I\fslash{\partial}\psi
+\frac{1}{2}\bar{\lambda}_{\sigma}[(\bar{\psi}\psi)^{2}\!-\!(\bar{\psi}\gamma_{5}\psi)^2]  
\!-\!\frac{1}{2}\bar{\lambda} _{\rm V}[(\bar{\psi}\gamma_{\mu}\psi)^2] \right\}\,.
\ee
Note that we could have also chosen to remove, e.~g., the vector-channel interaction term with the aid of Eq.~\eqref{eq:expFierz}
at the expense of getting the axial-vector interaction.
From a phenomenological point of view it is tempting to attach a physical meaning to, e.~g., the vector-channel
interaction and interpret it as an effective mass term for vector bosons~$V_{\mu}$ as done in mean-field studies of Walecka-type 
models~\cite{Walecka:1995mi}: $\bar{\lambda}_{\rm V}(\bar{\psi}\gamma_{\mu}\psi)^2\sim \bar{m}_{\rm V}^2 V_{\mu}V_{\mu}+\dots$.
However, the present analysis shows that one has to be careful to attach such a phenomenological interpretation to this term
since the Fierz transformations allow us to remove this term completely from the action, see also Sect.~\ref{subsec:bos}.

In this section we drop a possible momentum dependence of the four-fermion couplings. Thus, we only take into account
the leading term of an expansion of the four-fermion couplings in powers of the dimensionless external momenta~$|p_i|/k$, e.~g.
\be
\Gamma^{(2,2)}[\bar{\psi},\psi](p_1,p_2,p_3)\equiv\bar{\lambda}_{\rm V}(p_1,p_2,p_3)=\bar{\lambda}_{\rm V}(0,0,0) 
+ {\mathcal O} \left(\frac{|p_i|}{k}\right).\label{eq:expfc}
\ee 
In momentum space, the corresponding interaction term in the expansion of the 
effective action~\eqref{eq:NJLtruncBasic3} in terms of fermionic self-interactions then assumes the following form, see 
App.~\ref{sec:conv} for our conventions of the Fourier transformation:
\be
\Gamma_{\text{NJL}}\left[\bar{\psi},\psi\right] =  \ldots -
\frac{1}{2}\bar{\lambda}_{\rm V}
\prod_{i=1}^{3} \int \frac{d^4 p_i}{(2\pi)^4}
\bar{\psi}(p_1)\gamma _{\mu}\psi(p_2)\bar{\psi}(p_3)\gamma _{\mu}\psi(p_1\!-\! p_2\!+\! p_3)-\dots,\label{eq:FTlambdaV}
\ee
where $\bar{\lambda}_{\rm V} \equiv \bar{\lambda}_{\rm V} (0,0,0)$ and correspondingly for the other four-fermion interaction 
terms in Eq.~\eqref{eq:NJLtruncBasic3}. Note that only three of the four four-momenta $p_1,\dots,p_4$ 
are independent due to momentum conservation.
We stress that we also apply this expansion at finite temperature~$T$, see Sect.~\ref{subsec:DefFT}. In this case, it then corresponds to an expansion in
powers of the dimensionless Matsubara modes $\nu_n/k=(2n+1)\pi T/k$ and $|\vec{p}|/k$. Thus, we assume that $T/k \ll 1$.

The approximation~\eqref{eq:expfc} does not permit a study of properties of bound states of fermions, such as meson 
masses in QCD, in the chirally broken regime; such bound states manifest themselves as momentum singularities 
in the four-fermion couplings in Minkowski space. Nonetheless, the point-like limit can still be a reasonable approximation for $T/k \ll 1$. In the 
chirally symmetric regime above the chiral phase transition it allows us to gain some insight into the question how
the theory approaches the regime with broken chiral symmetry in the ground state~\cite{Braun:2005uj,Braun:2006jd,Braun:2008pi}. 
In Sect.~\ref{subsec:bos} we shall discuss how the momentum dependence of the fermionic interactions can
be conveniently resolved in order to gain access to the mass spectrum in the regime with broken chiral symmetry.

Let us now compute the RG flow equations, i.~e. the so-called $\beta$ functions, for the four-fermion couplings in the point-like limit.
To this end, we compute the second functional derivative of the effective action with respect to the fields 
\be
\Phi\equiv\Phi(q):= \left(\begin{array}{c}
 {\psi (q)}\\
{\bar{\psi}^{\rm T}(-q)}\end{array}
\right)\quad\text{and}\quad \Phi ^{\rm T} \equiv \Phi ^{\rm T} (-q):=\left(\psi ^{\rm T} (-q),\bar{\psi} (q)\right)\,,
\ee
see also Eq.~\eqref{eq:funcder}, and evaluate it for {\it homogeneous} (constant) 
background fields $\bar{\Psi}$ and $\Psi$. In momentum space this means 
that we evaluate $\Gamma^{(1,1)}_{\rm NJL}$ at 
\be
\psi(p)=\Psi\,(2\pi)^4 \delta^{(4)}(p)\quad\text{and} \quad \bar{\psi}(p)=\bar{\Psi}\,(2\pi)^4 \delta^{(4)}(p)\,,\label{eq:fermBF}
\ee
where $\Psi$ and $\bar{\Psi}$ on the right-hand side denote the homogeneous background 
fields. Following
Eq.~\eqref{eq:PF}, we then split the resulting matrix into a field-independent part and a part which depends on $\Psi$ and 
$\bar{\Psi}$. To detail the derivation of flow equations of four-fermion interactions in a simple manner, 
we first restrict ourselves to the simplified ansatz~\eqref{eq:NJLtruncBasic} of the effective action. In this case, 
the so-called (regularized) propagator matrix ${\mathcal P}_k$ and the fluctuation matrix ${\mathcal F}_k$ read
\be
 {\cal P}_k= \left(\begin{array}{cc}
 0 &  -Z_{\psi} \fss{p}^{\rm T}(1+r_{\psi}) \\
 -Z_{\psi}\fss{p}(1+r_{\psi}) & 0 \end{array} \right)(2\pi)^4\delta^{(4)}(p-p^{\prime})\,\label{eq:simpleFmatrix}
\ee
and
\be
{\cal F}_k= \left( \begin{array}{cc}
  {\mathcal F}_{11} & {\mathcal F}_{12} \\
  {\mathcal F}_{21} & {\mathcal F}_{22}  \end{array}\right)(2\pi)^4\delta^{(4)}(p-p^{\prime})\,, \label{eq:flucmatrix}
\ee
respectively, where
\be
{\mathcal F}_{11} = -\bar{\lambda}_{\sigma} \left[ \bar{\Psi}^{\rm T}\bar{\Psi} -\gamma_5\bar{\Psi}^{\rm T}\bar{\Psi}\gamma_5\right]\,,\quad 
{\mathcal F}_{22} = -\bar{\lambda}_{\sigma} \left[ \Psi\Psi^{\rm T} -\gamma_5\Psi\Psi^{\rm T}\gamma_5\right] \,,\nn
\ee
\be
{\mathcal F}_{12} = -\bar{\lambda}_{\sigma}\left[ (\bar{\Psi}\Psi)-\gamma_5(\bar{\Psi}\gamma_5\Psi)+
\Psi\bar{\Psi}-\gamma_5\Psi\bar{\Psi}\gamma_5\right]^{\rm T}=-{\mathcal F}_{21}^{\rm T}\,.\nn
\ee
Since we evaluated $\Gamma^{(1,1)}_{\rm NJL}$ for constant 
fields, both ${\cal P}_k$ and ${\cal F}_k$ are diagonal in
momentum space. At this point it is not yet necessary to specify the regulator function exactly. 

\begin{figure}[t]
\begin{center}
\includegraphics[scale=1]{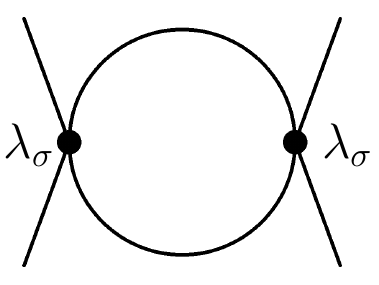}\hspace*{3cm}
\includegraphics[scale=1]{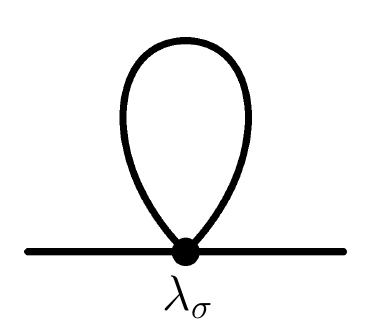}
\end{center}
\caption{Left diagram: $1$PI Feynman diagram associated with the $\lambda_{\sigma}^2$-term on the right-hand side of the RG flow
  equation~\eqref{eq:NJLbetasimple}. Note that our (functional) RG study includes resummations of this diagram to arbitrary 
  order in~$\lambda_{\sigma}$. Right diagram: $1$PI Feynman diagram associated with the RG running of the fermionic 
  wave-function renormalization~$Z_{\psi}$.}
\label{fig:feynman}
\end{figure}

The RG flow equation for~$\bar{\lambda}_{\sigma}$ can now be computed straightforwardly by comparing the coefficients of the four-fermion
interaction terms on the right-hand side of Eq.~\eqref{eq:flowexp} with the couplings included in our ansatz~\eqref{eq:NJLtruncBasic}.
From the fluctuation matrix ${\mathcal F}_k$ it is clear that only the term $\tr ({\mathcal P}_k^{-1}{\mathcal F}_k)^2\sim (\bar{\psi}O_i\psi)^2$ 
in Eq.~\eqref{eq:flowexp} contributes to the RG flow of the four-fermion coupling~$\bar{\lambda}_{\sigma}$. For this initial study, 
we simply take the four-fermion terms on the right-hand side of the flow equation ``as they appear" and ignore Fierz
transformations of these terms. We then find
\be
\beta_{\lambda_{\sigma}}\equiv\partial_{t}{\lambda}_{\sigma}=(2+2\eta_{\psi})\lambda_{\sigma} - 16 v_4\, l_{1}^{\rm (F),(4)}(0;\eta_{\psi})
\lambda_{\sigma}^2\,,
\label{eq:NJLbetasimple}
\ee
where $v_d^{-1}=2^{d+1}\pi^{d/2}\Gamma(d/2)$, i.~e. $v_4=1/(32\pi ^2)$.
Here, we have defined the dimensionless renormalized coupling
\be\label{eq:betasigmasimple2}
\lambda_{\sigma}=(Z_{\psi})^{-2} k^2 \bar{\lambda}_{\sigma}\,.
\ee
The so-called threshold function $l_{1}^{\rm (F),(d)}$ corresponds to a one-particle irreducible (1PI) Feynman diagram, 
see left diagram in Fig.~\ref{fig:feynman}, and describes the decoupling of massive and also thermal modes in case of finite-temperature
studies. Moreover, the regularization-scheme dependence is encoded in these threshold functions, see~App.~\ref{app:regthres} for 
their definitions.

\begin{figure}[t]
\begin{center}
\includegraphics[width=0.75\linewidth]{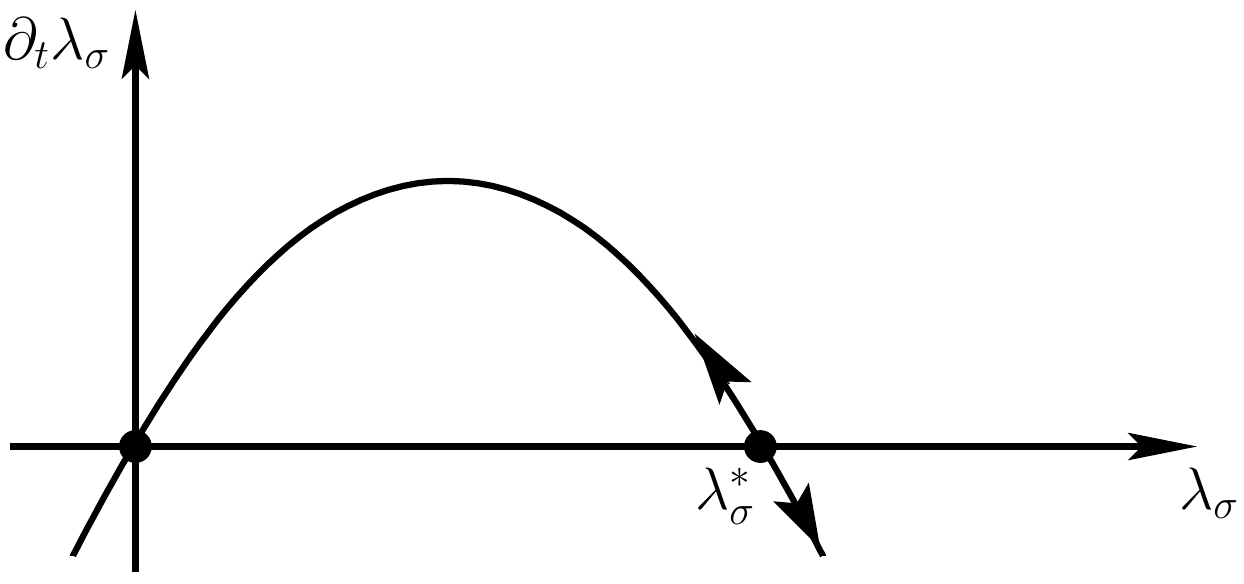}
\end{center}
\caption{Sketch of the $\beta_{\lambda_{\sigma}}$ function of the four-fermion interaction for 
zero temperature  (black/solid line). The arrows indicate the direction of the RG flow towards the infrared.}
\label{fig:parabola}
\end{figure}

In Fig.~\ref{fig:parabola} we show a sketch of the $\beta_{\lambda_{\sigma}}$-function for vanishing temperature.
Apart from a Gau\ss ian fixed point, $\lambda_{\sigma}^{\rm Gau\ss}=0$, we find a second non-trivial fixed 
point~$\lambda_{\sigma}^{\ast}$:
\be
\lambda_{\sigma}^{\ast} =\frac{1}{8v_4\,l_1^{{\rm (F)},(4)}(0;0)} + {\mathcal O}(\eta_{\psi}^{\ast})
\,.\label{eq:onechannelFP}
\ee
In the present leading-order approximation of the derivative expansion we have~$\eta_{\psi}\equiv 0$, see below.
We then find
\be
\lambda_{\sigma}^{\ast} =8\pi^2\label{eq:NJLFPOPTCUT}
\ee
for an optimized (linear) regulator function (for which~$l_1^{{\rm (F)},({\rm d})}(0;0)=2/d$) and
\be
\lambda_{\sigma}^{\ast} =4\pi^2\label{eq:NJLFPsc}
\ee
for the sharp cutoff (for which~$l_1^{{\rm (F)},({\rm d})}(0;0)=1$). It is instructive to have a closer look at Eq.~\eqref{eq:NJLbetasimple}.
This flow equation represents an ordinary differential equation which can be solved analytically for~$\eta_{\psi}=0$. Its solution
reads
\be
\lambda_{\sigma}(k)=\lambda_{\sigma}^{\rm UV}
\left[\left(\frac{\Lambda}{k}\right)^{\Theta}\left(1-\frac{\lambda^{\rm UV}_{\sigma}}{\lambda_{\sigma}^{\ast}}\right)+
\frac{\lambda^{\rm UV}_{\sigma}}{\lambda_{\sigma}^{\ast}}\right]^{-1}\,,
\label{eq:solfloweqse}
\ee
where
\be
\Theta:=-\frac{\partial (\partial_t \lambda_{\sigma})}{\partial \lambda_{\sigma}}\Big|_{\lambda_{\sigma}^{\ast}}\stackrel{(\eta_{\psi}\equiv 0)}{=}2\,.
\ee
In order to derive Eq.~\eqref{eq:solfloweqse}, it is convenient to expand the right-hand side of Eq.~\eqref{eq:NJLbetasimple}
about the fixed-point~$\lambda_{\sigma}^{\ast}$. The physical meaning of the so-called critical exponent~$\Theta$ will be discussed in more detail 
below. In Sect.~\ref{sec:NJLPL} we will then see that this exponent governs the scaling behavior of physical 
observables close to a quantum critical point.

For~$\lambda_{\sigma}^{\rm UV}=\lambda_{\sigma}^{\ast}$, we find that~$\lambda_{\sigma}(k)$ does not dependent on the 
RG scale~$k$ as it should be:~$\lambda_{\sigma}(k)=\lambda_{\sigma}^{\ast}$.
Choosing an initial value $\lambda_{\sigma}^{\rm UV} <\lambda_{\sigma}^{\ast}$  
at the initial UV scale $\Lambda$, the solution~\eqref{eq:solfloweqse} of the flow equation predicts 
that the theory becomes non-interacting in the 
infrared regime ($\lambda_{\sigma}\to 0$ for $k\to 0$), i.~e. chiral symmetry remains unbroken in this case, 
see Fig.~\ref{fig:parabola}. For $\lambda_{\psi}^{\rm UV}>\lambda_{\sigma}^{\ast}$, 
we find that the four-fermion coupling~$\lambda_{\sigma}$ increases rapidly and diverges eventually 
at a finite scale~$k_{\rm SB}$: $1/\lambda_{\sigma}(k_{\rm SB}) \to 0$. This behavior of the coupling and the associated fixed-point structure are tightly linked to the question 
whether chiral symmetry is broken in the ground state or not:  
The value of the non-trivial fixed-point can be considered as a critical value of the coupling which separates the chirally
symmetric regime and the regime with a broken chiral symmetry in the ground state. We shall discuss this in more detail
in Sects.~\ref{subsec:bos},~\ref{subsec:fourfermionSSB} and~\ref{subsec:NJLscaling}. 

In the derivation of the flow equation~\eqref{eq:NJLbetasimple} we have dropped contributions arising from
four-fermion interactions with different transformation properties, e.~g. a vector-channel interaction.
From the expansion~\eqref{eq:flowexp} of the flow equation, we can indeed read off that contributions to the flow of 
four-fermion couplings other than $\bar{\lambda}_{\sigma}$ might be generated, even though they
have not been included in the truncation~\eqref{eq:NJLtruncBasic}: the matrix multiplications on the 
right-hand side of Eq.~\eqref{eq:flowexp} mix the contributions from the propagator ${\mathcal P}_k$, 
which is proportional to $\gamma _{\mu}$, with the contributions from the field-dependent part $\mathcal{F}_k$:
\be
\bar{\lambda}_{\sigma}^2 \,\text{tr}\left\{\gamma _{\mu}\psi\bar{\psi}\gamma _{\mu} \psi\bar{\psi}\right\}
=-\bar{\lambda}_{\sigma}^2 (\bar{\psi}\gamma _{\mu}\psi)(\bar{\psi}\gamma _{\mu}\psi)\,.\label{eq:ffmotiv}
\ee
This term obviously contributes to the flow of the $\bar{\lambda}_{V}$-coupling.\footnote{At first glance, 
it seems possible that a term could arise in the calculation with opposite sign to the term in Eq.~\eqref{eq:ffmotiv},
so that both would cancel each other. We are aware of this and stress that Eq.~\eqref{eq:ffmotiv} should serve only as 
a motivation. As the full calculation of the NJL model shows (see below), not all terms which couple the RG flows of the different couplings 
drop out in the end.} Moreover, contributions of this type couple the flow equations of the various 
four-fermion interactions to one another. Thus, quantum fluctuations induce a vector-channel interaction, even though
we have not included such an interaction term initially. This observation explains why we need to include 
a basis which is complete with respect to Fierz transformations, such as in the effective action~\eqref{eq:NJLtruncBasic2}.
We stress that the effective action~\eqref{eq:NJLtruncBasic2} is closed in the sense that no 
contributions to four-fermion interactions, which are {\it not} covered by the truncation, are generated in the RG flow: any other 
pointlike four-fermion interaction compatible with the underlying symmetries of the theory can be written in terms of the interactions
included in these effective actions by means of Fierz transformations.

In the point-like limit the RG flow of the four-fermion coupling is completely decoupled
from the RG flow of fermionic $n$-point functions of higher order. For example, $8$-fermion interactions do not contribute
to the RG flow of the coupling $\bar{\lambda}_{\sigma}$ in this limit. Using the one-loop structure of the Wetterich equation, 
this statement can be proven diagrammatically: it is not possible to construct a one-loop diagram with only for external legs 
out of fermionic $n$-point functions ($n>4$) which are compatible with the underlying chiral symmetry.

Up to now we have only discussed the running of a four-fermion coupling. We have not
yet discussed how to compute the running of the wave-function renormalization~$Z_{\psi}$.
In general, the flow equation for~$Z_{\psi}$ can be obtained from $\Gamma^{(1,1)}_{\rm NJL}$ 
evaluated for a spatially varying background field, 
\be
\psi(p)=\Psi\,(2\pi)^4 \delta^{(4)}(p+Q) \quad\text{and} \quad \bar{\psi}(p)=\bar{\Psi}\,(2\pi)^4 \delta^{(4)}(p-Q)\,,\nn
\ee
where $Q$ denotes the external momentum.\footnote{These choices correspond to plane waves in position space.}
The $Q$-dependent second functional derivative~$\Gamma^{(1,1)}_{\rm NJL}$ 
can still be split into a field-independent and a field-dependent part. However, the latter is no longer diagonal in momentum space.
The flow equations for the wave-function renormalizations can then be computed by comparing the 
coefficients of the terms bilinear in fermionic fields which appear on the right-hand side of Eq.~\eqref{eq:flowexp} with
the kinetic terms in the ansatz for the effective action. In our present approximation, 
we find that the RG running of~$Z_{\psi}$ is trivial, i.~e.
\be
\partial_t Z_{\psi}=0\,.\label{eq:ZpsiDisc}
\ee
Thus, the associated anomalous dimensions $\eta_{\psi}=-\partial_t \ln  Z_{\psi}$ is zero. In fact, this follows immediately
from the associated 1PI Feynman diagram, see diagram on the right in Fig.~\ref{fig:feynman}, which has only one internal fermion 
line.\footnote{The momentum of the ingoing and outgoing fermion (line) is identical, namely~$Q$ in our conventions. Due
to momentum conservation, the loop momentum integration is then independent of~$Q$. Recall that we consider the
four-fermion coupling in the point-like limit.}
In the following we therefore set the wave-function renormalization to one, $Z_{\psi}\equiv 1$, which implies~$\eta_{\psi}\equiv 0$.

Let us now turn to the effective action~\eqref{eq:NJLtruncBasic3}. The flow equations of the various couplings can be
derived along the same lines as the RG equation for the $\lambda_{\sigma}$-coupling detailed above. We find
\be
\partial_{t}{\lambda}_{\sigma}&=& 2\lambda_{\sigma}- 
8v_4\,  l_{1}^{\rm (F),(4)}(0;0) \left[ \lambda_{\sigma}^2 + 4\lambda_{\sigma}\lambda_{\rm V}+3\lambda_{\rm V}^2
\right]\,,\label{eq:NJLsig2ch}\\ 
\partial_{t}{\lambda}_{{\rm V}}&=& 2\lambda_{\rm V}-
4v_4\,  l_{1}^{\rm (F),(4)}(0;0) \left[
\lambda_{\sigma}^2 + 2\lambda_{\sigma}\lambda_{\rm V}+\lambda_{\rm V}^2
\right]\,,\label{eq:NJLv2ch}
\ee
where the dimensionless (renormalized) couplings are defined as
\be\label{eq:betasigmasimple1}
\lambda_{\sigma}= k^2 \bar{\lambda}_{\sigma}\,,\quad\text{and}\quad
\lambda_{\rm V}= k^2 \bar{\lambda}_{\rm V}\,.
\ee
In the derivation of the flow equations for~$\lambda_{\sigma}$ and~$\lambda_{\rm V}$ also terms
of the type $\sim (\bar{\psi}\gamma_{\mu}\gamma_5\psi)^2$ and
\be
\left[ (\bar{\psi}\sigma_{\mu\nu}\psi)^2 - (\bar{\psi}\sigma_{\mu\nu}\gamma_5 \psi)^2\right] \label{eq:tensorchannel}
\ee
appear. While the latter vanishes identically, see also App.~\ref{sec:dirac}, the former can 
be completely transformed into a scalar-pseudoscalar and vector-interaction channel with the
aid of the Fierz transformation~\eqref{eq:expFierz}. In fact, any four-fermion interaction term appearing
in the derivation of the flow equations for the present system can be unambiguously rewritten in terms of these two interaction
channels. Thus, the above RG flows are closed with respect to Fierz transformations. 
Due to Eq.~\eqref {eq:expFierz} we could have also used, e.~g., a scalar-pseudoscalar and an axial-vector 
interaction to describe the properties of our simplified theory without loss of physical information. The present
choice for a complete basis of four-fermion interactions is one of several possibilities.

Our flow equations for~$\lambda_{\sigma}$ and~$\lambda_{\rm V}$ agree with the results found in
Refs.~\cite{Aoki:1999dw,Jaeckel:2002rm}. The RG flow of the couplings~$\lambda_{\sigma}$ and~$\lambda_{\rm V}$ is governed by three 
fixed points ${\mathcal F}_i=(\lambda_{\sigma}^{\ast},\lambda_{\rm V}^{\ast})$ which are 
given by\footnote{One might naively expect four fixed points. In order to show that our system of flow equations has only
three fixed points, we redefine the $\lambda_{\sigma}$-coupling 
according to $\lambda_{\sigma}\to (\lambda_{\sigma}-\lambda_{\rm V})$. The flow equation for the vector coupling
then has only one fixed point which depends on the value of the (new) $\lambda_{\sigma}$-coupling. However, 
the $\beta$-function of the latter is now cubic in~$\lambda_{\sigma}$ when we evaluate it at the fixed-point value of the
$\lambda_{\rm V}$-coupling. Thus, the system indeed has only three fixed points.} 
\be
{\mathcal F}_1 \equiv{\mathcal F}_{\rm Gau\ss}=(0,0)\,,\quad 
{\mathcal F}_2=\left(
{3\zeta}\,,
{\zeta}
\right)\,,\quad
{\mathcal F}_3=\left(
-32\zeta\,,
16{\zeta}
\right)\,,
\label{eq:allchannelFP}
\ee
where
\be
\zeta = \frac{1}{32v_4 l_{1}^{\rm (F),(4)}(0;0)}\,.\label{eq:zetadef}
\ee
These fixed-points are of phenomenological importance. First of all, they might be related to (quantum)
phase transitions. Second, we can define sets of initial values for the RG flows of the 
couplings~$\lambda_{\sigma}$ and~$\lambda_{\rm V}$ for which we find condensate formation 
associated with (chiral) symmetry breaking in the IR, as we shall discuss in detail in the two subsequent
sections. The existence of such sets of initial conditions is not a generic feature of fermionic models but 
also appears in (chiral) gauge theories. In QCD and QED${}_3$,
four-fermion interactions are generated dynamically due to strong quark-gluon interactions, see
our discussion in Sect.~\ref{sec:gaugetheories}. 

We can classify the various fixed points according to their directions in the space spanned by the couplings. 
To this end, we first linearize the RG flow equations of the couplings near a fixed point:
\be
\partial_t\lambda_{i}=\sum_{j}B_{ij}(\lambda_j-\lambda_j^{\ast}) + \dots\,, \quad\text{where}\quad
B_{ij}=\frac{\partial_t \lambda_{i}}{\partial \lambda_j}\Bigg|_{\lambda_{i}=\lambda_{i}^{\ast}}\,
\label{eq:Bdef}
\ee
and $i,j\in\{\sigma,{\rm V}\}$. We refer to $B$ as the {\it stability} matrix. The two eigenvectors~$\vec{v}_i$
and eigenvalues~$\Theta^{(i)}$ (critical exponents)
of this matrix essentially determine the RG evolution near a fixed point:\footnote{In other words, the critical exponents
are simply the zeroes of the (characteristic) polynomial $\det (B+\Theta\mathbbm{1})$.}
\be
\partial_t \vec{v}_i=B\cdot \vec{v}_i =: -\Theta^{(i)} \vec{v}_i\,.\label{eq:genthetadef}
\ee
The solution of the RG flow in the fixed-point regime is then given by 
\be
\lambda_i = \lambda_i^{\ast} + \sum_j c_j \left(\vec{v}_j\right)_i \left(\frac{k_0}{k}\right)^{\Theta^{(j)}}\,. 
\ee
Here, the $c_j$'s define the initial conditions at the scale~$k_0$. From the
solution of the linearized flow it becomes apparent that positive critical exponents, $\Theta^{(j)}>0$,
correspond to RG relevant, i.~e. infrared repulsive, directions. On the other hand, negative
critical exponents~$\Theta^{(j)}<0$ correspond to RG irrelevant, i.~e. infrared attractive, directions.
The classification of marginal directions associated with vanishing critical exponents 
requires to consider higher orders in the expansion about the fixed point.
\begin{figure}[t]
\begin{center}
\includegraphics[width=0.75\linewidth]{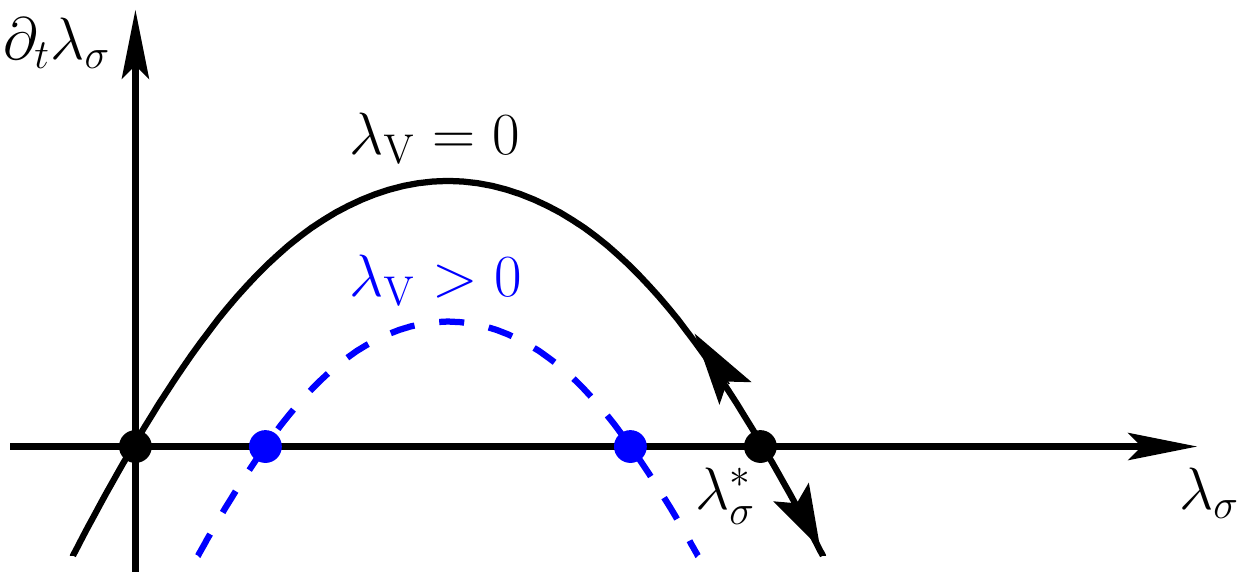}
\end{center}
\caption{Sketch of the $\beta_{\lambda_{\sigma}}$-function of the four-fermion interaction for
$\lambda_{\rm V}=0$ (black line) and~$\lambda_{\rm V} >0$ (blue/dashed line). 
The arrows indicate the direction of the RG flow towards the infrared.}
\label{fig:parabola2}
\end{figure}

Using the flow equations~\eqref{eq:NJLsig2ch} and~\eqref{eq:NJLv2ch}, we find that
the Gau\ss ian fixed point has two IR attractive directions; the eigenvalues are
 $\Theta_{{\mathcal F}_1}\equiv \Theta_{\rm Gau\ss }=\{-2,-2\}$.
The fixed points ${\mathcal F}_2$ with $\Theta_{{\mathcal F}_2}=\{2,-5/2\}$
and~${\mathcal F}_3$ with $\Theta_{{\mathcal F}_3}=\{2,-10)\}$
have both one IR attractive and one IR repulsive direction.
We would like to add that the fixed-point values of the four-fermion couplings are {\it not} universal quantities as 
the dependence of their RG flows on the threshold function indicates. However, the statement about the mere existence of these
fixed points is {\it universal}, because the regulator-dependent factor $l_1^{{\rm( F),(d)}}(0;0)$ is a positive number for any regulator. 
Moreover, the critical exponents~$\Theta_i$ themselves are universal. The latter
can be indeed related to the exponents associated with (quantum) phase transitions, as we shall discuss in Sect.~\ref{subsec:NJLscaling}.
Therefore the accuracy of the critical exponents can be used to measure the quality of a given truncation as has been done in 
the context of scalar field theories, see e.~g. Refs.~\cite{Tetradis:1993ts,Litim:2002cf,Bervillier:2007rc,Delamotte:2007pf,Benitez:2009xg}.
In a pragmatic sense, the computation of critical exponents 
allows us to estimate how well the dynamics close to a phase transition are captured within our ansatz for the effective action.

Let us conclude our discussion with a comparison of the RG flows~\eqref{eq:NJLsig2ch} and~\eqref{eq:NJLv2ch}
obtained from a complete basis of four-fermion interactions with the RG flow equation~\eqref{eq:NJLbetasimple}
from our single-channel approximation. We immediately observe that setting $\lambda_{\rm V}\to 0$ in Eq.~\eqref{eq:NJLsig2ch}
does not yield the flow equation~\eqref{eq:NJLbetasimple}. Thus, the values of the non-trivial fixed point of this coupling
are not identical but differ by a factor of two.\footnote{The RG flow also generates axial-vector channel interactions
which, in the present (Fierz) basis of four-fermion interactions, reduce the prefactor of the term~$\propto \lambda_{\sigma}^2$
in Eq.~\eqref{eq:NJLsig2ch} by a factor of two.} 
For a finite~$\lambda_{\sigma}$, 
we find that the vector-channel interaction is dynamically generated 
due to quantum fluctuations even if we have initially set the vector-channel interaction to zero. In fact, a finite 
$\lambda_{\sigma}$-coupling shifts the parabola associated with the $\beta$-function of the coupling $\lambda_{\rm V}$, 
and vice versa, see Fig.~\ref{fig:parabola2}. Thus, the $\lambda_{\sigma}$-coupling
can potentially induce critical behavior in the vector-channel, i.~e. a diverging four-fermion coupling. We shall
discuss this in more detail in Sect.~\ref{subsec:fourfermionSSB} after we have clarified the physical meaning 
of diverging four-fermion couplings in the subsequent section.

\subsection{Bosonization and the Momentum Dependence of Fermion~Interactions}\label{subsec:bos}

In this section we study the NJL model with one fermion species
in a partially bosonized form. Partial bosonization of fermionic theories is a well-established concept which
makes use of the so-called {\it Hubbard-Stratonovich} transformation~\cite{Hubbard:1959ub,Stratonovich}.
The advantage of a partially bosonized formulation of NJL-type models over their purely fermionic formulation is that
it allows us to include the momentum dependence of four-fermion interactions in a simple manner. 
Therefore it opens up the possibility to study conveniently the mass spectrum of a theory which emerges from the spontaneous
breakdown of its underlying symmetries, e.~g. the chiral symmetry. As a bonus, it
relates the {\it Ginzburg-Landau} picture of spontaneous symmetry breaking, as known from statistical
physics, with dynamical bound-state formation in strongly-interacting fermionic theories. 

In the following we derive the RG flow equations for the partially bosonized version of this theory and 
discuss dynamical chiral symmetry breaking. In particular, we explain the mapping of  the (partially) bosonized equations
onto the RG equations of the four-fermion couplings in the purely fermionic description of our model. This finally
allows us to relate the fixed-point structure of the purely fermionic formulation to spontaneous (chiral) symmetry breaking.

The generating functional $Z$ reads\footnote{For convenience, we do not display the source terms for the fields here
and in the following.}
\be
Z&\propto&\int{\mathcal D}\psi{\mathcal D}\bar{\psi}\,\E ^{-S[\bar{\psi},\psi]
}\label{eq:ZNJL1}
\ee
with the action
\be
S[\bar{\psi},\psi]&=&\int d^4 x
\left\{Z_{\psi}\bar{\psi} \I\fslash{\partial}\psi
+\frac{1}{2}\bar{\lambda}_{\sigma}[(\bar{\psi}\psi)^{2}-(\bar{\psi}\gamma_{5}\psi)^2]\right.\nn\\
& &\quad\qquad\qquad\qquad\qquad\quad -\frac{1}{2}\bar{\lambda} _{\rm V}[(\bar{\psi}\gamma_{\mu}\psi)^2]
-\frac{1}{2}\bar{\lambda} _{\rm A}[(\bar{\psi}\gamma_ {\mu}\gamma_{5}\psi)^{2}]\bigg\}\,.\label{eq:NJLaction}
\ee
see also Eq.~\eqref{eq:NJLtruncBasic2}. As discussed in the previous section, this action is over-complete 
in the sense that only two of the three couplings $\bar{\lambda}_{\sigma}$, $\bar{\lambda} _{\rm V}$ and $\bar{\lambda} _{\rm A}$ 
are independent. We shall come back to this issue in the partially bosonized formulation below.

Our NJL model possesses a chiral symmetry, see Eq.~\eqref{eq:chiraltrafo}, which can be broken dynamically, if a finite 
vacuum expectation value $\langle\bar{\psi}\psi\rangle$ is generated. This is associated with
the {\em Nambu-Goldstone} theorem~\cite{Nambu:1961tp,Nambu:1961fr,Goldstone:1961eq,Goldstone:1962es} 
which relates a spontaneously broken continuous symmetry of a given theory 
to the existence of massless states in the spectrum. To apply this theorem to the present model,
we need to compute the vacuum expectation value of the commutator of the so-called chiral charge $Q_5$, which is 
the generator of the chiral symmetry transformations, and the composite field $\bar{\psi}\I\gamma_5\psi$:
\be
\langle \left[\I Q_5 ,\bar{\psi}\I\gamma_5\psi\right]\rangle\propto\langle \bar{\psi}\psi\rangle\quad
\text{with}\quad Q_5 =\frac{1}{2}\int d^3 x\,\bar{\psi}\gamma _{0}\gamma_5 \psi\,.\label{eq:commut}
\ee
We observe that the generator $Q_5$ does not commute with the field~$\bar{\psi}\I\gamma_5\psi$, if the 
vacuum expectation value of $\bar{\psi}\psi$ is finite. Thus, the chiral symmetry of our model can be indeed
broken spontaneously. Following the {\em Nambu-Goldstone} theorem this implies the existence of a {\it massless} 
pseudo-scalar {\em Nambu-Goldstone} boson in the channel of the composite 
field $\bar{\psi}\I\gamma_5\psi$. Since the action $S$ does not contain such a state, the massless state must be a 
bound state. We refer to this type of boson as a {\it pion} in the context of QCD, see Sect.~\ref{sec:njlgn}.
At this point we have traced the question of chiral symmetry breaking back to the existence of a finite expectation
value of the composite field $\bar{\psi}\psi$.

\begin{figure}[t]
\begin{center}
\includegraphics[scale=1]{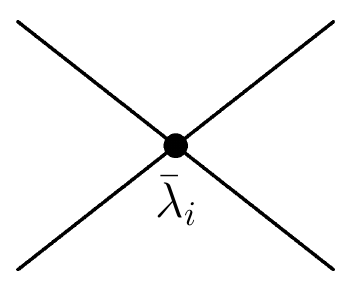}\hspace*{3cm}
\includegraphics[scale=1]{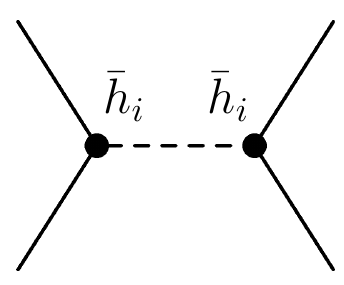}
\end{center}
\caption{Left diagram: Representation of a four-fermion interaction ($i\in\{\sigma,{\rm V},{\rm A}\}$) 
in the purely fermionic formulation of our NJL model. In general, the vertex
function has a non-trivial momentum dependence. Right diagram: The Feynman diagram
has the same external lines as the one on the left. However, the vertex function has been
replaced by a boson propagator (dashed line) which mediates the interaction. This illustrates the situation
in the partially bosonized version of the NJL model.
}
\label{fig:bosmed}
\end{figure}

Formally, we may introduce auxiliary fields in the path integral
by introducing an exponential factor into the 
integrand of the generating functional. This is known as a {\em Hubbard-Stratonovich}
transformation. To bosonize the scalar-pseudoscalar interaction channel we use
\be
{\mathcal N}\int {\mathcal D}\phi_1{\mathcal D}\phi_2 {\mathcal D}V_{\mu} {\mathcal D}A_{\mu}
\,\E^{-\int d^4 x\,\left\{
\frac{1}{2}\bar{m}^2_{\sigma}\vec{\phi}^{\,2}
+\frac{1}{2}\bar{m}_{\rm V}^2 V_{\mu}V_{\mu} 
+\frac{1}{2}\bar{m}_{\rm A}^2 A_{\mu}A_{\mu}
\right\}}&=&1\,,
\ee
where we have combined the scalar fields into the $O(2)$ vector $\vec{\phi}^{\,\rm T}=(\phi_1,\phi_2)$, where $\phi_1$ and $\phi_2$
are real-valued scalar fields.\footnote{Alternatively, we could have introduced a complex field $\varphi=(\phi_1+\I\phi_2)/\sqrt{2}$:
$\bar{m}^2_{\sigma}\varphi^{*}\varphi = \bar{m}^2_{\sigma} \vec{\phi}^{\,2}/2$.}
The fields $\phi_1$, $\phi_2$, $V_{\mu}$ and $A_{\mu}$ are auxiliary fields (which have no dynamics so far), $\mathcal N$ is a normalization factor, 
and the constants $\bar{m}_{\sigma}$, $\bar{m}_{\rm V}^2$ and $\bar{m}_{\rm A}^2$ remain arbitrary for the moment.
Multiplying the integrand of the generating functional with such a factor
leaves the Greens functions of the theory unchanged. We now shift the integration 
variables in the so-modified generating functional according to\footnote{For historical reasons 
we include a factor of $\sqrt{2}$ in the shift of the scalar fields; such a factor is usually present in studies of theories with only one fermion
species, such as QED, where one rather deals with a complex (charged) scalar field $\varphi=(\phi_1+\I\phi_2)/\sqrt{2}$.}
\be
&&\phi_1\rightarrow\phi_1 +\frac{\I\bar{h}_{\sigma}}{\sqrt{2}\bar{m}^2_{\sigma}}\left(\bar{\psi}\psi\right)\,,\quad %
\phi_2\rightarrow\phi_2 -\frac{\I\bar{h}_{\sigma}}{\sqrt{2}\bar{m}^2_{\sigma}}\left(\bar{\psi}\I\gamma_5\psi\right)\,,\nn\\
&& V_{\mu}\rightarrow V_{\mu} - \frac{\bar{h}_{\rm V}}{\bar{m}^2_{\rm V}}(\bar{\psi}\gamma_{\mu}\psi)   \,,\quad
A_{\mu} \rightarrow A_{\mu} - \frac{\bar{h}_{\rm A}}{\bar{m}^2_{\rm A}}(\bar{\psi}\gamma_{\mu}\gamma_5\psi)  \,.
\ee
where we have introduced the auxiliary constants $\bar{h}_{\sigma}$, $\bar{h}_{\rm V}$ and  $\bar{h}_{\rm A}$.
The new generating functional then reads
\be
Z&\propto&\int {\mathcal D}\phi_1{\mathcal D}\phi_2 {\mathcal D}V_{\mu} {\mathcal D}A_{\mu} {\mathcal D}\psi{\mathcal D}\bar{\psi}\,
\E ^{-S[\bar{\psi},\psi,\phi_1,\phi_2,V_{\mu},A_{\mu}]}
\ee
with the so-called partially bosonized action
\be
&& S[\bar{\psi},\psi,\phi_1,\phi_2,V_{\mu},A_{\mu}]=\int d^4 x\,\left\{ Z_{\psi}\bar{\psi}\I \fslash{\partial}\psi
+ \frac{1}{2}\bar{m}_{\sigma}^2\vec{\phi}^{\,2} 
+ \frac{\bar{h}_{\sigma}}{\sqrt{2}}\bar{\psi}(\vec{\tau}\cdot\vec{\phi})\psi \right.\nn\\
&& \qquad\qquad\qquad\; 
\left.+ \frac{1}{2}\bar{m}_{\rm V}^2 V_{\mu}V_{\mu}  
- \bar{h}_{\rm V} \bar{\psi}\fslash{V}\,\psi 
+ \frac{1}{2}\bar{m}_{\rm A}^2A_{\mu}A_{\mu} 
- \bar{h}_{\rm A} \bar{\psi}\gamma_{\mu}\gamma_{5}A_{\mu}\psi  \right\}.
\label{eq:bosaction}
\ee
Here, we have introduced $\vec{\tau}=(\I\cdot \mathbf{1},\gamma_5)$ in order to define a chirally invariant 
Yukawa interaction.
To obtain the (partially) bosonized action in this convenient form, we have exploited the fact that 
$\bar{m}_{\sigma}$, $\bar{m}_{\rm V}^2$ and $\bar{m}_{\rm A}^2$ as well as
$\bar{h}_{\sigma}$, $\bar{h}_{\rm V}$ and  $\bar{h}_{\rm A}$ are arbitrary parameters. To be specific, we have chosen
\be
\bar{\lambda} _{\sigma}\stackrel{!}{=}\frac{\bar{h}_{\sigma}^2}{2\bar{m}_{\sigma}^2}\,,\quad
\bar{\lambda} _{\rm V}\stackrel{!}{=}\frac{\bar{h}_{\rm V}^2}{\bar{m}_{\rm V}^2}\,\quad\text{and}\quad
\bar{\lambda} _{\rm A}\stackrel{!}{=}\frac{\bar{h}_{\rm A}^2}{\bar{m}_{\rm A}^2}\,.\label{eq:h2m2}
\ee
Instead of a four-fermion interactions, there are now Yukawa-type interactions and mass terms 
for the auxiliary fields. The interaction between the fermions is mediated by the bosonic fields $\vec{\phi}$,
$V_{\mu}$ and~$A_{\mu}$, see Fig.~\ref{fig:bosmed}. It is clear that the point-like approximation of the 
four-fermion interactions is only meaningful in the limit of large boson mass terms, $m_{i}^2\gg \Lambda^2 \geq p^2$ 
with $i\in\{\sigma,{\rm V},{\rm A}\}$, where~$p$ is a typical momentum of the composite boson.
We would like to emphasize that the Yukawa couplings and their associated mass terms 
are not independent in the pointlike limit: Only their ratio has a physical meaning which is due to our choice~\eqref{eq:h2m2}.

The new action~\eqref{eq:bosaction} is symmetric under a simultaneous chiral transformation
of the fermions and the composite fields. While the transformation of the fermionic fields is given by Eq.~\eqref{eq:chiraltrafo},
the scalar fields, which are of our particular interest, transform according to
\be
\vec{\phi} \;\mapsto\; 
\left(\begin{array}{cc}
\phantom{-} \cos (2\alpha) & \sin (2\alpha) \\
-\sin(2\alpha) &  \cos (2\alpha) \end{array} \right)\vec{\phi}\,,
\label{eq:O2trafo}
\ee
where $\alpha$ is an arbitrary rotation angle. The group of transformations defined by this matrix is simply 
the O($2$) rotation group. Note that the chiral transformation of the scalar fields follows
immediately from the transformation properties of $\bar{\psi}\psi$ and $\bar{\psi}\gamma_5\psi$. 

From the equations of motion of the bosonic field it follows that
\be
\!\!\!\!\!&&\frac{\delta S}{\delta \phi_1(x)}\!=\!\bar{m}_{\sigma}^2\phi_1(x)\!+\!\I\frac{\bar{h}_{\sigma}}{\sqrt{2}}\bar{\psi}\psi \!=\! 0\,,\quad
\frac{\delta S}{\delta \phi_2(x)}\!=\!\bar{m}_{\sigma}^2\phi_2(x)\!+\!\frac{\bar{h}_{\sigma}}{\sqrt{2}}\bar{\psi}\gamma_5\psi\! =\! 0\,,\\
\!\!\!\!\!&&\frac{\delta S}{\delta V_{\mu}(x)}\!=\!\bar{m}_{\rm V}^2 V_{\mu}(x)\!-\!\bar{h}_{\rm V}\bar{\psi}\gamma_{\mu}\psi\!=\! 0\,,\quad\!
\frac{\delta S}{\delta A_{\mu}(x)}\!=\!\bar{m}_{\rm A}^2 A_{\mu}(x)\!-\!\bar{h}_{\rm A}\bar{\psi}\gamma_{\mu}\gamma_5\psi\!=\! 0\,.
\ee
Inserting the solutions of these equations of motion into the action~\eqref{eq:bosaction} and using Eq.~\eqref{eq:h2m2},
we immediately recover the original action~\eqref{eq:bosaction}. From a phenomenological point of view, the
solutions of the equations of motion allow us to consider the bosonic fields as bound-states of the fermion fields.
In particular, they relate the vacuum expectation value of~$\bar{\psi}\psi$ to that 
of the scalar field~$\phi_1$, \mbox{$\langle \bar{\psi}\psi\rangle \sim \langle \phi_1\rangle$}.
From our consideration of the Nambu-Goldstone theorem, see Eq.~\eqref{eq:commut},
we then conclude that chiral symmetry breaking is indicated by a finite expectation value of the field~$\phi_1$.
Strictly speaking, chiral symmetry breaking is signaled by a finite expectation value of the field $\vec{\phi}$, where $|\langle \vec{\phi}\rangle|$ 
defines a circle in the two-dimensional space spanned by the fields~$\phi_1$ and~$\phi_2$. The points on 
this circle are related to each other by chiral transformations, i.~e. O($2$) rotations in the space of the chiral components, see Eq.~\eqref{eq:O2trafo}.
In any case, the fermions acquire a finite mass once the expectation value of $\phi_1$ is finite, as can
be seen from an evaluation of their equation of motion at a finite value of $\langle \phi_1\rangle$:
\be
\frac{\delta S}{\delta \bar{\psi}(x)} \Bigg|_{\tiny\begin{array}{l}
\phi_1=\langle \phi_1\rangle \\
\phi_2=0\end{array} }
=\left(\I  \fslash{\partial} + \I m_{\psi}\right)\psi(x)=0\,\quad
\text{with} \quad
m_{\psi}=\frac{\bar{h}_{\sigma}}{\sqrt{2}Z_{\psi}}\langle\phi_1\rangle\,.\label{eq:mpsidefsec3}
\ee
Thus, the fermions {\it condense} for $\langle \phi_1\rangle >0$ and their finite 
mass term $m_{\psi}$ signals the spontaneous breakdown 
of chiral symmetry in the physical ground state of the theory. Note that the chiral symmetry is still present
on the level of the microscopic theory.

The discussed (chiral) symmetry breaking pattern
plays an important role in various fields of research. In condensed-matter 
physics, such a mass term is associated with a fermion gap. In low-energy QCD, it plays the role of a constituent quark
mass. By contrast, in the electroweak theory it is considered as the (current) quark mass. 
The ground-state value~$\langle \phi_1\rangle$ does not necessarily have to be homogeneous, i.~e. spatially constant.
In fact, inhomogeneous condensates have been identified in some parts of the phase diagram of the Gross-Neveu model 
in~$d=2$ at finite temperature and large values of the chemical potential in the large-$\Nf$ limit~\cite{Thies:2003kk}; the
status for higher-dimensional fermionic models is a subject of ongoing work,  see e.~g. Refs.~\cite{Nickel:2009wj,Kojo:2009ha}.
In the following, however, we shall assume that the ground state of the theory is homogeneous.

Let us now critically discuss the partially bosonized form of our NJL model.
From its action we may already conclude that a non-trivial ground state of the theory, i.~e. a finite
expectation value $\langle \vec{\phi}\rangle$, is tightly linked to the sign of the term bilinear in the scalar fields:\footnote{Let us add a word
of caution on the meaning of~$\langle \vec{\phi}\rangle$ in the present work. For this discussion we put our theory in a finite volume~$V$. 
We then have~$\langle \vec{\phi}\,\rangle =0$, independent of our choice for the parameters~$\bar{h}_{\sigma}$ and~$\bar{m}_{\sigma}$.
Depending on our choice for~$\bar{h}_{\sigma}$ and~$\bar{m}_{\sigma}$, however, we may have~$\langle |\vec{\phi}|\rangle =0$.
In order to obtain a finite expectation value~$\langle \vec{\phi}\,\rangle$ in a finite volume, we may allow for a finite external source for, e.~g., 
the scalar field~$\phi_1$,
such that~$\langle \vec{\phi}^{\rm\, T}\rangle=(\langle \phi_1\rangle,0)$. In a field-theoretical model of ferromagnetism, such an external source
plays the role of an external magnetic field. Considering now the limit of a vanishing source term before taking the limit~$V\to\infty$, we encounter
$\langle \vec{\phi}\rangle =0$ also for~$V\to\infty$. Considering first the limit~$V\to\infty$ and then 
the limit of a vanishing source term, however, the
expectation value~$\langle \vec{\phi}^{\rm\, T}\rangle=(\langle \phi_1\rangle,0)$ remains finite in the infinite-volume limit, see e.~g. 
Ref.~\cite{Gasser:1987ah} for a detailed discussion. In our effective-action approach we {\it implicitly} take these two limits in the ``correct" order
to obtain a finite expectation value~$\langle \vec{\phi}\rangle$. In infinite volume, this is ensured
by expanding the effective action 
about, e.~g., a finite value of~$\langle \vec{\phi}^{\rm\, T}\rangle=(\langle \phi_1\rangle,0)$. From now on, we shall
always assume that the two limits have been taken in such a way that~$\langle \vec{\phi}\rangle$ remains finite, provided that
the parameters~$\bar{h}_{\sigma}$ and~$\bar{m}_{\sigma}$ have been chosen accordingly. In our present case, the (continuous) chiral symmetry of the 
ground state is then said to be broken spontaneously.} 
$\bar{m}_{\sigma}^2< 0$ seems to necessarily imply that \mbox{$|\langle \vec{\phi}\rangle| >0$}. To establish this relation and to discuss issues related to
Fierz transformations, we briefly analyze the model in a mean-field approximation. In
fermionic models with more than one fermion species, this approximation is often referred to as the large-$\Nf$ 
approximation, see e.~g. Ref.~\cite{Braun:2010tt}. 

In the action~\eqref{eq:bosaction} the fermions appear only
as bilinears. Thus, they can be integrated out straightforwardly and we obtain a purely bosonic
effective action:
\be
\Gamma_{\rm MF}[\phi_1,\phi_2,V_{\mu},A_{\mu}]&=&\int d^4 x\,\left\{ 
\frac{1}{2}\bar{m}_{\sigma}^2\vec{\phi}^{\,2} 
+ \frac{1}{2}\bar{m}_{\rm V}^2 V_{\mu}V_{\mu}  
+ \frac{1}{2}\bar{m}_{\rm A}^2A_{\mu}A_{\mu} \right\}\nn\\
&& \quad - {\rm Tr}\ln \left[ Z_{\psi}\I \fslash{\partial}
+ \frac{\bar{h}_{\sigma}}{\sqrt{2}}(\vec{\tau}\cdot\vec{\phi})
- \bar{h}_{\rm V} \gamma_{\mu}V_{\mu} 
- \bar{h}_{\rm A} \gamma_{\mu}\gamma_{5}A_{\mu}
\right],\label{eq:bosaction2}
\ee
where 
\be
{\rm Tr}\,{\mathcal O}={\rm \tr}_{\rm D}\int d^4 x \langle x | {\mathcal O} | x\rangle\,.
\ee
Here, ${\rm \tr}_{\rm D}$ sums over Dirac indices. In general, this purely bosonic
action is highly nonlocal since the bosonic fields depend on the space-time coordinates. In the
following we neglect such a dependence and evaluate the bosonic action~\eqref{eq:bosaction}
for constant background fields~$\bar{\phi}_1$, $\bar{\phi}_2$, $\bar{V}_{\mu}$ and~$\bar{A}_{\mu}$. We also  
set the wave-function renormalization of the fermion fields to one, $Z_{\psi}=1$. This yields
the effective potential~$U$:
\be
\Gamma_{\rm MF}[\bar{\phi}_1,\bar{\phi}_2,\bar{V}_{\mu},\bar{A}_{\mu}] 
= \int d^4 x \, U(\bar{\phi}_1,\bar{\phi}_2,\bar{V}_{\mu},\bar{A}_{\mu})\,.
\ee
The (homogeneous) ground state can be obtained from a variation of the effective potential
with respect to the fields evaluated at the physical ground state. Assuming for simplicity that
the vacuum expectation values of the vector bosons and the axial-vector boson are zero,
we find\footnote{Here, we evaluate the so-called gap equation 
for $\langle\vec{\phi}^{\,\rm T}\rangle=(\langle \phi_1\rangle,0)$. Equivalent ground states with 
$\langle\vec{\phi}^{\,\rm T}\rangle=(\langle \phi_1\rangle\neq 0,\langle \phi_2\rangle\neq 0)$ are related to
this ground state by means of a chiral transformation.}
\be
\langle \phi_1\rangle=4\frac{\bar{h}^2_{\sigma}}{2\bar{m}^2_{\sigma}}\int \frac{d^4 p}{(2\pi)^4}
\frac{\langle \phi_1\rangle}{p^2 +\frac{1}{2}\bar{h}^2_{\sigma} \langle \phi_1\rangle ^2 }\,.\label{eq:gap}
\ee
We immediately realize that the prefactor on the right-hand side is directly related to the 
four-fermion coupling~$\bar{\lambda}_{\sigma}$, see Eq.~\eqref{eq:h2m2}.
Depending on the value of the four-fermion 
coupling~$\bar{\lambda}_{\sigma}=\bar{h}^2_{\sigma}/(2 \bar{m}^2_{\sigma})$,
Eq.~\eqref{eq:gap} has apart from a trivial solution, $\langle\phi_1\rangle =0$, a non-trivial
solution for~$\langle\phi_1\rangle$. Since the integral on the right-hand side of Eq.~\eqref{eq:h2m2} is divergent, we have to 
regularize it. We employ a sharp UV cutoff $\Lambda\gg m_{\psi}$
for the four-momenta which permits a comparison of our results with the ones
from the previous section. The fermion mass~$m_{\psi}$ is then determined by the following implicit equation: 
\be
\left({4\pi^2}\left(\frac{2\bar{m}^2_{\sigma}}{\bar{h}^2_{\sigma}\Lambda^2}\right)-1\right)
=\left(\frac{m_{\psi}^2}{\Lambda^2}\right)\ln \left(\frac{m_{\psi}^2}{\Lambda^2}\right) \,,
\label{eq:mpsiMF}
\ee
where~$m_{\psi}$ is defined in Eq.~\eqref{eq:mpsidefsec3}.
Thus, the fermions acquire a finite mass due to the spontaneous breakdown of
chiral symmetry, if we choose 
\be
\lambda_{\sigma}=\bar{\lambda}_{\sigma}\Lambda^2 = \frac{\bar{h}_{\sigma}^2\Lambda^2}{2\bar{m}_{\sigma}^2}
> 4\pi^2\,.
\ee
From this inequality we can read off a critical value for the dimensionless four-fermion
coupling {$\lambda_{\sigma}=\bar{\lambda}_{\sigma}\Lambda^2$}:
\be
\lambda_{\sigma}^{\rm crit.}=4\pi^2\,.\label{eq:critvalscsec3}
\ee
It is instructive to repeat this analysis with the optimized regulator function that we have also employed in
the previous section. Using Eqs.~\eqref{eq:effactoneloop} and~\eqref{eq:fermreg2}, the gap equation~\eqref{eq:gap}
assumes the following form:
\be
\langle \phi_1\rangle=4\frac{\bar{h}^2_{\sigma}}{2\bar{m}^2_{\sigma}}\int \frac{d^4 p}{(2\pi)^4}
\left[
\frac{\langle \phi_1\rangle}{p^2 +\frac{1}{2}\bar{h}^2_{\sigma} \langle \phi_1\rangle ^2 }
- \frac{\langle \phi_1\rangle}{\Lambda^2 +\frac{1}{2}\bar{h}^2_{\sigma} \langle \phi_1\rangle ^2 }
\right]\theta(\Lambda^2 -p^2)
\,.\label{eq:gapOPTCUT}
\ee
Note that the second term in the square brackets 
is absent when we choose a sharp cutoff. From Eq.~\eqref{eq:gapOPTCUT} we obtain 
\be
\lambda_{\sigma}^{\rm crit.}=8\pi^2\,.\label{eq:critvaloptsec3}
\ee

Our results for the critical value~$\lambda_{\sigma}^{\rm crit.}$  
can be identified with 
the value of the non-trivial fixed point of the four-fermion coupling, which we have computed
in the previous section, see e.~g. Eqs.~\eqref{eq:NJLFPOPTCUT}, \eqref{eq:NJLFPsc} and~\eqref{eq:allchannelFP}.
In fact, we find that the critical values given in Eqs.~\eqref{eq:critvalscsec3} and~\eqref{eq:critvaloptsec3} are identical to the fixed-point 
values given in Eqs.~\eqref{eq:NJLFPsc} and~\eqref{eq:NJLFPOPTCUT}, respectively.
The role of the critical value as a fixed
point becomes apparent from the fact that the theory is (strongly) interacting 
for~$\lambda_{\sigma} = \lambda_{\sigma}^{\rm crit.}$ but remains massless and ungapped 
on all scales. 

\begin{figure}[t]
\begin{center}
\includegraphics[width=0.45\linewidth]{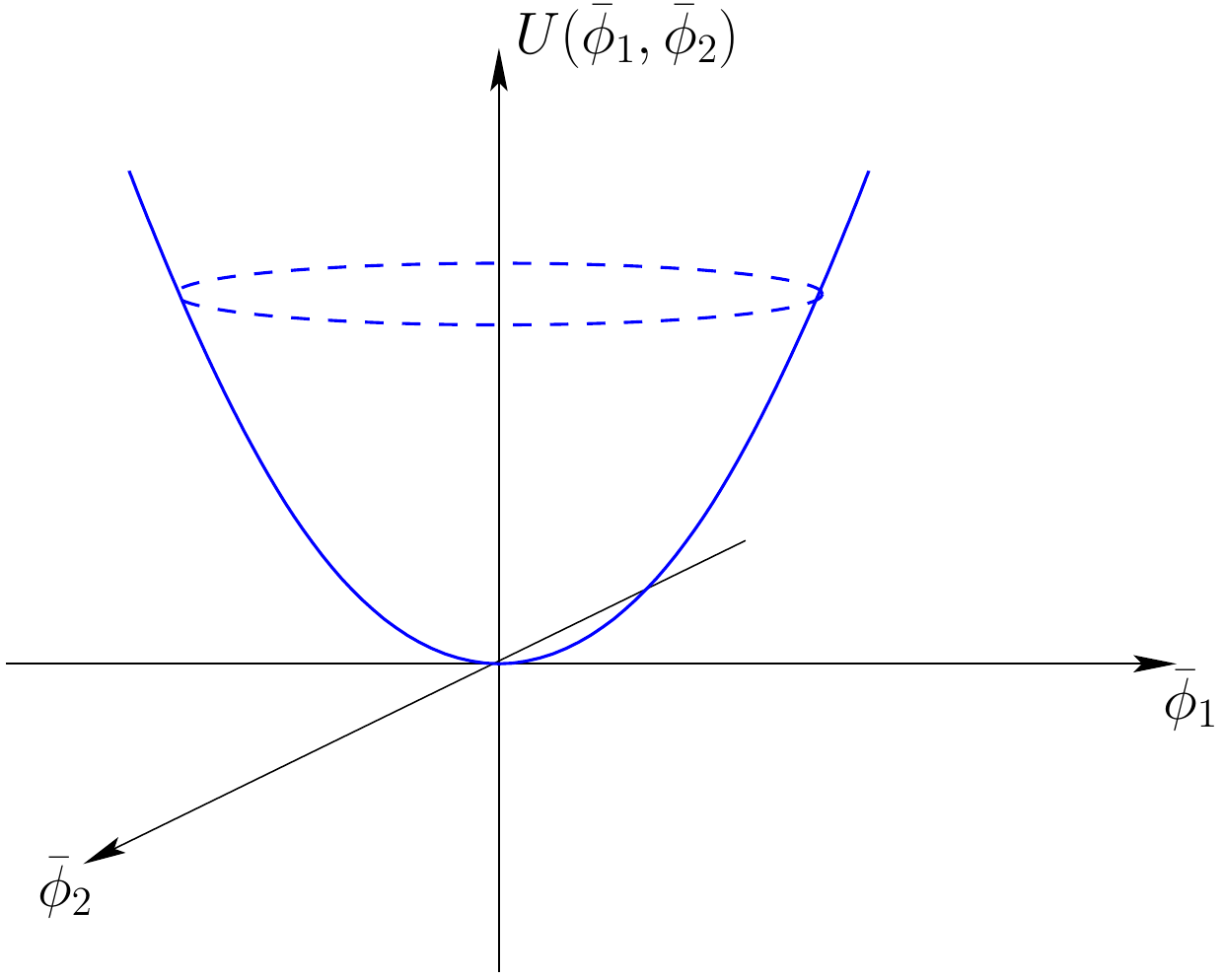}\hfill
\includegraphics[width=0.45\linewidth]{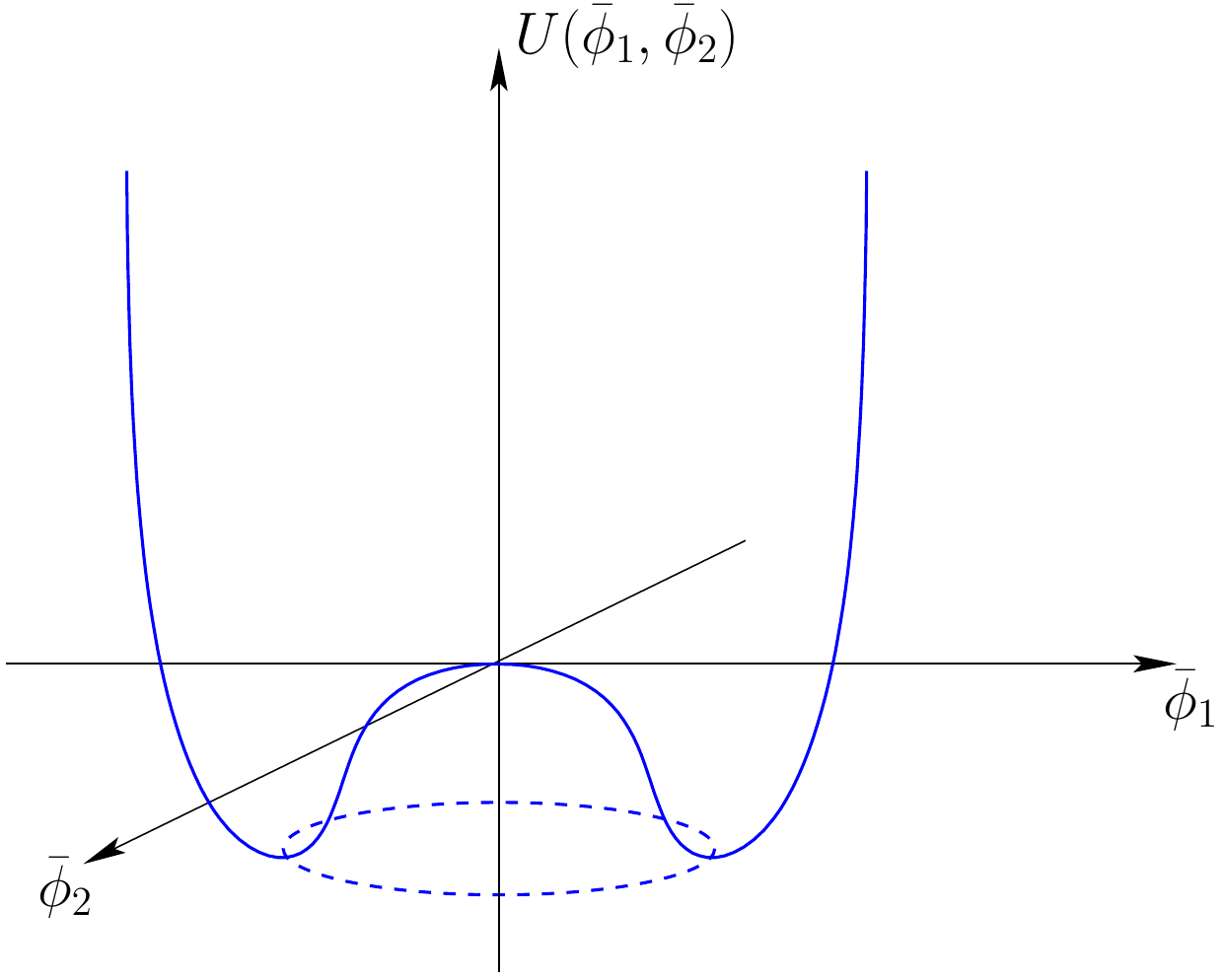}
\end{center}
\caption{Sketch of the mean-field order-parameter potential $U(\bar{\phi}_1,\bar{\phi}_2)$. The left panel depicts the shape of
the potential in the chirally symmetric phase 
($\langle \bar{\psi}\psi\rangle \sim |\langle \vec{\phi}\rangle| = 0$, \mbox{$\lambda_{\sigma} \leq \lambda_{\sigma}^{\rm crit.}$}).
The shape of the potential in a phase with broken chiral symmetry in the 
ground state ($|\langle\vec{\phi}\rangle| > 0$, $\lambda_{\sigma} > \lambda_{\sigma}^{\rm crit.}$) 
is shown in the right panel. }
\label{fig:effpot}
\end{figure}

In Fig.~\ref{fig:effpot} we show a sketch of the mean-field effective 
potential~$U(\bar{\phi}_1,\bar{\phi}_2,0,0)\equiv U(\bar{\phi}_1,\bar{\phi}_2,0,0)$ 
for~$\lambda_{\sigma} > \lambda_{\sigma}^{\rm crit.}$ (spontaneous breakdown of chiral
symmetry) and for~$\lambda_{\sigma} < \lambda_{\sigma}^{\rm crit.}$ (chirally symmetric phase).
This effective potential simply corresponds to a {\it Ginzburg-Landau} effective potential
for the order parameter fields. 

A word of caution on the parameter dependence of the model needs to be added here:
Our analysis indicates that the Yukawa couplings and the boson masses are not independent
input parameters of the partially bosonized theory at the UV scale~$\Lambda$. 
In fact, Eq.~\eqref{eq:mpsiMF} suggests that
the fermion mass, which is just one example for a physical observable, remains unchanged for
a fixed ratio~$\bar{h}_{\sigma}^2/(2\bar{m}_{\sigma}^2)$. As in the purely fermionic formulation,
this is indeed the case for~$d<4$. 
In $d=4$ space-time dimensions, however, the Yukawa coupling $\bar{h}$ is marginal. 
This then suggests that the partially bosonized theory in $d=4$ depends on 
two input parameters in contrast to $d<4$, see also Refs.~\cite{Braun:2009si,Braun:2010tt}.
We shall come back to this issue in Sect.~\ref{sec:njlgn} where we discuss two examples.

In Sect.~\ref{subsec:simpleex} we have discussed Fierz transformations in great detail in the context of
the purely fermionic formulation of our simple NJL model. One may wonder if the ambiguity arising 
from the possibility of Fierz transformations is still present in the partially bosonized formulation. As we have
discussed, the Fierz transformation~\eqref{eq:expFierz} states that only two of the three four-fermion
couplings are independent. Thus, one of these couplings, say~$\bar{\lambda}_{\rm A}$, can be chosen freely.
For example, we may choose~$\bar{\lambda}_{\rm A}=\gamma \bar{\lambda}_{\rm V}$ with~$\gamma\in {\mathbb{R}}$,
see Refs.~\cite{Jaeckel:2002rm,Jaeckel:2003uz}. Applying then the Fierz transformation~\eqref{eq:expFierz} to our
result for the critical coupling~$\lambda_{\sigma}^{\rm crit.}$ we observe that this quantity is not invariant
under such a transformation:
\be
\lambda_{\sigma}^{\rm crit.} \stackrel{\rm Fierz}{\longrightarrow}
\lambda_{\sigma}^{\rm crit.} - 2\gamma\lambda_{\rm V}
=4\pi^2 - 2\gamma\lambda_{\rm V}\,.
\ee
Thus, the critical value of the coupling depends on the arbitrary parameter~$\gamma$, if~$\lambda_{\rm V}$ 
is finite. As argued above, the critical couplings correspond to the value of the non-trivial fixed-point of the four-fermion
interactions which are of course not universal but depend on the renormalization scheme.
Therefore one might be tempted to not worry about this artificial $\gamma$-dependence.
However, it is fundamentally different from a scheme dependence since
the $\gamma$-dependence can be removed within any scheme. In particular, 
this so-called Fierz ambiguity affects not only the values of the critical couplings but also 
physical observables, such as the phase boundary at finite temperature or the 
mass spectrum of the theory. This can be readily seen from the corresponding change in the self-consistency 
equation~\eqref{eq:mpsiMF}:
\be
\left(\frac{4\pi^2}{\lambda_{\sigma}+2\gamma\lambda_{\rm V}} -1\right)
=\left(\frac{m_{\psi}^2}{\Lambda^2}\right)\ln \left(\frac{m_{\psi}^2}{\Lambda^2}\right) \,.
\ee
We stress that it is {\it not} 
possible to resolve this ambiguity within a mean-field approach~\cite{Jaeckel:2002rm}. 
Due to this, mean-field results of fermionic models will always be tainted with an uncertainty.

Since mean-field studies underly many investigations of phases of
strongly-interacting theories, ranging from condensed-matter physics to QCD, 
we briefly discuss the relevance of the Fierz ambiguity for these type of studies. 
Usually, the strategy for studying phases, e.~g. finite-temperature phase transitions, in 
NJL-type models is as follows: first, one chooses a physically 
motivated parameter set at zero temperature. Then, one uses the {\em same} 
initial conditions for a study of the model at finite temperature. Naively, one might now argue
that the Fierz ambiguity is of no relevance since it might always be possible to adjust 
the initial conditions such that the IR physics at zero temperature
remains unchanged. However, our present analysis already suggests 
that, e.~g., a vector-like coupling might be generated due to quantum and also thermal fluctuations,
see Sects.~\ref{subsec:fourfermionSSB} and~\ref{subsec:DefFT}. 
The finite-temperature dynamics might therefore be significantly altered for different
sets of parameters even though they have been adjusted to yield the same IR physics
at zero temperature.

\begin{figure}[t]
\begin{center}
\includegraphics[scale=1]{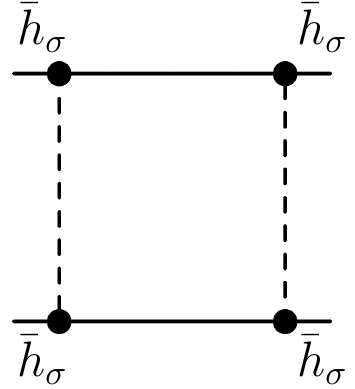}
\end{center}
\caption{Box diagram $\sim \bar{h}_{\sigma}^4$ which contributes to the RG flow of the four-fermion coupling. Dashed lines denote  
boson propagators while straight lines depict fermion propagators.}
\label{fig:boxbos}
\end{figure}

A further comment on the present bosonization procedure is in order.
In the remainder of this section, we shall follow the strategy to bosonize the four-fermion interaction at a given 
UV scale $\Lambda$. Below this scale, we then consider only operators in the RG flow which arise from the 
bosonization at~$k=\Lambda$. We stress that this is an approximation, since the four-fermion interactions 
are generated again in the RG flow, after each infinitesimal RG step~$\delta k$. This can be traced
back to the existence of finite Yukawa couplings. For example, the Yukawa coupling $\bar{h}_{\sigma}$ re-generates
the four-fermion coupling $\bar{\lambda}_{\sigma}$ in the RG flow due to the presence of so-called box diagrams,
see Fig.~\ref{fig:boxbos}. Thus, the $\beta$ functions of the four-fermion couplings are non-zero even though
we have removed these couplings from the action at the initial scale~$\Lambda$. The right-hand side of the RG flow 
equations of the four-fermion couplings then assumes the following form:
\be
\partial_t \bar{\lambda}_{i}  \sim \sum_{m,n} c_{mn}^{(i)}\bar{h}_{m}^2 \bar{h}_{n}^2 \,,
\ee
where $i,m,n\in\{\sigma,{\rm V},{\rm A}\}$ and the $c_{mn}^{(i)}$'s are scale-dependent functions depending on the boson masses.
The box diagrams shrink to point-like diagrams corresponding to four-fermion interactions in the case of large boson masses.
Of course, the running of the four-fermion interactions generated by the Yukawa couplings
is then expected to be small. However, it may become significant if the 
bosons become light, e.~g. close to a phase boundary. In any case, the re-generation
of the four-fermion couplings is problematic from a field-theoretical point of view, 
as one now encounters bosonic as well as purely fermionic operators at the scale $\Lambda -\delta k$. 
A so-called re-bosonization technique has been proposed to cure this problem by 
performing a bosonization of the newly generated four-fermion interactions in each RG 
step~\cite{Gies:2001nw,Gies:2002hq,Gies:2002kd,Pawlowski:2005xe,Gies:2006wv,Floerchinger:2009uf,Floerchinger:2010da}, 
to put it sloppily. This method has been successfully applied to QED and one-flavour 
QCD at zero temperature~\cite{Gies:2001nw,Gies:2002kd,Gies:2002hq} and finite temperature~\cite{Braun:2008pi}.
In addition to the phenomenological importance of this re-bosonization technique, it is of great importance
from a field-theoretical point of view. In fact, the discussed Fierz ambiguity, which is present in the partially bosonized 
description of the NJL model, can only be removed using this technique~\cite{Jaeckel:2002rm}.

Let us make one more point by comparing the NJL model and its bosonized version in more detail. Our
analysis of the partially bosonized theory allows us to define a sufficient and necessary criterion for 
the detection of the onset
of (chiral) symmetry breaking in the purely fermionic formulation:
\begin{itemize}
\item[(i)] Necessary condition: a four-fermion coupling must exceed its critical value to approach
an IR regime with broken (chiral) symmetry.
\item[(ii)] Sufficient criterion: a rapidly increasing (divergent) four-fermion coupling $\lambda$ 
signals the onset of (chiral) symmetry breaking:
\be
\frac{1}{\lambda}\rightarrow 0\quad\Leftrightarrow\quad 
\epsilon=\frac{\bar{m}^2}{Z k^2}\to 0\,,\nn
\ee
where $\epsilon$ and~$Z$ denote the dimensionless renormalized boson mass 
and the corresponding bosonic wave-function renormalization, respectively.  
In other words, a divergent four-fermion coupling signals a change in the sign of the coefficient of the
term quadratic in the fields in the Ginzburg-Landau effective potential.
\end{itemize}
Note that the necessary criterion is equivalent to the sufficient criterion for vanishing temperature only, see Sect.~\ref{subsec:DefFT}.
At finite temperature, the two criteria are no longer equivalent. This can be ultimately traced back to the fact that we only have
pseudo fixed-points in the RG flow once we introduce a dimensionful (external) parameter into the theory. This
so-called pseudo fixed-points then depend on, e.~g., the ratio of the temperature~$T$ and the RG scale~$k$.

Of course, the divergence of a four-fermion coupling at a finite scale $k_{\rm SB}$ is an artifact of the
point-like approximation employed in our studies of the purely fermionic formulation.
It can be resolved by taking into account (some of) the momentum dependence of the four-fermion coupling,
see also below. In any case, the scale $k_{\rm SB}$ at which $1/\lambda(k_{\rm SB})=0$ 
sets the scale for the IR observables. Thus, $k_{\rm SB}$ can be considered as the UV cutoff of
a low-energy theory describing the physics in the regime with broken chiral symmetry.

It is worth noting that, in general, the onset of (chiral) symmetry breaking is not simply indicated by a single rapidly growing 
four-fermion coupling. It is rather signaled by a strong RG running of various different four-fermion couplings associated with 
competing channels. This is due to the fact that a given rapidly increasing four-fermion coupling potentially entails
a strong running of other four-fermion couplings, see, e.~g., the set of flow 
equations~\eqref{eq:NJLsig2ch} and~\eqref{eq:NJLv2ch}. In a purely fermionic formulation it may therefore be 
a highly non-trivial task to resolve the symmetry-breaking patterns of a given theory. A partially bosonized formulation
might then be more beneficial.

Our discussion shows that both the purely fermionic formulation in a point-like approximation and the partially
bosonized formulation come with advantages and disadvantages from a technical as well as from a phenomenological
point of view. The purely fermionic formulation is very convenient to study the phase diagram of theories
in the (chirally) symmetric regime without suffering from Fierz ambiguities. From such a highly controlled study
of the symmetric phase we can then determine the regimes in parameter space in which the underlying symmetries are 
broken in the ground state of the theory. In particular, 
the purely fermionic description is well-suited for analytic studies. On the other hand, the 
partially bosonized ansatz allows us to resolve momentum dependences
of fermionic self-interactions in a simple manner and therefore permits a study of the formation of 
condensates and the mass spectrum of a given theory. However, the Fierz ambiguity is more difficult to resolve
in this picture. A fixed-point study in the purely fermionic formulation of the theory may therefore provide
important guidance for the construction of suitable low-energy models.

To further illustrate the relation (and equivalence) of the purely fermionic formulation and the partially
bosonized formulation we study the RG flow of the following action:
\be
\Gamma_k[\bar{\psi},\psi,\phi_1,\phi_2]
&=& \int d^4 x\,\Big\{ Z_{\psi}\bar{\psi}\I \fslash{\partial}\psi
+ \frac{\bar{h}_{\sigma}}{\sqrt{2}}\bar{\psi}(\vec{\tau}\cdot\vec{\phi})\psi \nn\\
&& \qquad\qquad\qquad +\,\frac{1}{2}Z_{\sigma}(\partial_{\mu}\vec{\phi})^2
+ \frac{1}{2}\bar{m}_{\sigma}^2\vec{\phi}^{\,2} 
+ \frac{1}{8}\bar{\omega}_{\sigma}\vec{\phi}^{\,4} \Big\}\,,\label{Eq:HSTAction}
\ee
where $Z_{\sigma}$ denotes the wave-function renormalization of the boson field. The
coupling~$\bar{\omega}_{\sigma}$ measures the strength of the four-boson interaction.
The boundary conditions of the flow equations at the initial RG scale~$\Lambda$ read
\be
\lim_{k\to \Lambda} Z_{\sigma} &=& 0\,,\label{eq:bc1}
\\
\lim_{k\to \Lambda} Z_{\psi} &=& 1\,,\label{eq:bc2}
\\
\lim_{k\to \Lambda} \bar{\omega}_{\sigma} &=& 0\,.\label{eq:bc3}
\ee
These conditions together with the identity
\be
\bar{\lambda}_{\sigma}=\frac{\bar{h}^2_{\sigma}}{2\bar{m}^2_{\sigma}}
\ee
allow us to map the ansatz~\eqref{Eq:HSTAction} onto the action~\eqref{eq:NJLtruncBasic} at the initial RG scale.
As our study of the RG flow of the action~\eqref{Eq:HSTAction} is just meant to highlight a few more aspects of the relation
of purely fermionic theories and their partially bosonized formulations, we drop the vector-interaction channel as well as the axial-vector
channel. We will also not make use of re-bosonization techniques. For a more complete study of the RG flow of 
partially bosonized theories based on the action~\eqref{eq:bosaction}, we refer the reader to Refs.~\cite{Jaeckel:2002rm,Jaeckel:2003uz}.

Let us now demonstrate that the partially bosonized formulation of our model allows us to conveniently resolve
the momentum dependence of four-fermion interactions. To this end, we consider the (quantum) equations of motion
of the fields~$\phi_1$ and~$\phi_2$ in momentum space, $(\delta \Gamma/\delta \phi_1(p))=0$ and~$(\delta \Gamma/\delta \phi_1(p))=0$:
\be
\phi_1(-p)&=&-\frac{\I}{Z_{\sigma}p^2 + \bar{m}_{\sigma}^2}\int \frac{d^4 q}{(2\pi)^4}
\frac{\bar{h}_{\sigma}}{\sqrt{2}}\left[\bar{\psi}(q)\psi(q-p)\right]\,,\label{eq:soleom1}\\
\phi_2(-p)&=&-\frac{1}{Z_{\sigma}p^2 + \bar{m}_{\sigma}^2}\int \frac{d^4 q}{(2\pi)^4}
\frac{\bar{h}_{\sigma}}{\sqrt{2}}\left[\bar{\psi}(q)\gamma_5 \psi(q-p)\right]\,,\label{eq:soleom2}
\ee
where we have set~$\omega_{\sigma}=0$ for simplicity. Now we insert these solutions for the scalar fields
into our ansatz~\eqref{Eq:HSTAction}. The Yukawa interaction term then assumes the following form in
momentum space:
\be
\!\!\!\! &&\frac{1}{2}\int\frac{d^4 p_1}{(2\pi)^4}\int\frac{d^4 p_2}{(2\pi)^4}\int\frac{d^4 p_3}{(2\pi)^4}\Big\{
\left[\bar{\psi}(p_1)\psi(p_2)\right]\frac{\bar{h}_{\sigma}^2}{Z_{\sigma}(p_1\!-\! p_2)^2 \!+\! \bar{m}_{\sigma}^2}
\left[\bar{\psi}(p_3)\psi(p_1\!-\! p_2\!+\! p_3)\right]\nn\\
&&\qquad\qquad\qquad\;
-\!\left[\bar{\psi}(p_1)\gamma_5\psi(p_2)\right]\frac{\bar{h}_{\sigma}^2}{Z_{\sigma}(p_1\!-\! p_2)^2 \!+\! \bar{m}_{\sigma}^2}
\left[\bar{\psi}(p_3)\gamma_5\psi(p_1\!-\! p_2\!+\! p_3)\right]
\!\!\Big\}. \label{eq:4psiMOMSPACE}
\ee
This corresponds to an s-channel approximation of the momentum dependence of the four-fermion vertex.
The expression in Eq.~\eqref{eq:4psiMOMSPACE} illustrates that the interaction between the fermions is mediated by the scalar fields
which act as exchange bosons (see Fig.~\ref{fig:bosmed}). We find that this term
simply reduces to the point-like four-fermion interaction term~$\sim\bar{\lambda}_{\sigma}$ included
in the action~\eqref{eq:NJLtruncBasic} for a large (renormalized)
boson mass, $m_{\sigma}^2 = \bar{m}^2_{\sigma}/Z_{\sigma} \gg \Lambda^2 \geq p^2$, see also Eq.~\eqref{eq:FTlambdaV}.
Thus, this analysis shows that the partially bosonized formulation allows us to resolve the momentum dependence of 
four-fermion interactions in a simple manner and that
the point-like approximation in our purely fermionic description in Sect.~\ref{subsec:simpleex}
is reasonable if the exchange bosons are heavy. On the other hand, the point-like approximation ultimately
breaks down in the limit of vanishing boson masses, 
e.~g. at the phase boundary and
in the phase with a spontaneously broken, continuous chiral symmetry where we encounter massless
bosonic excitations.

The solutions of the equation of motion of the scalar fields, Eqs.~\eqref{eq:soleom1} and~\eqref{eq:soleom2}, 
show also that the term $\sim \vec{\phi}^{\,4}$ in the effective action~\eqref{Eq:HSTAction} corresponds to an $8$-fermion interaction term  
in the purely fermionic description. Due to the boundary condition~\eqref{eq:bc3} this term is generated dynamically and not 
adjusted by hand in our RG approach, 
see e.~g. Refs.~\cite{Berges:1997eu,Gies:2002hq,Braun:2008pi,Braun:2009si,Braun:2009gm}.
Thus, the value of the corresponding coupling at the initial RG scale $\Lambda$ does not represent an additional parameter
of the theory. In any case, this coupling only plays a prominent role in the (deep) IR regime with broken chiral symmetry where it 
accounts for the mass difference between the Nambu-Goldstone boson and the mass of the radial mode.

Of course, the discussion in this section is mainly meant to address technical issues of fermionic models. Nonetheless we would like 
to remind the reader at this point that, in particular, the (partially) bosonized actions~\eqref{eq:bosaction} and~\eqref{Eq:HSTAction} 
resemble models from various fields of physics, ranging from condensed-matter physics (e.~g. Gross-Neveu-type models and 
ultracold atomic gases) to high-energy physics. In fact, our model is closely related to the action of the so-called {\em quark-meson model}, 
see Sect.~\ref{sec:njlgn}. In hadron physics, the auxiliary fields play the role of the scalar meson 
and the pseudo-scalar Nambu-Goldstone modes, i.~e. the pions, whereas the fermions are the constituent quarks. 
The Yukawa-coupling $\bar{h}_{\sigma}$ then specifies the strength of the quark-meson interaction.

Let us now discuss the RG flow equations for the couplings specified in the action~\eqref{Eq:HSTAction} as well as their mapping
onto the RG flows of the purely fermionic formulation of this model, see Eq.~\eqref{eq:NJLtruncBasic}. The partially bosonized version
of our model has been studied in detail in Ref.~\cite{Braun:2009si} for zero and finite temperature. 
To simplify the comparison between the two formulations,
we shall employ a so-called covariant regulator function for the bosons and the fermions which we have already used in Sect.~\ref{subsec:simpleex}. 
In the chirally symmetric regime the flow equations can be computed along the lines of Sect.~\ref{subsec:simpleex}:  
we compute the second functional derivative of the effective action~\eqref{Eq:HSTAction} with respect to the boson and the fermion
fields and evaluate it for homogeneous fermionic and bosonic background fields, 
respectively. The second functional derivative of the effective action can then
be split into a background-field dependent part (fluctuation matrix) 
and a part which does not depend on the background fields (propagator matrix). A straightforward calculation then
yields the following flow equations for the couplings:
\be
\partial _t \epsilon_{\sigma} &=& (\eta_{\sigma}-2)\epsilon_{\sigma} - 8v_4 l_1^{(4)}(\epsilon_{\sigma};\eta_{\sigma}) \omega_{\sigma}
 +\,8v_4 l_{1}^{{\rm (F)},(4)}(0;\eta_{\psi}) h^2_{\sigma}\,,\label{eq:m2flowNJL} \\
\partial _t \omega_{\sigma}&=& 2 \eta_{\sigma}\omega_{\sigma} + 20 v_4 l_2^{(4)}(\epsilon_{\sigma};\eta_{\sigma}) \omega_{\sigma}^2 
-\,8v_4 l_{2}^{{\rm (F)},(4)}(0;\eta_{\psi}) h^4_{\sigma}\,,\label{eq:lflowNJL}
\\
\partial _t h_{\sigma}^2 &=& (\eta_{\sigma} +2\eta_{\psi})h_{\sigma}^2\,,
\label{eq:hflowNJL}
\ee
where we have introduced the dimensionless renormalized couplings 
$\epsilon_{\sigma}=\bar{m}^{2}_{\sigma}/(Z_{\sigma}k^2)$,  $\omega_{\sigma}=\bar{\omega}_{\sigma}/Z_{\sigma}^2$ and 
the Yukawa coupling $h_{\sigma}=\bar{h}_{\sigma}/(Z_{\sigma}^{1/2}Z_{\psi})$. The equations for the
anomalous dimensions~$\eta_{\sigma}=-\partial _t \ln Z_{\sigma}$ and $\eta_{\psi}=-\partial _t \ln Z_{\psi}$ 
read\footnote{In order to obtain the flow equations for the wave-function renormalizations, we have to evaluate the second
functional derivative of the effective action for spatially varying bosonic and fermionic background 
fields, see Sect.~\ref{subsec:simpleex}.} 
\be
\eta_{\sigma}&=& 4 v_4 h^2_{\sigma} m_{4}^{({\rm F}),(4)}(0;\eta_{\psi}) \,,\\
\eta_{\psi}&=& 2 v_4 h^2_{\sigma} m_{1,2}^{({{\rm FB}}),(4)} (0,\epsilon_{\sigma};\eta_{\psi},\eta_{\sigma}) \,. \label{eq:PB1etaPsi}
\ee
The threshold functions are defined in App.~\ref{app:regthres}. The functions~$l_n^{(4)}$ represent purely bosonic loops.
Note that the fermions are massless (ungapped) in the symmetric
regime since the vacuum expectation value~$\langle \vec{\phi}\rangle$ of the boson fields vanishes. 

From the flow equation for the Yukawa coupling we recover the one-loop result for $\epsilon_{\sigma}\to 0$, 
see Refs.~\cite{Gies:2001nw,Braun:2009si}:
\be
\partial _t h^2_{\sigma}=\frac{h_{\sigma}^4}{4\pi^2}\,.\label{eq:h2flowNJL1flavor}
\ee
The running of the Yukawa coupling is driven only by the anomalous dimensions in the symmetric regime. This is a peculiarity of the present
model with a continuous chiral symmetry but only one fermion species. In theories with more than one fermion species,
such as two-flavor QCD, or in theories with a discrete chiral symmetry, the running of the Yukawa coupling receives also
contributions from so-called triangle diagrams, see Sect.~\ref{sec:njlgn}. In these types of theories, 
it turns out  that the contributions from triangle
diagrams are subleading in a systematic expansion of the flow equations in powers of the inverse number of 
fermion species.\footnote{In the Gross-Neveu model the expansion parameter is the inverse number of fermion flavors, whereas
it is the inverse number of colors in QCD.}

We now have a closer look at the relation between the partially bosonized and the purely fermionic
formulation. To this end, we consider the RG flow of the ratio 
$h^2_{\sigma}/(2\epsilon_{\sigma})=(k^2/Z_{\psi}^2) \bar{h}^2_{\sigma}/(2\bar{m}^2_{\sigma})=\lambda_{\sigma}$ 
which can be obtained straightforwardly from the flow equations~\eqref{eq:m2flowNJL} and~\eqref{eq:hflowNJL}:
\be
\partial_t \lambda_{\sigma}\equiv \partial_t \left( \frac{h^2_{\sigma}}{2\epsilon_{\sigma}}\right)=
(2+2\eta_{\psi})\lambda_{\sigma} - 16v_4 l_{1}^{{\rm (F)},(4)}(0;\eta_{\psi})\lambda_{\sigma} ^2
+ 8v_4 l_1^{(4)}(\epsilon_{\sigma};\eta_{\sigma}) \frac{\lambda_{\sigma}\omega_{\sigma}}{\epsilon_{\sigma}}\,.\label{eq:h2m2flow}
\ee
We observe that the first two terms on the right-hand side of this flow equation
are identical to the terms appearing in the flow equation~\eqref{eq:NJLbetasimple}
which we computed in the point-like approximation in the purely fermionic formulation.  
In particular, we find that the flows are identical in the limit~$k\to\Lambda$ where we have~$\omega_{\sigma}=0$
and~$\epsilon_{\sigma}\to\infty$ according to the boundary condtions~\eqref{eq:bc1}-\eqref{eq:bc3}.
From this comparison we deduce that the partially bosonized
description indeed allows us to go conveniently beyond the point-like approximation employed 
in Sect.~\ref{subsec:simpleex}. To be more specific, 
we observe that the momentum dependence of the four-fermion vertex is effectively parameterized
by the wave-function renormalization~$Z_{\sigma}$ and the mass parameter~$\epsilon_{\sigma} \sim m^2_{\sigma}$.
As discussed above,~$\omega_{\sigma}$ parameterizes fermionic self-interactions of higher order. Thus, the contribution~$\sim\omega_{\sigma}$
on the right-hand side of Eq.~\eqref{eq:h2m2flow} indicates that the RG flows of fermionic interactions of higher order,
such as eight-fermion interactions, contribute to the RG flows of four-fermion interactions, if we go beyond the point-like approximation.
For scales $k<\Lambda$, the purely fermionic point-like description and 
the partially bosonized description are then no longer identical. From the 
flow equation of the ratio $h_{\sigma}^2/(2\epsilon_{\sigma})$, however, it follows that
the differences are quantitatively small in the (chirally) symmetric regime since the renormalized mass
of the bosons is large.

In addition, we find that the anomalous dimension~$\eta_{\psi}$ is non-zero in the partially bosonized formulation whereas it
is identically zero in the point-like approximation of the purely fermionic description. Again, the running
of~$Z_{\psi}$ is small in the symmetric regime due to the large (renormalized) boson mass.\footnote{
This follows immediately when we replace the four-fermion vertex 
in the diagram on the right in Fig.~\ref{fig:feynman} with two Yukawa vertices connected by a boson
propagator.} 

Our analysis of the partially bosonized NJL model reveals clearly that spontaneous (chiral) symmetry breaking can 
be studied within a purely fermionic formulation. Some information on the symmetries
of the ground-state is already encoded in the strength of the four-fermion couplings compared to their values at the non-trivial RG fixed point.
These values can be viewed as critical values for the four-fermion couplings: in this framework, the system approaches a phase with broken
chiral symmetry in the ground state, if the critical values are exceeded by the initial values of the four-fermion couplings at the UV scale. 

Let us highlight a subtlety concerning the existence of the non-trivial RG fixed point of the four-fermion coupling. 
From Eqs.~\eqref{eq:m2flowNJL}-\eqref{eq:hflowNJL} we deduce that we can only have fixed-points with~$(h_{\sigma}^{\ast})^2\equiv 0$. Considering the 
relation~$\lambda_{\sigma}= h^2_{\sigma}/(2\epsilon_{\sigma})$, this suggests that only a Gau\ss ian fixed point exists 
for the four-fermion coupling beyond the point-like approximation. In fact, the non-trivial fixed point of the 
four-fermion coupling in the present type of fermionic models\footnote{This includes QCD low-energy models,
see Sect.~\ref{sec:NJLLQCD}.} is most likely an 
artifact of the point-like approximation in~$d=4$. It is only present for a finite UV cutoff~$\Lambda$. 
In the spirit of of low-energy models, the UV cutoff of a fermionic model in~$d=4$ should therefore always be considered as an additional 
parameter of the theory.\footnote{In contrast to the present NJL model in~$d=4$, the limit~$\Lambda\to\infty$ is well-defined in, e.~g.,
the Gross-Neveu model in~$d=3$, see Sect.~\ref{sec:GNmodel}.}
For any finite value of~$\Lambda$, it is then still possible to define a critical value of the 
four-fermion coupling above which we have chiral symmetry breaking in the IR limit, 
also beyond the point-like approximation. 
In our discussion we refer to this critical coupling as a quantum critical point, even if we discuss fermionic models in~$d=4$.

The partially bosonized formulation also allows us to study the deep IR regime in which
the dynamics are governed by massless Nambu-Goldstone bosons. Of course, this regime can also be studied
with the aid of RG flow equations, see e.~g. 
Refs.~\cite{Jungnickel:1995fp,Berges:1997eu,Schaefer:1999em,Gies:2002hq,Braun:2008pi,Braun:2009si,Braun:2009gm}.
We have shown that the mass parameter $m^2_{\sigma}\sim \epsilon_{\sigma}$ assumes 
negative values in this regime. This behavior signals the existence of a finite vacuum expectation value of the field $\vec{\phi}$. 
For an RG study, it is then convenient to parametrize the flow of the effective action in terms of
the four-boson coupling $\omega_{\sigma}$ and the vacuum expectation value $\langle\vec{\phi}\rangle$
rather than the mass parameter~$\epsilon_{\sigma}$. The RG flow of~$\langle\vec{\phi}\rangle$ 
can be derived from the following condition:
\be
\frac{d}{dt}\left[\frac{\partial}{\partial{\vec{\phi}}^{\,2}}\left(
\frac{1}{2}\bar{m}^2_{\sigma}{\vec{\phi}}^{\,2} + \frac{1}{8}\bar{\omega}_{\sigma}{\vec{\phi}}^{\,4}
\right)\right]_{\langle\vec{\phi}\rangle}\stackrel{!}{=}0\,.
\ee
However, a discussion of the regime governed by spontaneous symmetry breaking 
is beyond the scope of this review. For a detailed study of the IR dynamics of this particular NJL model, 
we refer the reader to Refs.~\cite{Jaeckel:2002rm,Braun:2009si}. In the following we restrict our discussions to the RG flow of four-fermion
interactions in the point-like limit in the symmetric regime. 
As discussed, this already allows us to map the phase diagram of a given theory as a function of various couplings in a clean and very 
controlled way. This has been explicitly demonstrated in the context of gauge theories in Ref.~\cite{Gies:2005as},  
where the regularization-scheme independence of 
universal quantities has been found to hold remarkably well in the point-like limit.


%
\subsection{Spontaneous Symmetry Breaking and Fermion Interactions}\label{subsec:fourfermionSSB}
Now that we have discussed the relation between spontaneous symmetry breaking and the
fixed-point structure of 
four-fermion interactions, we are in a position to compute the zero-temperature phase diagram
of the simple NJL model with only one species. To this end, we employ the flow equations~\eqref{eq:NJLsig2ch} 
and~\eqref{eq:NJLv2ch} as obtained from the Fierz-complete ansatz~\eqref{eq:NJLtruncBasic3} for the effective action.

In the previous section we have found that the non-trivial fixed-point of the four-fermion coupling in 
a single-channel approximation can be identified with the critical value which needs
to be exceeded in order to have chiral symmetry breaking in the IR limit. In such a single-channel approximation
this non-trivial fixed point is IR repulsive, as discussed in Sect.~\ref{subsec:simpleex}. Since the fermionic 
interactions are not induced by, e.~g., fluctuations of gauge fields, we have to tune
the initial value $\lambda^{\rm UV}_{\sigma}$ of the four-fermion coupling such that $\lambda_{\sigma}^{\rm UV}>\lambda_{\sigma}^{\ast}$
in order to observe chiral symmetry breaking in the IR. The four-fermion coupling then increases rapidly and diverges
at a finite RG scale~$k_{\rm SB}$, $1/\lambda_{\sigma}(k_{\rm SB})\to 0$. 
The value of this scale is clearly scheme-dependent. In any case, $k_{\rm SB}$ sets
the scale for a given (chiral) low-energy observable~$\mathcal O$, e.~g. the chiral condensate or the 
phase transition temperature: 
\be
{\mathcal O}\sim k_{\rm SB}^{d_{\mathcal O}}\,,\label{eq:kcrOIntro}
\ee
where $d_{\mathcal O}$ is the canonical dimension of the observable. This observation is at the heart of our study of scaling
in the subsequent section and our analysis of chiral symmetry breaking in gauge theories. On the other hand, choosing
$\lambda_{\sigma}^{\rm UV}<\lambda_{\sigma}^{\ast}$, we find that the $\lambda_{\sigma}$-coupling approaches 
zero in the IR limit. Thus, the non-trivial fixed point separates a trivial (non-interacting) phase from a phase with 
broken chiral symmetry in the ground state. The initial value for the four-fermion coupling can be considered
as an external parameter which controls chiral symmetry breaking and triggers a quantum phase
transition at~$\lambda_{\sigma}^{\rm UV}=\lambda_{\sigma}^{\ast}$. From a thermodynamical point of view,
one can think of 
$\lambda_{\sigma}^{\rm UV}$ as a proxy for the temperature of the system, and of~$\lambda_{\sigma}^{\ast}$ 
as the critical temperature.

At this point we would like to remind the reader that the divergence of the 
four-fermion coupling at a finite scale $k_{\rm SB}$ is an artifact of the point-like approximation.
This divergence simply indicates the onset of the 
breaking of the chiral U($1$) symmetry of our model. Hence the divergence has 
a physical meaning. 
It can be resolved by taking into account some of the momentum dependence of the coupling $\lambda_{\sigma}$, 
as discussed in the previous subsection. Such an 
improvement of the truncation is indispensable if we are interested 
in the IR properties of the model or the order of the phase transition. However, 
for a computation of the phase boundaries at zero as well as at finite temperature, 
it is a good approximation to simply detect the divergence in the RG flow.

\begin{figure}[t]
\begin{center}
\includegraphics[width=0.65\linewidth]{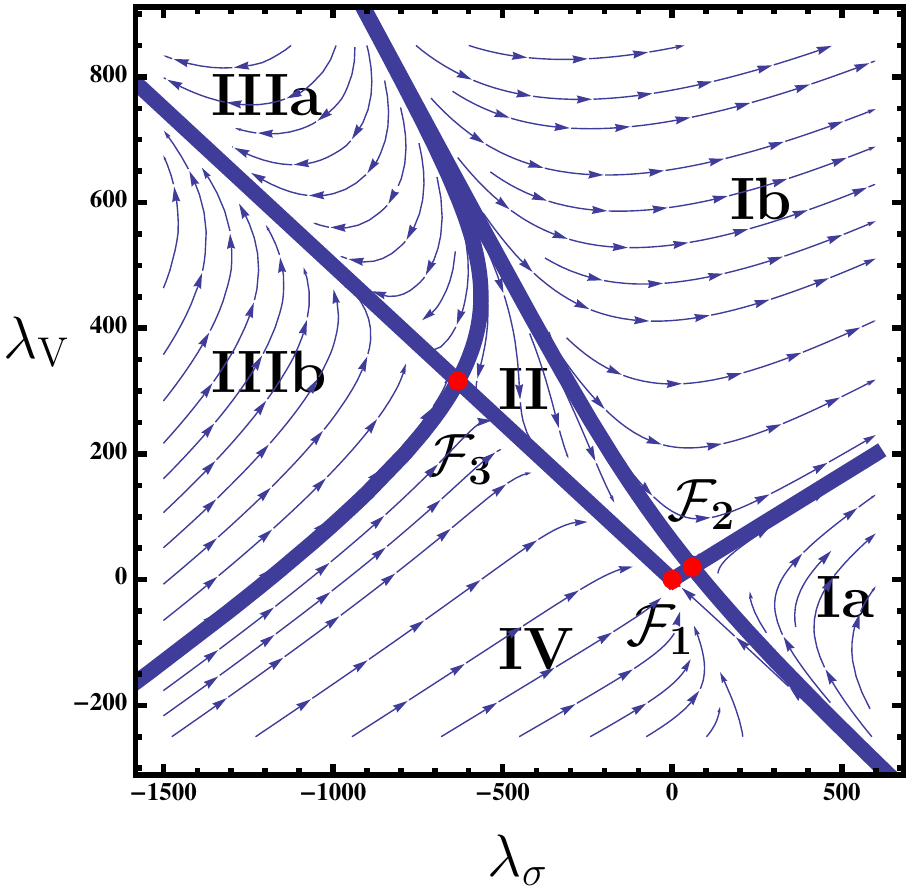}
\end{center}
\caption{Phase diagram of the NJL model with only one fermion species. The fixed points ${\mathcal F}_i$ are given
by red (light gray) dots. The separatrices (thick solid line) separate domains with different IR properties. The arrows of the various RG
trajectories (thin solid lines) indicate the direction of the flow towards the infrared.}
\label{fig:pd_njl}
\end{figure}

Let us now turn to a discussion of symmetry breaking within our Fierz-complete ansatz for the effective action.
To this end, we can indeed apply similar arguments as for the discussion of the single-channel approximation above.
However, the RG flow is now governed by the existence of fixed points in the two-dimensional plane spanned
by the couplings~$\lambda_{\sigma}$ and~$\lambda_{\rm V}$. As discussed in Sect.~\ref{subsec:simpleex},
we find three fixed points, see Eq.~\eqref{eq:allchannelFP}: one Gau\ss ian fixed point with two IR
attractive directions and two fixed points with both a repulsive and an attractive direction. The notion of a critical
coupling becomes more complicated in the present case. The role of the critical coupling in the single-channel approximation is
now essentially taken over by the so-called separatrices. 
In the $d_{\rm c}$-dimensional space spanned by the~$d_{\rm c}$ couplings of a given theory, 
separatrices can be viewed as $(d_{\rm c}\!-\!1)$-dimensional manifolds. These manifolds
separate disjunct domains of the RG flow. Fixed points are elements of the set of points defined by the separatrices.
This implies that a separatrix also determines the RG flow between two fixed points. In the present case, we have~$d_{\rm c}=2$ and
separatrices correspond to lines in the $(\lambda_{\rm V},\lambda_{\sigma})$-plane, see Fig.~\ref{fig:pd_njl}.

In Fig.~\ref{fig:pd_njl} we show the zero-temperature phase diagram of our model as obtained from an evaluation 
of the flow equations with an optimized regulator function. The fixed points~${\mathcal F}_i$ are given by red (light
gray) points. We observe that the separatrices (thick solid lines) allow us to divide the $(\lambda_{\sigma},\lambda_{\rm V})$-plane
into various distinct domains which we denote by roman numbers. Choosing a set of initial conditions in the regimes $\rm Ia, Ib, IIIa$ and~IIIb 
we find that the couplings increase rapidly and diverge at a finite scale~$k_{\rm SB}$ indicating the onset of chiral
symmetry breaking in the infrared.  
However, the four-fermion couplings in the domains $\rm Ia/b$ and~$\rm IIIa/b$ approach different points for $k\to k_{\rm SB}$.
While the RG flows in the regimes $\rm Ia$ and~$\rm Ib$ are attracted to the 
point
\be
{\mathcal F}_2^{\infty}:=\lim _{\alpha\to\infty}\alpha\cdot {\mathcal F}_2
\stackrel{\eqref{eq:allchannelFP}}{=}\lim _{\alpha\to\infty} (3\alpha\zeta,\alpha\zeta)
\,, 
\ee
the flows approach the point
\be
{\mathcal F}_3^{\infty}:=\lim _{\alpha\to\infty}\alpha\cdot {\mathcal F}_3
\stackrel{\eqref{eq:allchannelFP}}{=}\lim _{\alpha\to\infty} (-32\alpha\zeta ,\alpha\zeta )
\ee
for $k\to k_{\rm SB}$ in the regimes~$\rm IIIa$ 
and~$\rm IIIb$; the scheme-dependent quantity~$\zeta$ is defined in Eq.~\eqref{eq:zetadef}.
From a phenomenological point of view, 
the low-energy theories associated with the domains~$\rm Ia/b$ might therefore be
different, even though both correspond to theories with a broken chiral symmetry
in the ground state. To identify the IR properties of these theories, we would need to go beyond the point-like
approximation and include the momentum dependence of the fermionic interactions. However, 
this is beyond the scope of the present analysis. 

It is worth mentioning that the RG trajectories in the regimes $\rm Ib$ and $\rm IIIa$ originate from the very same point at infinity.
This point corresponds to a fixed point to which we refer as~${\mathcal F}_4$. This 
is the fixed point which seems to be missing in our analysis in Sect.~\ref{subsec:simpleex}. It has 
two IR repulsive directions. In a study with $N_{\rm f}$ fermions it indeed turns out that the two coordinates of ${\mathcal F}_4$
assume finite values, see e.~g. Refs.~\cite{Gies:2003dp,Gies:2005as,Braun:2006jd,Eichhorn:2011pc}.

In contrast to the regimes with a strongly-interacting IR limit, the dynamics in the regimes~$\rm II$ and~$\rm IV$ are governed
by the Gau\ss ian fixed point~${\mathcal F}_1$. These regimes represent the basin of attraction of the Gau\ss ian
fixed point. If we choose a set of initial conditions in these two domains, both couplings
are attracted to the Gau\ss ian fixed point and therefore vanish in the IR limit. Thus such initial conditions yield 
trivial, i.~e. non-interacting, IR theories.

From the phase diagram we can also read off that there exists no trajectory for which either of the
two couplings, namely~$\lambda_{\sigma}$ and~$\lambda_{\rm V}$, remains constant under RG transformations. For example,
a vector-channel interaction is always generated in the RG flow even if we set it to zero at the initial RG scale, provided
we choose a non-vanishing initial value for the $\lambda_{\sigma}$-coupling. The existence of a trajectory with 
$\lambda_{\rm V}\equiv 0$ (NJL-like trajectory) would imply
that one of the non-trivial fixed points was located on the $\lambda_{\sigma}$-axis of the phase diagram, with its IR repulsive
direction pointing towards the Gau\ss ian fixed point. 
 In Sects.~\ref{subsec:NJLSUNf} and~\ref{subsec:NJLUNf}, 
we further discuss 
whether an NJL-like trajectory exists in the limit of many fermion flavors. Note that this question is certainly an interesting 
and important question for, e.~g., the construction of low-energy QCD models in the limit of many colors. 
In a Fierz-complete study of the Thirring model in three dimensions, it 
has indeed been found that a pure Thirring-like trajectory exists in the limit of many fermion flavors. On this trajectory
we only have a running of the four-fermion interaction that is included in standard mean-field studies of the 
Thirring model~\cite{Gies:2010st}.

\subsection{A First Look at Scaling Behavior close to a Quantum Critical Point}\label{subsec:NJLscaling}
In statistical physics, one finds that completely different many-body systems show the same 
{\em quantitative} behavior near critical phase transitions, where long-range fluctuations 
are important. In the vicinity of these critical points, the behavior of a given theory is independent of 
its (microscopic) details and can be described completely by scaling relations and 
a small set of so-called critical exponents. 
This phenomenon is called {\em universality}. 
For example, finite-temperature phase transitions of the 3$d$ Ising model and SU($2$) Yang-Mills 
theory belong to the same universality class. 
Typical scaling relations near a critical point (e. g. a second-order phase transition) are of the form
\be
\xi\sim\left(\frac{T-T_{\text{cr}}}{T_{\text{cr}}}\right)^{-\nu}\quad\text{and}\quad 
G(x,0)\sim |x|^{2-d-\eta}\,, \label{eq:tbscalingrel}
\ee
where $\xi$ is the range of correlated fluctuations (correlation length) 
and $G$ denotes the two-point correlation function between fluctuations  at the origin and 
at space-time point $x$ at~$T=T_{\text{cr}}$. The universal critical exponents are $\nu$ and $\eta$, and 
$d$ denotes the number of Euclidean space-time dimensions.
The temperature is denoted by $T$ and the critical temperature by~$T_{\text{cr}}$.

The existence of universal behavior plays a very important role in modern physics. First, 
it allows us to classify theories according to their scaling behavior 
close to a phase transition which turns out to depend solely on the symmetry properties and the 
dimensionality of the critical system. Second, we may check numerical
data from studies of more involved theories with the aid of scaling functions as obtained
from an investigation of simpler models in the same universality class. Due to dimensional reduction at finite temperature
we can, for example, use scaling functions
from a simple three-dimensional scalar O$(4)$ model to analyze lattice QCD data for~$d=4$ 
close to the chiral phase transition, see e.~g. Refs.~\cite{Braun:2007td,Braun:2008sg,Ejiri:2009ac,Braun:2010vd,Engels:2011km}.

It is important to stress that such scaling laws and functions cannot be deduced from any fixed-order 
perturbation theory calculation, since there are inherently non-perturbative phenomena
underlying these laws.

In the following we discuss scaling behavior at zero temperature rather than at finite temperature. In
particular, we aim to understand the scaling behavior of fermionic theories at vanishing temperature close to 
a so-called {\it quantum critical point}. The associated quantum phase transition is driven purely by
the variation of a parameter other than the temperature. For example, a variation of the
number of fermion flavors may induce a quantum phase transition, as we shall discuss in
detail in Sect.~\ref{sec:gaugetheories}. In this section, however, we discuss only basic aspects of
the scaling behavior close to quantum phase transitions. This will be helpful for our investigations in the
remainder of this review.

\subsubsection{Power-law Scaling}\label{sec:NJLPL}
For simplicity, we start our discussion with an analysis of scaling in the single-channel 
approximation of the simple NJL model given in Eq.~\eqref{eq:NJLbetasimple}.
As discussed above in detail, the statement about the mere existence of the fixed-point~$\lambda_{\sigma}^{\ast}$,
see Eq.~\eqref{eq:onechannelFP}, is universal, even though its actual value is clearly scheme-dependent.
The control parameter for quantum critical behavior is given by the value~$\lambda_{\sigma}^{\rm UV}$
at the initial RG scale~$\Lambda$. In fact, choosing an initial value $\lambda_{\sigma}^{\rm UV} <\lambda_{\sigma}^{\ast}$  
at the UV scale $\Lambda$, we find that the theory becomes non-interacting in the
IR regime, see Fig.~\ref{fig:parabola}. For $\lambda_{\sigma}^{\rm UV}>\lambda_{\sigma}^{\ast}$, 
we find that the four-fermion coupling $\lambda_{\sigma}$ increases rapidly and diverges eventually 
at a finite scale~$k_{\rm SB}$, so that $1/\lambda_{\sigma}(k_{\rm SB})=0$. 
This indicates the onset of chiral symmetry breaking. Hence chiral symmetry breaking in the IR 
only occurs if we choose $\lambda_{\sigma}^{\rm UV}>\lambda_{\sigma}^{\ast}$. Since~$\lambda_{\sigma}^{\rm UV}$ 
distinguishes between two different phases in the IR limit, i.~e. in the long-range limit, we can 
consider~$\lambda_{\sigma}^{\ast}$  as a quantum critical point which divides the
model into two substantially different physical regimes.

The scale $k_{\rm SB}$ sets the scale for a given IR observable, see Eq.~\eqref{eq:kcrOIntro}. In the present case, this scale
can be directly obtained from the solution Eq.~\eqref{eq:solfloweqse} of the flow equation~\eqref{eq:NJLbetasimple}. 
Solving~$1/\lambda_{\sigma}(k_{\rm SB})=0$ for~$k_{\rm SB}$, we then find
\be
k_{\rm SB}=\Lambda \theta(\lambda_{\sigma}^{\rm UV} -\lambda_{\sigma}^{\ast})
\left(\frac{\lambda_{\sigma}^{\rm UV} -\lambda_{\sigma}^{\ast}}{\lambda_{\sigma}^{\rm UV}} \right)^{\frac{1}{|\Theta|}}
+{\mathcal O}(\eta_{\psi}^{\ast})\,,
\label{eq:lambdacr}
\ee
where $\theta(\dots)$ is the unit-step function. The critical exponent~$\Theta$ associated with the 
fixed point~$\lambda_{\sigma}^{\ast}$, which governs the critical behavior, is given by
\be
\Theta:=-\frac{\partial (\partial _t \lambda_{\sigma})}{\partial \lambda_{\sigma}}\Big|_{\lambda_{\sigma}^{\ast}} =2+2\eta_{\psi}^{\ast}
-2\lambda_{\sigma}^{\ast}
\frac{\partial \eta_{\psi}}{\partial \lambda_{\sigma}} \Big| _{\lambda_{\sigma}^{\ast}}
+ 16 v_4\, (\lambda_{\sigma}^{\ast})^2\,  \frac{\partial l_{1}^{\rm (F),(4)}(0;\eta_{\psi})}{\partial \lambda_{\sigma}}
\Big| _{\lambda_{\sigma}^{\ast}}\,,
\ee
see also Eq.~\eqref{eq:genthetadef}. This reduces to
\be
\Theta=2\,
\ee
for $\eta_{\psi}\equiv 0$, independent of the regulator function. 
Apparently an exact computation of the exponent~$\Theta$ requires to include corrections beyond the point-like limit. 

In any case, the value $\ksb$ scales with the distance of the initial value 
$\lambda_{\sigma}^{\rm UV}$ from the fixed-point value $\lambda_{\sigma}^{\ast}$. If this distance is increased,
the scale $\ksb$ increases and the values of physical (low-energy) observables~$\mathcal O$, such 
as the correlation length~$\xi$ or the fermion mass, increase in turn:
\be
{\mathcal O}\sim k_{\rm SB}^{d_{\mathcal O}}=  \Lambda^{d_{\mathcal O}} \theta(\lambda_{\sigma}^{\rm UV} -\lambda_{\sigma}^{\ast})
\left(\frac{\lambda_{\sigma}^{\rm UV} -\lambda_{\sigma}^{\ast}}
{\lambda_{\sigma}^{\rm UV}}
\right)^{\frac{d_{\mathcal O}}{|\Theta|}}\,.\label{eq:genscallaw}
\ee
Here, $d_{\mathcal O}$ is the canonical mass dimension of the observable~$\mathcal O$. From this expression
it is clear that the absolute values of the scheme-dependent quantities~$\lambda_{\sigma}^{\rm UV}$ 
and~$\lambda_{\sigma}^{\ast}$ are not physically relevant, but only their relative 
distance is.\footnote{In a phenomenological application of the model, the initial condition~$\lambda_{\sigma}^{\rm UV}$ 
is fixed by the values of a given set of IR observables.}
We stress again that the critical exponent $\Theta$ governs the long-range physics at the quantum critical
point~$\lambda_{\sigma}^{\ast}$. It is related to the 
standard correlation length exponent $\nu$ by $\nu=1/\Theta$.
In addition, it follows from the scaling law Eq.~\eqref{eq:genscallaw} that the quantum critical point is 
associated with a vanishing boson mass~$m\sim k_{\rm SB}$ and a diverging correlation length~$\xi\sim 1/m$ 
in the long-range limit. The critical exponent~$\eta$ in Eq.~\eqref{eq:tbscalingrel} is determined by the anomalous dimension
of the (bosonic) order-parameter field~$\sim \bar{\psi}\psi$, and the fixed-point value~$\lambda_{\sigma}^{\ast}$ plays the role
of the critical temperature.

One may wonder how the scaling behavior changes when we consider the Fierz-complete set of flow equations,
since we then have to deal with two couplings instead of one. For illustration, let us consider a situation in which
we have chosen a set of initial conditions in the domain~$\rm Ib$ of the phase diagram given in Fig.~\ref{fig:pd_njl}.
In this domain, the RG flows are attracted to the point~${\mathcal F}_2^{\infty}$:
\be
{\mathcal F}_2^{\infty}:=\lim_{\alpha\to\infty}\alpha\cdot {\mathcal F}_2\,.
\ee
For any initial condition~$(\lambda_{\sigma}^{\rm UV},\lambda_{\rm V}^{\rm UV})$ in this domain
we can then find a suitably chosen scale~$k_0<\Lambda$ for which the RG trajectory is close to the separatrix connecting the 
points~${\mathcal F}_2$ and~${\mathcal F}_2^{\infty}$. In other words, the point~$(\lambda_{\sigma}(k_0),\lambda_{\rm V}(k_0))$ 
lies close to this separatrix. Of course, the scale~$k_0$ depends strongly on the initial conditions at the 
scale~$\Lambda$. For scales~$k<k_0$, the direction of the RG flow is then essentially determined by the repulsive
direction, i.~e. the RG relevant direction, of the fixed point~${\mathcal F}_2$. Thus, our two-dimensional flow is effectively reduced
to a one-dimensional flow for~$k<k_0$ and we are left with the situation that we have discussed above in the single-channel
approximation. Consequently, we find the following scaling behavior of physical observables:
\be
{\mathcal O}\sim k_{\rm SB}^{d_{\mathcal O}} \sim k_0^{d_{\mathcal O}} 
{\left|\vec{\lambda}_{0}\right|^{-\frac{d_{\mathcal O}}{|\Theta|}}} {\left|\vec{\lambda}_{0} -\vec{\lambda}_{\ast} \right|^{\frac{d_{\mathcal O}}{|\Theta|}}}
\,, \label{eq:kcrgen}
\ee
where $\vec{\lambda}_{0}^{\,\rm T}:=(\lambda_{\sigma}(k_0),\lambda_{\rm V}(k_0))$ 
and~$\vec{\lambda}_{\ast}$ is determined by the coordinates of the fixed point~${\mathcal F}_2$. 
The exponent~$\Theta$ is given by the
critical exponent associated with the RG relevant direction of~${\mathcal F}_2$. Thus, the 
scale~$k_0$ now plays the role of the UV scale~$\Lambda$ in our single-channel approximation.

From the scaling relation~\eqref{eq:kcrgen}, we deduce that the scaling behavior at the quantum critical point
is still universal and fully determined by the critical exponent~$\Theta$.
Thus, the quantum critical behavior does {\it not} dependent on the initial conditions,
but only on the symmetries of the theory. The latter restrict the number of allowed interactions in the underlying action.
However, the absolute values of physical observables depend on the scale~$k_0$ and the 
distance~$|\vec{\lambda}_{0}-\vec{\lambda}_{\ast}|$. Both depend strongly on the choice of the initial 
conditions for~$k\to\Lambda$. The initial conditions are therefore of utmost importance for phenomenological 
applications.

Our discussion is obviously not restricted to initial conditions in the domain~$\rm Ib$ of the phase diagram
in Fig.~\ref{fig:pd_njl}. It can be easily adapted
to initial conditions in the domains $\rm Ia$, $\rm IIIa$ and~$\rm IIIb$. Of course, the present analysis of scaling
can be readily generalized to theories in $d$~dimensions with more than two interaction terms and symmetries
different from those of this model.

\subsubsection{Essential Scaling}\label{sec:ESNJL}
Next we discuss a very special behavior of a theory close to a quantum critical point, namely {\it essential} scaling.
This type of scaling plays a crucial role in our study of gauge theories in Sect.~\ref{sec:gaugetheories}. In the context
of gauge theories, such a scaling behavior is often referred to as 
Miransky scaling~\cite{Miransky:1988gk,Miransky:1996pd}. It has also been found 
in the context of specific 2-dimensional condensed-matter systems where it is known as
Berezinskii-Kosterlitz-Thouless (BKT) scaling~\cite{Berezinskii,Berezinskii2,Kosterlitz:1973xp}.

For this discussion we shall consider an extension of the presently employed NJL model.
To be specific, we shall assume that an extension of this model exists such that the RG flow equation for
the $\lambda_{\sigma}$-coupling remains unchanged, but that the one for the $\lambda_{\rm V}$-coupling receives
contributions from the extension of the model. These additional contributions will give rise to a
non-trivial, i.~e. interacting, IR fixed point for the $\lambda_{\rm V}$-coupling, even in the limit $\lambda_{\sigma}\to 0$.
The fixed-point value~$\lambda_{\rm V}^{\ast}$ can then be controlled by the additional parameters of the extended model.
This still describes a very general setup and may therefore apply to a large variety of theories. 
In particular, gauge theories in the limit of many fermion flavors are to some extent reminiscent of this setup. This will be discussed in 
Sect.~\ref{sec:gaugetheories}. There, the gauge coupling associated with the gauge vector bosons plays the 
role of the vector-channel interaction in our study.\footnote{The minimal coupling of the fermions and the gauge vector bosons
is given by a term~$\sim\bar{\psi} \gamma_{\mu}A_{\mu} \,\psi$. This type of interaction 
corresponds to the Yukawa-like interaction term~$\bar{\psi} \gamma_{\mu}V_{\mu}\,\psi$ which is present in the partially bosonized version of our 
simple NJL model, $\lambda_{\rm V} \sim h_{\rm V}^2$.}

In the following we ignore the running of the vector-like coupling and consider it as a {\it scale-independent} ``external"
parameter. The RG flow of this coupling is then governed by
\be
\partial _t \lambda_{\rm V}\equiv 0\,.\label{eq:lambdaVFP}
\ee
This corresponds to a case in which the $\lambda_{\rm V}$-coupling assumes a finite IR fixed point value~$\lambda_{\rm V}^{\ast}$.
In the vicinity of this fixed point, Eq.~\eqref{eq:lambdaVFP} may still be a reasonable approximation.\footnote{Note 
that~$\lambda_{\rm V}^{\ast}$ may depend on
other control parameters of the theory, such as the number of fermion flavors or an external magnetic field.}
The flow equation for the $\lambda_{\sigma}$-coupling reads
\be
\partial_{t}{\lambda}_{\sigma}= 2\lambda_{\sigma}- 
8v_4\,  l_{1}^{\rm (F),(4)}(0;0) \left[ \lambda_{\sigma}^2 + 4\lambda_{\sigma}\lambda_{\rm V}+3\lambda_{\rm V}^2
\right]\,,\label{eq:BKTsigma}
\ee
where~$v_4=1/(32\pi^2)$ and we have set $\eta_{\psi}\equiv 0$ for the sake of simplicity.
As illustrated in Fig.~\ref{fig:parabola2}, a variation of~$\lambda_{\rm V}$ allows to shift the fixed points of 
the $\lambda_{\sigma}$-coupling. In particular, a finite value of~$\lambda_{\rm V}$ turns 
the Gau\ss ian fixed point into an interacting fixed point. Moreover, we observe that a critical value 
for~$\lambda_{\rm V}$ exists for which the two fixed points of the scalar-pseudoscalar channel merge.
For~$\lambda_{\rm V}>\lambda_{\rm V}^{\rm crit.}$, the fixed points then annihilate each other and the RG flow
 is no longer governed by any (finite) real-valued fixed point. 
 From Eq.~\eqref{eq:BKTsigma} we find the following value for the critical coupling strength:
\be
\lambda_{\rm V}^{\rm crit.}=\frac{(2-\sqrt{3})}{8v_4 l_{1}^{\rm (F),(4)}(0;0)}=8\pi^2(2-\sqrt{3})\,.\label{eq:lvcritsec342}
\ee
Here, we have used an optimized regulator function to evaluate this expression; for this regulator, one 
finds~$l_1^{\rm (F),(4)}(0;0)=\frac{1}{2}$, see App.~\ref{app:covregfcts}.
Note that $\lambda_{\rm V}^{\rm crit.}$ is defined as the value of $\lambda_{\rm V}$ for which the 
right-hand side of Eq.~\eqref{eq:BKTsigma} has exactly one zero. 
Strictly speaking, there exist two solutions for~$\lambda_{{\rm V}}^{\rm crit.}$: \mbox{$0<\lambda_{{\rm V},1}^{\rm crit.}<\lambda_{{\rm V},2}^{\rm crit.}$}.
In the following we assume that we increase~$\lambda_{\rm V}$ by hand starting from~$\lambda_{\rm V}=0$. Therefore we can
exclude~$\lambda_{{\rm V},2}^{\rm crit.}$ from our considerations and set~$\lambda_{{\rm V}}^{\rm crit.}=\lambda_{{\rm V},1}^{\rm crit.}$,
where~$\lambda_{{\rm V},1}^{\rm crit.}$ is identical to the right-hand side of Eq.~\eqref{eq:lvcritsec342}.

For~$\lambda_{\rm V}>\lambda_{\rm V}^{\rm crit.}$, the $\lambda_{\sigma}$-coupling becomes a relevant
operator and increases rapidly towards the IR indicating the onset of chiral symmetry, even if
we choose~$\lambda_{\sigma}=0$ as initial condition at~$k=\Lambda$.
Thus, $\lambda_{\sigma}$ necessarily diverges at a finite RG scale~$k_{\rm SB}$ 
for~$\lambda_{\rm V}>\lambda_{\rm V}^{\rm crit.}$. 
Again, this scale sets the scale for chiral low-energy observables~$\mathcal O$.

In order to find the scaling behavior of the symmetry breaking scale~$k_{\rm SB}$ when~$\lambda_{\rm V}$
is varied by hand as a constant ``external" parameter, we have to solve the RG flow equation of the 
coupling~$\lambda_{\sigma}$. We find
\be
\ln k -\ln\Lambda=-\frac{2\arctan\left(\frac{2\lambda_{\rm V} -8\pi^2 + \lambda^{\prime}_{\sigma}}
{4\pi^2\delta(\lambda_{\rm V})}\right)}{\delta(\lambda_{\rm V})} 
\,\Bigg|^{\lambda_{\sigma}}_{\lambda_{\sigma}^{\rm UV}}\,,
\label{eq:mlambda_sol_NJL}
\ee
where we have again used an optimized regulator function for convenience. The
function~$\delta(\lambda_{\rm V})$ is given by
\be
\delta (\lambda_{\rm V})=\frac{1}{4\pi^2}\sqrt{3\lambda_{\rm V}^2 - 4(4\pi^2 - \lambda_{\rm V})^2}\,.
\ee
We shall only consider values of~$\lambda_{\rm V}$ such that~$\delta(\lambda_{\rm V})$ is real-valued, e.~g.~$\lambda_{\rm V}\geq \lambda_{{\rm V}}^{\rm crit.}$.
From Eq.~\eqref{eq:mlambda_sol_NJL}, we obtain $k_{\rm SB}$ by solving for the zero 
of $1/\lambda_{\sigma}(k)$, i. e. $1/\lambda_{\sigma}(k_{\rm SB})=0$:
\be
\ln k_{\rm SB} -\ln\Lambda = -\frac{\pi}{\delta(\lambda_{\rm V})} + {\rm const.}\,.\label{eq:m_general_NJL}
\ee
Here, we have chosen the initial conditions such that $\lambda_{\sigma}^{\rm UV}=\lambda_{\sigma}^{\rm max}$ where
$\lambda_{\sigma}^{\rm max}$ denotes the position of the maximum of the
$\beta_{\lambda_{\sigma}}$-function~\eqref{eq:BKTsigma}, i.~e., the peak of the parabola in
Fig.~\ref{fig:parabola2}. An expansion of Eq.~\eqref{eq:m_general_NJL} around $\lambda_{\rm V}^{\rm crit.}$ 
yields
\be
\ksb \propto \Lambda \theta(\lambda_{\rm V}-\lambda_{\rm V}^{\rm crit.}) 
\exp\left( {-\frac{\pi}{2\epsilon\sqrt{\lambda_{\rm V} - \lambda_{\rm V}^{\rm crit.}}  }}
\right)\,,
\label{eq:miransky_NJL}
\ee
where $\epsilon$ is simply a numerical factor,
\be
\epsilon =\frac{\sqrt[4]{3}}{2\pi}
\,,
\label{eq:defeps_NJL}
\ee
which in general depends on the details of the theory under consideration and 
is scheme dependent. In any case, the exponential (essential) scaling behavior of $k_{\rm SB}$ is {\it universal}
for $\lambda_{\rm V}$ close to $\lambda_{\rm V}^{\rm crit.}$.  Since the dynamically generated scale $k_{\rm SB}$
sets the scale for the low-energy sector, we expect that all IR observables
$\mathcal O$ with canonical mass dimension~$d_{\mathcal O}$ scale according to
\be
{\mathcal O}\sim k_{\rm SB}^{d_{\mathcal O}}\,.
\ee

A few comments on the history of the scaling law Eq.~\eqref{eq:miransky_NJL} are in order at this point.
In the context of QCD, the scaling law in Eq.~\eqref{eq:miransky_NJL} has been
first derived by Miransky~\cite{Miransky:1988gk,Miransky:1996pd}, but it has also
been found in the context of specific 2-dimensional condensed-matter
systems~\cite{Berezinskii,Berezinskii2,Kosterlitz:1973xp}. A way to derive the scaling
law~\eqref{eq:miransky_NJL} via an analysis of the RG flow of four-fermion operators has been recently
pointed out by Kaplan, Lee, Son and Stephanov~\cite{Kaplan:2009kr}.

The scaling behavior~\eqref{eq:miransky_NJL} is clearly distinguishable from the power-law 
behavior~\eqref{eq:lambdacr}. In particular, we observe that essential scaling is not governed
by the critical exponent~$\Theta$ associated with the IR repulsive direction at the fixed point~$\lambda_{\sigma}^{\ast}$.
To find exponential (essential) scaling behavior, we have 
assumed that the initial condition for the
four-fermion coupling $\lambda_{\sigma}$ at the UV scale~$\Lambda$ has been 
chosen such that $\lambda_{\sigma}^{\rm UV}$ is
smaller than the value of the IR repulsive fixed point~$\lambda_{\sigma}^{\ast}$, see
Fig.~\ref{fig:parabola2}.  We therefore stress that essential scaling behavior can only be
observed when~$\lambda_{\sigma}^{\rm UV}$ is chosen to be smaller than the 
corresponding value of the IR repulsive fixed point for a given value of the constant ``external" 
parameter $\lambda_{\rm V}$. Otherwise we expect power-law-like scaling behavior as discussed
above, see Eq.~\eqref{eq:lambdacr}. In our discussion of quantum critical behavior in 
gauge theories in Sect.~\ref{sec:gaugetheories},
we shall discuss a new kind of scaling behavior which can be considered as a 
superposition of power-law scaling and essential scaling.  In the present toy-model setup, 
this new type of scaling behavior emerges when we allow for an RG running of the~$\lambda_{\rm V}$-coupling.
\subsection{Deformations of Fermionic Theories}\label{subsec:DefTheories}
Deformations of a theory play a crucial role in physics. For example, studies
of fermionic models at finite temperature provide us with insights into the mechanisms of 
symmetry breaking as we expect them to have been at work in the early stage of the universe. Further
examples can be found in the context of condensed-matter physics:
Apparently, comprehension of so-called high-$T_{\rm c}$ superconductors requires an understanding of the
finite-temperature dynamics of strongly interacting fermions, see e.~g. Ref.~\cite{Honerkamp5}. Other important deformations are
given by the inclusion of terms which break explicitly the underlying symmetries. For example, explicit mass
terms for the fermions break the chiral symmetry. From the point of view of a Ginzburg-Landau effective potential, such terms
effectively play the role that an external magnetic field plays for a ferromagnet, e.~g. an Ising model. In the
theory of the strong interaction, such terms are of particular importance since they determine the so-called
current quark masses. Also, studies of fermionic theories in a finite volume are of interest and represent
a valuable deformation. Depending on the theory under consideration, the study 
of finite-size effects may help to make better contact between theoretical results and experimental
data, see e.~g. Ref.~\cite{Ku:2008vk}. On the other hand, finite-size effects are important from a field-theoretical
point of view. For example, Monte-Carlo simulations are necessarily performed in a finite volume. For the 
extrapolation to the continuum limit a rigorous understanding of the scaling behavior of a system with its volume
size is certainly required, see e.~g. Refs.~\cite{Braun:2004yk,Braun:2005fj,Braun:2005gy,Braun:2008sg,Braun:2010vd,Braun:2010pp} 
for RG studies of finite-size effects relevant for Monte-Carlo simulations. In any case, we start our
discussion of deformations with a first analysis of theories with many fermion flavors. To this end, we consider
two distinct extensions of our NJL model with one fermion species. In Sect.~\ref{subsec:NJLSUNf} we study
a NJL model with $\Nf$~fermions and a chiral ${\rm SU}(\Nf)_{\rm L} \otimes {\rm SU}(\Nf)_{\rm R}$ symmetry. An
extension with a chiral ${\rm U}(1)^{\otimes \Nf}$ symmetry is then discussed in Sect.~\ref{subsec:NJLUNf}. 
These studies of many-flavor physics should be considered as a warm-up for 
Sects.~\ref{sec:njlgn} and~\ref{sec:gaugetheories}.

\subsubsection{Many-Flavor Physics I: Chiral ${\rm SU}(\Nf)_{\rm L} \otimes {\rm SU}(\Nf)_{\rm R}$ Symmetry}\label{subsec:NJLSUNf}
For our first study of many-flavor physics we employ a NJL model with a (chiral) ${\rm SU}(\Nf)_{\rm L} \otimes {\rm SU}(\Nf)_{\rm R}$ symmetry.
We will encounter this type of chiral symmetry again when we discuss QCD and, in particular, when we analyze chiral symmetry
breaking in strongly-flavored gauge theories in Sect.~\ref{sec:gaugetheories}. QCD is in fact symmetric under 
${\rm SU}(\Nf)_{\rm L} \otimes {\rm SU}(\Nf)_{\rm R}$ transformations in the limit of massless quarks. Apart from its relevance 
for QCD phenomenology, this type of flavor symmetry appears in studies of chiral symmetry breaking in quantum gravity~\cite{Eichhorn:2011pc}.

In direct analogy to the simple NJLmodel with one fermion species,  the chiral charge~$Q_5$,
\be
Q_5^{a}=\int d^3x \bar{\psi}\gamma_0\gamma_5 T^{a} \psi\,,
\ee
does not commute with the composite 
fields~$\sim (\bar{\psi}\I\gamma_0\gamma_5 T^a\psi)\equiv\bar{\psi}_c\I\gamma_0\gamma_5 T^a_{cd}\psi_d$, 
if the vacuum expectation value~$\langle \bar{\psi}\psi\rangle$ is finite. In other words, the chiral symmetry
of such a model is broken dynamically when~$\langle \bar{\psi}\psi\rangle$ assumes a finite value.
Here, the $T^a$ denote the $(\Nf^2-1)$ generators of the group~${\rm SU}(\Nf)$ in the fundamental representation. 
For ${\rm SU}(2)$ these generators are related to the 
Pauli matrices, whereas they are related to the Gell-Mann matrices for~${\rm SU}(3)$. 
For a finite chiral condensate~$\langle \bar{\psi}\psi\rangle$ it then follows 
from the {\it Nambu-Goldstone} theorem that $(\Nf^2-1)$ massless states exist in the spectrum of the theory,
the {\it Nambu-Goldstone} bosons. In QCD, this yields a natural explanation for the existence of three
light (massless) hadrons, namely the pions, in the hadronic spectrum in the theory with two light (massless)
quark flavors.

We are again interested in an analysis of the fixed-point structure of the theory. To this end, we 
consider the effective action in leading order in the derivative expansion:
\be
\label{eq:NJLtruncBasicSUN}
\Gamma \left[\bar{\psi},\psi\right]=\int d^4 x
\left\{Z_{\psi}\bar{\psi} \I\fslash{\partial}\psi
+\frac{1}{2}\bar{\lambda}_{-}({\text{V--A}})  
+\frac{1}{2}\bar{\lambda} _{+}({\text{V+A}})
 \right\}\,,
\ee
where
\be
({\text{V--A}})&=&[(\bar{\psi}\gamma_{\mu}\psi)^{2} + (\bar{\psi}\gamma_{\mu}\gamma_{5}\psi)^2]\,,\nn\\
({\text{V+A}})&=&[(\bar{\psi}\gamma_{\mu}\psi)^{2} - (\bar{\psi}\gamma_{\mu}\gamma_{5}\psi)^2] \,.\nn
\ee
The flavor indices are contracted pairwise, $(\bar{\psi}{\mathcal O}\psi)\equiv (\bar{\psi}_i {\mathcal O}\psi_i)$.
This ansatz corresponds to the matter part employed in a study of chiral symmetry breaking
in quantum gravity~\cite{Eichhorn:2011pc}.

The effective action~\eqref{eq:NJLtruncBasicSUN} is 
complete with respect to Fierz transformations, i.~e. all other  ${\rm SU}(\Nf)_{\rm L} \otimes {\rm SU}(\Nf)_{\rm R}$
symmetric four-fermion interactions can be transformed into a linear combination of the terms included in our ansatz. Note that
\be
[(\bar{\psi}\psi)^{2}-(\bar{\psi}\gamma_{5}\psi)^2]\nn
\ee
is {\it not} invariant under ${\rm SU}(\Nf)_{\rm L} \otimes {\rm SU}(\Nf)_{\rm R}$ transformations. However,
\be
[(\bar{\psi}_i\psi_j)(\bar{\psi}_j\psi_i)-(\bar{\psi}_i\gamma_{5}\psi_j)(\bar{\psi}_j\gamma_{5}\psi_i)]\nn
\ee
is invariant, as can be seen from the following Fierz relation:
\be
({\text{V+A}}) \equiv [(\bar{\psi}\gamma_{\mu}\psi)^{2} - (\bar{\psi}\gamma_{\mu}\gamma_{5}\psi)^2] 
=-2[(\bar{\psi}_i\psi_j)(\bar{\psi}_j\psi_i)-(\bar{\psi}_i\gamma_{5}\psi_j)(\bar{\psi}_j\gamma_{5}\psi_i)]\,.\label{eq:expFierz2}
\ee
Of course, the effective action can be directly related to the action~\eqref{eq:NJLtruncBasic3} of our 
simple one-flavor model. Using the Fierz identity~\eqref{eq:expFierz2} , we find for $\Nf=1$ that
\be
\bar{\lambda}_{\sigma}=2(\bar{\lambda}_{-}-\bar{\lambda}_{+})\,\quad\text{and}\quad 
\bar{\lambda}_{\rm V}=-2\bar{\lambda}_{-}\,.\label{eq:SUNsimpleDic}
\ee

From the point of view of high-energy phenomenology it is instructive to relate the interaction
channels in Eq.~\eqref{eq:NJLtruncBasicSUN} to the interaction channel conventionally employed in 
so-called quark-meson models, namely the scalar-pseudoscalar channel, see Sect.~\ref{sec:NJLLQCD}. From the Fierz identities in
App.~\ref{sec:dirac} it follows that 
\be
({\text{V--A}}) &=&
\left[(\bar{\psi}_i\gamma_{\mu}\psi_j)(\bar{\psi}_j\gamma_{\mu}\psi_i) 
+ (\bar{\psi}_i \gamma_{\mu}\gamma_{5}\psi_j)(\bar{\psi}_j\gamma_{\mu}\gamma_{5}\psi_i)\right] \,
\label{eq:VmASUN}
\ee
and
\be
({\text{V+A}})
&=& -2 \left[(\bar{\psi}_i\psi_j)(\bar{\psi}_j\psi_i)-(\bar{\psi}_i\gamma_{5}\psi_j)(\bar{\psi}_j\gamma_{5}\psi_i)\right]\,\nn\\
&=& -4  [(\bar{\psi}T^{\alpha}\psi)^{2}-(\bar{\psi}\gamma_{5}T^{\alpha}\psi)^2] 
- \left(\frac{2}{\Nf}-1\right)[(\bar{\psi}\psi)^{2}-(\bar{\psi}\gamma_{5}\psi)^2] \nn\\
&\stackrel{\left(\Nf =2\right)}{=}& - [(\bar{\psi}\psi)^{2}- (\bar{\psi}\gamma_{5}\tau^{a}\psi)^2] - \dots\,, \label{eq:SUNsps}
\ee
with~$\tau^{a}$ being the Pauli matrices.
Here, we have dropped terms on the right-hand side of the last line in Eq.~\eqref{eq:SUNsps}.
However, we have included all terms in our flow equation study. This is most conveniently achieved
by working with the (V+A)-channel and (V--A)-channel. Note that $\alpha = 0,1,\dots,(\Nf^2 -1)$ and $a=1,2,\dots,(\Nf^2 -1)$. 
To derive Eq.~\eqref{eq:SUNsps}, we have used the Fierz transformation for flavor indices:\footnote{This identity can in principle be 
obtained from the general relation~\eqref{eq:genFierz}. The 
complete set of basis elements is now given by~$\{\frac{1}{2}\mathbbm{1},T^{1},T^{2},\dots\}$.} 
\be
\mathbbm{1}_{ab}\mathbbm{1}_{cd}
=2\left(\frac{2}{\Nf}-1\right)\mathbbm{1}_{ad}\mathbbm{1}_{cb}
+2\sum_{\alpha=0}^{\Nf^2-1}\left( T^{\alpha}_{ad}  \right) \left( T^{\alpha}_{cb}  \right)\,, \label{eq:flavorFT}
\ee
where $T^{0}:=(1/2)\mathbbm{1}$. 
We observe that the first term on the right-hand side of the last line in Eq.~\eqref{eq:SUNsps} corresponds
to the standard scalar-pseudoscalar channel included in low-energy models for 2-flavor 
QCD, see Sect.~\ref{sec:NJLLQCD} for a detailed discussion.  

Let us now discuss the RG flow equations and the associated fixed-point structure.
In our point-like approximation, it follows immediately that~$\eta_{\psi}=0$. Thus, we set $Z_{\psi}\equiv 1$ from now on. 
The flow equations for the two four-fermion couplings specified in the action~\eqref{subsec:NJLSUNf} read
\be
\partial_t \lambda_{-} &=& 2\lambda_{-} + 8 v_4 l_{1}^{\rm (F),(4)}(0;0)\left[ (\Nf-1)\lambda_{-}^2 + \Nf \lambda_{+}^2\right]
  \,,\\
\partial_t \lambda_{+} &=& 2\lambda_{+} +  8 v_4 l_{1}^{\rm (F),(4)}(0;0)\left[ 2\lambda_{+}^2 + 2\lambda_{+}\lambda_{-}(\Nf+1)\right]\,,
 \,
\ee
where $\lambda_{i}=Z_{\psi}^{-2}k^2\bar{\lambda}_i$. From this set of equations it is apparent that we only have three 
(finite) fixed points for~$\Nf=1$. For arbitrary $\Nf>1$, we find the following four fixed 
points~${\mathcal F}_i^{{\rm SU(\Nf)}}=(\lambda_{+}^{\ast},\lambda_{-}^{\ast})$:
\be
&&{\mathcal F}_1^{{\rm SU(\Nf)}}\equiv {\mathcal F}_{\rm Gau\ss }=(0,0)\,,\quad 
{\mathcal F}_2^{{\rm SU(\Nf)}}=\left(0,-\frac{8\zeta}{\Nf-1}\right)\,,\quad\nn\\
&&{\mathcal F}_3^{{\rm SU(\Nf)}}=\left(\frac{8\zeta}{2\Nf\!-\! 1}, -\frac{8\zeta}{2\Nf\!-\! 1} \right)\,,\quad
\!{\mathcal F}_4^{{\rm SU(\Nf)}}=\!\left(-\frac{8\zeta (\Nf+3)}{9\!+\! 5\Nf\! +\! 2\Nf^2},-\frac{8\zeta\Nf}{9\!+\! 5\Nf\! +\! 2\Nf^2}\right)\,,\nn
\ee
where $\zeta$ is defined in Eq.~\eqref{eq:zetadef}. From the stability matrix we obtain the critical exponents which allow
us to classify these fixed points. We find that the Gau\ss ian fixed point has two infrared-attractive directions. The second
fixed point has two infrared-repulsive directions whereas the third and the fourth fixed point have an infrared-attractive
as well as an infrared repulsive direction.
Using the relations given in Eq.~\eqref{eq:SUNsimpleDic}, the fixed-point values in the limit~$\Nf\to 1$ can be translated into
those of the NJL model with only one fermion species, see Eq.~\eqref{eq:allchannelFP}. This confirms that
the "missing" fourth fixed point in our one-flavor study is pushed to $(\lambda_{\sigma}\to -\infty,\lambda_{\rm V}\to\infty)$ 
for $\Nf\to 1$ from above, as indicated in Fig.~\ref{fig:pd_njl}.

Finally we would like to analyze this model in the limit of many flavors. In leading order in an expansion in powers
of $1/\Nf$ we find a Gau\ss ian fixed point~${\mathcal F}_1^{{\rm SU(\Nf)}}=(0,0)$ as well as three non-Gau\ss ian fixed points:
\be
{\mathcal F}_2^{{\rm SU(\Nf)}}=\left(0,-\frac{8\zeta}{\Nf}\right)\,,\quad 
{\mathcal F}_3^{{\rm SU(\Nf)}}=\left(\frac{4\zeta}{\Nf}, -\frac{4\zeta}{\Nf} \right)\,,\quad
{\mathcal F}_4^{{\rm SU(\Nf)}}=\left(-\frac{4\zeta}{\Nf}, -\frac{4\zeta}{\Nf} \right)\,.\nn
\ee
Note that the rescaled fixed-point couplings~$\Nf\cdot {\mathcal F}_i^{{\rm SU(\Nf)}}$ approach constant values
in the limit~$\Nf\to\infty$. The associated critical exponents in leading order in $1/\Nf$ are given by
\be
&&\Theta_1^{{\rm SU(\Nf)}}=\{-2,2\}\,,\quad
\Theta_2^{{\rm SU(\Nf)}}=\left\{2,2+\frac{8}{\Nf}\right\}\,,\nn\\ 
&&\Theta_3^{{\rm SU(\Nf)}}=\left\{2,-2-\frac{4}{\Nf}\right\}\,,\quad
\Theta_4^{{\rm SU(\Nf)}}=\left\{2,-2-\frac{4}{\Nf}\right\}\,.\nn
\ee
Thus, the critical behavior of such a theory at the quantum phase transition  
is modified when we vary the number of fermion flavors. Our analysis for $\Nf>1$ 
shows that an RG trajectory exists with \mbox{$\lambda_{+}\equiv 0$} which connects the fixed 
point~${\mathcal F}_2^{{\rm SU(\Nf)}}$ with the Gau\ss ian fixed point in the IR limit. However,
the interaction channel associated with the $\lambda_{-}$-coupling cannot be transformed into the 
phenomenologically relevant scalar-pseudoscalar channel, see Eqs.~\eqref{eq:VmASUN} and~\eqref{eq:SUNsps}.
On the contrary, a Fierz-complete analysis of the Thirring model in three dimensions provides evidence
that a pure Thirring-like trajectory indeed exists in the limit of many fermion flavors~\cite{Gies:2010st}. In any case,
these observations show that a careful analysis of fermionic theories in the large-$\Nf$ limit can provide useful information
for a controlled and systematic construction of effective theories for physically relevant systems.

\subsubsection{Many-Flavor Physics II: Chiral ${\rm U}(1)^{\otimes \Nf}$ Symmetry}\label{subsec:NJLUNf}

As a second example for fermionic models with many flavors we 
consider an NJL model with a continuous chiral ${\rm U}(1)^{\otimes \Nf}\equiv\otimes_{i=1}^{\Nf}{\rm U}(1)$ 
symmetry, i.~e. each flavor transforms independently under the chiral transformation given in Eq.~\eqref{eq:chiraltrafo}.
As a consequence, there is only one massless excitation in the spectrum for each (spontaneously broken)~U($1$) symmetry
rather than $(\Nf^2-1)$ massless excitations as in the case of a spontaneously broken 
chiral~${\rm SU}(\Nf)_{\rm L} \otimes {\rm SU}(\Nf)_{\rm R}$ symmetry. 
Thus, the number of fermions can in principle be arbitrarily increased
without changing the number of the Nambu-Goldstone bosons.
As we shall discuss in Sect.~\ref{sec:njlgn}, this chiral symmetry
is closely related to the one realized in the Gross-Neveu model with many flavors. Moreover, such a
flavor number dependence is faintly reminiscent of the dependence of QCD on the number
of colors. In fact, increasing the number of colors increases the number of fermions linearly but 
leaves the number of Nambu-Goldstone bosons unchanged. In Sect.~\ref{sec:gaugetheories}
we shall come back to QCD in the limit of many colors.

To be specific, let us consider the following ansatz for the effective action:
\be
\label{eq:NJLtruncBasicU1N}
\Gamma \left[\bar{\psi},\psi\right]&=&\int d^4 x
\left\{Z_{\psi}\bar{\psi} \I\fslash{\partial}\psi
+\frac{1}{2}\bar{\lambda}_{\sigma}[(\bar{\psi}\psi)^{2}-(\bar{\psi}\gamma_{5}\psi)^2]\right.\nn\\
& &\qquad\qquad\qquad\qquad\quad -\frac{1}{2}\bar{\lambda} _{\rm V}[(\bar{\psi}\gamma_{\mu}\psi)^2]
-\frac{1}{2}\bar{\lambda} _{\rm A}[(\bar{\psi}\gamma_ {\mu}\gamma_{5}\psi)^{2}]\bigg\}\,.
\ee
This ansatz has a continuous chiral ${\rm U}(1)^{\otimes \Nf}$ symmetry. It is 
even invariant under~U($\Nf$) flavor-transforma\-tions. However, it does not form a complete basis
of (point-like) four-fermion interactions with respect to the U($\Nf$) flavor-symmetry. For the latter, it is possible to show that 
the corresponding complete basis includes six distinct four-fermion interaction channels~\cite{HGPD}.
Nonetheless, the ansatz~\eqref{eq:NJLtruncBasicU1N} can be considered to be closed for our 
purposes in this section. By this, we mean that the RG flow of this 
model only generates four-fermion interactions which are included in the ansatz~\eqref{eq:NJLtruncBasicU1N}.
In other words, the ansatz~\eqref{eq:NJLtruncBasicU1N} forms an invariant subspace of operators with respect to
continuous chiral ${\rm U}(1)^{\otimes \Nf}$ transformations.

The RG flow equations for this model can be derived along the lines of Sect.~\ref{subsec:simpleex}. Since we restrict oureselves
to the point-like limit, the anomalous dimension vanishes, $\eta_{\psi}\equiv 0$, and we set~$Z_{\psi}\equiv 1$.
The flow equations for the three four-fermion couplings read:
\be
\partial _t \lambda_{\sigma} &=& 2\lambda_{\sigma} - 16 v_4 l_{1}^{\rm (F),(4)}(0;0)\left[ \Nf \lambda_{\sigma}^2 + 
2\lambda_{\sigma}\lambda_{\rm V}-\lambda_{\sigma}\lambda_{\rm A}\right]\,,\\
\partial _t \lambda_{\rm V} &=& 2\lambda_{\rm V} + 8 v_4 l_{1}^{\rm (F),(4)}(0;0)\left[ 
2\lambda_{\rm V}\lambda_{\rm A}-\lambda_{\sigma}\lambda_{\rm V} - (\Nf +1)\lambda_{\rm V}^2 \right]\,,\\
\partial _t \lambda_{\rm A} &=& 2\lambda_{\rm A} + 4 v_4 l_{1}^{\rm (F),(4)}(0;0)\left[ 3 \lambda_{\rm V}^2 
-\lambda_{\sigma}^2 + (1-2\Nf)\lambda_{\rm A}^2 \right. \nn\\
&& \qquad\qquad\qquad\qquad\qquad\qquad\qquad\qquad
\left. -2\lambda_{\rm V}\lambda_{\rm A} + 2\lambda_{\sigma}\lambda_{\rm A}
\right]\,.
\ee
As discussed in Sect.~\ref{subsec:simpleex}, for~$\Nf=1$
only two of the three couplings are independent, see also Eq.~\eqref{eq:expFierz}. Using 
\be
\lambda_{\sigma} \to \lambda_{\sigma}+2\lambda_{\rm A}
\quad\text{and}\quad
\lambda_{\rm V} \to \lambda_{\rm V} -\lambda_{\rm A}\,
\ee
in the above flow equations, we indeed recover the flow equations~\eqref{eq:NJLsig2ch} and~\eqref{eq:NJLv2ch} 
of the NJL model with one fermion species.

The fixed points and the associated critical exponents of our ${\rm U}(1)^{\otimes \Nf}$-symmetric NJL model can be
computed straightforwardly. We do not give the explicit expressions here, but we find
eight distinct fixed points for~$\Nf>1$. In leading order in an expansion in powers of~$1/\Nf$
the expressions simplify significantly. To be specific, we find the following eight
points~${\mathcal F}_i^{{\rm U(1)}}=(\lambda_{\sigma}^{\ast},\lambda_{\rm V}^{\ast},\lambda_{\rm A}^{\ast})$:
\be
\!\!\!\!\!\!\!\!\! &&
{\mathcal F}_1^{{\rm U(1)}}\!=\! (0,0,0)\,,\;
{\mathcal F}_2^{{\rm U(1)}}\! =\! \left( \frac{4\zeta}{\Nf},\frac{8\zeta}{\Nf},\frac{8\zeta}{\Nf}\right)\,,\;\nn               
{\mathcal F}_3^{{\rm U(1)}}\!=\! \left( \frac{4\zeta}{\Nf},\frac{8\zeta}{\Nf},0\right)
\,,\;
{\mathcal F}_4^{{\rm U(1)}}\!=\!\left( \frac{4\zeta}{\Nf},0,\frac{8\zeta}{\Nf}\right)\,,\nn\\
\!\!\!\!\!\!\!\!\! &&
{\mathcal F}_5^{{\rm U(1)}}\!=\!\!\left(0,\frac{8\zeta}{\Nf},\frac{8\zeta}{\Nf}\right)\,,\;   
{\mathcal F}_6^{{\rm U(1)}}\!=\!\left(0,\frac{8\zeta}{\Nf},0\right)\,,\; 
{\mathcal F}_7^{{\rm U(1)}}\!=\!\left(0,0,\frac{8\zeta}{\Nf}\right)
\,,\;
{\mathcal F}_8^{{\rm U(1)}}\!=\left( \frac{4\zeta}{\Nf},0,0\right)\,,\nn
\ee
where $\zeta$ is defined in Eq.~\eqref{eq:zetadef}. 

In contrast to our study of the ${\rm SU}(\Nf)_{\rm L} \otimes {\rm SU}(\Nf)_{\rm R}$ symmetric NJL model, we now
find an RG trajectory on which only the scalar-pseudoscalar interaction is non-vanishing. This trajectory 
connects the non-trivial fixed point ${\mathcal F}_8^{{\rm U(1)}}$ with the Gau\ss ian fixed point in the IR limit. This observation is 
of phenomenological interest. Generally speaking, we can find RG trajectories 
in the large-$\Nf$ limit such that condensation can only 
occur in one specific channel while the other channels remain trivial, i.~e. non-interacting. 
This type of observation 
may be used as a field-theoretical justification for models in which only one interaction channel has been taken
into account, e.~g. studies of Gross-Neveu models in the large-$\Nf$ limit. This will be discussed in more detail in 
Sect.~\ref{sec:GNmodel}.

\subsubsection{Finite Temperature}\label{subsec:DefFT}

Let us now study the dynamics of fermions at finite temperatures~$T$. To this end, we start with
an analysis of the symmetries of our Fierz-complete NJL model with one fermion species, see Eq.~\eqref{eq:NJLtruncBasic3}.

For $T>0$, the Euclidean time has a finite
extent $\beta=1/T$ and therefore Lorentz symmetry is broken explicitly.\footnote{We restrict our discussion to quantum field
theories in equilibrium. For related approaches to quantum field theories away from equilibrium we refer the reader to 
Refs.~\cite{Berges:2000ur,Aarts:2001qa,Berges:2004yj,Gasenzer:2008zz,Alkofer:2009zz}.} Thus, the time direction in our $d=4$ Euclidean
space-time is distinguished due to the presence of a heat-bath. In fact, we can introduce a velocity $n_{\mu}=(1,\vec{0})$ 
which defines the velocity of the heat-bath in the rest-frame of an observer. 
At finite temperature the most general ansatz for the effective action~$\Gamma_{\text{NJL}}$
compatible with the underlying symmetries of our model then reads\footnote{We only take into account four-fermion interactions.}
\be
\label{eq:NJLtruncBasic4}
\Gamma_{\text{NJL}}\left[\bar{\psi},\psi\right]&=&\int _0^{\beta} dx_0\int d^3 x
\left\{ Z_{\psi}^{\|} \bar{\psi} \I \gamma_0\partial_0  \psi + Z_{\psi}^{\perp}\bar{\psi} \I\gamma_i\partial_i \psi
+\frac{1}{2}\bar{\lambda}_{\sigma}[(\bar{\psi}\psi)^{2}-(\bar{\psi}\gamma_{5}\psi)^2]\right.\nn\\
& &\qquad\qquad\qquad\qquad -\frac{1}{2}\bar{\lambda}  ^{(0)}_{\rm V}[(\bar{\psi}\gamma_{0}\psi)^2]
-\frac{1}{2}\bar{\lambda} ^{\perp}_{\rm V}[(\bar{\psi}\gamma_{\mu}\psi)^2] \nn\\
&& \qquad -\frac{1}{2}\bar{\lambda} ^{(0)}_{\rm A}[(\bar{\psi}\gamma_ {0}\gamma_{5}\psi)^{2}]
-\frac{1}{2}\bar{\lambda}^{(0)} _{\rm T}[(\bar{\psi}\sigma_ {0\mu}\psi)^{2}-(\bar{\psi}\sigma_ {0\mu}\gamma_{5}\psi)^{2}]
\bigg\}\,,
\ee
where 
\be
\bar{\lambda} ^{(0)}_{\rm V}=\bar{\lambda} ^{\|}_{\rm V} -\bar{\lambda} ^{\perp}_{\rm V}\,,\quad
\bar{\lambda} ^{\perp}_{\rm V}\equiv \bar{\lambda}_{\rm V}\,,\quad
\bar{\lambda} ^{(0)}_{\rm A}=\bar{\lambda} ^{\|}_{\rm A} -\bar{\lambda} ^{\perp}_{\rm A}\,,\quad
\bar{\lambda} ^{(0)}_{\rm T}=\bar{\lambda} ^{\|}_{\rm T} -\bar{\lambda} ^{\perp}_{\rm T}\,.\nn\\
\ee
Note that both the kinetic term as well as the interaction terms in Eq.~\eqref{eq:NJLtruncBasic3} 
can be split up into contributions longitudinal~($\|$) and perpendicular ($\perp$) to the heat bath. We add that the action~\eqref{eq:NJLtruncBasic4}
can be deduced from the ansatz~\eqref{eq:NJLtruncBasic3} with the aid of the following relation:
\be
(\bar{\psi}{\mathcal O}_{\mu}\psi)^2\equiv (\bar{\psi}{\mathcal O}_{\mu}\psi)\mathbbm{1}_{\mu\nu}(\bar{\psi}{\mathcal O}_{\nu}\psi)
=(\bar{\psi}{\mathcal O}_{\mu}\psi)\left[  P_{\mu\nu}^{\|} + P_{\mu\nu}^{\perp}
\right] (\bar{\psi}{\mathcal O}_{\nu}\psi)\,,
\ee
where ${\mathcal O}_{\mu}$ stands for $\gamma_{\mu}$ and $\gamma_{\mu}\gamma_5$, respectively, and the heat-bath
projectors are defined as follows:
\be
P_{\mu\nu}^{\|} = n_{\mu}n_{\nu}\qquad\text{and}\qquad 
P_{\mu\nu}^{\perp} =  \mathbbm{1}_{\mu\nu} - P_{\mu\nu}^{\|}\,.
\ee
These projectors are orthogonal, $P^{\|} \cdot P^{\perp}=P^{\perp} \cdot P^{\|}=0$, and idempotent.
A further invariant at finite temperature emerges from the tensor-axialtensor channel
defined in Eq.~\eqref{eq:tensorchannel}. This channel is chirally symmetric but vanishes identically at zero
temperature, since 
\be
(\bar{\psi}\sigma_{\mu\nu}\psi)^2 = (\bar{\psi}\sigma_{\mu\nu}\gamma_5\psi)^2\,.\nn
\ee
However, the chirally symmetric combination
\be
[(\bar{\psi}\sigma_ {0\mu}\psi)^{2}-(\bar{\psi}\sigma_ {0\mu}\gamma_{5}\psi)^{2}]
\ee
is not identical to zero.

For convenience, we have introduced the 
couplings~$\bar{\lambda}_{\rm V}^{(0)}$, $\bar{\lambda}_{\rm A}^{(0)}$ and~$\bar{\lambda}_{\rm T}^{(0)}$
which effectively measure the difference between the couplings longitudinal and perpendicular to the heat bath and therefore allow for a simple mapping
between the couplings of the effective actions~\eqref{eq:NJLtruncBasic3} and~\eqref{eq:NJLtruncBasic4}.
To ensure comparability of our results for zero and finite temperature, we have to fix the boundary conditions 
for our couplings properly. To this end, it seems natural to fix the parameters (initial conditions) of our model at zero temperature, e.~g., by fitting them to a 
given set of low-energy observables. At finite temperature, we then use the
same set of initial conditions to compute, e.~g., the phase transition temperature. It is evident that such an approach requires $T/\Lambda\ll 1$.
Moreover, it naturally fixes the initial conditions for the new couplings~$\bar{\lambda}_{\rm V}^{(0)}$, $\bar{\lambda}_{\rm A}^{(0)}$ and~$\bar{\lambda}_{\rm T}^{(0)}$
as follows:
\be
\lim_{T/\Lambda\to 0} \bar{\lambda} ^{(0)}_{\rm V} = 
\lim_{T/\Lambda\to 0} \bar{\lambda} ^{(0)}_{\rm A} =
\lim_{T/\Lambda\to 0}\bar{\lambda} ^{(0)}_{\rm T} =0\,.
\ee

In order to study chiral symmetry restoration at finite temperature, we now have to derive the RG flow equations for the various four-fermion
couplings in our ansatz~\eqref{eq:NJLtruncBasic4}. 
To keep our discussion as simple as possible, we only take into account the $\bar{\lambda}_{\sigma}$-coupling and set all other couplings
to zero. This leaves us with the ansatz~\eqref{eq:NJLtruncBasic}. Of course, this ansatz is not complete with respect to Fierz transformations.
As in the zero-temperature case, however, the underlying mechanisms of chiral symmetry breaking can nevertheless be understood with such a simple ansatz.
The Fierz-complete study will be published elsewhere~\cite{BraunFT}.
\begin{figure}[t]
\begin{center}
\includegraphics[width=0.75\linewidth]{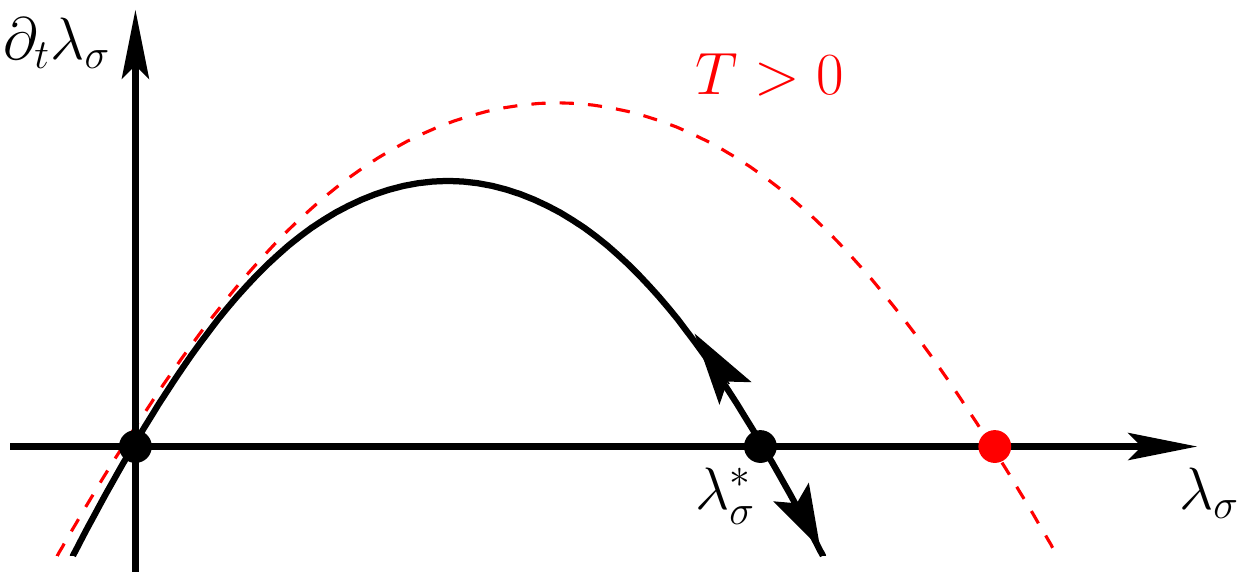}
\end{center}
\caption{Sketch of the $\beta_{\lambda_{\sigma}}$ function of the four-fermion interaction for  vanishing temperature  (black/solid line) and a given finite value of the 
temperature $T$ (red/dashed line). The arrows indicate the direction of the RG flow towards the infrared.}
\label{fig:parabolaT}
\end{figure}

The flow equation for the coupling~$\lambda_{\sigma}$ can in principle be derived along the lines of Sect.~\ref{subsec:simpleex}. Therefore
we only highlight the subtleties of the derivation. For studies of quantum field theories at finite temperature, it is convenient to
use a dimensionally reduced regulator function which only regularizes the spatial momenta but leaves the time-like momenta essentially
unconstrained~\cite{Braun:2003ii,Litim:2006ag,Blaizot:2006rj}. The (regularized) propagator matrix~${\mathcal P}_k$ 
then assumes the following form: 
\be
 {\cal P}_k= \left(\begin{array}{cc}
 0 &  -Z_{\psi}^{\|} \gamma_0^T \nu_n \! \! -Z_{\psi}^{\perp}\fss{\vec{p}}^{\,T}(1\!+\! r_{\psi}) \\
 - Z_{\psi}^{\|} \gamma_0 \nu_n \! -\! Z_{\psi}^{\perp}\fss{\vec{p}}(1\!+\! r_{\psi}) & 0 \end{array} 
 \right)\beta \delta_{n,n^{\prime}} (2\pi)^3 \delta^{(3)}(\vec{p}-\vec{p}^{\,\prime})\,,\nn
\ee
where the fermionic Matsubara frequencies are given by~$\nu_n=(2n+1)\pi T$. 

At this point we would like to add a word of caution concerning dimensionally-reduced (spatial) regulator functions, e.~g. 
Eq.~\eqref{eq:fermreg} in the appendix. 
From the propagator matrix it is apparent that such a regulator function necessarily breaks the O($d\!=\!4$) symmetry 
in the derivative terms of our truncation, i. e.~$\partial_t Z_{\psi}^{\parallel} \neq \partial_t Z_{\psi}^{\perp}$
even in the zero-temperature limit. In so-called local potential approximations,
this problem does not appear since the non-trivial momentum dependence of the propagators is 
neglected~\cite{Braun:2003ii,Schaefer:2004en,Litim:2006ag,Blaizot:2006rj,Braun:2009si}. In studies beyond the point-like limit, however, we would have to 
deal with the broken Poincare-invariance at vanishing temperature which arises due to the choice of our dimensionally reduced regulator 
function.\footnote{This issue does not occur if one applies a 4$d$ regulator function.}
In principle, one can solve this problem by taking care of the symmetry violating terms with the aid of 
the corresponding Ward identities. Equivalently, one can adjust the initial conditions for the RG flow equations 
such that one finds $Z_{\psi}^{\perp}=Z_{\psi}^{\parallel}$ for $k\to 0$ and $T\to 0$, see Ref.~\cite{Braun:2009si}.
From a field-theoretical point of view the adjustment of the initial conditions for a given truncation means nothing else than adding appropriate 
counter-terms, such that the theory remains Poincare-invariant for $k\to0$ and $T\to0$. At finite temperature, the breaking of the O($d$) symmetry in 
momentum space due to the regulator function is not problematic since this symmetry is 
broken anyway. In any case, the choice of a dimensionally reduced regulator function offers the possibility to perform the Matsubara sums in $1$PI 
diagrams analytically. This simplifies finite-temperature studies considerably and justifies our choice of such a regulator function.

Let us now discuss the flow equations at finite temperature.
In the present approximation, the fluctuation matrix~${\mathcal F}_k$ remains unchanged and is given by Eq.~\eqref{eq:simpleFmatrix}. 
To study effects of a finite temperature, it is convenient to consider the temperature~$T$ as an additional coupling and
define the corresponding dimensionless coupling~$\tau$ as
\be
\tau = \frac{T}{k}\,. 
\ee
Together with the propagator and the fluctuation matrix this yields the following set of flow equations:
\be
\beta_{\lambda_{\sigma}}\equiv \partial_{t}{\lambda}_{\sigma}&=&(2+2\eta_{\psi}^{\perp})\lambda_{\sigma} - 16 v_3\, l_{1}^{\rm (F),(4)}(\tau,0,0;\eta_{\psi},\hat{z}_{\psi})
\lambda_{\sigma}^2\,,
\label{eq:NJLbetasimpleT}\\
\partial_t \tau &=& -\tau\,.\label{eq:NJLtauflowsec353}
\ee
where $v_3=1/(8\pi^2)$. Note that the prefactor of the second term on the right-hand side of Eq.~\eqref{eq:NJLbetasimpleT} 
differs from the corresponding one in Eq.~\eqref{eq:NJLbetasimple} since
we use a dimensionally reduced regulator function; the definition of the corresponding (thermal) threshold function can be found in App.~\ref{app:regthres}.  
The (dimensionless) renormalized coupling~$\lambda_{\sigma}$ is given by
\be
\lambda_{\sigma}=(Z_{\psi}^{\perp})^{-2} k^2 \bar{\lambda}_{\sigma}\,.
\ee
In addition, we have $\eta_{\psi}^{\|,\perp}=-\partial_t \ln Z_{\psi}^{\|,\perp}$ and $\hat{z}_{\psi}=Z_{\psi}^{\|}/Z_{\psi}^{\perp}$. The flow of the
latter reads
\be
\hat{\eta}_{\psi}=-\frac{\partial_t \hat{z}_{\psi}}{\hat{z}_{\psi}}=\eta_{\psi}^{\|}-\eta_{\psi}^{\perp}\,.
\ee
Recall that at vanishing temperature the wave-function renormalizations have to satisfy the boundary condition~$\hat{z}_{\psi}=1$ for~$k\to 0$ 
to render the IR limit Poincare-invariant.

Next, we turn to a discussion of the fixed-point structure at finite temperature in the point-like limit, i.~e. $\eta_{\psi}^{\|,\perp}\equiv 0$.
Apart from a Gau\ss ian fixed point, we find a pseudo fixed-point $\lambda_{\sigma}^{\ast}(\tau)$ 
at which the right-hand side of the flow equation is zero:\footnote{At finite temperature $T$, the fixed-point value depends on the 
dimensionless temperature $\tau=T/k$. We shall refer to it as a pseudo fixed point since its value has an intrinsic dependence on the 
RG scale~$k$.}
\be
\lambda_{\sigma}^{\ast}(\tau) =\frac{1}{8v_3\, l_1^{{\rm (F)},(4)}(\tau,0,0;0,1)  } \,.
\ee
For high temperatures the fermions are screened due to the absence of a thermal zero mode. This is encoded in the 
large-$\tau$ behavior of the (fermionic) threshold function. Independent of the regularization 
scheme,\footnote{The associated perturbative Feynman diagram (diagram on the left in Fig.~\ref{fig:feynman}) 
has two internal fermion lines, yielding a factor~$\tau^{-2}$. The regulator insertion splits one of the two
internal lines apart and provides an extra factor~$\tau^{-1}$.} we have $l_1^{{\rm (F)},(4)} \sim (k/T)^{-3}$.
As an immediate consequence we find $\lambda_{\sigma}^{\ast} \sim (T/k)^3$. In other words, the value of the
pseudo fixed-point is pushed to larger values for increasing~$\tau=T/k$, see also Fig.~\ref{fig:parabolaT}. 

\begin{figure}[t]
\begin{center}
\includegraphics[width=0.65\linewidth]{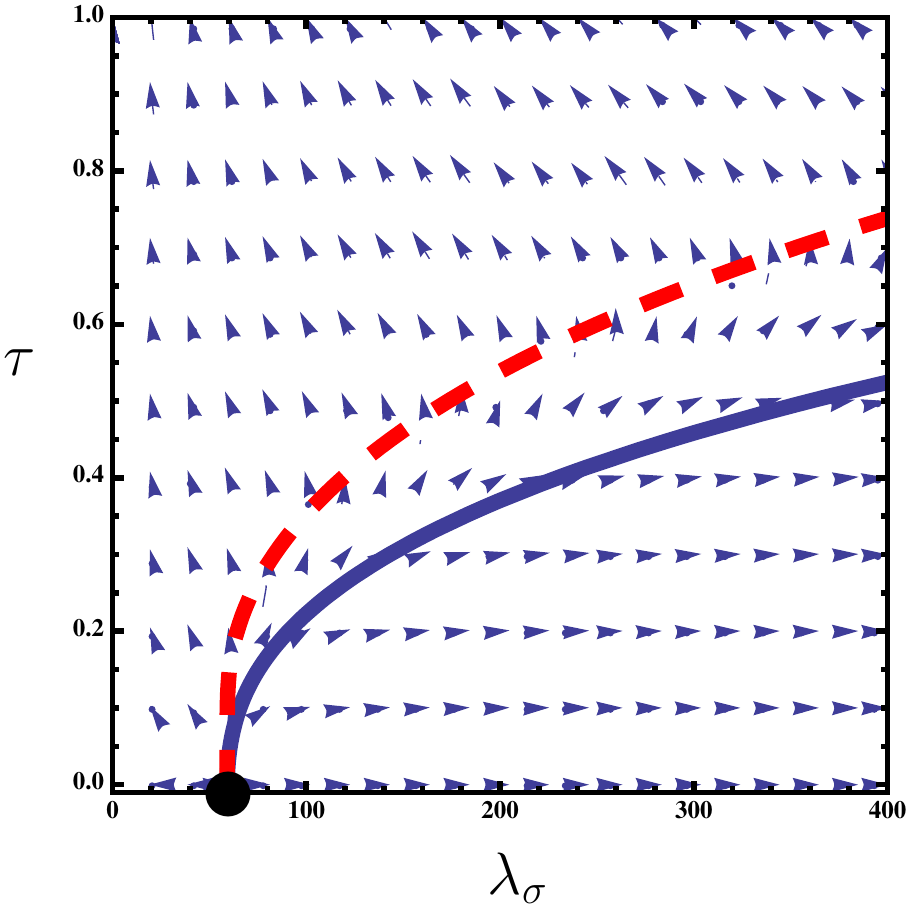}
\end{center}
\caption{RG flow of in the plane spanned by the four-fermion coupling~$\lambda_{\sigma}$ and the 
dimensionless temperature~$\tau=T/k$. The blue (straight) line depicts the separatrix~$\lambda_{\sigma}^{\rm sep.}(\tau)$, whereas
the red (dashed) line represents the line of pseudo fixed-points~$\lambda_{\sigma}^{\ast}(\tau)$.
Here, the separatrix has been obtained by solving the
flow equations~\eqref{eq:NJLbetasimpleT} and~\eqref{eq:NJLtauflowsec353} with suitably fine-tuned initial conditions.
For a given value of~$\tau$, the theory approaches a non-interacting IR limit for initial conditions~$\lambda_{\sigma}^{\rm UV}>\lambda_{\sigma}^{\rm sep.}(\tau)$,
whereas the system flows into an IR limit with broken chiral symmetry in the ground state for~$\lambda_{\sigma}^{\rm UV}<\lambda_{\sigma}^{\rm sep.}(\tau)$.
The dot represents the quantum critical point~$(\lambda_{\sigma}^{\ast}(0),0)$.
The arrows indicate the direction of the RG flow towards the infrared. 
 }
\label{fig:finiteTsep}
\end{figure}

Let us now assume that we have fixed the physical IR observables at zero temperature by 
choosing~$\lambda_{\sigma}^{\rm UV}>\lambda_{\sigma}^{\ast}(\tau\!=\! 0)$. This implies that we have
chosen the initial condition for the four-fermion interaction such that chiral symmetry is broken in the IR limit. Recall that the
fixed point~$\lambda_{\sigma}^{\ast}(\tau\!=\! 0)$ can be viewed as a quantum critical point.
Since the value of the (pseudo) fixed-point increases with increasing $\tau=T/k$, the rapid 
increase of the four-fermion coupling towards the IR is effectively slowed down and the RG flow may even change its 
direction in the $(\lambda_{\sigma},\tau)$-plane, see Fig.~\ref{fig:finiteTsep} for an illustration.
Due to this behavior of the pseudo fixed-point $\lambda_{\sigma}^{\ast}(\tau)$, it is already clear that 
for a fixed initial value $\luv$ a critical temperature $\Tc$ 
exists above which the $\lambda_{\sigma}$-coupling does not diverge but tends to zero for~$k\!\to\! 0$.
From a phenomenological point of view, such a behavior is indeed expected for high temperatures: The fermions 
become effectively  stiff degrees of freedom due to their 'Matsubara' mass $\sim T$ and chiral symmetry is 
restored.

With the aid of the pseudo fixed-point~$\lambda_{\sigma}^{\ast}(\tau)$ we can formulate a {\it sufficient} criterion for 
chiral symmetry breaking at finite temperature. First, we note that for a given initial condition~$\lambda_{\sigma}^{\rm UV}>\lambda_{\sigma}^{\ast}(\tau\!=\!0)$ 
the pseudo fixed-point~$\lambda_{\sigma}^{\ast}(\tau)$ determines a dimensionless 
temperature~$\tau_{\ast}$  through
\be
\lambda_{\sigma}^{\ast}(\tau_{\ast})=\lambda_{\sigma}^{\rm UV}\,.
\ee
Since the $\beta_{\lambda_{\sigma}}$-function is strictly positive for $\lambda_{\sigma}< \lambda_{\sigma}^{\ast}(\tau)$,
it follows immediately that it is {\it sufficient} to choose~$\tau>\tau_{\ast}$ to restore chiral symmetry for a given 
fixed value of~$\lambda_{\sigma}^{\rm UV}$. 
Thus, $T_{\ast}=\Lambda \tau_{\ast}$ defines a strict upper bound for the chiral phase transition temperature~$\Tc$.

It is important to stress that~$\lambda_{\sigma}^{\ast}(\tau)$ does not define a separatrix in the space spanned by the 
coupling~$\lambda_{\sigma}$ and~$\tau$. From the above discussion it is clear that~$\lambda_{\sigma}^{\ast}(\tau)$ only represents a strict
upper bound for the separatrix~$\lambda_{\sigma}^{\rm sep.}(\tau)$, see also Fig.~\ref{fig:finiteTsep}:
\be
\lambda_{\sigma}^{\ast}(\tau) \geq \lambda_{\sigma}^{\rm sep.}(\tau)\,.
\ee

For a given initial condition~$\lambda_{\sigma}^{\rm UV}>\lambda_{\sigma}^{\ast}(\tau\!=\!0)$, 
we can now define a second (dimensionless) temperature~$\tau_{\rm sep.}$ via
\be
\lambda_{\sigma}^{\rm sep.}(\tau_{\rm sep.})=\lambda_{\sigma}^{\rm UV}\,.
\ee
Due to the very definition of a separatrix in coupling space, this allows us to define a {\it necessary} criterion for chiral symmetry breaking (restoration)
at finite temperature: choosing~$\tau < \tau_{\rm sep.}$ ($\tau > \tau_{\rm sep.}$) for a given fixed UV coupling~$\lambda_{\sigma}^{\rm UV}$,
the theory approaches necessarily a regime with broken (restored) chiral symmetry in the IR limit. For a given value of the UV cutoff~$\Lambda$,
the quantity~$\tau_{\rm sep.}$ can then be translated into a physical temperature~$T_{\rm sep.}=\Lambda\tau_{\rm sep.}$. Moreover, we 
have~$\tau_{\rm sep.}\leq \tau_{\ast}$.

Strictly speaking, even the temperature~$T_{\rm sep.}$ defines only an upper bound for the actual
chiral phase transition temperature~$\Tc$ since it is only sensitive to an emergence of a condensate on intermediate 
momentum scales, but insensitive to the fate of the condensate in the deep IR due to fluctuations of the bosonic 
modes, e.~g. Nambu-Goldstone modes. In fact, the phase transition temperature decreases when one goes beyond
the point-like limit, e.~g., by taking into account bosonic loops in the RG flow to resolve the momentum dependence of the 
fermionic vertices~\cite{Braun:2009si}. In particular, the Nambu-Goldstone bosons play a prominent role at a thermal phase transition,
since they tend to restore the chiral symmetry, while fermions tend to build 
up a condensate and thereby break the symmetry of the ground state. In contrast to the bosonic fields, however, the anti-periodic boundary conditions 
of the fermion fields in Euclidean time direction lead to a suppression of the associated modes in the vicinity of the phase transition
and above. We shall come back to this in Sect.~\ref{sec:TQPTQCD}, 
where we discuss the thermal phase transition in QCD low-energy models.

In Fig.~\ref{fig:finiteTsep} we show the RG flow of our theory in the plane spanned by the coupling~$\lambda_{\sigma}$ and the dimensionless
temperature~$\tau$.
It is instructive to analyze the RG flow in this plane with the aid of the matrix~$\tilde{B}$:
\be
\tilde{B}=
\begin{pmatrix}
\frac{\partial (\partial_t \lambda_{\sigma})}{\partial \lambda_{\sigma}} & \frac{\partial (\partial_t \lambda_{\sigma})}{\partial \tau}   \\
\frac{\partial (\partial_t \tau)}{\partial \lambda_{\sigma}} & \frac{\partial (\partial_t \tau )}{\partial\tau}  
\end{pmatrix}
\,.\nn
\ee
Evaluated at the non-Gau\ss ian zero-temperature fixed point $(\lambda_{\sigma}^{\ast}(\tau=0),\tau=0)$, the matrix~$\tilde{B}$ corresponds to the stability matrix~$B$
defined in Eq.~\eqref{eq:Bdef}. Following our discussion in Sect.~\ref{subsec:simpleex}, we find that we have two IR repulsive (RG relevant) directions at the 
non-Gau\ss ian fixed point. The eigenvectors of~$B$ are simply given by~$\vec{v}_1=(1,0)$ and~$\vec{v}_2=(0,1)$. The associated critical exponents of~$B$ are given
by $\Theta_1=2$ and~$\Theta_2=1$, respectively. Note that the functional form of the separatrix in the $(\lambda_{\sigma},\tau)$-plane is determined
by the largest critical exponent, namely~$\Theta_1$, see Fig.~\ref{fig:finiteTsep}. Close to the quantum critical point~$(\lambda_{\sigma}^{\ast}(0),\tau\!=\! 0)$,
we therefore expect that the phase transition temperature~$\Tc$ scales according to
\be
\Tc\sim\Lambda\theta( \lambda_{\sigma}^{\rm UV} -\lambda_{\sigma}^{\ast}   )  \left(\frac{\lambda_{\sigma}^{\rm UV} -\lambda_{\sigma}^{\ast}}
{\lambda_{\sigma}^{\rm UV}}
\right)^{\frac{1}{|\Theta_1|}}\,,
\ee
see also Eq.~\eqref{eq:genscallaw} and Sect.~\ref{sec:TQPTQCD}. Corrections to this scaling behavior then arise from the subleading exponent~$\Theta_2$.

Away from the fixed point, we can use the eigenvectors of the matrix~$\tilde{B}$ to study the direction of the RG flow towards the IR. 
Let us first consider the eigenvectors of~$\tilde{B}$ for increasing~$\lambda_{\sigma}$ but fixed (dimensionless) temperature~$\tau>0$. 
For $\lambda_{\sigma}\gg \lambda_{\sigma}^{\ast}(\tau)$, the
eigenvectors become independent of~$\tau$: $\vec{v}_1=(1,0)$ and~$\vec{v}_2=(c\lambda_{\sigma},1)$ with  a constant $c>0$. Thus, the angle between the two directions 
shrinks to zero and we are effectively left with a one-dimensional RG flow pointing into $\lambda_{\sigma}$-direction, i.~e. $(1,0)$-direction.
At the Gau\ss ian fixed-point we find that the eigenvectors are given by~$\vec{v}_1=(1,0)$ and~$\vec{v}_2=(0,1)$ with 
exponents~$\Theta_1=-2$ and~$\Theta_2=1$, respectively. Thus, the flow is driven by the $(0,1)$-direction close to the Gau\ss ian fixed point. In accordance with this 
observation, we find that the flow is pushed into the $(0,1)$-direction ($\tau$-direction) for fixed~$\lambda_{\sigma}$ and increasing~$\tau$. 
For sufficiently large values of~$\tau$, the RG flows
are therefore attracted by the 'thermal' fixed-point~${\mathcal F}_T^{\infty}:=\lim_{\alpha\to \infty} (0,\alpha)$. 
On the other hand, the RG flows are attracted by the line of fixed points~${\mathcal F}_{\lambda}^{\infty}:=\lim_{\alpha\to \infty}(\alpha,\tau)$
for a given~$\lambda_{\sigma}^{\rm UV} < \lambda_{\sigma}^{\rm sep.}(\tau)$. Overall, this
is simply another way of saying that a finite temperature tends to restore chiral symmetry, whereas increasing the coupling tends to break the 
chiral symmetry of the ground state.

Finally we would like to comment on the effect of other four-fermion interaction channels compatible with the underlying symmetries of our model.
Qualitatively, the effect of a finite temperature is the same in a Fierz-complete ansatz. 
To be specific, we have shown that the (modulus) of the the pseudo fixed-point value increases with increasing~$\tau$. For the Fierz-complete
phase diagram\footnote{Note that the study of the phase diagram in Fig.~\ref{fig:pd_njl} is only Fierz-complete for~$T=0$. We only use it here to
illustrate our point concerning the effect of a Fierz-complete basis of four-fermion interactions. This is justified since the difference between
couplings longitudinal and transversal to the heat bath is parametrically suppressed for $T/k <1$ anyway. Note also that $\Tc/\ksb \lesssim 1$
in our present study.
}
in Fig.~\ref{fig:pd_njl} this means that the fixed-points~${\mathcal F}_2$ and~${\mathcal F}_3$ are pushed away 
from the Gau\ss ian fixed point~${\mathcal F}_1$ for increasing~$\tau$. Hence the domain~II increases for increasing~$\tau$ while the domains~Ia/b 
and~IIIa/b effectively shrink. Since the domains~II and~IV represent the basin of attraction of the Gau\ss ian fixed point, we conclude
that a chiral phase transition temperature~$\Tc$ must exist for any initial condition lying in the domains~Ia/b and~IIIa/b. Thus, the very general
mechanisms of dynamical chiral symmetry breaking remain unchanged. However, we expect that the prediction for the actual value of the
(chiral) phase transition temperature changes when a Fierz-complete ansatz is considered. 

\subsubsection{Mass-like Explicit Symmetry Breaking}\label{subsec:DefESB}
Let us now study another phenomenologically highly important deformation of fermionic theories, namely the inclusion of 
an explicit mass term~$\sim \bar{\psi}\bar{m}_{\psi}\psi$. For example, in QCD such terms correspond to the so-called current quark masses.
From a field-theoretical point of view, the deformation of a theory with an explicit mass term is also of great importance 
since it can be used as a control parameter to study systematically the scaling behavior of physical observables in finite volumes
and at finite temperatures, see e.~g. Refs.~\cite{Fisher:1971ks,Fisher:1972zza,Goldenfeld:1992qy,Zinn-Justin:2002ru,Braun:2007td}. 
From this scaling behavior the critical exponents at the phase transition can be read off, which then allows to assign the theory 
under consideration to a certain universality class.
Such an approach is indeed applied in the context of Monte-Carlo simulations to determine the QCD universality class, see e.~g. Ref.~\cite{Karsch:2010ya}.

In the following we shall discuss only general aspects of RG flows of fermionic theories with an explicit mass term. 
In our discussion we leave aside other possible deformations such as finite temperature and finite volume. As in our study
of finite-temperature effects in the previous section, we shall consider only a scalar-pseudosclar interaction channel to keep
our discussion of the physical mechanisms as simple as possible. However, a more quantitative study would again require the
consideration of a Fierz-complete set of four-fermion interactions. To be specific, we consider a theory with only one fermion species and
employ the following effective action:\footnote{The imaginary unit factor "i" in front of the mass term~$\bar{m}_{\psi}$ appears due to our conventions in Euclidean space-time.}
\be
\Gamma\left[\bar{\psi},\psi\right]&=&\int d^4 x
\left\{ \bar{\psi}\left(Z_{\psi} \I\fslash{\partial} + \I \bar{m}_{\psi}\right) \psi
+\frac{1}{2}\bar{\lambda}_{\sigma}(\bar{\psi}\psi)^{2}
-\frac{1}{2}\bar{\lambda}_{\text{PS}}(\bar{\psi}\gamma_{5}\psi)^2
\right\}\,. \label{eq:fermactmass}
\ee
This ansatz represents a straightforward generalization of the effective action~\eqref{eq:NJLtruncBasic}.
Since we allow for an explicit fermion mass term in our ansatz, the chiral symmetry is broken explicitly and 
the couplings~$\bar{\lambda}_{\sigma}$ and~$\bar{\lambda}_{\text{PS}}$ will in general be different. 
We assume that the UV cutoff~$\Lambda$ has been chosen such that $\bar{m}_{\psi}^{\rm UV}/\Lambda \ll 1$ and fix
the initial conditions for the various couplings at the UV scale $\Lambda$ as follows:
\be
\lim_{\bar{m}_{\psi}^{\rm UV}/\Lambda\to 0} \bar{\lambda}_{\sigma} = 
\lim_{\bar{m}_{\psi}^{\rm UV}/\Lambda\to 0} \bar{\lambda} _{\text{PS}} = \lambda_{\sigma}^{\rm UV}\,\qquad\text{and}\qquad
\lim_{\bar{m}_{\psi}^{\rm UV}/\Lambda\to 0}\bar{m}_{\psi} =\bar{m}_{\psi}^{\rm UV}\,.
\ee
Thus, we are
left with only two input parameters for our simple model, namely~$\bar{\lambda}_{\sigma}^{\rm UV}$ and~$\bar{m}_{\psi}^{\rm UV}$. In QCD, the latter parameter
plays the role of the current quark mass.

It is also possible to consider a partially bosonized version of the fermionic action~\eqref{eq:fermactmass}. This yields two
different Yukawa couplings and corresponding mass terms, and a linear term for the composite field~$\phi_1 \sim \bar{\psi}\psi$ appears.
The net effect of this term 
is to stretch and tilt the order-parameter potential shown in Fig.~\ref{fig:effpot} into the direction associated with the field~$\phi_1$. As a result,
we have a finite vacuum expectation value~$\langle \phi_1\rangle$ on all scales and the ground state is no longer degenerate.
Of course, it is also possible to deal with such an order-parameter potential in RG flows. As is well known, 
a linear symmetry breaking term remains unchanged in the RG flow~\cite{Zinn-Justin:2002ru}. Therefore the
usual strategy is to evolve the potential without a symmetry breaking term. Explicit symmetry breaking
is then taken into account after the quantum fluctuations have been integrated out on all scales \cite{Jungnickel:1995fp,Berges:1997eu,Schaefer:1999em}.
Alternatively, it is also possible to include the explicit symmetry breaking in the (partially) bosonized RG flows
which is particularly convenient for studies of finite-volume effects in quantum field 
theories~\cite{Braun:2004yk,Braun:2005fj,Braun:2005gy,Braun:2007td,Braun:2008sg,Braun:2010pp}.

Returning to the purely fermionic formulation, we first note that the propagator matrix now includes a mass term~$\bar{m}_{\psi}$:
\be
 {\cal P}_k= \left(\begin{array}{cc}
 0 &  -Z_{\psi} \fss{p}^T(1+r_{\psi}) -\I \bar{m}_{\psi} \\
 -Z_{\psi}\fss{p}(1+r_{\psi}) + \I   \bar{m}_{\psi} & 0 \end{array} \right)(2\pi)^4\delta^{(4)}(p-p^{\prime})\,.\nn
\ee
In the following we shall employ a covariant regulator function ($d$-dimensional regulator) for convenience. The fluctuation
matrix in Eq.~\eqref{eq:flucmatrix} remains unchanged except for the fact that the coupling~$\bar{\lambda}_{\text{PS}}$ is attached
to the terms depending on~$\gamma_5$, whereas the coupling~$\bar{\lambda}_{\sigma}$ is attached to the terms independent of~$\gamma_5$.
Since we allow for a term~$\sim \bar{\psi}\psi$ in our ansatz for the effective action, the Feynman diagram shown on the right in Fig.~\ref{fig:feynman} also
contributes to the RG flow of the effective action on all scales and yields a mass renormalization.\footnote{Strictly speaking, our RG approach includes
resummations of both diagrams in Fig.~\ref{fig:feynman} as well as combinations thereof.} Within the present truncation of the effective action  
there are nevertheless no contributions to the RG flow of the wave-function renormalization~$Z_{\psi}$ in the point-like limit, which we shall consider
from now on. For~$Z_{\psi}\equiv 1$ and~$\eta_{\psi}=0$, the flow equations then read
\be
\partial _t \epsilon_{\psi} &=& -2\epsilon_{\psi} - 8\left[ 3\lambda_{\sigma}-\lambda_{\text{PS}} \right] v_4\, b_1^{{\rm (F)},(4)}(\epsilon_{\psi};0)\,, \label{eq:epsi}\\
\partial_t \lambda_{\sigma} &=& 2\lambda_{\sigma} - 8\left[ \lambda_{\sigma}^2 + \lambda_{\sigma}\lambda_{\text{PS}}   \right] v_4\, \hat{l}_1^{{\rm (F)},(4)}(\epsilon_{\psi};0)\nn\\
&& \qquad\qquad\qquad\quad - 8\left[ \lambda_{\text{PS}}^2 - \lambda_{\sigma}\lambda_{\text{PS}}   \right] v_4\, \tilde{l}_1^{{\rm (F)},(4)}(\epsilon_{\psi};0)
\,,\\
\partial_t \lambda_{\text{PS}} &=& 2\lambda_{\text{PS}} - 8\left[ \lambda_{\text{PS}}^2 
+ \lambda_{\sigma}\lambda_{\text{PS}}   \right] v_4\, \hat{l}_1^{{\rm (F)},(4)}(\epsilon_{\psi};0)\nn\\
&& \qquad\qquad\qquad\quad - 8\left[ \lambda_{\text{PS}}^2 + 3\lambda_{\sigma}\lambda_{\text{PS}}   \right] v_4\, \tilde{l}_1^{{\rm (F)},(4)}(\epsilon_{\psi};0)\label{eq:lpspsi}
\,,
\ee
where the renormalized dimensionless fermion mass is given by
\be
\epsilon_{\psi}=\frac{\bar{m}_{\psi}^2}{k^2}\,,
\ee
and the renormalized dimensionless four-fermion couplings are defined as $\lambda_{\sigma}= k^2 \bar{\lambda}_{\sigma}$ 
and~$\lambda_{\text{PS}}=k^2 \bar{\lambda}_{\text{PS}}$. The various threshold functions are defined in App.~\ref{app:regthres}. Here we only
note that
\be
l_1^{{\rm (F)},(4)}(\epsilon_{\psi};\eta_{\psi})=\hat{l}_1^{{\rm (F)},(4)}(\epsilon_{\psi};\eta_{\psi}) + \tilde{l}_1^{{\rm (F)},(4)}(\epsilon_{\psi};\eta_{\psi})\,
\ee
and $l_1^{{\rm (F)},(4)}(0;0)=\hat{l}_1^{{\rm (F)},(4)}(0;0)$. This implies~$\tilde{l}_1^{{\rm (F)},(4)}(0;0)=0$.
The function $b_1^{{\rm (F)},(4)}$ behaves as
\be
 b_1^{{\rm (F)},(4)}(\epsilon_{\psi};0) \sim \epsilon_{\psi}
\ee
for $\epsilon_{\psi}\ll 1$, whereas $\hat{l}_1^{{\rm (F)},(4)}$ and~$\tilde{l}_1^{{\rm (F)},(4)}$ are constant for 
small $\epsilon_{\psi}$. Moreover, we have
\be
 b_1^{{\rm (F)},(4)}(\epsilon_{\psi};0) \sim \frac{1}{\epsilon_{\psi}}\,, \qquad
 \hat{l}_1^{{\rm (F)},(4)} \sim -\frac{1}{\epsilon_{\psi}^2}\,\qquad\text{and}\qquad
 \tilde{l}_1^{{\rm (F)},(4)} \sim  \frac{1}{\epsilon_{\psi}^2}
\ee
for $\epsilon_{\psi} \gg 1$. With the aid of these identities it is straightforward to show that the flow equations for~$\lambda_{\sigma}$
and~$\lambda_{\text{PS}}$ are identical in the limit~$\epsilon_{\psi}\!\to\! 0$. Thus, we have $\lambda_{\sigma}\equiv \lambda_{\text{PS}}$
in the chirally symmetric limit, as it should be. Moreover, the flow equations of these coupling  reduce to the flow equation~\eqref{eq:NJLbetasimple}
in this limit.

From our RG equations~\eqref{eq:epsi}-\eqref{eq:lpspsi} we read off that a strong four-fermion interaction induces a
strong increase in the fermion mass. This is in accordance with our expectations, since strong fermion self-interactions
signal the onset of chiral symmetry breaking which is associated with a finite fermion mass (gap). From the 1PI Feynman diagrams
and the associated threshold functions, it is clear that a large fermion mass suppresses quantum corrections and
therefore prevents the couplings and the mass from growing further. Nevertheless it is in general not possible
to ``stabilize" the RG flows in the point-like approximation
 with the aid of the evolving fermion mass term, such that the four-fermion interactions remain finite on all scales. We only find
that for a given value of $\lambda_{\sigma}^{\rm UV}$ a critical initial value~$\epsilon_{\psi}^{\rm cr.}\!=\! (m_{\psi}^{\rm cr.}/\Lambda)^2$ 
exists, such that the four-fermion interactions remain finite on all scales for~$\epsilon_{\psi}^{\rm UV}>\epsilon_{\psi}^{\rm cr.}$. 
For $\epsilon_{\psi}^{\rm UV} < \epsilon_{\psi}^{\rm cr.}$
the four-fermion couplings still diverge at a finite scale~$\ksb$, provided we choose the initial conditions~$\lambda_{\sigma}^{\rm UV}$
to be larger than the fixed-point~$\lambda_{\sigma}^{\ast}$ of the chirally symmetric theory ($\epsilon_{\psi}^{\rm UV}\to 0$), see Eq.~\eqref{eq:onechannelFP}. 
To put it sloppily, the critical value~$\epsilon_{\psi}^{\rm cr.}$ plays a role roughly similar to that of the critical (dimensionless) temperature. More precisely,
the functional dependence of~$\epsilon_{\psi}^{\rm cr.}$ on~$\lambda_{\sigma}^{\rm UV}$ can be viewed as a critical line similar to the separatrix 
discussed in our finite-temperature studies in the previous section. We indeed find that~$\epsilon_{\psi}^{\rm cr.}$ increases monotonously 
with increasing~$\lambda_{\sigma}^{\rm UV}$. In other words, larger masses are required to screen strong self-interactions.

From a phenomenological point of view, $\epsilon_{\psi}^{\rm cr.}$ distinguishes between a phase with a 
trivial explicit chiral symmetry breaking in the IR limit and a phase with spontaneous chiral symmetry breaking associated with a non-trivial momentum dependence
of the fermionic self-interactions. In fact, a large initial value~$\epsilon_{\psi}^{\rm UV}>\epsilon_{\psi}^{\rm cr.}$ suppresses the fermion self-interactions and
therefore keeps the theory from approaching criticality. From a field-theoretical point of view, the deformation of a fermionic theory with
an explicit mass term might be still useful to guide lattice studies of conformal phases in gauge 
theories~\cite{Dietrich:2010yw,DelDebbio:2010jy,DelDebbio:2010ze}, see also Sect.~\ref{sec:gaugetheories}.
Whereas functional approaches (RG and Dyson-Schwinger equations) allow to study explicitly the chiral limit ($\epsilon_{\psi}^{\rm UV}\!\to\! 0$) of such 
theories, lattice simulations usually involve explicit mass terms for the fermions and therefore require a controlled
extrapolation to the chiral limit. 

Finally we remark that once one has resolved the momentum-dependence of the fermionic vertices, the explicit
mass term can indeed be used to ``stabilize" the RG flows to study IR observables, even for (arbitrarily) small values of~$\epsilon_{\psi}^{\rm UV}$. This
was found in the context of condensed-matter physics~\cite{PTP.112.943,Honerkamp5}, where it was shown explicitly that the point-like
limit and the limit~$\epsilon_{\psi}^{\rm UV}\to 0$ do not commute in general.
 
%



%
\section{Non-relativistic Quantum Field Theories} \label{sec:nrtheories}
As first examples of strongly interacting fermionic field theories we consider non-relativistic many-body systems.
However, we do not aim at a quantitative study of these systems. We only present very general arguments concerning
universality in non-relativistic Fermi gases and the existence of inhomogeneous phases in such systems. This allows us
to apply our field-theoretical discussion in the previous section to phenomenologically relevant systems, such as 
ultracold atomic gases and nuclear physics. In Sect.~\ref{sec:coldgases}, we show that the non-Gau\ss ian fixed-point of the 
four-fermion coupling plays a decisive role in ultracold atomic gases. 
In fact, the coupling strength of fermions in two different
hyperfine states can be tuned in experiments by means of an external magnetic field~\cite{Feshbach}. 
This strong experimental control opens up the possibility to gain 
deep insights into the mechanisms of quantum many-body phenomena, such as {\it Bose-Einstein} (BEC)
condensation and {\it Bardeen-Cooper-Schrieffer} (BCS) superfluidity,
and to benchmark different approaches to strongly interacting field theories.

In principle, non-relativistic many-body problems can be studied  
by simply solving the {\it Schr\"odinger} equation. 
However, exact solutions of the {\it Schr\"odinger} equation are difficult (or even impossible) to find for most of the problems, 
making approximations necessary. Therefore a path-integral approach to many-body problems might be promising since it allows us to employ 
complementary approximation schemes and gain insights into the dynamics of the system from a different perspective.
In the spirit of this review, we shall restrict our discussion to path-integral approaches in what follows.

Before we actually discuss RG flows of non-relativistic systems, it is instructive to compare actions which 
describe the dynamics of relativistic and non-relativistic theories, respectively. For non-relativstic fermions with spin $1/2$, we consider an action consisting of 
only a one-body part (kinetic term) and a two-body part (interaction term). Whether it is justified to neglect 
higher $n$-body interactions depends on the system under consideration and needs to be carefully analyzed. 
For example, this
might be a reasonable approximation for the description of a dilute gas of ultracold atoms.
In nuclear physics, on the other hand, it has been found that three-body interactions have to be taken into account
to compute accurately ground-state properties of nuclei, see e.~g. Refs.~\cite{Hagen:2007ew,Otsuka:2009cs,Friman:2011vm}.
In any case, it is sufficient for our purposes to consider 
the following ansatz for the action in $d+1$ Euclidean space-time dimensions:
\be
&& S[\psi^{\dagger},\psi] = \sum_{\sigma} \int d\tau\int d^d x\, \psi^{\dagger}_{\sigma}(\tau,\vec{x}) 
\left( \partial _{\tau} - \Delta  \right) \psi_{\sigma}(\tau,\vec{x})\nn\\
&& \qquad
+\,\frac{1}{2}\sum_{\sigma,\sigma^{\prime}}  \int d\tau\!\int d^dx\! \int d^dy \,
\psi^{\dagger}_{\sigma}(\tau,\vec{x}) \psi^{\dagger}_{\sigma^{\prime}}(\tau,\vec{y})
U(\vec{x},\vec{y})
\psi_{\sigma^{\prime}}(\tau,\vec{y}) \psi_{\sigma}(\tau,\vec{x})\,,\label{eq:nractiongen}
\ee
where $\Delta$ denotes the Laplace operator and the indices~$\sigma$,~$\sigma^{\prime}$ 
refer to the spin components of the two-component Grassmann-valued 
spinor~$\psi^{T}=(\psi_{\uparrow},\psi_{\downarrow})$. For example, we can think of these two
components as two different hyperfine-states of the atoms in ultracold gases.
As usual, Hermitian conjugation is defined
as~$\psi^{\dagger}=(\psi^{\ast})^{T}$. For convenience, we have set $\hbar=1$ and~$2m=1$, where $m$
denotes the mass (parameter) of the fermions. The interaction potential is given by the function~$U$. Prominent examples are 
the Coulomb potential or a contact interaction potential.

Comparing the non-relativistic action~\eqref{eq:nractiongen} with a simple action describing 
relativistic fermions, see e.~g. Eq.~\eqref{eq:NJLtruncBasic}, we immediately observe that we have
an~O($d$) symmetry in the kinetic term in Eq.~\eqref{eq:NJLtruncBasic}
owing to Poincare invariance. In non-relativistic theories, on the other hand, we have Galilei invariance and
the (canonical) mass dimension of space- and time-like coordinates is different, see App.~\ref{app:units} for our conventions. This 
implies that the dimension of the interaction potential~$U$ and the four-fermion coupling 
in Eq.~\eqref{eq:NJLtruncBasic} is also different. On the other hand, both the relativistic as well as the
non-relativistic theory are invariant under continuous~U($1$) transformations. For the non-relativistic case, the 
symmetry transformations are given by
\be
\psi_{\uparrow} \mapsto \E^{\I\alpha}\psi_{\uparrow}\,,\qquad
\psi_{\uparrow} ^{\ast}\mapsto \E^{-\I\alpha}\psi_{\uparrow} ^{\ast}\,,\quad
\psi_{\downarrow} \mapsto \E^{\I\alpha}\psi_{\downarrow}\,,\qquad
\psi_{\downarrow} ^{\ast}\mapsto \E^{-\I\alpha}\psi_{\downarrow} ^{\ast}\,.\label{eq:symNR}
\ee
This symmetry reflects particle number conservation.
The corresponding transformation for the relativistic case is given in Eq.~\eqref{eq:U1reltrans}. 

In the subsequent section we study universality in non-relativistic Fermi gases. To this end, we consider
a special choice for the interaction potential, namely a contact interaction.  
In Sect.~\ref{sec:DFTRG} we then discuss an
RG approach to density functional theory (DFT) which gives direct access to the density and opens up the
possibility to conveniently resolve inhomogeneities of the ground state of (self-bound) many-body systems, such as nuclei.

\subsection{Cold Atomic Quantum Gases}\label{sec:coldgases}
While particles approach the classical limit for high temperatures, their quantum nature becomes
important for low temperatures~$T$, where the thermal wavelength~$\sim 1/\sqrt{T}$ becomes much larger
than the interparticle distance. 
In the past fifteen years it has become possible to achieve low temperatures in ultracold Fermi gases and
study quantum many-body phenomena, such as Bose-Einstein condensation and BCS 
superfluidity in great detail~\cite{Anderson14071995,PhysRevLett.75.1687,PhysRevLett.75.3969,DeMarco10091999}.
Different experimental setups can be used to study quantum effects of atomic gases 
at low temperatures. Here, we shall consider a dilute gas consisting of fermions in two different hyperfine states 
which interact resonantly. For example, such a situation can be achieved in experiments with ${}^{6}\text{Li}$ atoms.
From now on, we refer to the fermions in the two different hyperfine states as spin-up and spin-down fermions. 
For simplicity, we shall assume that
the number of fermions in these two spin states is identical which implies that the corresponding chemical potentials
are identical. However, we would like to point out that spin-polarized Fermi gases have also attracted 
a lot of attention in the past few years, both from the experimental~\cite{Zwierlein,Partridge} 
as well was from the theoretical side, see e.~g. Refs.~\cite{CombMora,Stoof,Stoof2,Ku:2008vk}.

Let us now discuss the interaction of the fermions in such experiments in more detail. The two-body interaction potential~$U$ of two
fermions at position $\vec{x}$ and~$\vec{y}$ is short-range repulsive and long-range attractive. A prominent and simple example for such a 
potential is a hard-core square-well potential with range~$R$: 
\be
U(\vec{x},\vec{y}) \to \infty\,
\quad\text{for}\quad 0\leq |\vec{x}-\vec{y}| \leq R_{\rm a}\,,
\quad U(\vec{x},\vec{y})= -U_0
\quad\text{for}\quad R_{\rm a}<|\vec{x}-\vec{y}| \leq R\,, \nn
\ee
and
\be
U(\vec{x},\vec{y})= 0
\quad\text{for}\quad |\vec{x}-\vec{y}|>R\,, \nn
\ee
where $U_0$ and $R_{\rm a}$ are positive constants characterizing the specific type of the atoms.
In a situation where the interparticle distance~$r$ is much larger than the range~$R$ of the interaction (limit of a dilute gas; 
atom density $n\sim 1/r^3$), the details of the interaction potential~$U$ are of no importance and we may simply approximate it 
by a $\delta$-function:
\be
U(\vec{x},\vec{y}) \approx \bar{\lambda}_{\psi} \delta^{(d)} (\vec{x}-\vec{y})\,,\label{eq:plnrpot}
\ee
where~$\bar{\lambda}_{\psi}$ defines the (bare) four-fermion coupling. 
Note that only fermions in different hyperfine states can interact via such a potential due to the Pauli exclusion
principle. It should be also stressed that our approximation of using a contact interaction potential causes UV divergences. 
To obtain the "true" effective interaction potential in the limit of a dilute Fermi gas, these divergences have to be removed by, e.~g., introducing
a UV cutoff~$\Lambda \gg n^{1/3}\sim 1/r$. From a phenomenological point of view, this UV cutoff acquires a physical meaning and
can be viewed as the {\it Bohr} radius of the atoms in cold gases. As we shall discuss below, the dynamics of the system can become effectively
independent of this parameter, i.~e. it is possible to remove this parameter by taking the limit~$\Lambda\to\infty$.

The (dimensionless) renormalized four-fermion coupling $\lambda_{\psi}$ can be directly related to an
experimentally accessible control parameter, namely the s-wave scattering length~$a_{\rm s}$ of the atoms. 
This can be readily seen from a study of the $2\to 2$ scattering process.
To be more specific, one needs to compute the contributions to the T-matrix arising from the $2\to 2$ scattering process.
These contributions can be summed up analytically and then be related to the 
s-wave scattering amplitude~\cite{Sakurai:2011zz}.
The leading order of an expansion of the
scattering amplitude in the external momentum (i.~e. the relative momentum of the atoms) 
is constant and defines minus the s-wave scattering length~$a_{\rm s}$. 
In the limit of low temperatures $1/\sqrt{T}\gg r$ and low densities $n\sim 1/r^3$, the
typical momenta of the atoms are indeed small (low-energy limit) and the wave-function describing the 
scattered atoms is therefore mainly dominated by s-wave states.
Overall, such an analysis yields a relation between the dimensionless renormalized four-fermion coupling $\lambda_{\psi}$
and the s-wave scattering length~$a_{\rm s}$:
\be
\lambda_{\psi}=  \frac{8\pi\Lambda}{\frac{1}{a_{\rm s}}-c_{\rm s}\Lambda}
\,, \label{eq:lpsiSL}
\ee
where~$\Lambda$ denotes the UV cutoff.
The constant~$c_{\rm s}>0$ depends on the employed 
regularization scheme. For example, we have~$c_{\rm s}=2/\pi$ for the sharp UV cutoff.

At this point a comment is in order concerning the expansion of the scattering amplitude in the (external) momentum.
The coefficient of the quadratic term in this expansion determines the so-called effective 
range~$r_{\rm e}$ of the interaction potential which allows us to distinguish between narrow and broad Feshbach resonances. 
While the effective range~$|r_{\rm e}|$ is much larger than~$R$
in the vicinity of {\it narrow} Feshbach resonance,\footnote{Near a narrow Feshbach resonance we have~$r_{\rm e} < 0$
and $|r_{\rm e}| > R$.} we have~$r_{\rm e} \sim R$ close to a {\it broad} Feshbach resonance. However, we have $a_{\rm s}\to\infty$ in both cases.
As we shall see below, these two types of Feshbach resonances are associated with two distinct fixed points of the theory.

In experiments, it is possible to tune the scattering length~$a_{\rm s}$ by hand to arbitrarily large values 
with the aid of an external magnetic field~$B$.
In fact, one finds the following relation between~$a_{\rm s}$ and a magnetic field~$B$ that couples to the magnetic moment
of the fermions~\cite{Feshbach}:
\be
a_{\rm s} \simeq a_{\rm bg}\left (1- \frac{\Delta_{\rm W}}{B-B_0} \right)\,,\label{eq:SLB}
\ee
where $a_{\rm bg}$ is the (background) scattering length away from the resonance limit and~$\Delta_{\rm W}$
defines the width of the resonance; $B_0$ is the value of the magnetic field at which the resonance occurs.
From Eqs.~\eqref{eq:lpsiSL} and~\eqref{eq:SLB} it follows that the 
(effective) interaction strength can be tuned to arbitrarily large values by varying the external magnetic field~$B$. 
The limit of large scattering length~$|a_{\rm s}|$
defines a {\it universal} regime (unitary regime), provided the range~$R$ of the interaction potential as well as the effective 
range~$r_{\rm e}$ are much smaller than the interparticle distance~$r$:
\be
0 \leftarrow\frac{1}{|a_{\rm s}|}\ll \frac{1}{r}\sim n^{\frac{1}{3}} \ll \frac{1}{r_{\rm e}}\sim \frac{1}{R}\,.
\ee
Here, $n$ is the atom density. Since $r/R\gg 1$ and $r/r_{\rm e}\gg 1$, the theory 
depends on only a single parameter, namely the density.\footnote{In this case, the scattering amplitude is proportional to $\I/q$, where $q$ denotes the external 
momenta. Therefore this limit is also known as the {\it unitary} limit.} 
Thus, the dynamics of the theory is independent of the details of the interaction (potential).
For a narrow Feshbach resonance ($r_{\rm e} > R$), the details of the interaction potential become important and the theory 
depends on more than one parameter. We shall return to the case of a narrow Feshbach resonance at the end of this section.

From our discussion it becomes already apparent that the s-wave scattering length $a_{\rm s}$ plays a prominent role in 
ultracold Fermi gases. Away from the Feshbach resonance for $B>B_0$, where $a_{\rm s}$ is positive and small, 
the interaction between the fermions is strongly attractive. Two fermions can form a tight bosonic bound
state (bosonic molecule) with a binding energy~$E_{\rm b} \sim 1/a_{\rm s}^2$. 
For sufficiently low temperatures and $a_{\rm s}>0$, we therefore expect Bose-Einstein 
condensation which is associated with a spontaneous breakdown of the U($1$) symmetry of the 
theory.\footnote{Note that the interaction between the tightly bound bosonic molecules is effectively repulsive.} We shall therefore refer to this regime 
as the BEC regime.

In the regime where $|a_{\rm s}|$ is small but $a_{\rm s}<0$, the interaction of atoms with opposite spin is weakly attractive. Therefore
it is not possible to form tightly bound bosonic molecules. However, pairs of fermions with opposite spin and opposite momenta 
can form bound states, so-called Cooper pairs, 
since the Fermi surface is unstable to pairing in this case~\cite{Bardeen:1957mv}. These pairs can be viewed as bosonic bound states with
a large spatial extent, i.~e. they are highly-localized in momentum space. 
These bound states can also condense at sufficiently low temperatures and form a superfluid macroscopic
state. The existence of this superfluid state is associated with a broken U(1) symmetry in the ground state of the theory. 
In the following we refer to this regime as the BCS regime.

At the Feshbach resonance $B=B_0$, i.~e. $|a_{\rm s}|\to\infty$, we encounter resonant Cooper pairs. The binding energy of these
states becomes arbitrarily small and we are left with spatially delocalized bound-states.

Let us now discuss the fixed-point structure of an ultracold Fermi gas in $d=3$ space dimensions 
close to a broad Feshbach resonance. To be specific, we consider the following 
ansatz for the effective action:\footnote{Our ansatz follows directly from the action~\eqref{eq:nractiongen}
by employing the interaction potential given in Eq.~\eqref{eq:plnrpot}.}
\be
\Gamma [\psi^{\dagger},\psi]=  \int d\tau\int d^3 x\, \left\{ \psi^{\dagger}
\left( Z^{\|}_{\psi} \partial _{\tau} - Z_{\psi}^{\perp}\Delta -\mu \right) \psi 
+\,\frac{1}{2}\bar{\lambda}_{\psi}(\psi^{\dagger}\psi) (\psi^{\dagger}\psi)
\right\}\,,\label{eq:UGaction}
\ee
where $\mu$ denotes the chemical potential of the fermions. The wave-function renormalizations~$Z_{\psi}^{\|}$ and~$Z_{\psi}^{\perp}$
are in general different. We would like to stress that
we solely consider the continuum limit in this section. This implies that we study the system in the infinite-volume limit. The particle
number is then not well-defined but only the (fermion) density. 

Ultracold Fermi gases have been studied extensively within the functional
RG approach. The phase diagram of symmetric Fermi gases at zero and finite temperature has been studied in 
Refs.~\cite{Diehl:2007ri,Diehl:2007th,Bartosch2009b,Floerchinger:2008qc,Floerchinger:2009pg}; for reviews, 
see Refs.~\cite{Diehl:2009ma,Scherer:2010sv}. The phase transition in a non-relativistic Bose gas
(regime with a small positive s-wave scattering length)
has been studied in great detail in
Refs.~\cite{Blaizot:2003vz,Blaizot:2004qa,Blaizot:2008xx}. Note that the latter studies have triggered the development of novel techniques
to resolve the momentum dependence of correlation functions in non-perturbative RG flows~\cite{Blaizot:2005xy,Blaizot:2005wd,Blaizot:2006vr}.
Moreover, spin-polarized Fermi gases have also been studied with Wilsonian-type RG flows~\cite{Stoof2,Schmidt:2011zu}.
Here, we do not aim at a quantitative study of the phase diagram of ultracold gases. 
In the spirit of this review, we are
rather interested in a simple analysis of the fixed-point structure of the four-fermion coupling~${\lambda}_{\psi}$. As it will turn out, such an analysis is already
sufficient to understand the experimentally observed universality in these systems.

The flow equation for the~$\bar{\lambda}_{\psi}$-coupling can be derived along the same lines as the flow equations for the
four-fermion couplings of the NJL model in Sect.~\ref{subsec:simpleex}. We only add that in the present case
it is convenient to define a generalized field vector~$\Phi^{\rm T}=(\psi^{\rm T},\psi^{\ast})$. 
The regularized propagator matrix (in $\Phi$-space) then reads\footnote{Recall that the mass-dimension of time-like 
and space-like momenta are different.}
\be
 {\cal P}_k\!=\! \left(\begin{array}{cc}
 0 &   \I p_0\! -\! (\vec{p}^{\,2} \! -\! \mu) \!-\! k^2 r_{\psi}({\mathcal Z}) \\
  \I p_0  \!+\! (\vec{p}^{\,2} \!-\!\mu) \!+\! k^2 r_{\psi}({\mathcal Z}) & 0 \end{array} 
 \right) (2\pi)^4 \delta (p_0\!-\! p_0^{\,\prime})\delta^{(3)}(\vec{p}\!-\!\vec{p}^{\,\prime})\,\nn
\ee
with ${\mathcal Z}=(\vec{p}^{\,2}-\mu)/k^2$. Here, we have chosen a spatial regulator for convenience. A possible choice for the shape
function~$r_{\psi}$ is given in Eq.~\eqref{eq:nr_cutoff_shape}. Since it is suffices to consider the point-like limit for our
purposes, we can set the wave-function renormalizations equal to one. The fluctuation matrix~${\mathcal F}$ can be derived
straightforwardly from the action~\eqref{eq:UGaction}. Using the regulator shape function~\eqref{eq:nr_cutoff_shape},
we find the following set of flow equations:
\be
\partial_t \lambda_{\psi}&=& \lambda_{\psi} + \frac{8}{6} v_3 \,l(\tilde{\mu}) \lambda_{\psi}^2\,, \label{eq:lambdapsiNR}
\\
\partial_t \tilde{\mu} &=& -2 \tilde{\mu}\,,
\ee
where $v_3=1/(8\pi^2)$. The dimensionless four-fermion coupling is defined as $\lambda_{\psi}=k\bar{\lambda}_{\psi}$.
The dimensionless chemical potential is given by $\tilde{\mu}=\mu/k^2$. Moreover, we have 
\be
l(\tilde{\mu})=(1+\tilde{\mu})^{\frac{3}{2}}\theta ( 1+\tilde{\mu}) - (\tilde{\mu}-1)^{\frac{3}{2}} \theta(\tilde{\mu}-1)\,.\label{eq:lfctNR}
\ee

Analogous to our studies of spontaneous symmetry breaking in the NJL model in Sect.~\ref{sec:example},
the onset of U($1$) symmetry breaking in the present case is signaled by the fact that the four-fermion coupling diverges at 
a finite scale~$\ksb$, i.~e.~$1/\lambda_{\psi}(\ksb)=0$. Recall that in experiments the 
spontaneous breakdown of the U(1) symmetry is associated with a superfluid behavior of the system.

Let us now analyze the fixed-point structure of an ultracold Fermi gas. We begin with a study of the UV limit of the theory, 
i.~e. we first consider~$k^2\gg |\mu|$. This limit can be viewed as the "vacuum" limit associated with the two-body scattering problem~\cite{Diehl:2007ri}.
The flow equation for the $\lambda_{\psi}$ coupling simplifies considerably for $k^2\gg |\mu|$:
\be
\partial_t \lambda_{\psi}&=& \lambda_{\psi} + \frac{8}{6} v_3 \lambda_{\psi}^2\,.
\ee
For the sharp cutoff, this flow equation reads
\be
\partial_t \lambda_{\psi}&=& \lambda_{\psi} + 2 v_3 \lambda_{\psi}^2\,.
\ee
Independent of the regularization scheme, we find two fixed points. The Gau\ss ian fixed point is IR attractive (UV repulsive), whereas the non-Gaussian fixed 
point~$\lambda_{\psi}^{\ast} < 0$ is IR repulsive (UV attractive). For the regulator function~\eqref{eq:nr_cutoff_shape}, the 
non-Gau\ss ian fixed point is given by
\be
\lambda_{\psi}^{\ast} = - 6\pi^2\,.
\ee
For the sharp cutoff, on the other hand, we find
\be
\lambda_{\psi}^{\ast} = - 4\pi^2\,.
\ee
From a comparison of these fixed-point values with the relation~\eqref{eq:lpsiSL}, we conclude
that the non-Gau\ss ian fixed point can be identified with the renormalized four-fermion coupling 
in the limit $a_{\rm s}\to\infty$ (for a broad Feshbach resonance). In other words, the {\it experimentally} observed universal
behavior of ultracold Fermi gases in the limit~$a_{\rm s}\to\infty$ 
is tightly linked to the existence of the non-Gau\ss ian fixed point~$\lambda_{\psi}^{\ast}$. This fixed point can be viewed
as a quantum critical point since the choice for the initial value~$\lambda_{\psi}^{\rm UV}$ relative to~$\lambda_{\psi}^{\ast}$ 
distinguishes between two distinct regimes in the IR limit, namely a strongly interacting superfluid phase and a weakly interacting
phase with restored~U($1$) symmetry.
The Gau\ss ian fixed point, on the other hand, can be associated with the limit of a narrow Feshbach 
resonance, as we shall see below.
\begin{figure}[t]
\begin{center}
\includegraphics[width=0.65\linewidth]{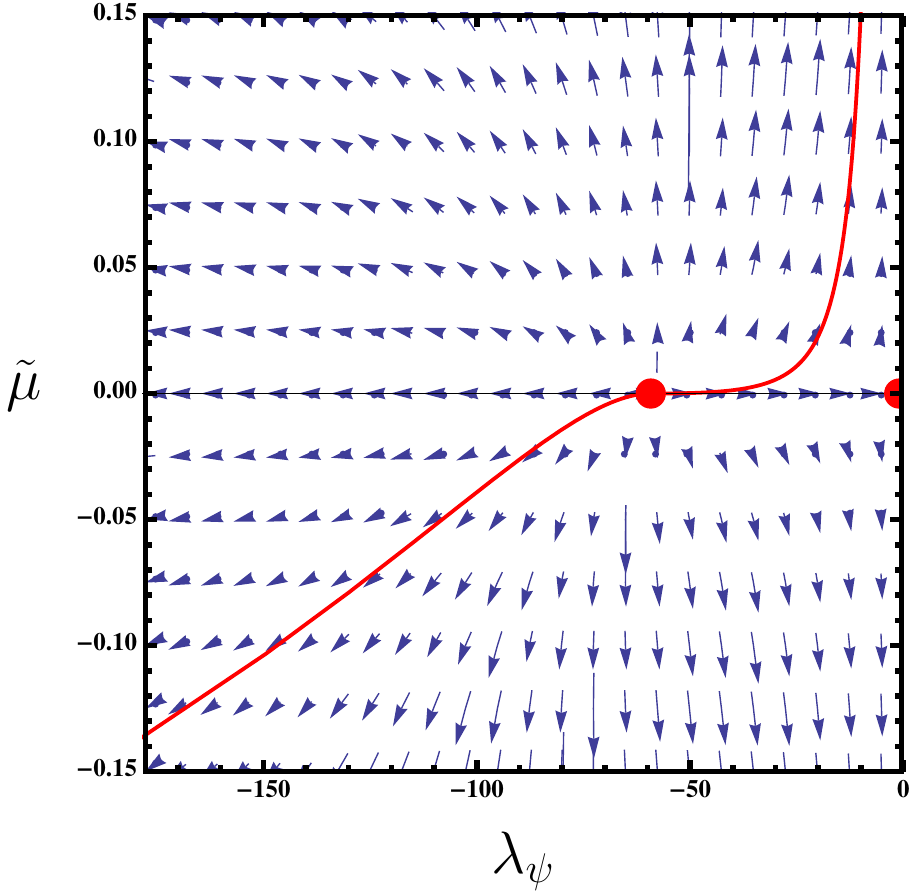}
\end{center}
\caption{RG flow of an ultracold Fermi gas in the plane spanned by the four-fermion coupling~$\lambda_{\psi}$ and the 
dimensionless chemical potential~$\tilde{\mu}$. The red (straight) line depicts the separatrix 
which separates the regime with a dynamically broken U(1) symmetry in the IR from a weakly interacting IR regime.
The dots represent the Gau\ss ian and the non-Gau\ss ian fixed point, respectively. The latter can be viewed as quantum critical point, see main text.
The arrows indicate the direction of the RG flow towards the infrared.}
\label{fig:BECBCSsep}
\end{figure}

In Fig.~\ref{fig:BECBCSsep} we show the RG flow of an ultracold Fermi gas in the plane spanned by the dimensionless 
four-fermion foupling~$\lambda_{\psi}$ and the dimensionless chemical potential~$\tilde{\mu}=\mu/k^2$. The red line
depicts the separatrix~$\lambda_{\psi}^{\rm sep.}(\tilde{\mu})$, 
which separates a weakly interacting IR regime from a strongly interacting regime associated with
spontaneous U(1)-symmetry breaking. By definition, we have~$\lambda_{\psi}^{\rm sep.}(\tilde{\mu}\!=\! 0)=\lambda_{\psi}^{\ast}$.
For large positive values of~$\tilde{\mu}$, the separatrix tends to
zero,~$\lambda_{\psi}^{\rm sep.}(\tilde{\mu})\sim \tilde{\mu}^{-1/2}$. On the other hand, it approaches the asymptote~$\tilde{\mu}=-1$ for large
negative values of~$\lambda_{\psi}$. 

Let us now discuss the (physical) meaning of the initial value~$\lambda_{\psi}^{\rm UV}$ at the scale~$\Lambda$.
Choosing $\lambda_{\psi}^{\rm UV} < \lambda_{\psi}^{\rm sep.}(\tilde{\mu})$, 
we find that the four-fermion coupling increases rapidly and diverges eventually at a finite scale~$\ksb$. This indicates the breakdown
of the U($1$)~symmetry of the ground state, see Fig.~\ref{fig:BECBCSsep}.
From Eq.~\eqref{eq:lpsiSL}, on the other hand, it follows that
\be
\Delta_{\psi}:=\frac{1}{\lambda_{\psi}^{\ast}} - \frac{1}{\lambda_{\psi}^{\rm UV}} \sim -\frac{1}{a_{\rm s}\Lambda}\,, \label{eq:rellpsiaNR}\,
\ee
where we have tacitly assumed that~$\lambda_{\psi}^{\rm UV}$ is of the order of the fixed-point 
value~$\lambda_{\psi}^{\ast}$. Since the s-wave scattering length~$a_{\rm s}$ is directly related to the applied (external) magnetic field~$B$,
see Eq.~\eqref{eq:SLB}, it follows that a variation of~$\lambda_{\psi}^{\rm UV}$ corresponds to a variation of~$B$. Recall that 
the latter is an experimentally controllable parameter. 

From the action~\eqref{eq:UGaction} we deduce that a negative chemical potential has a similar effect as a fermion mass. In other
words, a negative chemical potential suppresses the fermion propagation and the dynamics of the 
system is rather governed by bosonic bound states of two fermions.  
The binding energy of these bosons
is twice the chemical potential of the fermions~\cite{Diehl:2005an}. We conclude that the low-energy limit 
for~$\lambda_{\psi}^{\rm UV} < \lambda_{\psi}^{\rm sep.}(\tilde{\mu}_{\rm UV})$ (i.~e. positive s-wave scattering length)
and~$\tilde{\mu}_{\rm UV}=\mu/\Lambda^2<0$
defines a superfluid phase of bosonic bound states: these bosons macroscopically 
occupy the ground-state and form a Bose-Einstein condensate.\footnote{This is signaled by a finite expectation value of the order-parameter potential.}

Now we turn to the limit of infinite s-wave scattering length. Here, the initial condition for the four-fermion coupling
is given~$\lambda_{\psi}^{\rm UV}=\lambda_{\psi}^{\ast}$. From Fig.~\ref{fig:BECBCSsep} it follows immediately that
condensation can then only occur for~$\tilde{\mu}>0$. For~$\mu/\Lambda^2 \ll 1$,
we can approximate~$l(\tilde{\mu})$ by~$l(\tilde{\mu})\approx 1+\frac{3}{2}\tilde{\mu}$. This
allows us to solve the flow equation for the four-fermion coupling~\eqref{eq:lambdapsiNR} analytically. 
For~$\lambda_{\psi}^{\rm UV}=\lambda_{\psi}^{\ast}$ we then find
that the symmetry breaking scale~$\ksb$ is solely determined by the density, as it should be in the universal regime:
\be
\ksb \sim \sqrt{\mu} \sim n^{\frac{1}{3}}\,.
\ee
Here, we have dropped a scheme-dependent constant of proportionality. As discussed in detail for the NJL model in Sect.~\ref{subsec:bos},
the scale~$\ksb$ sets the scale for all low-energy observables, e.~g. the fermion gap or the critical temperature.
We conclude that all physical low-energy observables are fully determined by our choice for the density~$n$.

Finally we discuss the regime defined by a positive chemical potential and~$\lambda_{\psi}^{\rm UV}\neq\lambda_{\psi}^{\ast}$.
The flow equation for the $\lambda_{\psi}$-coupling can again be solved analytically and assumes a simple form, provided we approximate
the function~$l(\tilde{\mu})$ by $l(\tilde{\mu})\approx 1 + 3\sqrt{\tilde{\mu}}$; the limit $\tilde{\mu}=0$ 
as well as the limit $\tilde{\mu}\gg 1$ are correctly reproduced by this approximation. The solution for~$\lambda_{\psi}$ then reads
\be
\frac{1}{\lambda_{\psi}(k)} = \frac{1}{\lambda_{\psi}^{\ast}} - \left(\frac{\Lambda}{k}\right)\Delta_{\psi}
-\frac{1}{2\pi^2}\sqrt{\frac{\mu}{k^2}}\ln\left(\frac{k}{\Lambda}\right)\,.
\ee
From this expression we can derive the symmetry breaking scale~$\ksb$. For \mbox{$\ksb/\Lambda\ll 1$}, we find
\be
\ksb =\Lambda \exp\left({-\frac{2\pi^2 \Lambda \Delta_{\psi}}{\sqrt{\mu}}}\right)\sim \Lambda 
\exp
\left( 
-\frac{c_{\text{\tiny BCS}}}{|a_{\rm s}| n^{\frac{1}{3}}}
\right)\,,
\ee
where we have used Eq.~\eqref{eq:rellpsiaNR} and~$a_{\rm s}<0$; $c_{\text{\tiny BCS}}$ is a positive constant.
Thus, $\ksb$ scales exponentially in the BCS regime. Moreover, we observe
that~$\ksb$ depends only on the dimensionless quantity~$a_{\rm s}n^{1/3}$. With decreasing $a_{\rm s}n^{1/3}$, the scale~$\ksb$ decreases exponentially. 
As an immediate consequence, we expect that all low-energy observables (fermion gap, phase transition temperature, $\dots$) 
depend only on the value of~$a_{\rm s}n^{1/3}$ for~$a_{\rm s}<0$
and change exponentially when~$a_{\rm s}n^{1/3}$ is varied. 
This was first observed by Bardeen, Cooper and Schrieffer~\cite{Bardeen:1957mv} and then further investigated 
by Gorkov and Melik-Barkhudarov~\cite{Melik}. 
We stress that $\ksb$ does not scale exponentially on the BEC side for 
increasing~$1/(a_{\rm s}n^{\frac{1}{3}})$. In fact, it is well known that the phase transition temperature of a Bose gas is finite, even
in the non-interacting limit.

We would like to point out that the transition from a BCS-type superfluid to a BEC-type
superfluid at~$1/|a_{\rm s}| \to 0$ should not be confused with a quantum phase transition. On the contrary, the U(1) symmetry of the
ground state is spontaneously broken in the (deep) IR on both sides of the Feshbach resonance. Thus, the
transition from positive to negative s-wave scattering length rather corresponds to a smooth crossover between two phenomenologically different
superfluid regimes. 

We have seen that the analysis of the fixed point structure and the RG flows in the U(1) symmetric regime already allows us 
to understand many aspects of ultracold Fermi gases in the limit of a broad Feshbach resonance. Moreover, such an analysis provides insights into the 
dynamics away from the Feshbach resonance. Let us close this section by discussing the limit
of a {\it narrow} Feshbach resonance. To this end, it is convenient to consider the partially bosonized form of the effective
action~\eqref{eq:UGaction}. Following our discussion in Sect.~\ref{subsec:bos}, 
this form can be obtained by introducing a complex scalar field~$\varphi \sim (\bar{h}_{\varphi}/\bar{m}^2_{\varphi})(\psi_{\uparrow}\psi_{\downarrow})$ 
into the path integral by means of a Hubbard-Stratonovich transformation. The complex field~$\varphi$ describes the dynamics of  
the bound states; the prefactors $\bar{h}_{\varphi}$ and~$\bar{m}_{\varphi}$ are {\it a priori}
constants at our disposal and are chosen such that the four-fermion interaction term is canceled,~$\bar{\lambda}_{\psi}=-\bar{h}_{\varphi}^2/\bar{m}_{\varphi}^2$, 
see also Sect.~\ref{subsec:bos}. Due to the Hubbard-Stratonovich transformation, the (classical) action now includes a Yukawa-type interaction 
term~$\sim\bar{h}_{\varphi}$
and a mass term~$\bar{m}_{\varphi}$ for the composite field~$\varphi$:
\be
S = \int d\tau \!\int d^3 x\, \Big\{ \psi^{\dagger}
\left( \partial _{\tau}  -  \Delta -\mu \right) \psi 
+  \bar{m}_{\varphi}^2 \varphi^{\ast}\varphi
- \bar{h}_{\varphi}\left[ \varphi^{\ast}\psi_{\uparrow}\psi_{\downarrow}\! -\! \varphi\, \psi_{\uparrow}^{\ast}\psi_{\downarrow}^{\ast}\right]
\label{eq:YukNR}
\Big\}\,.
\ee
Due to quantum corrections, the Yukawa interaction generates kinetic terms for the bosonic fields in the RG flow,
\be
Z_{\varphi}^{\|}\varphi^{\ast}\partial_{\tau}\varphi
\quad\text{and}\quad
Z_{\varphi}^{\perp}\varphi^{\ast}\Delta\varphi\,, 
\ee
even if these terms have been set to zero at the initial RG scale~$\Lambda$. The (non-trivial) momentum-dependence of the four-fermion vertex 
is to some extent encoded in these kinetic terms. Close to a narrow Feshbach resonance, the details of the interaction (potential)
of the atoms matter. The wave-function renormalizations~$Z_{\varphi}^{\|,\perp}$ should then be rather considered as parameters of the theory,
which in general assume finite values even at the initial scale~$\Lambda$, see also below.

In leading order in a derivative expansion, the anomalous dimensions associated with the wave-function renormalizations~$Z_{\varphi}^{\|,\perp}$
are essentially given by a purely fermionic 1PI diagram in the U($1$) symmetric regime. Thus, we have
\be
\eta_{\varphi}^{\|,\perp} = c_{\varphi} \bar{h}_{\varphi}^2\,,\label{eq:etaflowNR}
\ee
where $c_{\varphi}$ is a positive constant~\cite{Diehl:2009ma}.
Contributions from 1PI diagrams with one internal fermion line and one internal boson line are suppressed
in the symmetric regime due to the large (renormalized) boson mass, at least in the limit of a broad 
Feshbach resonance.
For the same reason, the running of the fermionic wave-function renormalizations is subleading.
Finally, the RG flow equation of the Yukawa coupling also assumes a simple form
in the symmetric regime, since it is only driven by the anomalous dimensions of the bosonic 
fields:
\be
\partial_t h^2_{\varphi} = (\eta^{\perp}_{\varphi}-1)h^2_{\varphi}\,,
\ee
where $h_{\varphi}^2=\bar{h}_{\varphi}^2/(Z_{\varphi}^{\perp}k)$, see Ref.~\cite{Diehl:2009ma}. 
Note that we are free to choose either~$Z_{\varphi}^{\|}$ or~$Z_{\varphi}^{\perp}$ to renormalize the Yukawa coupling. 
Since we have $\lambda_{\psi}\equiv -h_{\varphi}^2/m_{\varphi}^2$, 
we conclude that a non-trivial fixed-point of the four-fermion coupling~$\lambda_{\psi}$ 
requires that $\eta^{\perp}_{\varphi}=1$. From Eq.~\eqref{eq:etaflowNR} it is then clear that~$(h_{\varphi}^{\ast})^2 > 0$.  

In our study of the NJL model with one species in Sect.~\ref{subsec:bos},  we have also found that  the Yukawa coupling is only 
driven by the anomalous dimensions of the bosonic fields.\footnote{Recall that the partially bosonized 
version of our NJL model with one fermion species can be rewritten in terms of
a complex scalar field~$\varphi=(\phi_1+\I\phi_2)/\sqrt{2}$. The associated Yukawa interaction term
in Eq.~\eqref{eq:bosaction} is then given by~$\sim\bar{h}_{\sigma}\left[ \varphi \bar{\psi}_{\rm R}\psi_{\rm L} - \varphi^{\ast}\bar{\psi}_{\rm L}\psi_{\rm R}\right]$,
where the left- and right-handed fermions are defined as~$\psi_{\rm L,R}=(\mathbbm{1}\pm\gamma_5)\psi$. This reformulation 
of the Yukawa interaction term in the NJL model with one fermion species is clearly reminiscent of the Yukawa interaction given in Eq.~\eqref{eq:YukNR}.
Note that the mass dimension of the Yukawa coupling in relativistic and non-relativistic theories in $d=4$ space-time dimension 
is different, see also App.~\ref{app:units}.}
However, the Yukawa coupling is marginal in the NJL model in $d=4$ space-time
dimensions. Strictly speaking, the non-trivial fixed-point of the four-fermion coupling in the NJL
model in~$d=4$ is an artifact of our point-like approximation and is only there for a finite UV cutoff~$\Lambda$. In the present case of a non-relativistic theory, the
non-trivial four-fermion fixed-point also exists in the limit~$\Lambda\to\infty$ beyond the point-like approximation. In this respect, our non-relativistic model should
be rather compared to an NJL model with one fermion species in $d=3$ space-time dimensions, where the Yukawa coupling is also a relevant coupling. We shall come
back to this when we discuss the partially bosonized form of the Gross-Neveu model in $d=3$ space-time dimensions in Sect.~\ref{sec:GNPB}.

For the NJL model we have shown in Eq.~\eqref{eq:4psiMOMSPACE} that the partially bosonized form of the action allows us
to conveniently resolve momentum dependences of four-fermion vertices. A corresponding expression can be derived for the 
present non-relativistic theory. In any case, we conclude that the  
Gau\ss ian fixed point of the four-fermion coupling~$\lambda_{\psi}\sim h_{\varphi}^2/m_{\varphi}^2$ is associated with the Gau\ss ian fixed-point of the 
Yukawa coupling~$h_{\varphi}$. Since~$\bar{\lambda}_{\psi}\sim a_{\rm s}$, it also follows that
the width of the resonance is directly related to the value of the Yukawa coupling at the initial RG scale, see Eq.~\eqref{eq:SLB}. 
The narrow resonance limit is therefore approached with~$\bar{h}_{\varphi}^2\to 0$, corresponding to the Gau\ss ian fixed-point
of the four-fermion coupling.

As discussed above, {\it universality} in the 
limit of a broad Feshbach resonance means that the RG flow is governed by the non-trivial fixed-point.
In the partially bosonized formulation of the action~\eqref{eq:UGaction}, this is the case, if 
the initial Yukawa coupling $(\bar{h}^{\rm UV}_{\varphi})^2$ is chosen to be reasonably close to the fixed-point~$(h^{\ast}_{\varphi})^2\Lambda$. 
Close to the Gau\ss ian fixed-point, the theory can be
treated perturbatively~\cite{Diehl:2005an}. However, the details of the
underlying interaction potential (e.~g. the effective range~$r_{\rm e}$) now become important. These are encoded in the 
momentum dependence of the four-fermion vertex. 
Therefore we expect that a description of an ultracold Fermi gas close to the narrow Feshbach resonance in general depends on more than one parameter.
For a detailed discussion of this fixed point, we refer the reader to Ref.~\cite{Diehl:2007ri}.

Up to this point, we have mainly discussed the (UV) fixed-point structure of a non-relativistic Fermi gas. Although such a study already provides
us with important information about the properties of the theory, the phenomenologically relevant IR observables are not accessible. For a consideration
in the continuum limit, a derivative expansion of the partially bosonized action allows us to conveniently gain access to IR 
observables~\cite{Diehl:2007ri,Diehl:2007th,Floerchinger:2008qc,Floerchinger:2009pg,Diehl:2009ma,Scherer:2010sv}. 
However, a derivative expansion becomes inefficient when
we are interested in, e.~g., density profiles of finite systems. In the next section, we discuss
an RG approach which treats the density as an effective degree of freedom and therefore provides an alternative approach for a study of 
inhomogeneities of the ground state of strongly interacting fermionic theories.

\subsection{Excursion: Density Functional Theory and the Renormalization Group}\label{sec:DFTRG}

In general, the ground-state density of a given finite many-body problem is inhomogeneous, even in the case of a non-interacting
system. For example, the ground-state density~$n_{\rm gs}(\vec{x})$ of $N$~fermions in a harmonic oscillator potential is clearly not uniform: 
it varies rapidly for $|\vec{x}| \lesssim \ell _{\text{HO}}\sim 1/\sqrt{w}$ and approaches 
zero exponentially for $|\vec{x}| \gg \ell _{\text{HO}}$, where~$\omega$ is the oscillator frequency and~$\ell _{\text{HO}}$ denotes the 
length scale set by the oscillator potential. For increasing~$N$ the effective extent of the ground-state wave-function 
increases and inhomogeneities of the ground-state 
density are washed out. In the limit~$N\to\infty$, we effectively approach the continuum limit and the density becomes uniform.

We emphasize that it depends on the details of the theory under consideration, 
whether the continuum limit is approached rapidly for increasing particle number~$N$, see e.~g. Refs.~\cite{Ku:2008vk,Lee:2010qp,Endres:2011er,BraunDSMMS}
for explicit studies of this issue.
However, the ground-state of a theory might also be inhomogeneous  
in the continuum limit. For example, this can be the case in a (weakly) interacting theory 
if the chemical potentials of the spin-up and the spin-down fermions are
different. It was shown by Fulde and Ferrell and independently by Larkin and Ovchinnikov that only Cooper pairs with
a finite center-of-mass momentum can be formed if the difference in the chemical potentials is large~\cite{Fulde}. The 
finite center-of-mass momentum then renders the ground state inhomogeneous.
We would like to point out that the existence of inhomogeneous phases is not bound to non-relativistic systems, a stable
inhomogeneous ground state can also occur in relativistic theories. This has been explicitly shown for the high-density phase
of the Gross-Neveu model in $d=1\!+\! 1$ space-time dimensions~\cite{Thies:2003kk}.

In the following we are particularly interested in strongly-interacting many-body systems away from the continuum limit. In particular, we aim at a study
of a finite system of fermions interacting via a non-local interaction which is repulsive at short range and attractive at long range. Prominent examples
for such systems are nuclei, but also ultracold trapped Fermi gases fall into this class of systems. Indeed, the density profiles of protons and neutrons in (heavy) 
nuclei are reminiscent of the density profiles of the spin-up and spin-down fermions in spin-polarized Fermi gases. 
In order to study the ground-state properties of such systems, density functional theory (DFT) has proven to be useful.

Let us begin with a brief discussion of the underlying principles of DFT. To this end, we consider the following action:
\be
&& S[\psi^{\dagger},\psi] = \sum_{\sigma} \int d\tau\int d^d x\, \psi^{\dagger}_{\sigma}(\tau,\vec{x}) 
\left[ \partial _{\tau} -  \Delta + V(\vec{x}) \right] \psi_{\sigma}(\tau,\vec{x})\nn\\
&&\quad +\,\frac{1}{2}\sum_{\sigma,\sigma^{\prime}}  \int d\tau\!\int d^dx\! \int d^dy \,
\psi^{\dagger}_{\sigma}(\tau,\vec{x}) \psi^{\dagger}_{\sigma^{\prime}}(\tau,\vec{y})
U(\vec{x},\vec{y})
\psi_{\sigma^{\prime}}(\tau,\vec{y}) \psi_{\sigma}(\tau,\vec{x})\,,\label{eq:actionDFT}
\ee
where~$V(\vec{x})$ is a (background) potential. DFT is based on the famous 
{\it Hohenberg-Kohn theorem}~\cite{Hohenberg:1964zz}. For a given interaction potential~$U$, this theorem states that 
there exists a one-to-one correspondence between the ground-state density and the potential~$V(\vec{x})$ (up to an additive constant),
 at least for non-degenerate ground states. This implies that the ground-state density (uniquely)
determines the ground-state wave-function of the $N$-body problem under consideration. The latter can therefore be considered 
as a functional of the ground-state density. Moreover, the expectation value of any physical observable 
is determined by a unique functional of the ground-state density. In particular, this is true for the ground-state energy of the system and implies the existence
of an energy density functional~$E[n]$. Following the {\it Rayleigh-Ritz theorem}, the ground-state energy~$n_{\rm gs}(\vec{x})$ 
can then be obtained by minimizing~$E[n]$ with respect to the density:
\be
E_{\rm gs}=\min_n E[n]\,.
\ee
Moreover, it can be shown that the energy density functional in the limit of vanishing external potential~$V$,
the so-called {\it Hohenberg-Kohn} functional~$E_{\rm HK}$, is {\it universal} for a given interaction potential~$U$:
\be
E_{\rm HK}[n]=E[n] - \int d^d x\, n(\vec{x})V(\vec{x})\,.
\ee
These considerations can be generalized to the case
of degenerate ground states.\footnote{Here, we leave aside a discussion of the issue of $V$-representability.} 

Originally, DFT has been invented to compute efficiently ground-state energies of atoms. In recent years, however, DFT has also been successfully employed 
in condensed-matter physics and for studies of ground-state properties of (heavy) nuclei~\cite{Dobaczewski:2001ed,Stoitsov:2003pd,Bender:2003jk}. 
For reviews and introductions to DFT approaches in nuclear physics, we refer the reader to Refs.~\cite{Furnstahl:2007xm,Drut:2009ce}. 
At this point we would like to highlight an important difference between DFT studies of atoms and nuclei. In studies of the ground-state energy of atoms, the center-of-mass 
momentum of the system is essentially carried by the nucleus. This allows us to easily subtract the center-of mass energy of the system. 
The electrons in the atomic shell can then essentially be considered as an electron gas with vanishing center-of-mass motion in an external
Coulomb potential. In studies of ground-state
properties of nuclei, however, 
the center-of-mass energy cannot be simply subtracted since the constituents ,
namely the protons and neutrons, have almost identical masses. Therefore an accurate computation of ground-state energies (binding energies) 
of nuclei is generically
spoilt by the finite center-of-mass energy of the system. Fortunately, these contributions become smaller when the number of nucleons
increases. Nonetheless, a systematic computation of the corrections to the ground-state energies of nuclei due
to the center-of-mass motion of the system is inherently difficult~\cite{Engel,Jennings}.

The Hohenberg-Kohn theorem can be viewed as a starting point for an efficient description of many-body problems. However, the theorem
does not provide a recipe for the computation of the Hohenberg-Kohn functional. Similar to the effective action in conventional quantum field theory,
the Hohenberg-Kohn functional consists of infinitely many terms.\footnote{As we shall see below, the Hohenberg-Kohn functional is indeed
closely related to the effective action.}
Therefore it is in general not possible to write down the exact Hohenberg-Kohn functional for a given many-body problem. This implies that
an ansatz for this functional is required in order to compute the ground-state energy of a given many-body problem. 
The simplest approximation to the energy density functional
is the so-called local density approximation (LDA). This approximation can be obtained straightforwardly from the density dependence of the 
ground-state energy~$E_{\rm gs}(n)$ of the associated uniform many-body 
problem,\footnote{For example, LDA requires the knowledge of the ground-state energy of a uniform electron gas when we are interested in a 
computation of the ground-state energies of atoms. We add that the computation of ground-state energies of uniform systems is already a highly
non-trivial problem.}
where~$E_{\rm gs}(n)=V\epsilon_{\rm gs} (n)$ ($n={\rm const.}$).
The energy-density functional in LDA then corresponds to the coordinate-space integral of $\epsilon_{\rm gs}(n(\vec{x}))$. It is 
possible to show that LDA represents the lowest order in a derivative expansion of the exact
energy-density functional~\cite{Leeuwen}. Such an approximation might be justified in systems with weakly varying 
densities,\footnote{We stress that LDA should by no means be confused with the so-called local potential approximation (LPA) which
represents the lowest order in an expansion of the effective action in derivatives of the fields.} 
such as ultracold Fermi gases with a large number of atoms in an isotropic trap~\cite{Mueller1}.

Let us now make contact to the effective action approach to quantum field theories which underlies most of our studies in the present 
review.\footnote{For conciseness, 
we shall skip many details in the following. For reviews and introductions on this subject 
matter, we refer the reader to, e.~g., Refs.~\cite{Polonyi:2001uc,Furnstahl:2007xm,Drut:2009ce}.} To this end, we consider the following path integral
\be
Z [J]
=\int  \mathcal{D}\psi^{\dagger}  \mathcal{D}\psi\, \E ^{-S[\psi^{\dagger},\psi] + \sum_{\sigma} \int _0^{\beta} d\tau\int d^d x J_{\sigma}(\tau,\vec{x})
( \psi^{\dagger}_{\sigma}(\tau,\vec{x}) \psi_{\sigma}(\tau,\vec{x}))
}\equiv \E ^{W[J]}\,, \label{eq:DFTpath}
\ee
where the action~$S$ is defined in Eq.~\eqref{eq:actionDFT} and, for convenience, we assume that the Euclidean-time direction is compactified. 
In order to fix the particle number in a study of a finite many-body problem, we have
essentially two options: First, one can introduce chemical potentials into the path integral to fix the numbers of the various particle species. 
Second, one does not include chemical potentials into the path integral
but fixes the particle numbers by choosing appropriate boundary conditions for the equations of motion~\cite{Puglia2003145}. In the following
we shall follow the latter approach to fix the particle number since it turns out to be more convenient for studies of finite many-body problems.

In contrast to the conventional textbook approach to quantum field theories, we have  
coupled the external source~$J$ in Eq.~\eqref{eq:DFTpath} to a term which is bilinear in the fermion fields and can be viewed as a composite bosonic degree of freedom. 
As usual, we may define the classical field~$\rho _{\sigma}(\tau,\vec{x})$ as the (functional) derivative of~$W[J]$ with respect to the 
corresponding source~$J_{\sigma}(\tau,\vec{x})$:
\be
\rho _{\sigma}(\tau,\vec{x}) = \frac{\delta W[J]}{\delta J_{\sigma}(\tau,\vec{x})}\,.
\ee
Apparently, these classical fields are related to the particle densities.
Similar to the derivation of the 1PI effective action, we can now define a 2PPI effective action as follows:\footnote{Here, ``2PPI" stands for
``two-particle point-irreducible". A 2PPI diagram is a 1PI diagram that cannot be split into two by cutting two internal lines
attached to the same vertex.
To put it sloppily, this type of effective actions arises generically from the path-integral when one couples a local source term to a term bilinear in the fields.
For a more general discussion of the properties of 2PPI effective actions and 
RG flow equations thereof, we refer the reader to Ref.~\cite{Pawlowski:2005xe}.}
\be
\Gamma[\rho] = \sup_{\{J_{\sigma}\}} \left\{ -W[J] + \sum_{\sigma} \int _0 ^{\beta} d\tau \int d^d x\, J_{\sigma}(\tau,\vec{x}) \rho_{\sigma}(\tau,\vec{x})
\right\}\,.
\ee
The effective action~$\Gamma[\rho]$ determines the dynamics of the many-body
problem under consideration and
should be compared with the energy-density functional mentioned above in the context of the standard Hohenberg-Kohn DFT 
formalism. The exact equivalent of the energy density functional as introduced by Hohenberg and Kohn can be derived along these lines by employing
time-independent sources~$J_{\sigma}(\vec{x})$, see Refs.~\cite{PTP.92.833,1997cond.mat..2247V,PhysRevB.66.155113,Puglia2003145}. 

As in the conventional 1PI formalism, the effective action~$\Gamma[\rho]$ does
not dependent on the sources~$J_{\sigma}$, i.~e. $(\delta\Gamma[\rho]/\delta J_{\sigma})=0$.
Let us now consider the first functional derivative of~$\Gamma[\rho]$ with respect to the classical field~$\rho_{\sigma}$:
\be
\frac{\delta\Gamma[\rho]}{\delta \rho_{\sigma} (\tau,\vec{x})}=J_{\sigma}(\tau,\vec{x})\,.\label{eq:gsDFT}
\ee
The ground-state configuration~$\rho_{\sigma,\rm gs}$ 
is determined by this equation in the limit~$J_{\sigma}\to 0$. We add that this equation can be also viewed as the 
equation of motion of the composite degree of freedom~$\rho_{\sigma}$. 
From the solutions~$\rho_{\sigma,\rm gs}(\tau,\vec{x})$ of Eq.~\eqref{eq:gsDFT}, we obtain the 
(time-independent) ground-state density~$n_{\rm gs}(\vec{x})$:
\be
{n}_{\rm gs}(\vec{x}) = \frac{1}{\beta}\int _0^{\beta}d\tau\, \rho_{\rm gs}(\tau,\vec{x})\,.
\ee
The ground-state energy~$E_{\rm gs}$ can be obtained from an evaluation of the effective action~$\Gamma[\rho]$ at the ground-state~$\rho_{\rm gs}$:
\be
E_{\rm gs}=\lim_{\beta\to\infty}\frac{1}{\beta}\Gamma[\rho_{\rm gs}]\,.
\ee
We would like to add that the {\it universality} of the Hohenberg-Kohn functional~$E_{\rm HK}$ can be easily proven in
the effective action approach~\cite{Schwenk:2004hm}. It follows from the fact that background potential can be absorbed
into source term~$J_{\sigma}$ by a simple shift, $J_{\sigma}\to J_{\sigma}+V$. Exploiting this observation, we find
\be
\Gamma[\rho] =\Gamma_{\rm HK}[\rho]+ \sum_{\sigma}\int _0 ^{\beta} d\tau \int d^dx\, V(\vec{x})\rho_{\sigma}(\tau,\vec{x})\,,
\ee
where~$\Gamma_{\rm HK}[\rho]=\Gamma_{V=0}[\rho]$. Thus, the functional~$\Gamma_{\rm HK}[\rho]$ depends only
on our choice for the interaction potential but not on the background potential~$V$.

As mentioned above, the computation of the effective action~$\Gamma[\rho]$ for a given theory can be inherently difficult. Following
Refs.~\cite{PhysRevB.66.155113,Schwenk:2004hm,BSP}, we now present an RG flow equation which allows for a systematic computation of the
density functional~$\Gamma[\rho]$. Since this equation can be essentially derived along the lines of the Wetterich equation (see Sect.~\ref{sec:rg}), we shall be brief here.
To keep our discussion as simple as possible, we consider a system of $N$~spinless fermions. However, the derivation of the flow equation
is not bound to such a theory but 
can be straightforwardly generalized to other non-relativistic theories~\cite{Schwenk:2004hm}. To be specific, we consider an action of the following form: 
\be
&& S_{\gamma}[\psi^{\ast},\psi] = \int_0^{\beta} d\tau\int d^d x\, \psi^{\ast}(\tau,\vec{x}) 
\left[ \partial _{\tau} - \Delta +V_{\gamma}(\vec{x}) \right] \psi (\tau,\vec{x})\nn\\
&& \quad +\,\frac{\gamma}{2} \int _0^{\beta}\! d\tau\!\int \! d^dx\! \int _0^{\beta}\!d\tau^{\prime}\!\int\! d^dx^{\prime} \,
\psi^{\ast}(\tau,\vec{x})  \psi(\tau,\vec{x})
U(\tau,\tau^{\prime};\vec{x},\vec{x}^{\prime})
\psi^{\ast}(\tau^{\prime},\vec{x}^{\prime})
\psi (\tau^{\prime},\vec{x}^{\prime})\,.\label{eq:actionDFTRG}\nn
\ee
For convenience, we have reordered the fermion fields in the second term compared to Eq.~\eqref{eq:actionDFT}.
The parameter~$\gamma \in [0,1]$ denotes a dimensionless control parameter. For~$\gamma=0$, the two-body interaction potential~$U$
is turned off and we are left with an exactly soluble problem. For~$\gamma=1$, the potential~$U$ is fully turned on.
We shall assume that the interaction potential~$U$
is attractive at long range and repulsive at short range, such that the system can potentially be self-bound for~$V_{\gamma}\to 0$. 
In particular, we shall assume that the Fourier transform of the interaction potential~$U$ falls off sufficiently rapidly for large momenta
to avoid the occurrence of UV divergences. 
Examples for such a two-body potential are nucleon-nucleon interaction potentials. 
The one-body potential~$V_{\gamma}$ is at our disposal. However, it needs to be
chosen such that the $N$-body system has a finite extent~$\ell _{\rm V}$ for~$\gamma=0$, where the interaction term vanishes identically. 
This length scale~$\ell _{\rm V}$ sets a momentum scale~$1/\ell_{\rm V}$ which screens (potentially existing) IR divergences.
Moreover, our choice for~$V_{\gamma}$ partially defines the many-body problem under consideration. For a study of ground-state 
properties of nuclei, one may choose~$V_{\gamma}=\frac{1}{4}(1-\gamma)\omega^2 \vec{x}^2$ where~$\omega$ is the oscillator frequency, 
see Ref.~\cite{Schwenk:2004hm}. For a study of
trapped Fermi gases, on the other hand, one may choose a $\gamma$-independent potential~$V_{\gamma}=\frac{1}{4}\omega^2 \vec{x}^2$. (Note that
we have set $2m=1$ in our conventions.) 

Now it is straightforward to derive the effective action~$\Gamma_{\gamma}[n]$ associated with 
the (classical) action~$S_{\gamma}$. By taking the derivative of~$\Gamma_{\gamma}[\rho]$
with respect to~$\gamma$ we find the following flow equation~\cite{Schwenk:2004hm,BSP}: 
\be
\partial_{\gamma} \Gamma_{\gamma}[\rho] &=&(\partial_{\gamma} V_{\gamma}) \cdot \rho
+\frac{1}{2}  \rho\cdot U \cdot \rho  + \frac{1}{2}\,{\rm Tr}\, U\cdot \left( \frac{\delta^2 \Gamma[\rho]}{\delta \rho\delta \rho}\right)^{-1}\,,
\label{eq:DFTflow}
\ee
where the dot indicates a product in Euclidean space-time, i.~e. 
\be
A\cdot B\equiv \int _0^{\beta}d\tau\int d^dx A(\tau,\vec{x})B(\tau,\vec{x})\,.
\ee
This flow
equation governs the flow from the non-interacting system at~$\gamma=0$ to the (strongly) interacting system at~$\gamma=1$. 
In the terminology of many-body physics, the second term can be identified as the so-called Hartree term. The third term on the right-hand side
depends on the density-density correlator $\delta^2\Gamma[\rho]/(\delta \rho\delta \rho)$ and includes all higher order corrections to the 
effective action, e.~g. the so-called Fock term.

Note that the derivation
of the flow equation~\eqref{eq:DFTflow} does not require that the (effective) interaction strength is small. Moreover, this flow equation does not rely on an approximation
scheme, such as a gradient expansion. In fact, the flow equation~\eqref{eq:DFTflow} is exact, if we only allow for a two-body
interaction potential in the underlying action~$S_{\gamma}$. A generalization of this flow equation to include the effects of 
higher $n$-body operators is straightforward. In any case, 
it is in general not possible to solve the flow equation~\eqref{eq:DFTflow} exactly. Therefore, a systematic approximation scheme is required, such as an expansion of the
effective action~$\Gamma_{\gamma}[\rho]$ about the ground state~$\rho_{\rm gs}$. This would correspond to a vertex expansion in the terminology of
quantum field theory.

In order to obtain the initial condition of the flow equation~\eqref{eq:DFTflow} at~$\gamma =0$ for a given many-body problem, 
we need to solve the non-interacting $N$-body problem
defined by the action~$S_{\gamma}$ evaluated at \mbox{$\gamma=0$}. The effective action~$\Gamma_{\gamma=0}[\rho]$ associated with 
this exactly soluble $N$-body problem then determines the initial condition of the RG flow. For illustration, let us assume that we would like to compute
the ground-state energies of so-called Alexandrou-Negele nuclei~\cite{Alexandrou:1988jg}, i.~e. self-bound systems in~$d=1\!+\! 1$ Euclidean space-time dimensions
consisting of $N$~spinless fermions interacting
via a specific choice for a long-range attractive and short-range repulsive potential~$U$. A convenient and appropriate choice for 
the background potential~$V$ is then a harmonic oscillator potential~\cite{Schwenk:2004hm}:~$V_{\gamma}(\vec{x})=(1/4)(1-\gamma)\omega^2 x^{2}$. 
Thus, the initial condition at~$\gamma=0$ corresponds to a simple oscillator potential in which the $N$ lowest 
lying states are filled. In other words, the initial 
condition is simply given by a one-dimensional shell model ("mean-field" approximation). By lowering the control parameter~$\gamma$, i.~e. 
by solving the flow equation~\eqref{eq:DFTflow}, we gradually remove the background potential~$V$ and turn on the interaction potential~$U$. In the spirit of the RG,
removing the background potential corresponds to lowering the RG scale which is given by the inverse of the $\gamma$-dependent oscillator
length~$\ell _{\rm HO}\sim ((1\!-\!\gamma)^{\frac{1}{4}}\omega^{\frac{1}{2}})^{-1}$. For~$\gamma\to 1$, the background-potential is removed
and we are left with the fully interacting system. A quantitative study of Alexandrou-Negele nuclei with the presented RG approach to DFT is 
on its way~\cite{BSP}; preliminary results have been presented in Refs.~\cite{BraunRBRG,SchwenkDFTRG}.

Let us close this section by stating that the presented RG approach to DFT is promising for a study of finite
many-body problems since an expansion of the theory in terms
of the density scales favorably to large systems, i.~e. systems with many constituents. Since the presented RG-inspired approach 
allows for an ab-initio calculation of ground-state
properties of strongly interacting many-body systems, it might be promising tool for studies of ground-state properties
of (heavy) nuclei from microscopic interactions.



%
\section{Gross-Neveu and Nambu-Jona-Lasinio-type Models}\label{sec:njlgn}
In this section we apply the techniques discussed and developed in Sect.~\ref{sec:example}
to specific examples of relativistic quantum field theories, namely the Gross-Neveu model and QCD low-energy models.
While Gross-Neveu-type models play a prominent role in the description of (ferromagnetic) superconductors,
QCD low-energy models aim to describe the equation of state of hadronic matter under extreme conditions.
Recently, the Gross-Neveu model in two space-time dimensions has attracted a lot of attention in the high-energy physics
community, since its finite-temperature phase diagram can be computed analytically in the limit of many fermion flavors 
and shows an intriguing phase structure exhibiting inhomogeneous phases at high densities~\cite{Thies:2003kk}. 
Whether such phases also exist as stable ground states beyond this limit is still not clear.
Since relatives of the 
Gross-Neveu model, namely Nambu-Jona-Lasinio-type models, underlie the construction of QCD low-energy models,
it is tempting to speculate whether such inhomogeneous phases are also present in the phase diagram of $\,$3+1-dimensional field
theories, such as QCD~\cite{Nickel:2009wj,Kojo:2009ha}. Even though these are certainly interesting 
questions, we shall restrict ourselves in the following to much simpler issues which arise in studies of Gross-Neveu models 
and QCD low-energy models.

After a brief discussion of the Gross-Neveu model and its symmetries in Sect.~\ref{sec:GNmodelSYM}, we study the fixed-point 
structure of its purely fermionic formulation in Sect.~\ref{sec:GNfermionic}. The partially bosonized formulation and
quantum critical behavior are then analyzed in detail
in Sect.~\ref{sec:GNPB} in the large-$\Nf$ limit and in Sect.~\ref{sec:QPTGN} in next-to-leading order in the $1/\Nf$-expansion.
In addition to a discussion of quantum criticality, we highlight the differences between the derivative 
expansion and the $1/\Nf$-expansion of the effective action. 
We shall keep our discussion as general as possible. However, explicit numerical solutions are presented only 
for the 2+1-dimensional case, if not stated otherwise.
As a bonus, we relate quantum criticality to the issue of
non-perturbative renormalizability of quantum field theories in Sect.~\ref{sec:QCAS}. 

In Sect.~\ref{sec:NJLLQCD} we study aspects of QCD low-energy models in $d\!=\!3\!+\!1$ space-time
dimensions. To this end, we begin in Sect.~\ref{sec:LowQCDF} with 
a discussion of the Fierz ambiguity in QCD low-energy models and study the fixed-point structure of a
Fierz-complete low-energy model in the limit of many colors, the so-called large-$\Nc$ limit.
This limit corresponds to the limit of many-flavors
in the Gross-Neveu model. Quantum and thermal phase transitions in low-energy QCD models are then discussed
in Sect.~\ref{sec:TQPTQCD}, where we mainly focus on the large-$\Nc$ limit. In addition, we comment on the widely used so-called local potential
approximation (LPA) and its relation to the $1/\Nc$-expansion. This includes a discussion of the effects of the
next-to-leading order corrections in a $1/\Nc$-expansion.

In addition to the discussion of physical aspects of the Gross-Neveu model and QCD low-energy models, our goal is 
to highlight the field-theoretical differences and the similarities in the description of these models.
The latter allow for a cross-fertilization of QCD and condensed-matter physics.

\subsection{Gross-Neveu Model and Quantum Criticality}\label{sec:GNmodel}
\subsubsection{Gross-Neveu Model}\label{sec:GNmodelSYM}
The Gross-Neveu model represents a quantum field theory of $\Nf$ massless
(relativistic) fermion flavors in $d$ space-time dimensions which allows to study dynamical chiral symmetry
breaking. This model is related to the Peierls-Froehlich model and models for ferromagnetic (relativistic) superconductors, see 
e.~g. Refs.~\cite{PSSB:PSSB2221030242,Schnetz:2004vr}. Moreover, the Gross-Neveu model has attracted
a lot of attention in high-energy physics, since the finite-temperature phase boundary in $d=2$
can be studied analytically, at least in the limit of many flavors, see e.~g. Ref.~\cite{Thies:2003kk}. 

The classical action of the Gross-Neveu model reads
\be
\label{eq:fermionic_action}
S_{\text{GN}}[\bar{\psi},\psi]
=\int d^d x \left\{\bar{\psi}\mathrm{i}\fslash{\partial}\psi+ \frac{1}{2}\bar{\lambda}_{\sigma}(\bar{\psi}\psi)^2
\right\}\,,
\ee
where $(\bar{\psi}\psi)\equiv \bar{\psi}_i\psi_i$ and~$i=1,\dots,\Nf$.
The model depends on a single parameter which is the coupling constant $\bar{\lambda}_{\sigma}$ 
with mass dimension $2-d$. In $d=2$, the model is asymptotically free and perturbatively renormalizable, 
as the Gau\ss ian fixed point is UV attractive. In this respect the Gross-Neveu model
is reminiscent of QCD. On the other hand, $d=4$ turns out to be a marginal case, see our discussion of the NJL model in Sect.~\ref{subsec:bos}.
In the following we shall restrict ourselves mainly to the case $d\!=\! 3$ for which we shall employ a four-component 
(reducible) representation for the Dirac $\gamma$-matrices,~$\gamma_{\mu}=\{\gamma_0,\gamma_1,\gamma_2\}$, 
see App.~\ref{sec:dirac} for our conventions.

The action of the Gross-Neveu model is invariant under global U($\Nf$)
transformations of the fermion fields. This implies that the model is invariant under
$\text{U}(1)^{\otimes \Nf}$ transformations, i.e. the associated U($1$)-charge is conserved
in each flavor-sector separately.
The matrix $\gamma_{35}=\I \gamma_3\gamma_5$ further realizes 
a U$^{35}$($1$) symmetry in each flavor sector:
\be
\label{eq:vector_trafo}
\bar{\psi}_j \mapsto \bar{\psi}_j\mathrm{e}^{-\mathrm{i}\alpha\gamma_{35}},\,\quad
\psi_j \mapsto\mathrm{e}^{\mathrm{i}\alpha\gamma_{35}}\psi_j\,,
\ee
where~$\alpha$ denotes the ``rotation" angle.
In addition to these two symmetries, the Gross-Neveu model is invariant under 
discrete $\mathbbm{Z}_{2}^{5}=\{\mathbbm{1}_{4},\gamma_{5}\}$ 
chiral transformations acting on all flavors simultaneously: 
\be
\label{eq:chiral_trafo}
\bar{\psi} \mapsto -\bar{\psi}\gamma_{5}\,,\quad
\psi \mapsto \gamma_{5}\psi\,.
\ee
Note that a similar symmetry transformation involving $\gamma_3$ can be understood as a
combination of $\mathbbm{Z}_{2}^{5}$ and U$^{35}$($1$) transformations.

In Sect.~\ref{sec:GNPB} we shall see that the chiral symmetry of the Gross-Neveu model can be associated with a $\mathbbm{Z}_2$ 
symmetry for the order parameter. As we shall also discuss below, the infrared regime of the theory is governed by
dynamical chiral symmetry breaking, provided the {\it only} parameter of the model,
namely $\bar{\lambda}_{\sigma}$, is adjusted accordingly.

\subsubsection{Fermionic Formulation}\label{sec:GNfermionic}

Let us start with an analysis of the fermionic fixed-point structure of the Gross-Neveu
model. To this end, we consider only a point-like four-fermion
interaction. Our ansatz for the effective action then reads
\be
\Gamma_{\text{GN}}[\bar{\psi},\psi]=
\int d^d x\,\left\{Z_{\psi}\bar{\psi}\mathrm{i}\fslash{\partial}\psi+
 \frac{1}{2} {\bar{\lambda}_{\sigma}}(\bar{\psi}\psi)^2\right\}\,.
\label{eq:eff_action_fermion_GN}
\ee
As we have discussed in detail in Sect.~\ref{sec:example}, 
this ansatz can be viewed as a derivative expansion of the
effective action, with the leading-order defined by a constant $Z_{\psi}$. 
A running wave-function renormalization then corresponds
to the next-to-leading order approximation. Aside from derivatives,
further fermion interaction channels and higher-order interactions compatible with the
symmetries can be taken into account. Recall that $Z_{\psi}$ is indeed constant
when we treat the four-fermion interaction in the point-like limit, see discussion of
Eq.~\eqref{eq:ZpsiDisc}. Thus, we set $Z_{\psi}\equiv 1$ in the following.

The RG flow equation for the four-fermion coupling 
can be derived along the lines of Sect.~\ref{subsec:simpleex}. We find 
\be
\partial _t \lambda_{\sigma} = (d - 2)\lambda_{\sigma} - 4(\Nf d_{\gamma} \!- \! 2 )
 v_d\,l_1^{({\rm F}),(d)}(0;0)\, \lambda_{\sigma}^2\,,
\label{eq:GNfourfermionflow}
\ee
where
\be
\lambda_{\sigma} =  k^{d-2} \bar{\lambda}_{\sigma}\,,
\ee
and $d_{\gamma}=4$ denotes the dimension of the Dirac algebra. 
For the sake of simplicity, we have dropped contributions from further
(possibly fluc\-tuation-induced) interaction channels as well as dependences on
the Fierz basis, see discussion in Sect.~\ref{subsec:simpleex} and 
Refs.~\cite{Jaeckel:2002rm,Jaeckel:2003uz,Gies:2009da}. Nevertheless, we expect this
to be a reasonable approximation for two reasons: first, we find that the considered
scalar interaction channel does not generate other interaction 
channels.\footnote{Note that this does not mean that the ansatz~\eqref{eq:eff_action_fermion_GN} is closed
with respect to Fierz transformations. In fact, other four-fermion interactions compatible
with the symmetries may generate contributions to the flow of the coupling~$\lambda_{\sigma}$.}
Second,  we are particularly interested in a study of the Gross-Neveu model in the large-$\Nf$ limit
which we consider as a controlled starting point for the construction of an effective (low-energy) theory for this model.
This means we assume that a pure Gross-Neveu-type RG trajectory exists in the large-$\Nf$ limit,
i.~e. a trajectory on which only the $\lambda_{\sigma}$-coupling is finite but all other four-fermion
couplings are identical to zero.
In Sect.~\ref{subsec:NJLUNf} we have shown that a fermionic model with similar symmetry properties
as the Gross-Neveu model
indeed exhibits a trajectory which corresponds to a pure Gross-Neveu-type trajectory in our present
study. However, further studies are needed to confirm the existence of such a trajectory in the
Gross-Neveu model.

Apart from a Gau\ss ian fixed point, we find a second non-trivial fixed point for the coupling $\lambda_{\sigma}$ which is given by 
\be
\lambda^{\ast}_{\sigma}=\frac{d - 2}{4 (\Nf d_{\gamma}\! - \! 2)v_d\, l_1^{({\rm F}),(\rm d)}(0;0)}\,.
\ee
For the optimized regulator function defined in Eq.~\eqref{eq:fermreg} we find 
\be
\lambda^{\ast}_{\sigma}=\frac{d(d-2)}{8(\Nf d_\gamma\!-\! 2) v_d}\stackrel{(d=3)}{=}\frac{3\pi^2}{(4\Nf\!-\! 2)}\,.\nn
\ee
Recall that the fixed-point value is a non-universal quantity as indicated by
the regulator dependence. Nevertheless, the mere existence of the fixed
point is a universal statement. 

For a sketch of $\partial_t \lambda_{\sigma}$ as 
a function of $\lambda_{\sigma}$ we refer to Fig.~\ref{fig:parabola}. 
As discussed in detail in Sect.~\ref{sec:example}, the choice for the
initial condition $\lambda_{\sigma}^{\rm UV}$ relative to the fixed-point $\lambda_{\sigma}^{\ast}$
distinguishes between two different
phases in the long-range limit: for $\lambda_{\sigma}^{\rm UV}<  \lambda_{\sigma}^{\ast}$, 
we approach a trivial phase in the IR limit, whereas we run into a phase with broken
chiral symmetry in the ground state for $\lambda_{\sigma}^{\rm UV}>  \lambda_{\sigma}^{\ast}$.
Thus, the fixed-point~$ \lambda_{\sigma}^{\ast}$ can be considered as 
a quantum critical point which divides the model into two physically different regimes.

The scale for a given IR observable $\mathcal O$ 
is set by the scale $k_{\rm SB}$ at which $1/\lambda_{\sigma}(k_{\rm SB}) \to 0$. 
In our study of power-law scaling in Sect.~\ref{sec:NJLPL}, we have discussed how 
this scale can be obtained from the flow equation of the four-fermion coupling. We proceed along these
lines to compute~$k_{\rm SB}$. From the flow equation~\eqref{eq:GNfourfermionflow} we then find
\be
k_{\rm SB}=\Lambda \theta(\lambda_{\sigma}^{\rm UV}-\lambda_{\sigma}^{\ast})
 \left(\frac{ \lambda_{\sigma}^{\rm UV}-\lambda_{\sigma}^{\ast}}{\lambda_{\sigma}^{\rm UV}}\right)^{\frac{1}{|\Theta|}},\label{eq:scalingGN} 
\ee
where the critical exponent $\Theta$ is given by
\be
  \Theta=- \frac{\partial (\partial_t \lambda_{\sigma})}{\partial \lambda_{\sigma}}\Big|_{\lambda_{\sigma}^{\ast}}
 = d-2\,.
\label{eq:fermThetaGN}
\ee
Thus, relation~\eqref{eq:scalingGN} determines how a given  
physical IR observable scales when $\lambda_{\sigma}^{\rm UV}$ is varied. In particular, 
we observe that the exponent~$\Theta$ governing the scaling behavior does not depend on~$\Nf$ in 
the present approximation.

Let us conclude with a word of caution concerning the derivative expansion in the fermionic truncation. In this simple approximation, the fixed point seems to
exist with similar properties in any dimension $d>2$, in particular also in~$d=4$ and beyond. This conclusion changes when composite bosonic degrees
of freedom are taken into account in the subsequent section. In the purely fermionic description, the bosonic degrees of
freedom correspond to specific nonlocal interactions or momentum-structures in the fermionic vertices. These are not properly resolved in a derivative
expansion. As $d=4$ is a marginal case, the conclusions for fermionic theories in $d=4$ may depend on the details of the interaction and the algebraic
structure of a given model.

\subsubsection{Partial Bosonization and the Large-$\Nf$ Limit}\label{sec:GNPB}
We now study the fixed-point structure of the Gross-Neveu model by employing its (partially) bosonized version.
We follow closely our discussion of (partial) bosonization in Sect.~\ref{subsec:bos}
We relate our findings to the purely fermionic description, study the
large-$\Nf$ limit analytically and show how corrections beyond the mean-field approximation can be systematically taken into account. 
In particular, we demonstrate how the quantum critical behavior of the model is affected by the inclusion of momentum dependences and corrections
beyond the large-$\Nf$ limit.
Note that the partially bosonized formulation has been used to study various aspects of the Gross-Neveu
model, such as the phase structure at zero and finite temperature and density~\cite{Hands:1995jq,Thies:2003kk,Castorina:2003kq}.

A partially bosonized description of the Gross-Neveu model is appealing from a field-theoretical point of view as it also forms the basis for an expansion in powers 
of~$1/\Nf$. In addition, it allows us to systematically resolve parts of the momentum dependence of the vertices by
means of a derivative expansion. As we shall see below, these two expansion
schemes are {\it not} identical and should therefore not be confused with each
other. For our study we use the following ansatz for the effective action:
\be
\Gamma[\bar{\psi},\psi,\sigma]=\int d^d x \left\{\frac{{1}}{2}Z_{\sigma}(\partial_{\mu}\sigma)^2+
Z_{\psi}\bar{\psi} \mathrm{i}\fslash{\partial}\psi + \mathrm{i}\bar{h}_{\sigma}\bar{\psi}\sigma \psi 
+ \frac{1}{2}\bar{m}_{\sigma}^2 \sigma ^2 + \frac{1}{8}\bar{\omega}_{\sigma}\sigma^4
\label{eq:PBgammaGN}
\right\}\,,
\ee
where we allow all couplings and wave-function renormalizations to be scale dependent. Note that we drop possibly generated higher-order
bosonic self-interaction terms for the sake of simplicity. As we have discussed in Sect.~\ref{subsec:bos}, cf. Eq.~\eqref{Eq:HSTAction},
the kinetic term of the boson field allows us to go beyond the local approximation of simple mean-field theory. In terms of the fermionic language, this kinetic
term corresponds to a specific momentum dependence in the scalar interaction channel of the four-fermion coupling. Thus, it allows us to conveniently resolve (part of) the
momentum dependence of the associated four-point function. As we shall see below, this
term and the associated wave-function renormalization receive contributions to leading order in the large-$\Nf$ approximation. 
Therefore the large-$\Nf$ approximation corresponds to the following choice for the 
initial conditions of $Z_{\sigma}$ and $Z_{\psi}$:
\be
&& \lim _{k\to\Lambda} Z_{\sigma}= 0\,,\qquad \partial_t Z_{\sigma}\neq 0\,, 
\qquad \lim_{k\to\Lambda}  Z_{\psi}= 1\,,\qquad \partial_t Z_{\psi}\equiv 0\,.
\ee
This choice for the initial conditions exemplifies nicely the difference between
large-$\Nf$ and derivative expansions, as we include next-to-leading order corrections in terms of a
derivative expansion in the bosonic sector but treat the fermionic sector in the leading-order approximation. 

As detailed in Sect.~\ref{subsec:bos}, the choice for the initial conditions allows us to map the partially
bosonized action~\eqref{eq:PBgammaGN} exactly onto the purely fermionic ansatz~\eqref{eq:eff_action_fermion_GN} 
at the initial RG scale~$\Lambda$. In fact, we fix the bosonized Yukawa-type model to the fermionic
Gross-Neveu model by a suitable choice of initial conditions for $\Gamma$ at
the scale $\Lambda$: 
\be
\lim_{k\to \Lambda} Z_{\sigma} = 0\,,\quad
\lim_{k\to \Lambda} Z_{\psi} = 1\,,\quad
\lim_{k\to \Lambda} \bar{\omega}_{\sigma} = 0\,.  \label{eq:BCGN}
\ee
Thus, the renormalized boson mass $m_{\sigma}=\bar{m}_{\sigma}/Z_{\sigma}^{1/2}$ at the UV cutoff
$\Lambda$ becomes much larger than $\Lambda$ and renders the boson propagator
essentially momentum independent. This so-called compositeness condition for the bosonic
formulation can be considered as a locality condition at the UV scale for
the four-fermion coupling $\lambda_{\sigma}$ in the purely fermionic formulation of the
model.

Since we are interested in the quantum critical behavior associated with the fixed-point structure\footnote{More precisely, quantum 
critical behavior in the Gross-Neveu model is associated with the UV fixed-point structure, see also Sect.~\ref{sec:QCAS}.} of 
the partially bosonized Gross-Neveu model, we anticipate that a possible non-Gau\ss ian fixed point occurs in the
symmetric regime. Therefore, we only need to study the RG flow in the
symmetric regime with vanishing vacuum expectation value for the $\sigma$ field. 
The flow equations for the Yukawa coupling~$h$ as well as the anomalous
dimensions $\eta_{\sigma}=-\partial_t \ln Z_{\sigma}$ and
\mbox{$\eta_{\psi}=-\partial_t \ln Z_{\psi}$} read\footnote{These flow
equations agree with those derived in Refs.~\cite{Hofling:2002hj,Gies:2009hq}.}
\be
\partial_{t}h^2_{\sigma}  &=&  (d-4 + 2\eta_{\psi}+\eta_{\sigma})h^2_{\sigma}  
+ 8 v_d\,h^4_{\sigma} \, l_{1,1}^{\rm (FB),(d)}(0,\epsilon_{\sigma};\eta_{\psi},\eta_\sigma)\,, \label{eq:hflowGN}\\ 
\eta_{\sigma}  &=&  8 \Nf
\frac{d_{\gamma}v_d}{d}h^2_{\sigma}\,m_{4}^{\rm(F),(d)}(0;\eta_{\psi})\,, \label{eq:etasigmaGN}\\ 
\eta_{\psi}  &=& 8 \frac{v_{d}}{d}h^{2}_{\sigma}\,m_{1,2}^{\rm (FB),(d)}(0,\epsilon_{\sigma};\eta_{\psi},\eta_{\sigma})\,,\label{eq:etapsi_flowGN}
\ee
where $\epsilon_{\sigma} = \bar{m}_{\sigma}^2/(Z_{\sigma}k^2)$, $\omega_{\sigma}=\bar{\omega}_{\sigma}/(Z_{\sigma}^2k^{(d-4)})$
and $h^2_{\sigma}=\bar{h}^2_{\sigma}/(Z_{\sigma}Z_{\psi}^2 k^{(d-4)})$.
The definition of the threshold functions can be found in App.~\ref{app:regthres}. The RG flow equations
of the bosonic self-interactions are given by  
\be
\partial _t \epsilon_{\sigma} &=&  (\eta_{\sigma}-2)\epsilon_{\sigma} - 6v_d\, l_1^{(d)}(\epsilon_{\sigma};\eta_{\sigma}) \omega_{\sigma}
 +\,4\Nf d_{\gamma} v_d\, l_{1}^{\rm (F),(d)}(0;\eta_{\psi}) h^2_{\sigma}\,,\label{eq:m2flowGN}\\
\partial _t \omega_{\sigma}&=& (d-4 + 2 \eta_{\sigma})\omega_{\sigma} + 18 v_d l_2^{(d)}(\epsilon_{\sigma};\eta_{\sigma}) \omega_{\sigma}^2 
-\,8\Nf d_{\gamma} v_d\, l_{2}^{\rm (F),(d)}(0;\eta_{\psi}) h^4_{\sigma}\,.\label{eq:lflowGN}
\ee
From these flow equations we deduce that 1PI diagrams with at least one inner bosonic line do not contribute
to the flow of the couplings in the large-$\Nf$ limit. As a consequence, $\eta_{\sigma}$ is non-vanishing in
leading order in the $1/\Nf$ expansion, whereas the fermionic anomalous dimension is zero.
We emphasize that the large-$\Nf$ counting is very different from that in scalar O($N$) models, where
the anomalous dimensions are zero to leading order at large~$N$, see e.~g. Ref.~\cite{ZinnJustin:2002ru}.
This already makes clear that such a large-$N$ expansion should not be confused
with the large-$\Nf$ expansion in the Gross-Neveu model. In fact, 
we always have to deal with an ${\rm O}($1$)\simeq {\rm Z}(2)$ symmetric order-parameter potential in the Gross-Neveu model
due to its underlying symmetries.

At this point it is instructive to make contact to expansion schemes employed in the context of high-energy physics.
The parameter $\Nf$ counting the number of flavors in the Gross-Neveu model 
should then be compared to the number of colors in QCD.  Thus, a large-$\Nc$ expansion in QCD (models) corresponds to a large-$\Nf$ 
expansion in the Gross-Neveu model.\footnote{For an analysis of the role of corrections beyond the large-$\Nc$
approximation for the thermodynamics of QCD low-energy models we refer to Sect.~\ref{sec:NJLLQCD} 
and~Ref.~\cite{Braun:2009si}.} On the other hand, the number of fermion flavors in QCD is
directly related to the number of Nambu-Goldstone modes and therefore to the symmetry properties
of the order-parameter potential. In fact, we expect\footnote{Note that QCD exhibits a (continuous) chiral $\rm{SU}(\Nf)_{\rm L}\otimes \rm{SU}(\Nf)_{\rm R}$ symmetry,
in contrast to the Gross-Neveu model, see also Sect.~\ref{sec:NJLLQCD} and Sect.~\ref{sec:gaugetheories}.}
 that QCD falls into the ${\rm O}(\Nf^2)$ universality class~\cite{Pisarski:1983ms}.

\begin{figure}[t]\center
\includegraphics[scale=1]{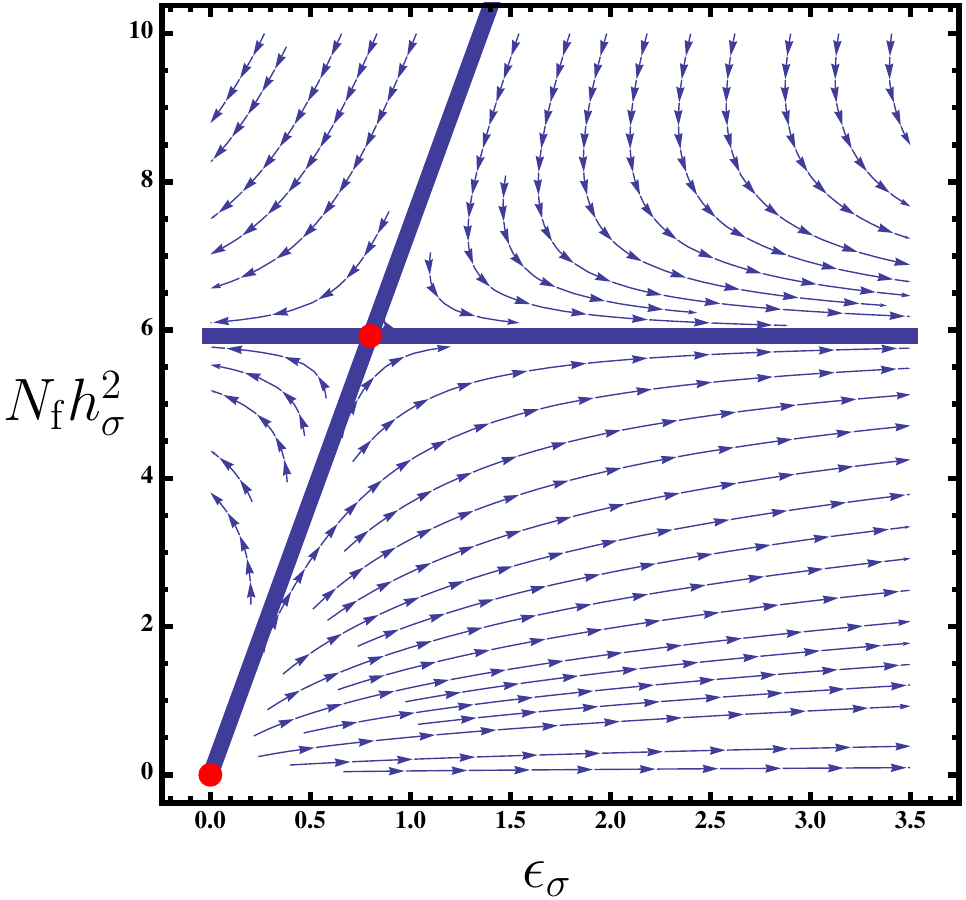}
\caption{\label{fig:loflowGN} RG flow of the partially bosonized Gross-Neveu model 
in leading order in the $1/\Nf$-expansion in the plane spanned by $\Nf h^2_{\sigma}$ and~$\epsilon_{\sigma}$, see Ref.~\cite{Braun:2010tt}.
The arrows indicate the direction of flow towards the  infrared.}
\end{figure}

Let us now begin with an analysis of the Gross-Neveu model in the large-$\Nf$ limit. In this limit, the flow equations
simplify considerably:
\be
\partial _t \epsilon_{\sigma} &=&  (\eta_{\sigma}-2)\epsilon_{\sigma}
 +\,4\Nf d_{\gamma} v_d\, l_{1}^{\rm (F),(d)}(0;\eta_{\psi}) h^2_{\sigma}\,,\label{eq:m2flowGNLN}\\
\partial _t \omega_{\sigma}&=& (d-4 + 2 \eta_{\sigma})\omega_{\sigma} 
-\,8\Nf d_{\gamma} v_d\, l_{2}^{\rm (F),(d)}(0;\eta_{\psi}) h^4_{\sigma}\,,\label{eq:lflowGNLN}\\
\partial_{t}h^2_{\sigma}  &=&  (d-4 + 2\eta_{\psi}+\eta_{\sigma})h^2_{\sigma}\,,  
\label{eq:hflowGNLN}\\ 
\eta_{\sigma}  &=&  8 \Nf
\frac{d_{\gamma}v_d}{d}h^2_{\sigma}\,m_{4}^{({\rm F}),(d)}(0;\eta_{\psi})\,, \label{eq:etasigmaGNLN}\\ 
\eta_{\psi}  &=& 0\,.\label{eq:yuk_flowGNLN}
\ee
Fixed points of the theory can be identified as the zeroes of these $\beta$ functions. 
Of course, we have a Gau\ss ian fixed point with all couplings vanishing. As the RG flows for the bosonic couplings
decouple in the large-$\Nf$ limit, a non-trivial fixed point requires~\mbox{$h_{\sigma,\ast}\neq0$}. 
From this it follows that
\be
\eta_{\sigma}^{\ast}=4-d\,, \label{eq:etaNf}
\ee 
independent of the RG scheme.
While this close relation between the dimensionality and the bosonic anomalous dimension here is an artifact of the large-$\Nf$ expansion, a
similar relation exists in gravity for the graviton anomalous dimension at the
(interacting) fixed point as a consequence of background gauge invariance~\cite{Reuter:1996cp}.
Similar sum rules are known for Yukawa theories with chiral symmetries~\cite{Gies:2009da}. Such a sum rule for a corresponding fixed point is also
responsible for the universality of the BCS-BEC crossover in the broad resonance limit of ultracold Fermi gases, see 
Sect.~\ref{sec:coldgases} and Ref.~\cite{Diehl:2007th}. In the present case, this rule simply determines the value of the Yukawa fixed-point coupling:
\be
(h_{\sigma}^{\ast})^2=\frac{1}{\Nf}\left(\frac{d}{8d_{\gamma}v_{d}}\right)\frac{(4-d)}{m_{4}^{({\rm F}),(d)}(0;0)}
=\frac{1}{\Nf}\left(\frac{d}{d_{\gamma}v_{d}}\right)\frac{(d-4)(d-2)}{(8-6 d)}\,,
\label{eq:hFPGN}
\ee
where we have used the optimized regulator function defined in Eq.~\eqref{eq:fermreg} to evaluate the threshold function~$m_{4}^{({\rm F}),(d)}(0;0)$.
Note that the fixed point is interacting for $2<d<4$ and merges with the Gau\ss ian fixed point of the Yukawa coupling in $d=4$, see also Fig.~\ref{fig:loflowGN}.
Also the fixed-point values for the bosonic mass parameter and couplings can be given analytically. For the optimized regulator we find
\be
\epsilon_{\sigma}^{\ast}= -\frac{8d(d-4)(d-2)}{d_{\gamma}(8-6d)(2-d)}\,,\qquad 
\omega_{\sigma}^{\ast}= \frac{1}{\Nf}\frac{8d^2(d-4)^2(d-2)^2}{d_{\gamma}(8-6d)^2 (4-d)v_d}\,.
\ee
Thus, the fixed-point values for the bosonic mass parameter as well as the four-boson self-interaction
are non-vanishing. In a purely fermionic formulation of the model, higher bosonic self-interactions, such as 
the four-boson interaction, correspond to higher (non-local) fermionic self-interactions. In any case, we find that the fixed-point structure in
$2<d<4$ for $\Nf\to\infty$ is not identical to the effective action~\eqref{eq:eff_action_fermion_GN} 
at the initial RG scale~$\Lambda$ but involves operators of higher order. 
Recall our discussion in Sect.~\ref{subsec:bos} where we have shown that bosonic self-interaction terms~$\sim \sigma^{2n}$ can be viewed 
as momentum-dependent (non-local) fermionic self-interaction terms~$\sim (\bar{\psi}\psi)^{2n}$.

We add that the fixed-point values 
of all bosonic self-interactions of the form~$\sigma^{2n}$ can be computed analytically in the limit~$\Nf\to\infty$
and that they are non-vanishing with their sign determined by~$(-1)^n$. It turns out that the resulting alternating
series for the full fixed-point (effective) potential, 
\be
u^{\ast}(\sigma^2)={\rm const.} + \frac{1}{2}\epsilon_{\sigma}^{\ast}\sigma^2 + \frac{1}{8}\omega_{\sigma}^{\ast}\sigma^4 + \dots
\,,\nn
\ee
can be resummed analytically, yielding $u^{\ast}(\sigma^2)\sim (\sigma)^{3/2}$ for large~$\sigma$, 
see Ref.~\cite{Braun:2010tt}. 

From a phenomenological point of view the scale-invariant fixed-point action describes a theory of massless fermions
interacting via the exchange of scalar bosons. On the other hand, we have seen that the purely fermionic formulation of the Gross-Neveu model depends only on
a single input parameter, namely the four-fermion coupling at the initial RG scale.
In the following we show that the partially bosonized theory still has predictive power. It will turn out that the deviations of the partially bosonized 
fixed-point theory from the classical GN action are only due to IR irrelevant operators which neither modify the predictive power of the Gross-Neveu model 
nor the leading-order scaling behavior at the quantum critical point.

The number of physical parameters is determined by the number of RG relevant directions corresponding to the number
of positive critical exponents. As discussed in Sect.~\ref{subsec:simpleex}, the latter can be computed from
the stability matrix:
\be
B=
\begin{pmatrix}
\frac{\partial (\partial_t h^2_{\sigma})}{\partial h^2_{\sigma}} & \frac{\partial (\partial_t h^2_{\sigma})}{\partial \epsilon_{\sigma}} & \frac{\partial (\partial_t h^2_{\sigma})}{\partial \omega_{\sigma}} &  \\
\frac{\partial (\partial_t \epsilon_{\sigma})}{\partial h^2_{\sigma}} & \frac{\partial (\partial_t \epsilon_{\sigma})}{\partial \epsilon_{\sigma}} & 
\frac{\partial (\partial_t \epsilon_{\sigma})}{\partial \omega_{\sigma}} &  \\
\frac{\partial (\partial_t \omega_{\sigma})}{\partial h^2_{\sigma}} & \frac{\partial (\partial_t \omega_{\sigma})}{\partial \epsilon_{\sigma}} & \frac{\partial (\partial_t \omega_{\sigma})}{\partial \omega_{\sigma}} &  
\end{pmatrix}
_{h_{\sigma}^{\ast},\epsilon_{\sigma}^{\ast},\omega_{\sigma}^{\ast}}
\,.
\ee
More precisely, the zeroes of the (characteristic) polynomial $\mathrm{det}(B+\Theta\mathbbm{1})$ of the stability matrix $B$ 
represent the critical exponents at the non-Gau\ss ian fixed point. In the present case the polynomial reads
\be
\mathrm{det}(B+\Theta\mathbbm{1})=\left( 4-d+\Theta \right)\left( 2-d +\Theta \right)\left( 4-d +\Theta \right)\,.
\ee
In $d=3$ we thus have one positive critical exponent associated with an RG relevant direction and two negative critical exponents associated with RG irrelevant directions.
As it should be, the exponent of the RG relevant direction is identical to the the one found in the purely fermionic formulation, see Eq.~\eqref{eq:fermThetaGN}. 
In~$d=4$ the situation is substantially different. There, we have two marginal directions and one relevant direction. 
This suggests that the Gross-Neveu model in~$d=4$ may depend on more than one input parameter. In any case, we observe that the critical exponents
are independent of~$\Nf$.

Even if we take into account bosonic self-interactions~$\sigma^{2n}$ of arbitrarily high order, we still have 
only a single RG relevant direction at the non-Gau\ss ian fixed point. All additional directions are RG irrelevant,
see Ref.~\cite{Braun:2010tt}. Thus, the IR physics of the model depends on only a single parameter as stated above
in the case of the purely fermionic formulation.

At the Gau\ss ian fixed point the critical exponents coincide with the mass dimension of the
Yukawa coupling and the bosonic couplings, reproducing simple perturbative power counting. In total, one finds three relevant RG
directions and one marginal RG direction in $d=3$.  Note that the Gau\ss ian fixed point in the purely fermionic flow in Sect.~\ref{sec:GNfermionic}
translates into a diverging {dimensionless} renormalized boson mass, $\epsilon_{\sigma}^{1/2}\sim \lambda_{\sigma}^{-1/2}$.  

At this point it is instructive to make the contact with the purely fermionic description more explicit and 
deduce the fixed point of the four-fermion coupling from the fixed point values of the Yukawa
coupling and the bosonic mass parameter~$\epsilon_{\sigma}$. We find 
\be
\lambda_{\sigma}^{\ast}=\frac{(h_{\sigma}^{\ast})^2}{\epsilon_{\sigma}^{\ast}}=\frac{1}{\Nf}\left(\frac{d}{d_{\gamma}v_{d}}\right)\frac{d-2}{8}
\stackrel{(d=3)}{=}\frac{3\pi^2}{4\Nf}
\ee
for the optimized regulator, which agrees with our findings in Sect.~\ref{sec:GNfermionic}. As we have discussed in Sect.~\ref{subsec:bos}, the 
flow equation of the four-fermion interaction $\lambda_{\sigma}$ can be reconstructed
from the flow of $h^2_{\sigma}/\epsilon_{\sigma}$:
\be
\partial_t \left(\frac{h^2_{\sigma}}{\epsilon_{\sigma}}\right)&=& (2 +  2\eta_{\psi})\left(\frac{h^2_{\sigma}}{\epsilon_{\sigma}}\right)
-4\Nf d_{\gamma}v_d\, l_{1}^{\rm (F),(\rm d)}(0;\eta_{\psi})\left(\frac{h^2_{\sigma}}{\epsilon_{\sigma}}\right)^2
 \nn\\ 
&&
 \;\;  + 6v_d\, l_1^{(d)}(\epsilon_{\sigma};\eta_{\sigma}) \omega_{\sigma}  \left(\frac{h^2_{\sigma}}{\epsilon^2_{\sigma}}\right) 
 \!+\! 8v_d\, l_{1,1}^{\rm  (FB),(d)}(0,\epsilon_{\sigma};\eta_{\psi},\eta_{\sigma})  \left(\frac{h^4_{\sigma}}{\epsilon_{\sigma}}\right). \label{eq:GNlsBLN}
\ee
Using Eq.~\eqref{eq:BCGN} and $\lambda_{\sigma}=h^2_{\sigma}/\epsilon_{\sigma}$ as well as 
\be
l_{1,1}^{\rm  (FB),(d)}(0,\epsilon_{\sigma};\eta_{\psi},\eta_{\sigma})\quad
\stackrel{ (\epsilon_{\sigma}\gg 1)}{\longrightarrow}\quad
\frac{1}{\epsilon_{\sigma}}\, l_{1}^{\rm (F),(d)}(0;\eta_{\psi})\,, \label{eq:lFtolFB}
\ee
we recover the flow equation~\eqref{eq:GNfourfermionflow} found in the purely fermionic formulation.
In the large-$\Nf$ limit, the flow equation~\eqref{eq:GNlsBLN} simplifies considerably. We find
\be
\partial_t \left(\frac{h^2_{\sigma}}{\epsilon_{\sigma}}\right)=
 (d - 2)\left(\frac{h^2_{\sigma}}{\epsilon_{\sigma}}\right) - 4\Nf
d_{\gamma}  v_d\,l_1^{\rm ({F}),(d)}(0;0)\, \left(\frac{h^2_{\sigma}}{\epsilon_{\sigma}}\right)^2\,.
\ee
We observe that the flow equations
for $\lambda_{\sigma}$ and $h^2_{\sigma}/\epsilon_{\sigma}$ are indeed identical. In turn, also the scale~$k_{\rm SB}$, which sets 
the scale for low-energy observables, must be the same in both cases. Recall that $\eta_{\psi}=0$ in this limit.

Due to the equivalence of $\lambda_{\sigma}$ and $h^2_{\sigma}/\epsilon_{\sigma}$, the quantum critical point found in the purely fermionic
formulation is also present in the partially bosonized theory for $\Nf\to\infty$, as it should be the case. 
As can be seen from the scaling law~\eqref{eq:scalingGN}, this quantum critical
point is associated with a vanishing boson mass, i.e. a diverging correlation length, in the long-range limit. The scaling
behavior of physical observables close to this point is governed by the critical exponent associated with the RG relevant direction.

Let us conclude our large-$\Nf$ analysis with a word of caution on the widely used LPA in which
the running of the wave-function renormalizations are neglected. If we ignored the running of the wave-function renormalization of the
bosonic field in the present case, the model would depend on more than one physical parameter and this additional 
dependence would be artificial. 
To be more specific, let us consider the mass spectrum of the theory in the regime with broken chiral symmetry and assume that we have
already fixed the mass of the fermions. Using the definition of the masses and the flow equations of the couplings, we find that the
(dimensionless) renormalized boson mass $m_{\sigma}$ in the broken regime can be written in terms of the (renormalized)
fermion mass $m_{\psi}$:
\be
m^2_{\sigma} =\omega_{\sigma} \sigma_0^2
\sim Z_{\sigma}^{-1}\bar{h}^2_{\sigma}(\bar{h}^2_{\sigma}\bar{\sigma}_0^2)\sim 
Z_{\sigma}^{-1}\bar{h}^2_{\sigma} m_{\psi}^2\,,\label{eq:GNmfms}
\ee
where $\sigma_0$ is the expectation value of the bosonic field in the regime
with broken chiral symmetry in the ground state,\footnote{Note that our conventions are such that~$\epsilon_{\sigma}=m_{\sigma}^2/k^2$ 
and~$\epsilon_{\psi}=m_{\psi}^2/k^2$.} see also Fig.~\ref{fig:effpot} for illustration. Neglecting the running of $Z_{\sigma}$,
i.~e. setting $Z_{\sigma}={\rm const.}$ as done in the LPA, we observe that the boson mass does not only depend 
on a single physical parameter, but on two parameters independently, namely the fermion mass and
the (bare) Yukawa coupling. In contrast, taking the running of $Z_{\sigma}\sim \bar{h}^2_{\sigma}$ into account, the value of the 
boson mass is fixed solely in terms of the fermion mass, in agreement with the fixed-point analysis. 

While this argument might be altered in $d=4$ space-time dimensions, where the Yukawa coupling is marginal, 
it is true for the Gross-Neveu model (as well as the Nambu-Jona-Lasinio model) 
in any dimension~$d$ in which the flow equation for $Z_{\sigma}$ is non-vanishing even at 
leading order in an expansion in $1/\Nf$. Therefore the flow of $Z_{\sigma}$ has to be taken into account in 
a systematic and consistent expansion of the flow equations in powers of $1/\Nf$. To be
specific, the flow of the bosonic self-interactions Eqs.~\eqref{eq:m2flowGN} and~\eqref{eq:lflowGN} incorporates 
fluctuations at next-to-leading order in $1/\Nf$ due to the presence of the bosonic 
loop. However, for a systematic and consistent study of the effects of corrections beyond 
the large-$\Nf$ expansion, the flow of $Z_{\sigma}$, $Z_{\psi}$ as well as of the Yukawa 
coupling needs to be taken into account.

\subsubsection{Quantum Critical Behavior Beyond the Large-$\Nf$ Limit}\label{sec:QPTGN}
Let us now discuss quantum critical behavior beyond the large-$\Nf$ limit. Beyond the large-$\Nf$ approximation, 
bosonic fluctuations play an important role. As an immediate consequence, a new fixed point for $h_{\sigma}\equiv 0$ arises for the flow 
of the effective potential in $2<d<4$. In the limit $\Nf\to 0$ we are left with a scalar $\mathbbm{Z}_2$ theory. This limit
is equivalent to the limit $h_{\sigma}\to 0$ and therefore this fixed point of the purely
bosonic theory is nothing but the Wilson-Fisher fixed point which describes critical phenomena in the Ising universality class. 

The non-Gau\ss ian fixed point of the full Gross-Neveu system can now be understood as stemming from the leading large-$\Nf$ terms discussed
above and the bosonic fluctuations inducing a Wilson-Fisher fixed point. Depending on the value of $\Nf$, the non-Gau\ss ian Gross-Neveu fixed
point interpolates between the large-$\Nf$ fixed point for $\Nf\to \infty$ and the Wilson-Fisher fixed point in the (formal) limit \mbox{$\Nf\to0$}. For the
latter, the functional RG has proven to be a useful quantitative tool for describing non-perturbative critical phenomena, see 
e.~g. Refs.~\cite{Tetradis:1993ts,Berges:2000ew,Litim:2001hk,Litim:2002cf,Bervillier:2007rc,Benitez:2009xg,Litim:2010tt}.

In the following we repeat the large-$\Nf$ analysis with the full set of flow equations for~$d=3$ at next-to-leading order of the derivative expansion,
see Eqs.~\eqref{eq:hflowGN}-\eqref{eq:lflowGN}. In particular for our numerical study of the critical behavior in $d=3$, we restrict ourselves to this set
of flow equations evaluated for the linear regulator. This implies that we only consider the RG flows of bosonic self-interaction terms up to order~$\sigma^4$.
We do not take into account corrections arising from higher bosonic self-interactions terms. However,
we expect that our analysis is sufficient to analyze the effect of corrections beyond the large-$\Nf$ limit as well as of
the next-to-leading order terms in the derivative expansion. For a more quantitative study of the critical exponents including,
effects of bosonic self-interactions up to 22nd order in $\sigma$, we refer the reader to Refs.~\cite{Braun:2010tt,Hofling:2002hj}.
  
In the chirally symmetric regime, a nontrivial fixed point in the Yukawa coupling requires that
the following inequality is satisfied:
\begin{equation}
d-4 + 2\eta_{\psi}^{\ast}+\eta_{\sigma}^{\ast}<0\,,\label{eq:ineq}
\end{equation}
for $\Nf<\infty$. This holds because the second term of the Yukawa flow Eq.~\eqref{eq:hflowGN} is
strictly positive for all admissible values of the anomalous dimensions
$\eta_\sigma,\eta_\psi \lesssim \mathcal{O}(1)$. For instance, in $d=3$, the
sum of the anomalous-dimension terms is always slightly smaller than one, see
Tab.~\ref{tab:GNfp}. The inequality becomes an equality in the large-$\Nf$ limit, see Eq.~\eqref{eq:etaNf}.
\begin{table*}[t]\center
\begin{tabular}{p{50pt}||p{50pt}|p{50pt}|p{50pt}|p{50pt}p{0pt}}
\centering $\Nf$ & \centering $\eta_{\sigma}^{\ast}$ & \centering $\eta_{\psi}^{\ast}$ & \centering $\Nf\,(h_{\sigma}^{\ast})^{2}$ & \centering $\epsilon_{\sigma}^{\ast}$ & \\ \hline\hline
\centering $2$ & \centering $0.7598$ & \centering $0.0320$ & \centering $4.5576$ & \centering $0.3966$ & \\ 
\centering $4$ & \centering $0.8869$ & \centering $0.0138 $ & \centering $5.2812$ & \centering $0.5831$ & \\ 
\centering $6$ & \centering $0.9267$ & \centering $0.0087$ & \centering $5.5068$ & \centering $0.6534$ & \\
\centering $8$ & \centering $0.9458$ & \centering $0.0063$ & \centering $5.6152$ & \centering $0.6894$ & \\
\centering $10$ & \centering $0.9571$ & \centering $0.0050$ & \centering $5.6788$ & \centering $0.7113$ & \\
\centering $12$ & \centering $0.9644$ & \centering $0.0041$ & \centering $5.7205$ & \centering $0.7260$ & \\
\centering $50$ & \centering $0.9917$ & \centering $0.0009 $ & \centering $5.8746$ & \centering $0.7821$ & \\
\centering $100$ & \centering $0.9958$ & \centering $0.0004 $ & \centering $5.8983$ & \centering $0.7911$ & \\ \hline 
\centering $\infty$ & \centering 1 & \centering 0 & \centering $5.9218$ & \centering $0.8$  
\end{tabular}
\caption{Non-Gau\ss ian fixed-point values of the {\it universal} anomalous dimensions
 and the (non-universal) fixed-point values of the couplings~$h_{\sigma}^{\ast}$
  and~$\epsilon_{\sigma}^{\ast}$ for various flavor numbers $\Nf$ in \mbox{$d=3$}. The results have been obtained using the flow 
  equations~\eqref{eq:hflowGN}-\eqref{eq:lflowGN} evaluated for the linear
  regulator. In Monte-Carlo simulations~\cite{Karkkainen:1993ef}, $\eta_{\sigma,\ast}=0.754(8)$
has been found for $\Nf=4$ two-component fermions (corresponding to $\Nf=2$ in our setting).}
\label{tab:GNfp}
\end{table*}

In the following we use the optimized regulator functions given in Eqs.~\eqref{eq:bosreg2} and~\eqref{eq:fermreg2} to evaluate the flow equations. 
The fixed point values for the Yukawa coupling in $d=3$ are given in Tab.~\ref{tab:GNfp}. For increasing $\Nf$, the fixed point quickly 
approaches its large-$\Nf$ limit \eqref{eq:hFPGN}, whereas it decreases and (slowly) tends to zero for small $\Nf$, leaving us with the pure Wilson-Fisher fixed 
point of a pure scalar model~\cite{Braun:2010tt,Hofling:2002hj}. As the latter is known to exhibit a fixed-point potential in the
broken regime~\cite{Tetradis:1993ts,Berges:2000ew,Litim:2001hk,Litim:2002cf,Bervillier:2007rc,Benitez:2009xg}
with a non-vanishing expectation value of the scalar field $\sigma$, we expect
such a fixed-point potential to appear for small $\Nf$. For all integer values of $\Nf\geq 1$
Dirac (four-component) fermions, we still find fixed-point actions in the chirally symmetric regime, see also
Ref.~\cite{Hofling:2002hj}. As has also been pointed out in Ref.~\cite{Hofling:2002hj}, 
the fixed point seems to occur in the broken regime for the
Gross-Neveu model with one two-component fermion.\footnote{Note that two-component fermions have been used in Ref.~\cite{Hofling:2002hj}.}
This would correspond to $\Nf=1/2$ in our setting. 

Let us now turn our discussion to the universal critical exponents~$\Theta^{(i)}$.
In our conventions, $\Theta^{(1)}$ denotes the leading, i.~e. the largest, critical exponent. 
The critical exponents for $\Nf=2$ read
\be
\Theta^{(1)}=0.9928 \,,\qquad \Theta^{(2)}=-0.8687\,,\qquad \Theta^{(3)}=-1.5743\,.\nn
\ee
Our result for the critical exponent $\Theta^{(1)}$ agrees within 
errors with the result from corresponding Monte-Carlo (MC) simulations~\cite{Karkkainen:1993ef}, 
$1/\Theta^{(1)}_{\rm MC}=\nu_{\rm MC}\approx 1.00(4)$. For $\Nf=12$, we find the following values for the critical exponents:
\be
\Theta^{(1)}=0.9898 \,,\qquad \Theta^{(2)}=-0.9735\,,\qquad \Theta^{(3)}=-1.0701\,.\nn
\ee
From this we conclude that the critical exponents do depend on~$\Nf$ but converge rapidly to their values in the large-$\Nf$ limit, namely
$\Theta^{(1)}=1$, $\Theta^{(2)}=-1$ and~$\Theta^{(3)}=-1$.
This is also visible in the values of the anomalous dimensions and of the fixed points for the couplings, see Tab.~\ref{tab:GNfp}. 
Whereas the values of the fixed-point couplings are non-universal, the anomalous 
dimensions are universal and illustrate the inequality~\eqref{eq:ineq}.

As we have discussed above, the critical exponent $\Theta^{(1)}$ governs the long-range physics at the quantum critical
point associated with the fixed-point~$\lambda_{\sigma}^{\ast}\sim (h_{\sigma}^{\ast})^2/\epsilon^{\ast}_{\sigma}$ of the four-fermion coupling.
The exponent~$\Theta^{(1)}$ is related to the correlation length exponent $\nu$ by $\nu=1/\Theta^{(1)}$. Together with the 
scalar anomalous dimension and the corresponding scaling and hyperscaling relations, all thermodynamic exponents 
of the (quantum) phase transition are determined. The results presented here are in quantitative agreement with the functional RG study of
H\"ofling et al.~\cite{Hofling:2002hj}. 
The agreement with results from other methods such as~$1/\Nf$ expansions~\cite{Gracey:1993kc} and Monte Carlo
simulations is also satisfactory~\cite{Hands:1992be,Karkkainen:1993ef}. Discrepancies are mainly visible in the anomalous
dimensions for small $\Nf$, a feature familiar from scalar models. 

To summarize, the non-perturbative features of the Gross-Neveu model near the quantum critical point can be described well
by our functional RG approach, as the model interpolates between two well-accessible limits within this framework: the
large-$\Nf$ limit and the Wilson-Fisher fixed point in the Ising universality class for $\Nf\to 0$. 
The ansatz for the effective action employed in our study is still incomplete in the sense that is does not exhibit all possible
four-fermion terms compatible with the underlying symmetries of the model. However, we have argued that a Gross-Neveu-type
RG trajectory might exist in the large-$\Nf$ limit, similar to the corresponding NJL-type trajectory in our study 
in Sect.~\ref{subsec:NJLUNf}. In any case, our results suggest that the Gross-Neveu model depends only on a single parameter, 
even when we take into account corrections beyond the large-$\Nf$ limit. 
This might provide helpful information for a systematic study of the finite-temperature phase diagram of the Gross-Neveu model beyond the 
large-$\Nf$ approximation.

\subsubsection{Excursion: Quantum Criticality and Asymptotic Safety}\label{sec:QCAS}

Let us finally exploit our results for the Gross-Neveu model to discuss aspects of renormalizability in
quantum field theories. In the context of our model, this question can be related to
the question of quantum criticality. In fact, the Gross-Neveu model allows us to discuss
the issue of renormalizability in quantum gravity in a very pedagogic way, as pointed out in Ref.~\cite{Braun:2010tt}.

Renormalizability is often described as a technical cornerstone for the construction of admissible models in particle physics. 
Renormalization fixes physical parameters of a model to measured values of observable
quantities. A main physical meaning of renormalizability is the capability of a model to provide an accurate description 
of a physical system over a wide range of scales at which measurements can be performed. The set of
physical parameters, e.~g. mass parameters, measured at different scales then defines the renormalized trajectory in parameter space.
We have already introduced this idea in our basic discussion 
of the RG in Sect.~\ref{sec:rg}. 
If we demand for a specific model to provide a {\em fundamental} description of nature, the model
must be valid on all scales, in particular down to arbitrary short-distance scales, i.~e. large momentum 
scales. In turn, the renormalized trajectory must exist on all scales without developing singularities. 

The requirement of renormalizability can formally be verified and realized in perturbatively renormalizable theories in a weak coupling expansion. Here, all
free parameters of a model can be fixed to physical values and the renormalized trajectory can be constructed order-by-order in a perturbative
expansion. This strategy can be applied successfully to theories that exist 
over a wide range of scales, such as QED. However, the perturbative construction can even be applied on all 
scales if a theory is asymptotically free, i.~e. if the Gau\ss ian fixed point is UV attractive. Prominent examples are SU($\Nc$) Yang-Mills theories in $d=4$ 
and the theory of the strong interaction, namely~QCD.

However, renormalizability is by no means tied to a perturbative construction. Even though reliable non-perturbative information might be difficult to obtain, the
concept of renormalizability and the existence of a renormalized trajectory on all scales can 
be formulated rather generally within {\it Weinberg's asymptotic safety scenario}~\cite{Weinberg:1976xy}. 
To put it sloppily, asymptotic safety is the generalization of 
asymptotic freedom at the Gau\ss ian fixed point to the
case of a non-Gau\ss ian fixed point. By construction, a fixed point of the RG defines a point in parameter 
space where the system becomes scale invariant, RG trajectories that hit the fixed 
point towards the UV can be extended to arbitrarily high energy scales, thereby defining a 
fundamental theory; for reviews we refer to Refs.~\cite{Weinberg:1996kw,Rosten:2010vm}.
On the other hand, we have shown that non-Gau\ss ian fixed-points can be viewed as quantum critical points governing long-range physics.

While quantum criticality plays a crucial role in fermionic models of, e.~g., condensed-matter systems,  
the asymptotic safety scenario has recently become an important ansatz for quantizing gravity. In contrast to other approaches, this scenario is 
based on the standard gravitational degrees of freedom and also the quantization procedure proceeds in a rather standard fashion; for recent developments
we refer to Refs.~\cite{Reuter:1996cp,Dou:1997fg,Lauscher:2001ya,Litim:2003vp,Codello:2006in,
Machado:2007ea,Codello:2007bd,Eichhorn:2009ah,Groh:2010ta,Eichhorn:2010tb,Eichhorn:2011pc}

In our study of the purely fermionic formulation of the Gross-Neveu model, we have found
that the critical exponent $\Theta$ is positive for $d>2$ and determines the scaling behavior of physical IR 
observables. From a field-theoretical point of view this means that the Gross-Neveu coupling~$\lambda_{\sigma}$ 
corresponds to an RG relevant coupling being repulsed by the non-Gau\ss ian fixed point~$\lambda_{\sigma}^{\ast}$
towards the IR. However, we can also think of~$\lambda_{\sigma}$ as a coupling which is in the RG flow
attracted by the non-Gau\ss ian fixed point~$\lambda_{\sigma}^{\ast}$  
towards the UV. In our simplified study, this suggests that the Gross-Neveu model can be renormalized and 
extended as a fundamental theory over all scales on RG trajectories that emanate from the non-Gau\ss ian fixed point. 
As there is only one relevant direction, only one physical parameter has to be fixed (say, the
value of the coupling at a UV scale~$\Lambda$) in order to predict all physical quantities in the long-range limit. 
Thus, the Gross-Neveu model is asymptotically safe, i.~e. non-perturbatively renormalizable. This observation has also been confirmed by our study
of the partially bosonized Gross-Neveu model in Sect.~\ref{sec:QPTGN} where we have systematically included corrections beyond the large-$\Nf$ limit.
Our results indeed suggest that the Gross-Neveu model in $d=3$ is asymptotically safe for all $\Nf>0$. In any case, the 
present discussion illustrates the tight connection between quantum criticality and asymptotic safety.

Let us make one more point concerning the asymptotic safety scenario 
by considering the Gross-Neveu model at the Gau\ss ian fixed point. At the Gau\ss ian fixed point the critical exponent associated
with the coupling~$\lambda_{\sigma}$ is given by~$\Theta_{\rm Gau\ss }=2-d$. Thus, the Gau\ss ian fixed point
is IR attractive. In this case, the limit $\Lambda\to \infty$ can only be taken if the RG trajectory emanates
from the fixed point, but then the theory would be noninteracting on all scales and therefore ``trivial''. 
Within perturbation theory, one therefore concludes that the Gross-Neveu model is perturbatively {\it non-renormalizable}. 
Note that this conclusion remains unchanged also if the anomalous dimension is taken into account. In fact, we have
$\eta_\psi=0$ at the Gau\ss ian fixed point, implying that standard power-counting can only be modified logarithmically. 

In quantum gravity, there is strong evidence that an IR repulsive non-Gau\ss ian fixed point exists. It has been found that
this fixed point indeed exists in (simple) truncations based on derivative expansions in 
all $d>2$, see Refs.~\cite{Litim:2003vp,Fischer:2006at}. The (upper) critical dimension for the existence of this non-Gau\ss ian 
fixed point is~\mbox{$d=2$} as in the Gross-Neveu model.  It is interesting to speculate about a possible destabilization of
 the fixed point above another so far unknown critical dimension due to 
strong-coupling phenomena such as bound-state formation. A similar phenomenon has been observed in 
extra-dimensional Yang-Mills theories~\cite{Gies:2003ic},  where a non-Gau\ss ian fixed point exists in 
$d=4+\epsilon$ but is non-perturbatively destabilized for $\epsilon \gtrsim \mathcal{O}(1)$.

We would like to add that the underlying ideas of the asymptotic safety scenario allow to construct 
UV-complete scenarios for the matter sector of the standard model (as well as for toy models of the Higgs sector), 
see Refs.~\cite{Gies:2003dp,Gies:2009hq,Gies:2009sv}, and Ref.~\cite{Percacci:2009fh} for a study of the non-linear sigma model.

Our comparison of the Gross-Neveu model and quantum gravity shows that the property of asymptotic safety in the Gross-Neveu model is 
closely related to the occurrence of a quantum phase transition of second order separating a disordered phase from a phase
with broken chiral symmetry in the ground state. More generally, models with such quantum phase transitions of second order 
are guaranteed to be asymptotically safe. Whether the converse is true, i.e., whether asymptotically safe models always exhibit a physically 
relevant order-disorder quantum phase transition, is an interesting question for future studies.

\subsection{Nambu-Jona-Lasinio Models and QCD at Low Energies}\label{sec:NJLLQCD}

The Gross-Neveu model represents an effective theory for systems relevant in the context of condensed-matter physics. However,
we have already hinted that the Gross-Neveu model also shares certain aspects with the theory of the strong interaction, namely QCD. 
In this section we now discuss aspects of strongly interacting fermions in the context of QCD low-energy models. As we shall see below,
such models can be viewed as an extension of the simple NJL model discussed in Sect.~\ref{sec:example}. Of course, the construction of the
NJL model has been originally inspired by the mechanisms of superconductivity in condensed-matter physics~\cite{Nambu:1961tp,Nambu:1961fr}.
The application of these type of models to QCD exemplifies once more that the action itself does not completely 
specify a physical system. Only the (quantum) equations of motion
together with their boundary conditions fully determine the physical system under consideration. The ``universality" of the equations
of motion resulting from a given action allows for a cross-fertilization of seemingly different fields of physics,
e.~g. QCD and condensed-matter physics.

In Sect.~\ref{sec:LowQCDF} we discuss some facets of QCD and related low-energy models. Quantum and thermal
phase transitions are then discussed in Sect.~\ref{sec:TQPTQCD} with the aid of a particular low-energy model,
the so-called {\it quark-meson model}. The corresponding RG flow equations can be derived
along the lines of Sect.~\ref{sec:example}. In particular, we shall see that QCD low-energy models are closely
related to the NJL model with a continuous chiral ${\rm SU}(\Nf)_{\rm L}\otimes {\rm SU}(\Nf)_{\rm R}$ symmetry
as discussed in Sect.~\ref{subsec:NJLSUNf}.

\subsubsection{Low-energy QCD Models and the Fierz Ambiguity}\label{sec:LowQCDF}

As a prelude to Sect.~\ref{sec:gaugetheories} we briefly outline how QCD low-energy models can in principle be derived
from the QCD action functional. One of the ``ancestors" of QCD is the Gell-Mann-Zweig quark model
for hadrons ~\cite{GellMann:1964nj,Zweig:1981pd} which is based on the observation that hadrons can be classified according to
the structure of the (flavor) ${\rm SU}(\Nf=3)$ gauge group. Even though this model was able to explain the existence of some particles,
it was not possible to explain the existence of all hadronic resonances in a simple manner. For example, 
it was not possible to reconcile the existence of the baryonic resonance $\Delta^{++} ({\tiny \frac{3}{2} }^{+})$,
which is made up of three up-quarks in a spin-up state, with the Pauli Principle. 
In the 1970s, it was then shown that the shortcomings of the original quark model could be resolved 
by assigning a so-called color charge to the quarks~\cite{Fritzsch:1973pi,Nambu:1974zg,Bardeen:1976tm,Greenberg:1976ph}. 
Considering the problems of the original quark model with the resonance~$\Delta^{++} ({\tiny \frac{3}{2} }^{+})$,
this was a reasonable assumption.
Starting from the principle of gauge invariance, the existence of gauge bosons, the gluons, which carry combinations
of color and anti-color, was postulated as well. The gluons are massless and mediate the interaction between the
quarks, as the photons do in QED. In contradistinction to QED, however, the gauge bosons of the strong interaction
interact with each other, which explains the short range of the strong force. As a consequence, photons are represented by
Abelian gauge fields while gluons are described by non-Abelian gauge fields.

The action functional of QCD is approximately symmetric 
under continuous chiral flavor ${\rm SU}(\Nf)_{\rm L}\otimes {\rm SU}(\Nf)_{\rm R}$ transformations. As motivated above,
the quarks also carry a color charge resulting in a local ${\rm SU}(\Nc)$ gauge symmetry, where~$\Nc$ denotes the number
of colors. The simplest action compatible with these symmetry constraints reads
\be
S_{\rm QCD}=\int d^4 x\left\{ \frac{1}{4}F_{\mu\nu}^{a} F_{\mu\nu}^{a} + \bar{\psi}\left( \I \fslash{\partial} + \bar{g}\fslash{A}\right)\psi \right\}\,,
\label{eq:SQCD}
\ee
where $A_{\mu}\equiv A_{\mu}^{a}T^{a}_{ij}$ denotes the gauge fields and~$\bar{g}$ is the (bare) coupling constant; the $T^{a}_{ij}$
are the generators of the~${\rm SU}(\Nc)$ gauge group. Note that the fermions transform under the fundamental representation
of the ${\rm SU}(\Nc)$ color group, whereas the gluons transform under the adjoint representation. The parameters~$\Nc$ and~$\Nf$
can in principle be chosen freely. For low-energy QCD phenomenology, we have~\mbox{$\Nf=2$} (and~$\Nc=3$). In nature
the flavor symmetry is broken explicitly by so-called current quark mass terms for the fermions, see also Sect.~\ref{subsec:DefESB}. 
For a study of the thermal phase transition of QCD, however, we expect that the 
dynamics is essentially driven by the two lightest quark flavors, which we shall assume to be massless in the following. The masses of the 
other quarks are large compared to the relevant scale of the theory,\footnote{Strange quarks with mass $m_{\rm s}\approx 200\,{\rm MeV}$
might still affect the dynamics close to the chiral phase boundary. Here, we do not take these  effects into account for the sake
of simplicity.} e.~g. the (chiral) phase transition temperature or the condensate~$|\langle\bar{\psi}\psi\rangle|^{1/3}$ (associated with the
light quarks).

The QCD action functional possesses also a conformal symmetry which is broken in the quantum theory due to the generation of a mass gap, even 
in the limit~\mbox{$\Nf\to 0$}. Moreover, the axial~${\rm U}_{\rm A}(1)$ symmetry of the flavor group, which is present on the classical level, is broken
explicitly due to topologically non-trivial gauge configurations.
This is known as the chiral anomaly. We shall comment on this anomaly below.

We now discuss how the basic structure of QCD low-energy models emerges from the QCD action functional. 
The local gauge symmetry gives rise to an interaction between the quarks and the gluons. The details of the momentum-scale dependence
of the associated (renormalized) quark-gluon coupling~$g$ is of no relevance for the subsequent discussion. We only state
that it increases towards the IR for $\Nf < {\tiny \frac{11}{2}}\Nc$. In Sect.~\ref{sec:gaugetheories} we shall then discuss this issue in more detail.
In any case, the interaction between the quarks and the gluons induces quark self-interactions, e.~g. by two-gluon exchange, of the following form:
\be
\int d^4 x\, \bar{\lambda}_{\alpha\beta\gamma\delta} 
\bar{\psi}_{\alpha}\psi_{\beta}\bar{\psi}_{\gamma}\psi_{\delta}
\,,
\ee
where $\alpha,\beta\,\dots$ denote a specific set of collective indices including, e.~g., flavor and/or color indices. Thus, quark self-interactions
are gluon-induced in QCD, and the associated couplings are not fundamental parameters. In fact, QCD 
in the chiral limit (i.~e. limit of vanishing current quark masses) does depend only on a single parameter,
e.~g. the value of the (renormalized) coupling~$g$ at a given scale. Once we have fixed the coupling, all physical observables are fixed. This is
reminiscent of the situation in the Gross-Neveu model in $2\leq d<4$ which also depends on only a single input parameter, 
see Sect.~\ref{sec:GNmodel}.

Let us assume for the moment that it would be possible to integrate out the gauge degrees of freedom completely. 
The dynamics associated with the gauge fields would then be encoded in highly non-local fermionic self-interactions
of arbitrarily high order.\footnote{In studies of the Gross-Neveu model in the large-$\Nf$ limit the fermions are often integrated out
explicitly, yielding a highly non-local but purely bosonic action, as discussed in Sect.~\ref{subsec:bos}.} 
We assume further that the strength of these dynamically generated fermionic self-interactions can be related to a set of 
initial conditions for this purely fermionic theory at a given (UV) scale~$\Lambda_{\rm H}$ by means of an RG trajectory. 
These initial conditions are in principle fixed by the gauge dynamics at scales~$k\gtrsim \Lambda_{\rm H}$.
The scale~$\Lambda_{\rm H}$ can then be viewed as a UV cutoff for this (purely) fermionic effective low-energy theory. 
For (momentum) scales~$k\lesssim \Lambda_{\rm H}$ we may expect a description of QCD in terms of a purely fermionic effective field 
theory to be valid and convenient, at least for a study of some aspects of QCD, such as dynamical chiral symmetry breaking.

We add a comment on the scale~$\Lambda_{\rm H}$: This scale essentially divides QCD into two regimes, namely a perturbative
high-energy regime and a low-energy regime governed by dynamical mass generation. We therefore expect~$\Lambda_{\rm H}$ to be at least of
the order of the scale~$\ksb$ associated with chiral symmetry breaking, $\Lambda_{\rm H} \gtrsim k_{\rm SB}$, see also
Sects.~\ref{sec:ScalLEM} and~\ref{sec:RGYM}. Of course, the scale~$k_{\rm SB}$ is 
a scheme-dependent quantity. In the context of QCD low-energy models, however, $\Lambda_{\rm H}$ should be considered as an
additional (input) parameter which needs to be fixed by comparison to physical observables. 

Based on these considerations effective low-energy QCD models are usually constructed from a given set of four-fermion interactions, see e.~g. 
Refs.~\cite{Klevansky:1992qe} for a review.
Fermionic operators of higher order are usually dropped for the sake of simplicity. A commonly used ansatz for 
a low-energy effective theory for QCD with two massless quark flavors is given by\footnote{Note that 
this ansatz does not possess a~U${}_{\rm A}(1)$ symmetry, see discussion below. On the other hand, 
the ansatz~\eqref{equ::truncationQM0Nf3} has a~U${}_{\rm A}(1)$ symmetry.} 
\be
\Gamma_{\Nf=2} [\bar{\psi},\psi]&=&
\int d^4x \Bigg\{ \bar{\psi} {\rm i}\fslash{\partial} 
\psi 
+\frac{1}{2} \bar{\lambda}_{\sigma} \Big[
(\bar{\psi}\psi)^2 - (\bar{\psi}\tau^{\chi}\gamma_5 \psi)^2
\Big]\Bigg\}\,,
\label{equ::truncationQM0}
\ee
with $\tau^{\chi}$ being the Pauli matrices ($\chi=1,2,3$),
see e.~g. Refs.~\cite{Klevansky:1992qe,Alkofer:1995mv,Jungnickel:1995fp,Berges:1997eu,Schaefer:1999em,Berges:2000ew,Braun:2003ii,Fukushima:2003fw,Schaefer:2004en, %
Braun:2005fj,Ratti:2005jh,Sasaki:2006ww,Nakano:2009ps,Braun:2009si,Braun:2010vd,Skokov:2010wb,Herbst:2010rf,Skokov:2010uh}.
For three light (massless) quark flavors one may 
use~\cite{Vogl:1989ea,Klimt:1989pm,Klimt:1990ws,Klevansky:1992qe,Alkofer:1995mv,Schaefer:2008hk,Fukushima:2008wg,Hell:2009by}
\be
\Gamma_{\Nf=3} [\bar{\psi},\psi]&=&
\int d^4x \Bigg\{ \bar{\psi} {\rm i}\fslash{\partial}  
\psi 
+\frac{1}{2} \bar{\lambda}_{\sigma} \Big[
(\bar{\psi}T^{\chi}\psi)^2 - (\bar{\psi}T^{\chi}\gamma_5 \psi)^2
\Big]\Bigg\}\,,
\label{equ::truncationQM0Nf3}
\ee
where the $T^{\chi}$ denote the generators of the ${\rm SU}(\Nf)$ flavor group. 
In the three flavor case the generators are related to the so-called Gell-Mann matrices. 
In Eqs.~\eqref{equ::truncationQM0} and~\eqref{equ::truncationQM0Nf3}
color indices ($a,b,\dots$) are contracted pairwise, e.~g. $(\bar{\psi}\psi)\equiv \bar{\psi}_a \psi_a$. 
Moreover, we employ the following convention for the flavor indices ($\alpha,\beta,\dots$): 
$(\bar{\psi}T^{\chi}\gamma_5 \psi)^2 \equiv (\bar{\psi}_{\alpha}T^{\chi}_{\alpha\beta}\gamma_5 \psi_{\beta})(\bar{\psi}_{\gamma}T^{\chi}_{\gamma\delta}\gamma_5 \psi_{\delta})$.
From a phenomenological point of view this ansatz is obvious since it incorporates pions $\pi ^{\chi} \sim (\bar{\psi}\tau^{\chi}\gamma_5 \psi)$ as composite degrees of
freedom. This can be seen most easily from the partially bosonized formulation of this ansatz.\footnote{In the literature
the partially bosonized versions of the purely fermionic actions~\eqref{equ::truncationQM0} 
and~\eqref{equ::truncationQM0Nf3} are often referred to as the {\it quark-meson model}. We refer the reader to
Sect.~\ref{subsec:bos} for a detailed discussion of bosonization and to Sect.~\ref{subsec:NJLSUNf} for a discussion of 
symmetry breaking in theories with a chiral~${\rm SU}(\Nf)_{\rm L}\otimes {\rm SU}(\Nf)_{\rm R}$ symmetry.}
The pions arise as effective degrees of freedom in the low-energy limit of QCD 
due to the spontaneous breakdown of the chiral flavor symmetry. Since they
are the massless Nambu-Goldstone bosons of QCD, they dominate the long-range physics and are of utmost importance for an accurate
description of, e.~g., the dynamics close to the finite-temperature phase boundary.

The ansatz~\eqref{equ::truncationQM0} represents an effective two-flavor model for dynamical chiral symmetry breaking at intermediate 
scales of $k \lesssim \Lambda_{\rm H}$. It is important to stress that this model cannot predict the temperature dependence of physical observables 
exactly. By construction, it has neither gluons nor quark confinement. At moderate energy scales, below the
hadronic mass scale $\Lambda_{\rm H}$, unconfined {\it constituent quarks} appear instead of baryonic degrees of freedom. 
However, the low-energy couplings as derived from the partially bosonized version of this model are compatible with those of chiral perturbation 
theory~\cite{Jungnickel:1997yu,Jendges:2006yk}. 

From a phenomenological point of view the ansatz~\eqref{equ::truncationQM0} is well-justified  
for a description of two-flavor QCD at low energies. However, it is not complete with respect to Fierz transformations.
The most general ansatz for the effective action compatible with the underlying symmetries of QCD reads~\cite{Gies:2003dp}
\begin{eqnarray}
\Gamma_k [\bar{\psi},\psi]&=&
\int d^4x \Bigg\{ \bar{\psi} {\rm i}\fslash{\partial} 
\psi 
+\frac{1}{2} \Big[
  \bar\lambda_-(\text{V--A}) +\bar\lambda_+ (\text{V+A}) +\bar\lambda_\sigma (\text{S--P})\nn\\
  && \qquad\qquad\qquad\qquad\qquad\;  +\bar\lambda_{\text{VA}}
  [2(\text{V--A})^{\text{adj}}\!+({1}/{\Nc})(\text{V--A})] \Big]\Bigg\}\,.
\label{equ::truncationQM1}
\end{eqnarray}
The four-fermion interactions appearing here have been classified according to their 
color and flavor structure. Color and flavor singlets are
\begin{eqnarray}
(\text{V--A})&=&(\yb\gamma_\mu\psi)^2 + (\yb\gamma_\mu\gamma_5\psi)^2, \label{eq:VmAdef}
\\
(\text{V+A}) &=&(\yb\gamma_\mu\psi)^2 - (\yb\gamma_\mu\gamma_5\psi)^2 ,
\end{eqnarray}
where (fundamental) color ($i,j,\dots$) and flavor ($\chi,\xi,\dots$) indices are contracted pairwise, e.g., $(\yb\psi)\equiv (\yb_i^{\chi}
\psi_i^{\chi})$.  The remaining operators have non-singlet color or flavor structure,
\begin{eqnarray}
(\text{S--P})&=&\!(\yb^{\chi}\psi^{\xi})^2\!-(\yb^{\chi}\gamma_5\psi^{\xi})^2
\!\equiv\!
   (\yb_i^{\chi}\psi_i^{\xi})^2\!-(\yb_i^{\chi}\gamma_5\psi_i^{\xi})^2\!,\nonumber\\
(\text{V--A})^{\text{adj}}\!\!\!&=&\!(\yb \gamma_\mu T^a\psi)^2 
   + (\yb\gamma_\mu\gamma_5 T^a\psi)^2, \label{eq::colorflavor}
\end{eqnarray}
where $(\yb^{\chi}\psi^{\xi})^2\equiv \yb^{\chi}\psi^{\xi} 
\yb^{\xi} \psi^{\chi}$, etc., and $(T^a)_{ij}$ denote the generators of the gauge group in the fundamental representation.  
We stress that the set of fermionic self-interactions introduced in Eq.~\eqref{equ::truncationQM1} forms a complete basis. This means that any other
pointlike four-fermion interaction which is invariant under \mbox{$\textrm{SU}(\Nc)$} gauge symmetry and
$\textrm{SU}(\Nf)_{\rm L} \otimes \textrm{SU}(\Nf)_{\rm R}$ flavor symmetry can be related to those in \eqref{equ::truncationQM1} by means of Fierz
transformations. Here, we have dropped U${}_{\text{A}}(1)$-violating interactions induced by topologically
non-trivial gauge configurations. These interactions can in principle be parametrized by fermionic self-interactions of the form~$\sim (\bar{\psi}\psi)^{\Nf}$ and 
then be included straightforwardly in effective low-energy models~\cite{tHooft:1976fv,Shifman:1979uw,Shuryak:1981ff,Schafer:1996wv,Pawlowski:1996ch}.

The action~\eqref{equ::truncationQM1} represents a very general ansatz for an effective low-energy model.
The strategy for employing such a model is usually as follows: First, one uses a set of parameters and the 
UV cutoff~$\Lambda_{\rm H}$ to fit the values of low-energy observables at vanishing temperature and 
chemical potential, e.~g. the pion decay constant and the meson masses.\footnote{We stress that the limit~$\Lambda_{\rm H}\to \infty$ is
not meaningful here, in contrast to the Gross-Neveu model in $2\leq d<4$ where the limit~$\Lambda\to\infty$ is well-defined.}
Second, one then computes the phase boundary while keeping the parameters at the UV cutoff fixed. A shortcoming of these models
is apparent: The set of parameters used to fit a given set of low-energy observables is not necessarily unique. Even worse,
two sets of parameters, which both give the same results for the low-energy observables, do not
necessarily lead to the same results for the chiral phase boundary, see e.~g. Ref.~\cite{Stephanov:2007fk}.
Strictly speaking, the initial conditions for the four-fermion couplings are fixed by gluodynamics at high 
momentum scales~$p\sim k\gtrsim \Lambda_{\rm H}$.
However, this information is usually not available. Here, RG approaches may provide a systematic and consistent framework to derive 
these initial conditions from first principles QCD, see Sect.~\ref{sec:gaugetheories}.

The effective action~\eqref{equ::truncationQM1} naturally encompasses the ansatz~\eqref{equ::truncationQM0}. One may wonder whether
the action~\eqref{equ::truncationQM0} represents an "exact" limit of the action~\eqref{equ::truncationQM1}, e.~g. in the limit~$\Nc\to\infty$.
This would be indicated by, e.~g., the existence of an RG trajectory on which only the coupling included in Eq.~\eqref{equ::truncationQM0} is finite
and all other interaction channels vanish identically. In order to investigate this question we study the fixed-point structure 
of the Fierz-complete ansatz~\eqref{equ::truncationQM1}. The RG flow equations for the couplings can be derived straightforwardly 
along the lines of Sect.~\ref{subsec:simpleex}. In the point-like limit we then find (see Refs.~\cite{Gies:2003dp,Gies:2005as})
\begin{eqnarray}
\pat\lm
&=& 2\lm\!
 -8 v_4\lFna \Big\{-\Nf\Nc(\lm^2+\lp^2) + \lm^2\nn\\
&& \quad\qquad\qquad\qquad\qquad -2(\Nc+\Nf)\lm\lva
       +\Nf\lp\lsf + 2\lva^2 \Big\},
\\
\pat\lp &=&2 \lp\! 
-8 v_4 \lFna \Big\{ - 3\lp^2 - 2\Nc\Nf\lm\lp - 2\lp(\lm+(\Nc+\Nf)\lva)\nn\\
 && \qquad\qquad\qquad\qquad\qquad\qquad\qquad       + \Nf\lm\lsf
+ \lva\lsf
        +{\small\frac{1}{4}}\lsf{}^2 \Big\},
\\
\pat\lsf
&=& 
2\lsf
  -8 v_4 \lFna
     \Big\{ \, 2\Nc \lsf^2\!  -\! 2\lm\lsf\! - 2\Nf\lsf\lva\!\!
- \!6\lp\lsf\! \Big\},
\\
\pat\lva
&=& 2 \lva\!
 -8 v_4 \lFna
  \Big\{ - \!(\Nc+\Nf)\lva^2\! + 4\lm\lva\!
        - {\small\frac{1}{4}} \Nf \lsf^2\Big\}\,,
\end{eqnarray}
where $\lFna =\lF$ and 
the dimensionless couplings~$\lambda_{i}$ are defined as~$\lambda_{i}=k^2\bar{\lambda}_{i}$ with $i\in\{+,-,\sigma,{\rm VA}\}$.
Apart from a Gau\ss ian fixed point, this set of equations possesses 15 non-trivial fixed points. The values of these can be computed straighforwardly
from the flow equations. In the following, however, we restrict our analysis to the limit~$\Nc\to\infty$. Loosely speaking, this limit corresponds
to the large-$\Nf$ limit in the Gross-Neveu model discussed in Sect.~\ref{sec:GNmodel}.

From the (Dirac) Fierz transformations
given in App.~\ref{sec:dirac} we deduce that the channels associated with the couplings~$\lambda_{-}$ and~$\lambda_{\rm VA}$
cannot be directly transformed into a channel with a scalar-pseudoscalar Dirac structure as included in the commonly
employed action~\eqref{equ::truncationQM0}. On the other hand, the (V+A)-channel can be transformed into a channel with a scalar-pseudoscalar structure.
Thus, we are left with the (V+A)-channel and the (S--P)-channel. In the large-$\Nc$ limit
it then turns out that no non-trivial fixed point exists at which only the $\lambda_{+}$-coupling assumes a finite value. Moreover,
no non-trivial fixed point exists at which only the~$\lambda_{-}$- and~$\lambda_{\rm VA}$-coupling are zero. 
However, a non-Gau\ss ian fixed point exists at which only the $\lambda_{\sigma}$-coupling assumes a finite value.
In fact, the set of flow equations for the couplings~$\lambda_{-}$, $\lambda_{+}$, $\lambda_{\sigma}$ and $\lambda_{\rm VA}$
simplifies considerably in the limit~$\Nc\to\infty$ for $\lambda_{-}=\lambda_{+}=\lambda_{\sigma}=0$:
\be
\pat\lsf &=& 2\lsf -16\Nc v_4 \lFna \lsf^2 \,.\label{eq:LNQM}
\ee
The associated fixed point ${\mathcal F}_{\text{(S-P)}}^{\infty}=(\lambda_{-}^{\ast},\lambda_{+}^{\ast},\lambda_{\sigma}^{\ast},\lambda_{\rm VA}^{\ast})$ of the
full set of equations in this limit reads
\be
{\mathcal F}_{\text{(S-P)}}^{\infty}=\left(0,0,\frac{8\pi^2}{\Nc},0\right)\,, 
\ee
where we have used the optimized regulator function for illustration, \mbox{$\lF=\frac{1}{2}$}.

Let us have closer look at the (S--P)-channel and its relation to the the interaction channel included in the simplified
action~\eqref{equ::truncationQM0}. Similar to our study in Sect.~\ref{subsec:NJLSUNf}, we may use the Fierz transformation
for flavor indices Eq.~\eqref{eq:flavorFT} to obtain
\be
(\text{S--P})&=&\left(\frac{1}{2\Nf}-\frac{1}{4}\right)\left[(\bar{\psi}\psi)^2\! -\! (\bar{\psi}\gamma_5 \psi)^2\right]\! +\! 
\left[ (\bar{\psi}T^{\chi}\psi)^2\! -\! (\bar{\psi}T^{\chi}\gamma_5 \psi)^2\right]\,, \label{eq:nflavorint}
\ee
where $\chi =0,1,\dots, \Nf^2-1$. For convenience, we have defined $T^{0}=\frac{1}{2}\mathbbm{1}$ and 
the $T^{\chi}$ for $\chi \geq 1$ denote the generators of the ${\rm SU}(\Nf)$ flavor group in the fundamental representation. 
Note that the second term in Eq.~\eqref{eq:nflavorint} is invariant even under~U($\Nf$) transformations. For the phenomenologically
important case $\Nf=2$ we are left with
\be
(\text{S--P})&=&
\frac{1}{4}\left[ (\bar{\psi}\tau^{\chi}\psi)^2\! -\! (\bar{\psi}\tau^{\chi}\gamma_5 \psi)^2\right]\nn\\
&=& \frac{1}{2} \left[ (\bar{\psi}\psi)^2 \!-\! (\bar{\psi}\tau^{\chi}\gamma_5 \psi)^2\right] - 
\left[  \det\bar{\psi}(1+\gamma_5)\psi + \det\bar{\psi}(1-\gamma_5)\psi \right]\,, \label{eq:2flavorint}
\ee
with $\tau^{\chi}$ being the Pauli matrices, $T^{\chi}=\frac{1}{2}\tau^{\chi}$, see also Ref.~\cite{Klevansky:1992qe}.
The determinant is performed in flavor space. 

Apparently, the (S--P)-channel includes
the standard scalar-pseudoscalar interaction channel of commonly used effective low-energy models for QCD with two massless flavors,
see Eq.~\eqref{equ::truncationQM0}.
The second term in Eq.~\eqref{eq:2flavorint} has the same structure as a term associated with topologically
non-trivial gauge configurations that break the ${\rm U}_{\rm A}(1)$ symmetry of the 
theory~\cite{tHooft:1976fv,Shifman:1979uw,Shuryak:1981ff,Schafer:1996wv,Pawlowski:1996ch}. 
In our case, we have attached the same coupling to the first and the second term in Eq.~\eqref{eq:2flavorint}. This keeps the 
${\rm U}_{\rm A}(1)$ symmetry intact. In an even more general approach we could allow for an explicit ${\rm U}_{\rm A}(1)$ breaking term by, e.~g., 
replacing the (S--P)-channel in Eq.~\eqref{equ::truncationQM1} as follows:
\be
\bar{\lambda}_{\sigma}\, \text{(S--P)} \longrightarrow \bar{\lambda}_{\sigma}\, \text{(S--P)} - 
\left(\bar{\lambda}_{\sigma} - \bar{\lambda}_{ {\rm  top.}}\right) \left[  \det\bar{\psi}(1+\gamma_5)\psi + \det\bar{\psi}(1-\gamma_5)\psi \right]\,.
\ee
In any case, our analysis suggests that in the large-$\Nc$ limit
a separatrix in coupling-constant space exists on which only the coupling included in Eq.~\eqref{equ::truncationQM0} assumes a finite value.
The fixed point on this separatrix can be considered as a quantum critical point in analogy to our study of the 
Gross-Neveu model. The initial condition for the scalar-pseudoscalar coupling being smaller or larger than
the value of the fixed point then distinguishes between two different phases in the long-range limit (IR limit):
a non-interacting phase and a strongly-interacting phase in which the dynamics is governed by the pions.
In practice, the initial condition for the scalar-pseudoscalar coupling
is adjusted to 
fit the values of low-energy observables. In the subsequent section we shall now employ the effective action~\eqref{equ::truncationQM0}
to study quantum and thermal phase transitions.

\subsubsection{Phase Transitions in QCD Low-energy Models}\label{sec:TQPTQCD}

We now study quantum\footnote{Concerning the issue of the existence of a non-trivial fixed-point of the four-fermion coupling 
associated with
a quantum critical point in~$d=4$ space-time dimensions, we refer the reader to our discussion in Sects.~\ref{subsec:bos} and~\ref{sec:coldgases}.
}
and thermal phase transitions with the effective action~\eqref{equ::truncationQM0}.
According to our analysis in the previous section this ansatz
can be considered as a controlled starting point for an analysis of the QCD matter sector in the large-$\Nc$ limit.\footnote{In (full) QCD, we have
a ${\rm U}_{\rm A}(1)$ symmetry. However, this symmetry is anomalously broken due to the presence of topologically
non-trivial gauge configurations. The widely-used definition of the two-flavor quark-meson model, see Eq.~\eqref{equ::truncationQM0},
does not have such a ${\rm U}_{\rm A}(1)$ symmetry. This becomes apparent from the fact that 
the interaction term in the ansatz~\eqref{equ::truncationQM0} essentially arises from  
the ${\rm U}_{\rm A}(1)$-invariant expression Eq.~\eqref{eq:2flavorint} by simply dropping the contributions from the determinants.}
As mentioned above,
the partially bosonized action corresponding to the purely fermionic action~\eqref{equ::truncationQM0} is widely known as
the {\it quark-meson model}. 

Despite the shortcomings of this low-energy model, we believe that its study can shed some light on the mechanisms 
underlying chiral symmetry breaking in QCD. While the actual mechanism in (full) QCD may be different due to the presence of color interactions, 
the approach employed here gives a possible explanation for the scaling behavior observed
in the low-energy observables, as far as they relate to the mechanisms of chiral symmetry breaking in an effective low-energy 
description of QCD by means of light Nambu-Goldstone particles.

We start our discussion of chiral symmetry breaking at zero and finite temperature 
with an analysis of the fixed-point structure of the purely fermionic formulation of the quark-meson model, see Eq.~\eqref{equ::truncationQM0}.
In this two-flavor model, 
the spontaneous breakdown of chiral symmetry gives rise to the existence of three massless Nambu-Goldstone bosons. 
In the following, we restrict ourselves to the case~$\Nf=2$ but keep the number of colors~$\Nc$ as an arbitrary (control) parameter.
Of course, the ansatz~\eqref{equ::truncationQM0} is not complete with respect to Fierz transformations. 
In particular, we do not take into account additional four-fermion operators which arise at finite temperature due to the broken Poincare invariance, 
see Sect.~\ref{subsec:DefFT}. In the large-$\Nc$ limit, however, the 
scalar-pseudoscalar channel defines an RG trajectory on which all other four-fermion couplings are identical to zero, see Sect.~\ref{sec:LowQCDF}. 

In our study we follow closely the steps detailed in Sects.~\ref{subsec:simpleex} and~\ref{subsec:DefFT}. This means that we
drop a possible momentum dependence of the four-fermion coupling, $\lambda_{\sigma}(|p|\ll k)$ which implies~$\eta_{\psi}=0$.
This approximation does not permit a study of properties of the system, such as the meson 
mass spectrum, in the chirally broken regime; for example, mesons manifest themselves as momentum singularities 
in the four-fermion couplings. As discussed in detail in Sect.~\ref{subsec:bos}, the point-like limit can still be a reasonable approximation in the 
chirally symmetric regime above the chiral phase transition. It allows us to gain some insight into the question how
the theory approaches the regime with broken chiral symmetry in the ground state~\cite{Braun:2005uj,Braun:2006jd,Braun:2008pi}. 
Since we are simply interested in mapping the phase diagram in the plane spanned by the temperature and 
the initial condition for the coupling~$\lambda_{\sigma}$, the point-like limit still represents a reasonable approximation.

At this point we would also like to remind the reader that in the point-like limit 
the RG flow of the four-fermion coupling, which signals the onset of chiral symmetry breaking, is completely decoupled
from the RG flow of fermionic $n$-point functions of higher order. For example, $8$-fermion interactions do not contribute
to the RG flow of the coupling $\bar{\lambda}_{\sigma}$ in this limit. 

Using the ansatz~\eqref{equ::truncationQM0} we obtain the RG flow equation for the dimensionless renormalized 
four-fermion coupling $\lambda_{\sigma}=k^2\bar{\lambda}_{\sigma}$ in the point-like limit:\footnote{Note that the factor
in front of~$\lambda_{\sigma}^2$ differs from the one in Eq.~\eqref{eq:LNQM} for two reasons: first, we have used
a 3$d$ regulator function. Second, we started from the action~\eqref{equ::truncationQM0} to derive Eq.~\eqref{eq:lpsi_flowQM}.
This action is not complete with respect to Fierz transformations. However, this only changes the (non-universal) fixed-point value
but not the (universal) critical exponents in the large-$\Nc$ limit.}
\be
\beta_{\lambda_{\sigma}}\equiv\partial_t \lambda_{\sigma} = 2\lambda_{\sigma} 
-16  (2\Nc + 1)v_3 l_{1}^{\rm (F),(4)}(\tau,0)\,\lambda_{\sigma}^2\,,
\label{eq:lpsi_flowQM}
\ee
where $v_3=1/(8\pi^2)$. Note that the coupling $\lambda_{\sigma}$ depends on the dimensionless temperature $\tau=T/k$. 
Here, we have used a 3$d$ (optimized) regulator function which is convenient for an analysis of chiral symmetry breaking at finite temperature.
The definition of the regulator and the temperature-dependent threshold function~$l_{1}^{\rm (F),(4)}$ can be found in App.~\ref{app:regthres}.
Recall that the wave-function renormalizations longitudinal and transversal to the heat bath are constant in leading order in the derivative expansion, 
i.~e.~$\eta_{\psi}^{\|,\perp}=0$.

Let us now discuss the fixed-point structure of the coupling $\lambda_{\sigma}$. Apart from a Gau\ss ian 
fixed point we have a second non-trivial fixed point. The
value of this fixed point depends on the dimensionless temperature $\tau$, see also Fig.~\ref{fig:parabolaT}.
At vanishing temperature we find
\be
\lambda_{\sigma}^{\ast}=\frac{1}{8(2\Nc +1)v_3 l_1^{({\rm F}),(4)}(0,0,0)} =\frac{6\pi^2}{(2\Nc+1)} \label{eq:QMfpT0}
\ee
for the non-Gau\ss ian fixed point. For illustration we have evaluated the threshold function
$l_1^{\rm (F),(4)}$ for the 3$d$ optimized regulator, see Eq.~\eqref{eq:fermreg}. Note that the rescaled fixed-point coupling $\Nc\lambda_{\sigma}^{\ast}$ 
approaches a constant value for $\Nc\to\infty$: $\Nc\lambda_{\sigma}^{\ast} \to 3\pi^2$.

Let us briefly recall the physical observations of Sect.~\ref{sec:example} which are of relevance here:
First of all, we stress that the fixed-point value $\lambda_{\sigma}^{\ast}$ is not a universal quantity, as its dependence on
the threshold function indicates. However, the statement about the mere existence of the 
fixed point is universal. Choosing an initial value $\lambda_{\sigma}^{\rm UV} < \lambda_{\sigma}^{\ast}$  
at the initial UV scale $\Lambda_{\rm H}$, we find that the theory becomes non-interacting in the
infrared regime ($\lambda_{\sigma}\to 0$ for $k\to 0$), see Fig.~\ref{fig:parabolaT}. For $\lambda_{\sigma}^{\rm UV}>\lambda_{\sigma}^{\ast}$ 
we find that the four-fermion coupling $\lambda_{\sigma}$ increases rapidly and diverges eventually 
at a finite scale~$k_{\rm SB}$. This behavior indicates the onset of chiral symmetry breaking
associated with the formation of a quark condensate and the emergence of massless Nambu-Goldstone bosons,
namely the pions. Hence chiral symmetry breaking in the IR only occurs if we choose $\lambda_{\sigma}^{\rm UV}>\lambda_{\sigma}^{\ast}$
and~$\lambda_{\sigma}^{\ast}$ can be viewed as a quantum critical point.

The scale $k_{\rm SB}$ at which $1/\lambda_{\sigma}(k_{\rm SB})=0$ sets the scale for a given IR observable~$\mathcal O$:
\be
{\mathcal O}\sim k_{\rm SB}^{d_{\mathcal O}}\,,\label{eq:lambdacrOQM}
\ee
where $d_{\mathcal O}$ is the canonical mass dimension of the observable $\mathcal O$. At vanishing temperature
the scale $k_{\rm SB}$ can be computed analytically from Eq.~\eqref{eq:lpsi_flowQM}. 
Analogously to our derivation of Eq.~\eqref{eq:lambdacr}, we find
\be
k_{\rm SB}=\Lambda_{\rm H} \theta(\lambda_{\sigma}^{\rm UV} -\lambda_{\sigma}^{\ast})
\left(\frac{\lambda_{\sigma}^{\rm UV} -\lambda_{\sigma}^{\ast}}{\lambda_{\sigma}^{\rm UV}} \right)^{\frac{1}{\Theta}}\,,
\label{eq:lambdacrQM}
\ee
where the critical exponent $\Theta$ is independent of~$\Nc$ in the present approximation:
\be
\Theta=-\frac{\partial (\partial_t \lambda_{\sigma})}{\partial \lambda_{\sigma}}\Big|_{\lambda_{\sigma}^{\ast}}=2\,.
\ee 
Thus, the critical value $k_{\rm SB}$ scales with the distance of the initial value 
$\lambda_{\sigma}^{\rm UV}$ from the fixed-point value $\lambda_{\sigma}^{\ast}$. For increasing $\lambda_{\sigma}^{\rm UV}$
the scale $k_{\rm SB}$ increases and, in turn, the values of low-energy observables, such 
as the pion decay constant $f_{\pi}$ and the chiral phase transition temperature~$T_{\chi}$, increase.

From now on, let us assume that we have fixed $\lambda_{\sigma}^{\rm UV}>\lambda_{\sigma}^{\ast}$ at $T=0$. 
The value of $\lambda_{\sigma}^{\rm UV}$ then determines the scale $k_{\rm SB}\equiv k_{\rm SB}(\lambda_{\sigma}^{\rm UV})$ 
which is related to the values of the low-energy values. For a study of the effects of a finite temperature 
we then leave our choice for $\lambda_{\sigma}^{\rm UV}$ unchanged. This ensures comparability of the results at zero and finite 
temperature for a given theory defined by the choice for $\lambda_{\sigma}^{\rm UV}$ at zero temperature.

At finite temperature we still have a Gau\ss ian fixed point. Moreover, we find a pseudo fixed-point $\lambda_{\sigma}^{\ast}(\tau)$ 
for arbitrary values of $\tau =T/k$ at which the right-hand side of the flow equation is zero:
\be
\lambda_{\sigma}^{\ast}(\tau) =\frac{1}{8(2\Nc +1)v_3 l_1^{{\rm (F)},(4)}(\tau,0,0)  } \,.
\ee
We would like to remind the reader that the pseudo fixed-point $\lambda_{\sigma}^{\ast}(\tau)$ is not necessarily 
an element of the separatrix in the plane spanned by the coupling~$\lambda_{\sigma}$ and 
the dimensionless temperature $\tau$. However, we have~$\lambda_{\sigma}^{\ast}(\tau)\geq \lambda_{\sigma}^{\rm sep.}(\tau)$
for a given value of $\tau$, where $\lambda_{\sigma}^{\rm sep.}(\tau)$ defines the separatrix in $(\lambda_{\sigma},\tau)$-space,
see our detailed discussion in Sect.~\ref{subsec:DefFT}. Therefore the analytically accessible value of the 
pseudo fixed-point~$\lambda_{\sigma}^{\ast}(\tau)$ is well-suited for a discussion of the basic mechanisms of 
chiral symmetry breaking at finite temperature.

For high temperatures $T\gg k$ we find $\lambda_{\sigma}^{\ast} \sim (T/k)^3$. 
Since the value of the (pseudo) fixed-point 
increases with increasing $\tau=T/k$, the rapid increase of the four-fermion coupling towards the IR ($k\to 0$) is effectively slowed 
down and may even change its direction in the plane spanned by the coupling~$\lambda_{\sigma}$ and~$\tau$,
see Figs.~\ref{fig:parabolaT} and~\ref{fig:finiteTsep}. This behavior of the pseudo fixed-point $\lambda_{\sigma}^{\ast}(\tau)$ already 
suggests that for a fixed initial value $\lambda_{\sigma}^{\rm UV}$ a critical temperature $T_{\chi}$ 
exists above which the four-fermion coupling does not diverge but approaches zero in the IR. 
Such a behavior is indeed expected for high temperatures since the quarks become effectively 
stiff degrees of freedom due to their thermal mass $\sim T$ and chiral symmetry is restored.
As discussed in Sect.~\ref{subsec:DefFT}, the equation~$\lambda_{\sigma}^{\ast}(\tau_{\ast})=\lambda_{\sigma}^{\rm UV}$
defines a strict upper bound for the critical temperature~$T_{\chi}$ for a given value of~$\lambda_{\sigma}^{\rm UV}$ and~$\Lambda_{\rm H}$,
i.~e. $T_{\chi}\leq T_{\ast}=\tau_{\ast}\Lambda_{\rm H}$.
Moreover, the equation~$\lambda_{\sigma}^{\rm sep.}(\tau_{\rm sep.})=\lambda_{\sigma}^{\rm UV}$ defines an even stronger upper bound
for the critical temperature: $T_{\chi}\leq T_{\rm sep.}\leq T_{\ast}$. The actual phase transition temperature~$T_{\chi}$
might be smaller than~$T_{\rm sep.}$ due to fluctuations of the Nambu-Goldstone bosons in the IR close to the phase boundary.
However, we have $T_{\chi}\equiv T_{\rm sep.}$ for~$\Nc\to\infty$ since bosonic fluctuations are parametrically 
suppressed in this limit. 

\begin{figure}[t]
\begin{center}
\includegraphics[scale=0.8]{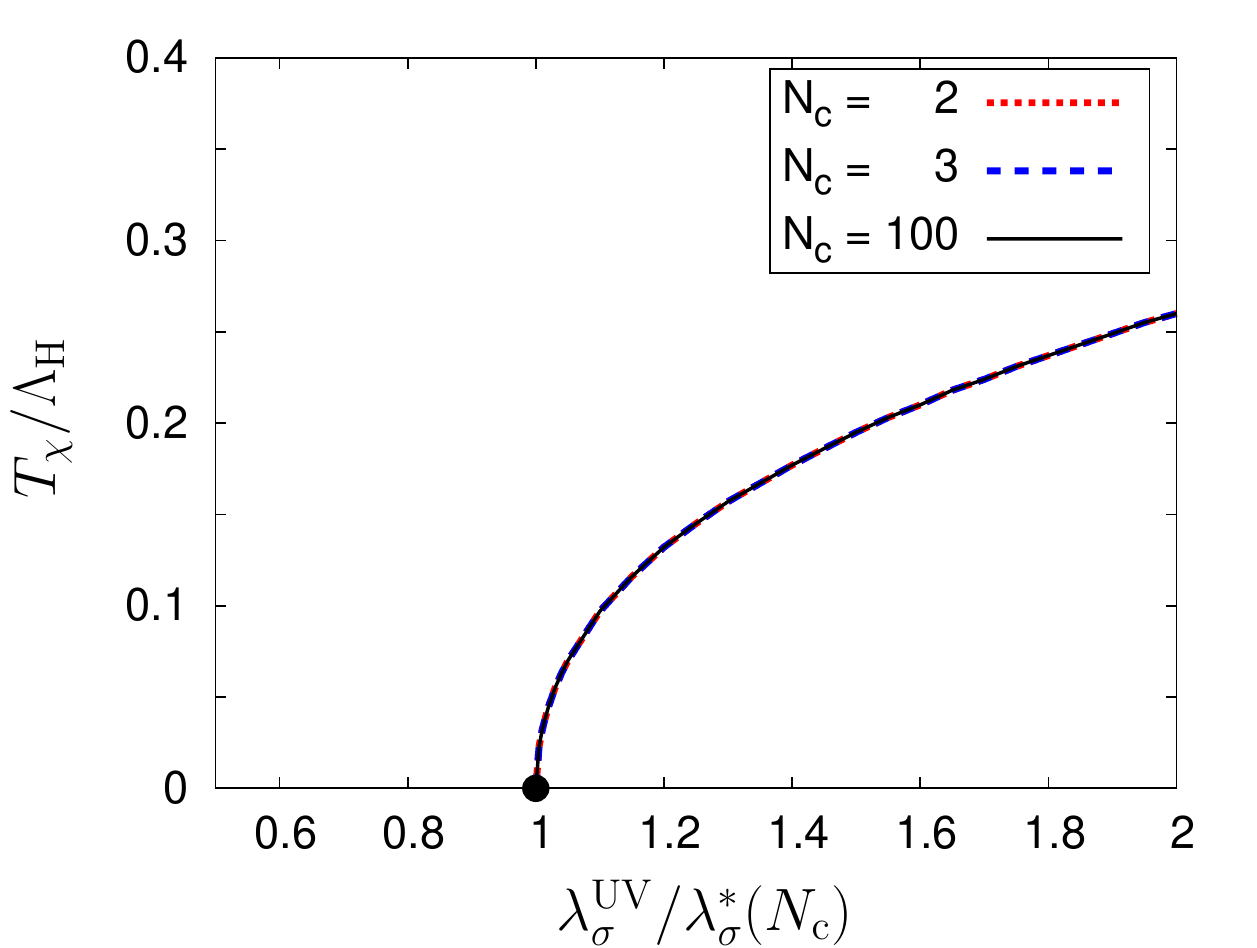} 
\end{center}
\caption{Phase diagram for two massless quark flavors 
and $\Nc=2,3,100$ colors in the plane spanned by the temperature and the rescaled 
coupling $\lambda_{\sigma}^{\rm UV}/\lambda_{\sigma}^{\ast}(\Nc)$. The 
lines depict our results for the phase boundary for various values of~$\Nc$.
For $T>T_{\chi}$ the theory remains in the chirally symmetric phase on all skales~$k\leq \Lambda_{\rm H}$.
For $T=0$ and increasing $\lambda_{\sigma}^{\rm UV}$ the system undergoes 
a quantum phase transition at $\lambda_{\sigma}^{\rm UV}/\lambda_{\sigma}^{\ast}(\Nc)=1$.
Note that $\Tc$ as a function of $\lambda_{\sigma}^{\rm UV}/\lambda_{\sigma}^{\ast}(\Nc)$ does not depend 
on~$\Nc$ in the present approximation.
}
\label{fig:pdQM}
\end{figure}

To illustrate our analytic results we have studied the RG flow of the four-fermion coupling $\lambda_{\sigma}$ 
for finite~$\Nc$ numerically. Following the discussion in the previous paragraph, the phase transition temperature $T_{\chi}$ 
is defined to be the smallest temperature for which~$\lambda_{\sigma}$ remains finite in the infrared limit $k\to 0$. 
In Fig.~\ref{fig:pdQM} we present the phase diagram for two massless quark flavors and various values of~$\Nc$ 
in the plane spanned by the temperature and the UV coupling $\lambda_{\sigma}^{\rm UV}(\Nc)$. 
For the UV cutoff $\Lambda_{\rm H}$ defining the range of validity of our model, we have chosen $\Lambda_{\rm H}=1\,\text{GeV}$.
In accordance with our analytic findings we observe that there is only a chirally symmetric phase 
for~$\lambda_{\sigma}^{\rm UV}<\lambda_{\sigma}^{\ast}$. 
Increasing $\lambda_{\sigma}^{\rm UV}$ above $\lambda_{\sigma}^{\ast}$, the system undergoes a quantum phase
transition at~$\lambda_{\sigma}^{\ast}$. We expect 
that the chiral phase transition temperature~$\Tc$ increases monotonously with~$\lambda_{\sigma}^{\rm UV}>\lambda_{\sigma}^{\ast}$ 
according to
\be
T_{\chi} \sim k_{\rm SB}\,. 
\ee
Therefore we expect  that the scaling behavior
of this chiral observable is determined by the exponent~\mbox{$\Theta=2$}, see Eq.~\eqref{eq:lambdacrQM}. 
In fact, the numerical data can be fitted to this analytic estimate:
\be
\frac{T_{\chi}}{\Lambda_{\rm H}}\approx 0.316 
\left( \frac{\lambda_{\sigma}^{\rm UV}-\lambda_{\sigma}^{\ast}}{\lambda_{\sigma}^{\rm UV}} \right)^{0.498}\,,
\ee
which is in good agreement with the expected behavior.
For the fit we have used the results for~$\Tc/\Lambda_{\rm H}$ for 11 equidistant values of~$\lambda_{\sigma}^{\rm UV}/\lambda_{\sigma}^{\ast}$
between~$\lambda_{\sigma}^{\rm UV}/\lambda_{\sigma}^{\ast}=1.0$ and~$\lambda_{\sigma}^{\rm UV}/\lambda_{\sigma}^{\ast}=1.01$.
Note that our estimate for the phase transition temperature 
does not depend on~$\Nc$ if we keep the initial condition~$\lambda_{\sigma}^{\rm UV}/\lambda_{\sigma}^{\ast}(\Nc)$ fixed.
This illustrates that the values of (chiral) observables in the large-$\Nc$ limit depend only on the relative distance of the initial 
condition~$\lambda_{\sigma}^{\rm UV}$ from the fixed-point value~$\lambda_{\sigma}^{\ast}(\Nc)$, as indicated by the
scaling behavior of the critical scale~$k_{\rm SB}$ in Eq.~\eqref{eq:lambdacrQM}.

Let us now relate our model study of the phase diagram in the $(T,\lambda_{\sigma}^{\rm UV})$-plane to QCD.
First of all, the initial condition $\lambda_{\sigma}^{\rm UV}$ is not a free parameter in QCD but originally generated by quark-gluon interactions at high (momentum) 
scales. In a given regularization scheme, the value of $\lambda_{\sigma}^{\rm UV}$ can therefore in principle be related to the value of the 
gauge coupling at some scale (e.~g. the $\tau$ mass scale), see Sect.~\ref{sec:gaugetheories} and Refs.~\cite{Gies:2005as,Braun:2005uj,Braun:2006jd,Braun:2008pi}.
We would like to point out that neither the value of the gauge coupling at some scale nor the value of
$\lambda_{\sigma}^{\rm UV}$ on a given RG trajectory are physical observables. Nevertheless, their values can be related to 
physical low-energy observables, e.~g. the pion decay constant or the quark condensate. For example, this has been explicitly illustrated in Ref.~\cite{Braun:2011fw} by
means of partial bosonization. In fact, the value of $\lambda_{\sigma}^{\rm UV}$
determines the scale for IR observables, as explicitly demonstrated for the phase transition 
temperature~$\Tc$. Since we expect that the (constituent) quark mass~$m_{\psi}$ also scales as~$m_{\psi}\sim \ksb$ in leading order, it follows that~$\Tc \sim m_{\psi}$.
In the following, however, we are not interested in a quantitative analysis of the relation of~$\lambda_{\sigma}^{\rm UV}$ to low-energy observables 
but rather in very general aspects of QCD low-energy models. 

Within our point-like approximation of the fermionic vertices it is not possible 
to predict the order of the finite-temperature phase transition. 
In particular, our present approximation
does not allow to determine the associated critical exponents. However, we expect our model to fall into 
the~${\rm O}(\Nf^2)$ universality class at the critical temperature. This can be understood most easily in terms of the partially bosonized version
of our model:\footnote{Recall that at finite temperature the Poincare invariance of the theory is broken explicitly. Therefore 
the wave-function renormalizations longitudinal and transversal to the heat-bath
obey a different RG running. We neglect this difference for our general discussion in this section. In general,  this is at least 
a reasonable approximation since it has indeed been found in Ref.~\cite{Braun:2009si} that the difference is small at low temperatures 
and only yields mild corrections to, e.~g., the thermal mass of the bosonic degrees of freedom for intermediate temperatures~$T\gtrsim\Tc$.
Moreover, we identify the wave-function renormalizations and the couplings of the Nambu-Goldstone modes and the so-called
radial $\sigma$-mode. This is justified for our general discussion concerning universality at finite temperature.}
\be
&&\Gamma_k[\bar{\psi},\psi,\sigma,\{\pi^{\chi}\}]
= \int d^4 x\,\Big\{ Z_{\psi}\bar{\psi}\I \fslash{\partial}  \psi 
+ \frac{1}{2}Z_{\sigma}\left(\partial_{\mu}\sigma\right)^2  +\frac{1}{2}Z_{\sigma}\left(\partial_{\mu}\pi^{\chi}\right)^2 \nn\\
&& \qquad\qquad\qquad\quad
+ {\rm i}\bar{h}_{\sigma}\bar{\psi}(\sigma + {\rm i}\tau^{\chi}\pi^{\chi}\gamma_5)\psi 
+\frac{1}{2}\bar{m}^2_{\sigma}(\sigma^2+\vec{\pi}^{\,2}) + \frac{1}{8}\bar{\omega}_{\sigma}(\sigma^2+\vec{\pi}^{\,2})^2  \label{Eq:HSTActionQM}
\Big\}\,,
\ee
with $\chi=1,\dots,\Nf^2-1$, $\vec{\pi}^{\,2}:=\pi^{\chi}\pi^{\chi}$, a Yukawa coupling $\bar{h}_{\sigma}\in \mathbb{R}$ and the boundary conditions
\be
\lim_{k\to \Lambda_{\rm H}} Z_{\sigma}=0\,,\quad
\lim_{k\to \Lambda_{\rm H}} Z_{\psi} = 1\,,\quad
\lim_{k\to \Lambda_{\rm H}} \bar{\omega}_{\sigma} &=& 0\,.\label{eq:bcQM}
\ee
These boundary conditions together with the identity
\be
\bar{\lambda}_{\sigma}=\frac{\bar{h}^2_{\sigma}}{\bar{m}^2_{\sigma}}\label{eq:hstmapQM}
\ee
allow us to map the ansatz~\eqref{Eq:HSTActionQM} onto the fermionic action~\eqref{equ::truncationQM0} at the initial UV scale $\Lambda$. 
In this picture, we assume that the bosons are composites of fermions and do not carry an internal charge, e.~g. color or flavor: $\sigma\sim (\bar{\psi}\psi)$ and 
$\pi ^{\chi} \sim (\bar{\psi} \tau^{\chi}\gamma_5\psi)$. The interaction between the fermions is mediated by the 
Nambu-Goldstone bosons, namely the pion fields~$\pi^{\chi}$.
The expectation value of the $\sigma$-field is proportional to the pion decay constant $f_{\pi}$. 
This is a consequence of the {\em Goldberger-Treiman relation}, which results from a detailed study of the axial-vector 
currents of the bosonized action~\cite{Goldberger:1958tr}.

At this point we mention a subtlety concerning the mapping of the partially bo\-sonized formulation onto the purely fermionic formulation
of our model. First of all, the identity~\eqref{eq:hstmapQM} suggests that we have 
only one input parameter in the present study, namely the value of the ratio~$(\bar{h}_{\sigma}^{\rm UV}/\bar{m}_{\sigma}^{\rm UV})$ 
at the UV scale~$\Lambda_{\rm H}$. In $d=4$ space-time dimensions, however, 
the Yukawa coupling $\bar{h}_{\sigma}$ is marginal.
This suggests that the partially bosonized theory in $d=4$ depends on 
two input parameters in contrast to $d=3$, see also our discussion of the Gross-Neveu model in $d=3$ in Sect.~\ref{sec:GNmodel}.
Nonetheless the critical scale~$k_{\rm SB}$ depends only on the 
ratio~$(\bar{h}_{\sigma}^{\rm UV}/\bar{m}_{\sigma}^{\rm UV})$ in leading order in an expansion in powers of $1/\Nc$ and receives only small 
corrections from the next-to-leading order, see also below. On the other hand, 
the ratio of IR observables, such as the ratio of the $\sigma$-mass and the constituent quark mass, 
depends on both parameters, see Eq.~\eqref{eq:GNmfms}. In practice, this means that the ratio~$(\bar{h}_{\sigma}^{\rm UV}/\bar{m}_{\sigma}^{\rm UV})$ 
and the Yukawa coupling~$\bar{h}_{\sigma}^{\rm UV}$ should be considered as independent input parameters at the UV scale~$\Lambda$. 

As we have seen in Sect.~\ref{subsec:bos}, the partially bosonized formulation of a fermionic model allows us to conveniently compute
the order parameter, i.~e. the chiral condensate, and in principle also its temperature dependence. The latter can be used to determine
the order of the phase transition and to relate the initial condition~$\lambda_{\sigma}^{\rm UV}$ to physical observables.
The gap equation for the expectation value of the $\sigma$-field in the mean-field approximation can be obtained
along the lines of the derivation of Eq.~\eqref{eq:gap}. From the effective action~\eqref{Eq:HSTActionQM} with $Z_{\sigma}=0$ and~$\bar{\omega}_{\sigma}=0$,
we find
\be
\langle \sigma \rangle=8\Nc \left(\frac{\bar{h}_{\sigma}^2}{\bar{m}_{\sigma}^2}\right) 
\!T\!\sum_{n=-\infty}^{\infty}\int \!\frac{d^3 p}{(2\pi)^3}\left[  \frac{\langle \sigma \rangle}{\nu _n ^2\! +\! \vec{p}^{\,2}\! +\! \langle \sigma \rangle^2} -  \frac{\langle \sigma \rangle}{\nu _n ^2\! +\! \Lambda_{\rm H}^2\! +\! \langle \sigma \rangle^2} 
\right]\theta(\Lambda_{\rm H}^2\! -\! \vec{p}^{\,2})\,, \label{eq:QMgapEq}
\ee
where we have used Eq.~\eqref{eq:effactoneloop} and the optimized 3$d$ regulator function given in Eq.~\eqref{eq:fermreg}. 
We have chosen this regulator since it allows us to directly relate the initial conditions~$\lambda_{\sigma}^{\rm UV}$ in Fig.~\ref{fig:pdQM} 
to the expectation value~$\langle \sigma\rangle$. The Yukawa coupling has been absorbed into a 
redefinition of the bosonic field which is possible in the mean-field approximation. 
Using Eq.~\eqref{eq:QMgapEq} we find that $(\Lambda_{\rm H}^2 \bar{h}_{\sigma}^2/\bar{m}_{\sigma}^2)\approx 14.1$ yields a (constituent)
quark mass~$m_{\psi}\approx 0.3\,\text{GeV}$ for $T\!=\! 0$, $\Nc\!=\! 3$ and~$\Lambda_{\rm H}\!=\! 1\,\text{GeV}$. 
Now we can either read off the corresponding phase transition temperature~$\Tc$ from Fig.~\ref{fig:pdQM} or we can compute~$\Tc$ with
the gap equation~\eqref{eq:QMgapEq}. It is reassuring that we find~$\Tc\approx 0.183\,\text{GeV}$ either way.
Moreover, the phase transition turns out to be of second
oder in this simple large-$\Nc$ approximation. 

Concerning the phase diagram in the $(T,\lambda_{\sigma}^{\rm UV})$-plane, we expect that corrections to the large-$\Nc$
approximation arising from boson fluctuations will lower the phase transition temperature. This implies that the $\Nc$-independence of the phase boundary is lifted.
To be specific,  we expect that the exact phase boundary approaches the large-$\Nc$ phase
boundary from below. Since fluctuations of the bosons are parametrically suppressed for~$\Nc\to\infty$, 
the phase boundary shown in Fig.~\ref{fig:pdQM} simply represents the large-$\Nc$ phase boundary.
Note that $1/\Nc$-corrections to the gap equation~\eqref{eq:QMgapEq} 
can be conveniently taken into account within our RG framework, see e.~g. Refs.~\cite{Berges:1997eu,Schaefer:1999em,Braun:2003ii,Braun:2011fw}. 
However, a detailed analysis of such corrections is beyond the scope of the present work. We only state 
that the results of these studies agree qualitatively with the simple estimates presented here.

Returning to the universal critical behavior of the theory at the thermal phase transition, we find
that the Nambu-Goldstone bosons, namely the pions, tend to restore the chiral symmetry while the fermions tend to build 
up a condensate and thereby break the symmetry of the ground state. However, the anti-periodic boundary conditions for 
the fermions in Euclidean time direction lead to a suppression of the fermionic modes in the vicinity and above the 
phase transition due to the absence of a zero mode. 

The critical temperature~$\Tc$ of our model can be viewed to be the temperature at which
all $\Nf^2$ bosonic modes are exactly massless in the limit~$k\to 0$. In this limit the (dimensionless) extent~$1/\tau=k/T$ 
of the Euclidean time direction becomes arbitrarily small. This means that the wave-length of these bosonic modes
becomes much larger than the extent of the Euclidean time direction.
Thus, the dynamics close to the thermal phase transition is effectively described by 
an ${\rm O}(\Nf^2)$ scalar theory in $d=3$ dimensions:
\be
\Gamma_k[\sigma,\{\pi^{\chi}\}]
= \int d^3 x\,\left\{ 
 \frac{1}{2}Z_{\sigma}^{\perp}\left(\partial_{i}\sigma\right)^2 + \frac{1}{2}Z_{\sigma}^{\perp}\left(\partial_{i}\pi^{\chi}\right)^2
+\frac{1}{2}\bar{m}^2_{\sigma}(\sigma^2 + \vec{\pi}^{\,2}) + \dots \label{Eq:HSTActionO4}
\right\}\,,
\ee
where~$i=1,2,3$.
This three-dimensional scalar field theory possesses a quantum critical point similar to the one
in our fermionic theories at vanishing temperature. This critical point divides the theory into two physically
distinct regimes in the IR limit, namely one with a spontaneously broken O($\Nf^2$) symmetry in the ground-state and one
with a restored O($\Nf^2$) symmetry. The critical exponents associated with this quantum critical point govern
the scaling behavior of physical observables at the thermal phase transition of the quark-meson model.
The computation of these (thermal) critical exponents is beyond the scope of this review. However, we stress
that the Wetterich equation can indeed be used to determine these exponents accurately, see e.~g. 
Refs.~\cite{Tetradis:1993ts,Litim:2002cf,Delamotte:2007pf,Bervillier:2007rc,Benitez:2009xg,Litim:2010tt}.

In the present work we have essentially restricted ourselves to an analysis of QCD low-energy effective models in the large-$\Nc$ limit.
In this limit we expect the effective action~\eqref{equ::truncationQM0} to be a controlled starting point according to
our fixed-point analysis in the previous section. In principle, we have to take into account the Fierz-complete
basis of four-fermion interactions given in Eq.~\eqref{equ::truncationQM1} when we intend to go beyond the large-$\Nc$ limit. Moreover,
we would have to take into account further four-fermion operators at finite temperature due to the broken
Poincare invariance, see our discussion in Sect.~\ref{subsec:DefFT}. In a partially bosonized language this means that we have
to take into account the associated bosonic degrees of freedom, e.~g. vector bosons and axial-vector bosons. 
This already indicates that the determination of the order of the chiral phase transition in QCD and, in particular, the 
determination of the location of a possibly existing critical point at finite temperature and density is an inherently complicated
task. Note that  in our considerations  we have not even taken into account effects of topologically non-trivial gauge configurations
which break the ${\rm U_A}(1)$ symmetry. 

We would also like to add that our simple model description breaks down above the phase
transition. In fact, we expect gluonic degrees of freedom to become relevant at high temperatures. 
In order to cure this shortcoming, extensions of the presently studied low-energy model 
have been put forward in Refs.~\cite{Meisinger:1995ih,Pisarski:2000eq,Braun:2003ii,Mocsy:2003qw,Fukushima:2003fw,Megias:2004hj,Ratti:2005jh,
Sasaki:2006ww,Schaefer:2007pw,Hell:2008cc,Mizher:2010zb,Skokov:2010wb,Herbst:2010rf,Skokov:2010uh,Hell:2011ic}.

Despite all these shortcomings of this simple model, our study already reveals the basic mechanism of dynamical
chiral symmetry breaking. Moreover, we have shown in this and the previous section that a detailed analysis of the fixed-point structure 
is extremely useful since it provides us with important insights for a reliable construction of effective models.

Let us conclude our discussion of QCD low-energy models with a few words of caution concerning the LPA 
which is often used to include corrections beyond the large-$\Nc$ limit in studies of the 
partially bosonized action Eq.~\eqref{Eq:HSTActionQM}. To this end, we 
consider the mapping of the fermionic and partially bosonized theory in more detail. 
In the chirally symmetric regime the flow equations for the dimensionless renormalized couplings $\epsilon_{\sigma}=\bar{m}^{2}_{\sigma}/(Z_{\sigma}k^2)$, 
$\omega_{\sigma}=\bar{\omega}_{\sigma}/Z_{\sigma}^2$ and the Yukawa coupling $h_{\sigma}=\bar{h}_{\sigma}/(Z_{\sigma}^{1/2}Z_{\psi})$
are given by\footnote{We have set~$Z_{\psi}^{\|}=Z_{\psi}^{\perp}$ and~$Z_{\sigma}^{\|}=Z_{\sigma}^{\perp}$ for simplicity. This 
implies~$\eta_{\psi}^{\|}=\eta_{\psi}^{\perp}$ and~$\eta_{\sigma}^{\|}=\eta_{\sigma}^{\perp}$.}
\be
\partial _t \epsilon_{\sigma} &=& (\eta_{\sigma}- 2)\epsilon_{\sigma} - 12v_3\, l_1^{(4)}(\tau,\epsilon_{\sigma};\eta_{\sigma}) \omega_{\sigma} 
+32\Nc v_3\,  l_{1}^{\rm (F),(d)}(\tau,0,0;\eta_{\psi}) h^2_{\sigma}\,,\label{eq:m2flowQM}
\\
\partial _t \omega_{\sigma}&=& 2 \eta_{\sigma}\omega_{\sigma} +24v_3\, l_2^{(4)}(\tau,\epsilon_{\sigma};\eta_{\sigma}) \omega_{\sigma}^2 
- 64 \Nc v_3\,  l_{2}^{\rm (F),(4)}(\tau,0,0;\eta_{\psi}) h^4_{\sigma}\,,\label{eq:lflowQM}
\\
\partial _t h^2_{\sigma} &=& (\eta_{\sigma} +2\eta_{\psi})h^2_{\sigma}
  - 16v_3\, l_{1,1}^{\rm  (FB),(4)}(\tau,0,\epsilon_{\sigma};\eta_{\psi},\eta_{\sigma})  h^4_{\sigma} ,
\label{eq:hflowQM}
\ee
where $v_3=1/(8\pi^2)$. Again, we have employed 3$d$ regulator functions to derive these equations. The threshold functions
are defined in App.~\ref{app:regthres}. Note that the sign in front of the term~$\sim  h^4_{\sigma}$ in Eq.~\eqref{eq:hflowQM} differs
from the one in Eq.~\eqref{eq:hflowGN}. This change in the sign is due to the existence of the Nambu-Goldstone bosons in our QCD model; the
latter enter this
equation with the opposite sign compared to the radial mode ($\sigma$-mode). Moreover, we would like to point out that the 
term~$\sim  h^4_{\sigma}$ is not present in the NJL model with one fermion species and a continuous chiral symmetry. In this case, we have
one Nambu-Goldstone mode and one radial mode which cancel each other identically, see Eq.~\eqref{eq:h2flowNJL1flavor}.

As in our study of the Gross-Neveu model we can now study the RG flow of the ratio $h^2_{\sigma}/\epsilon_{\sigma}$ which can be obtained straightforwardly from
the flow equations~\eqref{eq:m2flowQM} and~\eqref{eq:hflowQM}. We obtain
\be
\partial_t \left(\frac{h^2_{\sigma}}{\epsilon_{\sigma}}\right) &=& (2 +  2\eta_{\psi})\left(\frac{h^2_{\sigma}}{\epsilon_{\sigma}}\right)
-32\Nc v_3\, l_{1}^{\rm (F),(4)}(\tau,0,0;\eta_{\psi})\left(\frac{h^2_{\sigma}}{\epsilon_{\sigma}}\right)^2
 \nn\\ 
&&
 +12v_3\, l_1^{(4)}(\tau,\epsilon_{\sigma};\eta_{\sigma}) \omega_{\sigma}  \left(\frac{h^2_{\sigma}}{\epsilon^2_{\sigma}}\right) 
\!-\!16v_3\, l_{1,1}^{\rm  (FB),(4)}(\tau,0,\epsilon_{\sigma};\eta_{\psi},\eta_{\sigma})  \left(\frac{h^4_{\sigma}}{\epsilon_{\sigma}}\right).
\label{eq:h2m2flowQM}
\ee
Using Eqs.~\eqref{eq:lFtolFB},~\eqref{eq:bcQM} and~\eqref{eq:hstmapQM},  
we recover the RG flow equation~\eqref{eq:lpsi_flowQM} of the four-fermion coupling~$\lambda_{\sigma}$. 
Thus, the partially bosonized and the purely fermionic description are indeed identical at the UV scale $\Lambda_{\rm H}$. 
Note that the prefactor of the term \mbox{$\sim h^2_{\sigma}/\epsilon_{\sigma}$} would turn out to be incorrect if we did not include the RG running of the
Yukawa coupling, as done in the standard LPA. In other words, a standard LPA does not incorporate
all terms associated with a systematic expansion of the flow equations in powers of~$1/\Nc$. This observation
agrees with our analysis of the Gross-Neveu model in Sect.~\ref{sec:GNmodel}. 



%
\section{Gauge Theories}\label{sec:gaugetheories}
\subsection{Gauge Theories with Few and Many Fermion Flavors - A Motivation}
Chiral gauge theories are of utmost importance for our understanding of the fundamental forces in nature. For example,
the strong interaction is mediated by the exchange of gluons, the gauge bosons of the theory of the strong interaction~(QCD).
To obtain a quantitatively and qualitatively consistent description of the generation of hadron masses in the early 
universe, a comprehensive understanding of chiral symmetry breaking in gauge theories is therefore mandatory. 

Strongly-flavored asymptotically free gauge theories, i.~e. gauge theories with many flavors, are currently very actively researched.
Two prominent examples are QCD with many (light) flavors and QED${}_3$, i.~e. QED in $d=2+1$ dimension.
While there is a long-standing interest in QED${}_3$ as an effective
theory for graphene~\cite{PhysRevLett.95.146801,PhysRevLett.96.256802,PhysRevB.79.165425}, 
QCD with many quark flavors has drawn a lot of attention in recent years. 
The reasons for this great interest in strongly-flavored gauge theories are manifold.
First, the number of (massless) fermions can be considered as 
an external parameter. Such gauge theories are then expected to exhibit a 
quantum phase transition from a chirally broken to a {\it conformal phase} when the number of 
fermion flavors is increased. Second, the understanding of strongly-flavored 
gauge theories underlies (walking) technicolor-like scenarios for the Higgs sector, see e.~g. 
Refs.~\cite{Weinberg:1979bn,Holdom:1981rm,Hong:2004td,Sannino:2004qp,Dietrich:2005jn,Dietrich:2006cm,Ryttov:2007sr,Antipin:2009wr,Sannino:2009za}.
Returning to the dynamics underlying the generation of hadron masses, 
a controllable deformation of real QCD (with two light flavors) can teach us important lessons about the 
underlying principles of chiral symmetry breaking in nature.

The phase structure of gauge theories with $\Nf$ fermions can indeed be rich, as simple considerations may already suggest. 
Due to the screening property of fermionic fluctuations in gauge theories, asymptotic freedom is lost for large $\Nf$. For
instance, an SU($\Nc$) gauge theory with $\Nf$ fermions is no longer asymptotically free (a.f.) for $\Nf>\Nf^{\text{a.f.}}:=\frac{11}{2}\Nc$. 
Another prominent fermion number, $\Nf^{\text{CBZ}}$, potentially exists and denotes the smallest flavor number for which an infrared 
fixed point  $g^2_{\ast}$ of the running gauge coupling in QCD still occurs. Consider the universal\footnote{While the one-loop coefficient
is independent of the scheme, the two-loop coefficient is only universal in mass-independent regularization schemes, e.~g. the $\overline{\text{MS}}$ scheme.} 
two-loop $\beta_{g^2}$ function of the gauge coupling $g^2$:
\be
\beta_{g^2}\equiv \partial_t g^2 = -\left(\beta_0 + \beta_1\left(\frac{g^2}{16\pi^2}\right) +\dots
\right)\frac{g^4}{8\pi^2}\, \label{eq:beta2loop}
\ee
with
\be
\beta_{0}=\frac{11}{3}\Nc-\frac{2}{3}\Nf\,\qquad\text{and}\qquad
\beta_{1}=\frac{34\Nc^3+3\Nf-13\Nc^2\Nf}{3\Nc}\,. \label{eq:betadef2loop}
\ee
One readily observes that this $\beta$-function exhibits a non-trivial fixed point for
 $\Nf>\Nf^{\text{CBZ}}$, the so-called Caswell-Banks-Zaks (CBZ) fixed point \cite{Caswell:1974gg}.
For SU(3), we have $\Nf^{\text{CBZ}}\simeq 8.05$ in the two-loop approximation, marking a sign change of
the coefficient~$\beta_1$.
With increasing~$\Nf>\Nf^{\text{CBZ}}$, 
the fixed-point value $g^2_{\ast}$ decreases and
a perturbative treatment of the theory therefore seems possible near $\Nf\lesssim\Nf^{\text{a.f.}}$.
Moreover, the existence of this fixed point suggests the
existence of a conformally invariant limit in the deep infrared~\cite{Banks:1981nn}. For decreasing $\Nf$, $g^2_{\ast}$~becomes
larger, suggesting the onset of chiral symmetry breaking.\footnote{Of course, it is well-known that the 
phenomenologically important case of two massless flavors 
exhibits chiral symmetry breaking in the IR limit.} 
The fermions then acquire a mass  and decouple from the dynamics of
the theory. This effect destabilizes the CBZ fixed point $g^2_{\ast}$ in the gauge sector of the theory. 
In this case, the IR limit of the theory is 
dominated by massless bosonic excitations, the Nambu-Goldstone modes, and the
spectrum of the theory is characterized by a dynamically generated mass gap. A similar reasoning also applies to QED${}_3$, see
e.~g. Refs.~\cite{Pisarski:1984dj,Appelquist:1986fd}.

\begin{figure*}[t]
\centering\includegraphics[scale=1.0,clip=true]{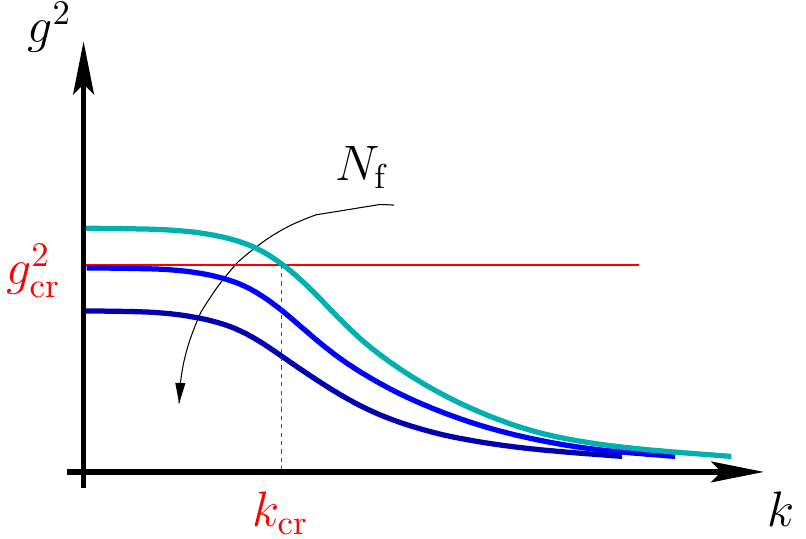}
\caption{Illustration of the IR running of the gauge coupling in comparison to the critical value of the gauge coupling~$g_{\text{cr}}^2$. 
  The figure has been taken from Ref.~\cite{Braun:2010qs}. Below the
  conformal window, $\Nf<\Nfcr$, the gauge coupling $g^2$ exceeds its critical value $g^2_{\rm cr}$ at the scale~$k=\kcr$,  triggering 
  chiral symmetry breaking. For increasing flavor number, the IR fixed-point 
  value $g_{\ast}^2$ becomes smaller than the critical value, indicating that the theory
  is inside of the conformal window. Note that $\kcr$ should not be confused with the chiral
  symmetry breaking scale~$\ksb$. In fact, we have~$\kcr \geq \ksb$, see discussion below.
}
\label{fig:sketch1}
\end{figure*}

Our considerations suggest the existence of a critical value~$g_{\text{cr}}^2$ of the gauge coupling which needs to be exceeded to trigger
chiral symmetry breaking. As a direct consequence, we expect the existence of a {\it quantum critical point} associated with a critical flavor number
$\Nf^{\text{CBZ}}\leq\Nfcr<\Nf^{\text{a.f.}}$ above which gauge theories approach a conformally invariant IR limit, see Fig.~\ref{fig:sketch1}. Thus, $\Nf$ serves as
a control parameter for a quantum phase transition. Note the difference to our studies of quantum critical behavior in the Gross-Neveu model and NJL-type models, 
where we have varied the initial values of the fermionic couplings directly to force the system to undergo a quantum phase transition.

Studies of the phase structure of strongly-flavored gauge theories have been performed with continuum methods as well as lattice 
simulations.  In QED${}_3$ many studies have provided estimates for $\Nfcr$ using Dyson-Schwinger equations
and resummation techniques~\cite{Pisarski:1984dj,Appelquist:1986fd,Appelquist:1988sr,Atkinson:1989fp,
  Pennington:1990bx,Curtis:1992gm,Burden:1990mg,Maris:1995ns,Gusynin:1995bb,Maris:1996zg,Fischer:2004nq}.
Since the dynamically generated mass is substantially smaller than the scale set by the gauge coupling, lattice simulations 
of QED${}_3$ with many flavors are inherently challenging~\cite{Dagotto:1989td,Hands:1989mv,Hands:2002dv,Hands:2004bh}.  The
phase structure of many-flavor QCD has also been studied with continuum
methods~\cite{Caswell:1974gg,Banks:1981nn,Kondo:1991yk,Miransky:1996pd,Appelquist:1996dq,
  Appelquist:1997dc,Schafer:1996wv,Velkovsky:1997fe,Appelquist:1998rb,%
  Harada:2000kb,Sannino:1999qe,Harada:2003dc,Gies:2005as,Braun:2005uj,Braun:2006jd,Terao:2007jm,%
  Poppitz:2009uq,Armoni:2009jn,Sannino:2009qc,Sannino:2009me} and lattice simulations
\cite{Kogut:1982fn,Gavai:1985wi,Fukugita:1987mb,Brown:1992fz,Damgaard:1997ut,%
  Iwasaki:2003de,Catterall:2007yx,Appelquist:2007hu,Deuzeman:2008sc,Deuzeman:2009mh,Appelquist:2009ty,%
  Fodor:2009wk,Fodor:2009ff,Pallante:2009hu,DeGrand:2010ba,Kusafuka:2011fd}. Recent results suggest in this
case that a conformal phase indeed exists, with a quantum phase transition
occurring around~$9\lesssim\Nf^{\text{cr}}\lesssim 13$.

Given the existence of such a quantum critical point in an asymptotically free gauge theory with $\Nf$ flavors, the question arises how the 
spectrum of the theory behaves when we approach this point from below. This question is tightly bound to the 
$\Nf$-dependence of the dynamically generated chiral symmetry breaking scale~$\ksb$.  It is
well-known from studies of Dyson-Schwinger equations in the rainbow-ladder approximation that physical observables, e.~g. 
the fermion condensate or the fermion mass, exhibit an exponential scaling close to $\Nfcr$, {\it provided that} the (momentum) scale
dependence of the gauge coupling can be 
neglected~\cite{Miransky:1984ef,Miransky:1985aq,Miransky:1988gk,Appelquist:1996dq,Chivukula:1996kg,Appelquist:1998xf}:
\be
m_{\psi}\propto \Lambda\theta(\Nfcr-\Nf)\exp\left(
  {-\frac{\pi}{2\epsilon\sqrt{|\alpha_1||\Nfcr
        - \Nf|} } 
}\right).\label{eq:1}
\ee
This type of scaling behavior has already been introduced in Sect.~\ref{sec:ESNJL}.
Here, $m_{\psi}$ denotes the dynamically generated fermion mass, 
and~$\Lambda$ denotes a suitably chosen UV scale.
The quantities $\epsilon$ and $\alpha_1$ in Eq.~\eqref{eq:1} 
are constants arising from the details of the theory and will be defined in Sect.~\ref{sec:MS}. 
This scaling behavior can be viewed as a generalization of essential Berezinskii-Kosterlitz-Thouless (BKT)
scaling~\cite{Berezinskii,Berezinskii2,Kosterlitz:1973xp} to higher-dimensional systems
\cite{Kaplan:2009kr}. We rush to add that the spectra of the different theories below and above $\Nfcr$ are substantially 
different. In particular, a construction of an effective low-energy theory in terms of light scalar fields
may no longer be possible above $\Nfcr$. 
In any case, the fact that essential scaling behavior may occur in various different
systems, ranging from specific 2-dimensional condensed-matter systems over QED${}_3$ to QCD, exemplifies once more
that a phenomenological and technical exchange between these seemingly different research fields offers great potential to gain
deep insights into the mechanisms of symmetry breaking in fermionic theories.

One may wonder how the essential scaling behavior~\eqref{eq:1} close to the quantum phase transition is altered
when the running of the gauge coupling is taken into account. We shall derive the corresponding modified
scaling laws for physical observables in this case, following the discussion in Ref.~\cite{Braun:2010qs}.
Our main arguments are based on very general considerations and involve only few assumptions about the 
fixed-point structure of the theory. The resulting scaling laws can be tested in QCD and QED${}_3$, for example 
with the aid of Monte-Carlo simulations.
In Sect.~\ref{sec:fewflavor} we begin our discussion of chiral symmetry breaking in gauge theories 
with the simple few-flavor case using the example of QCD. We explain the issue of scale fixing in gauge theories which arises when
one is interested in a meaningful comparison of theories with a different number of flavors. These ideas have been put forward
in Refs.~\cite{Braun:2006jd,Braun:2009ns} and have been used
to improve the parameter fixing in QCD model studies~\cite{Schaefer:2007pw}. In Sect.~\ref{sec:QCGT} we 
analyze the scaling behavior in gauge theories close to a quantum critical point on general grounds.
In Sect.~\ref{sec:MS}, we briefly repeat the arguments given in Sect.~\ref{sec:ESNJL} which
lead to an exponential scaling behavior at a quantum phase transition. In addition, we give the leading-order correction
to the exponential scaling behavior in gauge theories. In Sect.~\ref{sec:powerlaw}, we discuss 
power-law-like scaling behavior in gauge theories, which provides a strict upper bound for the $\Nf$-scaling of 
chiral low-energy observables.
In Sect.~\ref{sec:beyondmiransky}, we present
the {\it universal} corrections to the exponential scaling behavior~\eqref{eq:1} 
which arise due to the running of the gauge coupling. Moreover, we show that  power-law scaling and exponential scaling
arise as two different limits of one and the same RG flow. The consequences for low-energy observables from the existence of 
a nearby quantum critical point are explained in Sect.~\ref{sec:ScalLEM} with the aid of a simple low-energy model.
To illustrate our analytic findings, we review numerical results from non-perturbative RG studies of the 
scaling behavior in many-flavor QCD in Sect.~\ref{sec:SUNYM}.

For our discussion of many-flavor QCD in Sect.~\ref{sec:SUNYM}, we focus on the chiral phase transition, even though we 
also expect an impact of the confining nature of the theory on the properties of the system near criticality. However, since we work in the chiral limit, 
there is no good order parameter for confinement, in particular for many quark flavors. This implies that  for large~$\Nf$ nonanalyticities in the
correlation functions are rather dominated by the chiral degrees of freedom. 
For few-flavor QCD, on the other hand, we may indeed expect that the fixed-point
structure of the matter sector is significantly affected by the confining dynamics of the theory. In Sect.~\ref{sec:CSBCONF}, we show that at finite temperature
an interrelation of the confinement and chiral order parameter exists, at least for small~$\Nf$. This can be simply understood
in terms of the fixed-point structure of the matter sector.

An outlook is given in Sect.~\ref{sec:outlookYM},
including a discussion of the implications of our findings for other gauge theories, e.~g., with fermions in the adjoint representation.

\subsection{The Issue of Scale Fixing in Gauge Theories}
\label{sec:fewflavor}
To illustrate the issue associated with scale fixing in gauge theories, let us consider few-flavor QCD. 
In the limit of zero current quark masses (chiral limit), QCD depends only on one parameter, 
namely the gauge coupling $g$, see Eq.~\eqref{eq:SQCD}. In the quantum theory, the gauge coupling has to be fixed at a certain
momentum scale in terms of a renormalization condition. The RG finally trades the gauge coupling fixed at an arbitrary scale in for one
single parameter $\LQCD$ of mass dimension one. The latter sets the mass scale for all physical observables of the theory. In other words, all physical
observables respond trivially to a variation of $\LQCD$ according to their canonical mass dimension. In units of $\LQCD$, the theory is completely fixed. 

In order to discuss the dependence on quantities such as the flavor number, it is important to emphasize that a variation of the flavor number does not
correspond to a change of a parameter of the theory. It rather corresponds to changing the theory itself. In particular, there is no unique
way to unambiguously compare theories of different flavor number with each other, as different theories may have different scales~$\LQCD$. 

For example, it may seem natural to compare theories with different flavor numbers at fixed $\LQCD$ with each other. However, $\LQCD$ itself is not a
direct observable. Hence, such a comparison is generically spoilt with theoretical uncertainties. Moreover, $\LQCD$ is regularization-scheme
dependent which can affect comparisons between different theoretical methods, say, lattice and functional approaches. Another option could be a scale fixing in
the deep perturbative region, say, at the Z-boson mass scale by fixing~$g^2(M_{\rm Z})$. However, theories with different flavor numbers then exhibit a
different perturbative running, such that IR observables vary because of both high-scale perturbative as well as non-perturbative (RG) evolution. 

Following Ref.~\cite{Braun:2009ns}, we propose to choose a mid-momentum scale for the scale fixing, as the high-scale perturbative running is then separated from the more
interesting non-perturbative dynamics. For example, we may fix the theories at any $\Nf$ by keeping the running coupling at the $\tau$ mass scale fixed to
$g^2(m_\tau)/(4\pi)=0.322$. Even though also this choice is scheme dependent, these dependences should be subdominant, as they follow a perturbative
ordering. In general, fixing the scale via the coupling is a prescription which is well accessible by many non-perturbative methods. 
As an alternative of the described scale-fixing prescription, however, one might think of keeping the value of an IR observable
fixed for theories with different~$\Nf$, e.~g. the pion decay constant or the critical temperature. Of course, this is also possible,
e.~g. in lattice QCD simulations. We shall comment on possible difficulties arising from such a procedure in Sect.~\ref{sec:powerlaw}. 
\begin{figure}[t]
\begin{center}
\includegraphics[scale=0.5,clip=true]{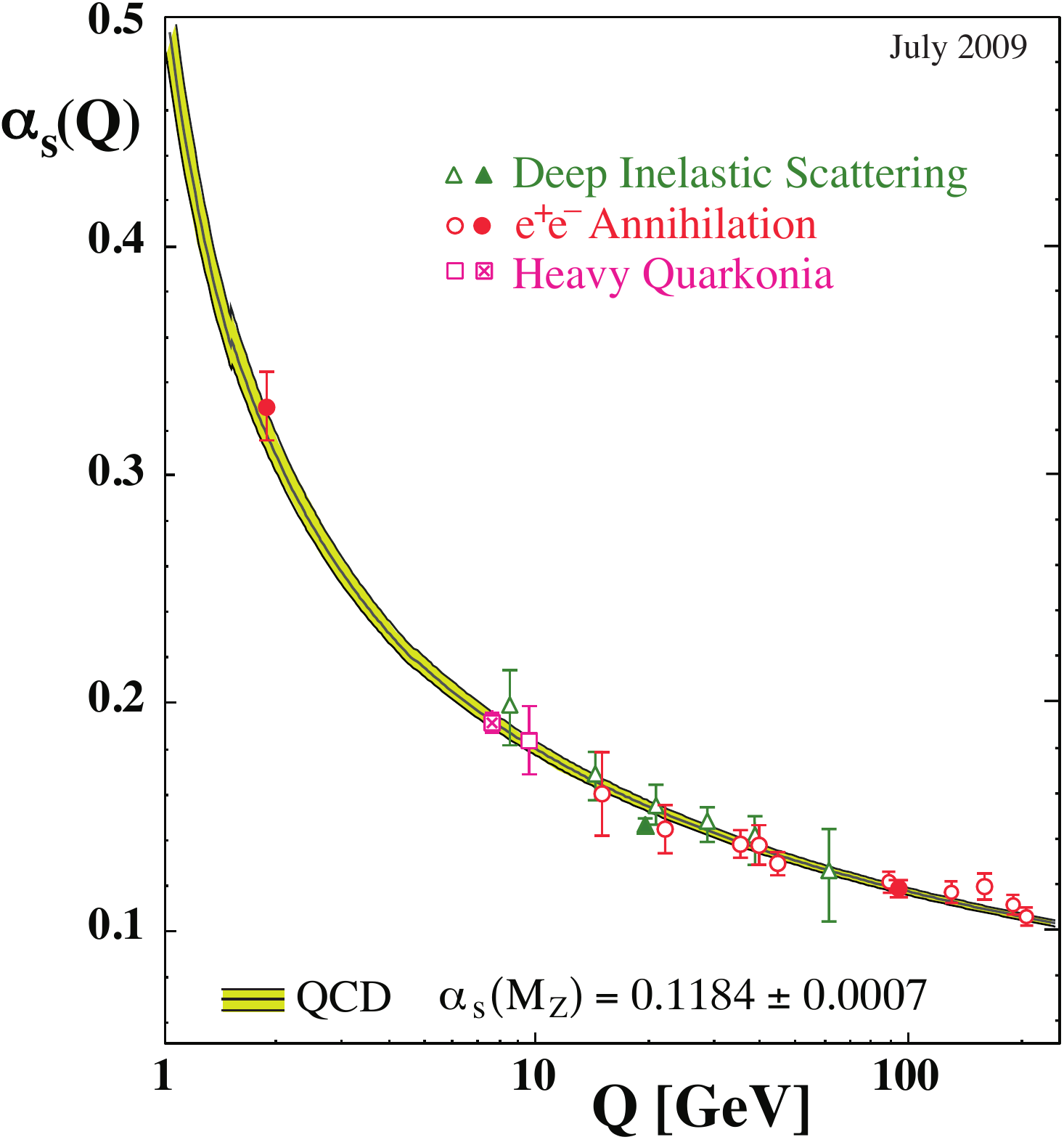}
\end{center}
\caption{Strong coupling~$\alpha_{s}= g^2/(4\pi)$ as a function of the momentum (transfer)~$Q$. 
The figure has been taken from~Ref.~\cite{Bethke:2009jm}.}
\label{fig:alphas}
\end{figure}

Let us now present a simple argument that illustrates how the $\Nf$-dependence of physical observables can be understood in the limit of small $\Nf$. As
already stated above, all IR observables such as the chiral phase transition temperature $\Tc$, the pion decay constant $f_\pi$,  the chiral condensate $\langle
\bar\psi\psi\rangle$, and model-dependent concepts such as the constituent quark mass, are proportional to $\LQCD$. On the one hand, the latter 
can be read off from the UV behavior of the running coupling, $g^2 (k)\sim 1/\ln (k/\LQCD)$ for large $k\sim Q$, see also Fig.~\ref{fig:alphas}.
On the other hand, the value of $\LQCD$
can be associated with the position of the Landau pole in perturbation theory.\footnote{Of course, this statement has to be taken with care, since
$\LQCD$ is a meaningful scale, whereas the Landau pole is simply an artifact of perturbation theory.}
In this simple reasoning, the artificial Landau pole in the one-loop $\beta_{g^2}$ function can be taken as an estimate for the scaling of physical observables. 
To be specific, we have
\be
\beta_{g^2}^{\text{1-loop}} \equiv \partial_t g^2 = - \beta_0\left(\frac{g^4}{8\pi^2}\right)\,,
\ee
where $\beta_0$ is defined in Eq.~\eqref{eq:betadef2loop}. The position of the Landau pole can then be read off from 
\be
0=\frac{1}{g^2(\LQCD)}=\frac{1}{g^2(\mu_0)} + \beta_0\, \ln \left(
\frac{\LQCD}{\mu_0}\right)\,, 
 \label{eq:LandauPole}
\ee
where $\mu_0$ denotes a perturbative scale, such as the $\tau$-mass scale $m_\tau$ or the Z-boson mass~$M_{\rm Z}$. 
Solving this equation for $\LQCD$ and expanding the result for small $\Nf$ leads us to
\be
  \LQCD &\simeq&\mu_0\, \E^{-\frac{1}{b_0
      g^2 (\mu_0)}} 
  \simeq \mu_0\, \E^{-\frac{24\pi ^2}{11\Nc g^2(\mu_0)}} \left(
  1-x \Nf + \mathcal O ((x \Nf)^2)\right)\,.
  \label{eq:LQCD1loopEst}
\ee
Choosing $\mu_0=m_\tau$, we find $x = \frac{48 \pi^2}{121 \Nc^2 g^2(\mu_0)} \simeq 0.107$ for $\Nc=3$. Two conclusions can immediately be
deduced from this expression: first, $\LQCD$ can be expanded in $\Nf$ and has a generically nonvanishing linear term. 
Second, for the present way of scale fixing the linear behavior should be a reasonable approximation 
for $\Nf\lesssim 4$, as the (dimensionless) expansion parameter $x$ is small in this regime.

Since $\LQCD$ sets the scale for all dimensionful IR observables, we are tempted to conclude that all IR observables scale linearly with $\Nf$ for
small $\Nf$ with the same proportionality constant $x$. Of course, this would be too simple since the dynamics which establishes the value of
the IR observables generically carries an $\Nf$ dependence as well. For example, the chiral symmetry-breaking dynamics depends on the number of light
mesonic degrees of freedom, which is an $\Nf$-dependent quantity. 
Nonetheless, it is reasonable to expect that in leading order (chiral) low-energy observables~${\mathcal O}$ indeed scale according to
\begin{equation}
{\mathcal O}= {\mathcal O}_0\left(\,1- x \Nf + \dots \,\right),
\label{eq:lowNfTc}
\end{equation}
where ${\mathcal O}_0$ is a dimensionful proportionality constant. For example, a linear dependence of the chiral phase transition
temperature~${\mathcal O}=\Tc$ on~$\Nf$ has been found in lattice simulations~\cite{Karsch:2000kv} and in studies with
functional RG methods~\cite{Braun:2005uj,Braun:2006jd}, see also Sect.~\ref{sec:PSFYM}. 
Note that Eq.~\eqref{eq:LQCD1loopEst}, which underlies Eq.~\eqref{eq:lowNfTc}, has led to a significant improvement of the 
parameter fixing in studies of so-called Polyakov-loop extended low-energy models~\cite{Schaefer:2007pw,Schaefer:2009ui,Herbst:2010rf}. 

Let us finally generalize our discussion to gauge theories other than QCD. Our scale-fixing prescription indeed also applies to other
theories in which dynamical chiral symmetry breaking is triggered by a running coupling that approaches a non-trivial IR fixed point.
To be more specific, we shall focus our discussion on strongly-flavored asymptotically free gauge theories, such as QCD with many flavors and ${\rm
QED}_3$. By asymptotic freedom, we refer here to the vanishing of the dimensionless renormalized coupling in the UV. 
In such theories, it seems natural to expect that the dependence of the coupling on the (momentum) scale modifies the exponential 
scaling behavior~\eqref{eq:1}. We will discuss this in detail in Sect.~\ref{sec:beyondmiransky}.

\subsection{General Aspects of Quantum Critical Behavior in Gauge Theories}\label{sec:QCGT}
\subsubsection{Miransky Scaling}\label{sec:MS}
In Sect.~\ref{sec:ESNJL} we have discussed essential/exponential scaling behavior in a simple NJL-type model. These
considerations can be straightforwardly generalized to gauge theories. The role of the vector-coupling in our simple 
model is now played by the squared gauge coupling~$g^2$. 
In gauge theories, exponential scaling behavior near a {\it quantum critical point} is also known as 
Miransky scaling~\cite{Miransky:1988gk,Miransky:1996pd}. 
\begin{figure}[t]
\begin{center}
\includegraphics[scale=0.95,clip=true]{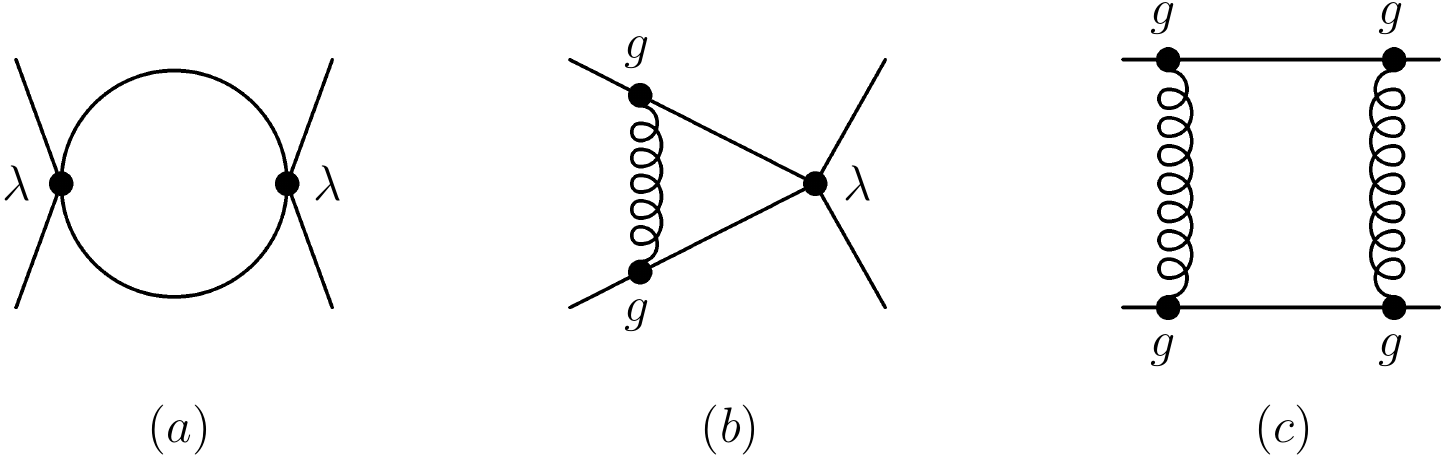}
\end{center}
\caption{Representation of the terms on the right-hand side of the RG flow
  equation~\eqref{eq:4psiflow} by means of 1PI Feynman diagrams, see Refs.~\cite{Braun:2006wu,Braun:2010qs}. 
  Our functional RG approach, see e.~g. Sect.~\ref{sec:SUNYM}, includes
    resummations of all diagram types including ladder-diagrams generated by type (b) and (c) as well as the corresponding crossed-ladder diagrams.  }
\label{fig:feynmanYM}
\end{figure}

The following analysis is by no means bound to QCD. To make this explicit, we shall keep our discussion as general as possible and
consider a very general class of theories where symmetry breaking and condensate formation is driven by fermionic self-interactions. Independently
of whether these interactions may be fluctuation-induced (as in QCD) or fundamental (as in beyond standard-model applications)
This class of theories can be parameterized by the following action:
\be
S_{\rm M} &=& \int d^dx \Big\{ \bar{\psi}({\rm i}\fslash{\partial}
+\bar{g}\fslash{A} 
)\psi 
 + \bar{\lambda}_{\alpha\beta\gamma\delta} 
\bar{\psi}_{\alpha}\psi_{\beta}\bar{\psi}_{\gamma}\psi_{\delta}
\Big\}\,,
\label{eq:miransky_ansatz}
\ee
where $\alpha,\beta\,\dots$ denote a specific set of collective indices including, e.~g., flavor and/or color 
indices. In general, we expect to have more than just one four-fermion interaction channel as it is indeed the case
in QCD, see Eq.~\eqref{equ::truncationQM1}. 

From the action~\eqref{eq:miransky_ansatz} we can derive the $\beta$ function of the dimensionless four-fermion coupling $\lambda$
in the point-like limit. It assumes the following simple form:
\be
\beta_{\lambda}\equiv\partial_t \lambda = (d-2)\lambda - a \lambda^2 -b\lambda g^2 -c g^4\,.
\label{eq:4psiflow}
\ee
The couplings $\lambda\sim\bar{\lambda}/k^{(d-2)}$ and $g\sim\bar{g}/k^{4-d}$ denote
dimensionless and suitably renormalized couplings. The quantities
$a$, $b$ and $c$ do not depend on the RG scale but may depend on control parameters, such as the number of fermion flavors~$\Nf$ or the number of 
colors~$\Nc$ in QCD. This $\beta_{\lambda}$ function can be directly compared to the flow equation~\eqref{eq:BKTsigma} of our toy model, where 
the role of~$g^2$ is played by the vector coupling.
Note that the coefficients $a$, $b$ and $c$ can depend implicitly on the RG scale as soon as we introduce a dimensionful external
  parameter, e.~g., temperature~$T$. However, the coefficients remain dimensionless since they depend only on the ratio $T/k$, see
   e.~g. Refs.~\cite{Braun:2005uj,Braun:2006jd}.
  
The various terms on the right-hand side of Eq.~\eqref{eq:4psiflow} can be understood in terms of perturbative
Feynman diagrams~\cite{Braun:2006wu}, see Fig.~\ref{fig:feynmanYM}. Note that we have dropped terms  in Eq.~\eqref{eq:4psiflow} which are 
proportional to the anomalous dimension~$\eta_{\psi}$ of the fermion fields. 
In contrast to purely fermionic theories, $\eta_{\psi}$ can be finite in gauge theories even in the point-like limit.
This is due to the existence of 1PI diagrams proportional to $g^2$ with one internal fermion line and one internal gauge boson 
line.\footnote{This is in close analogy to the partially bosonized formulations of purely fermionic models where $\eta_{\psi}$ is finite due
to the existence of a 1PI diagram with one internal fermion line and one internal boson line,
see e.~g. Eqs.~\eqref{eq:PB1etaPsi} and~\eqref{eq:etapsi_flowGN}.} In the following 
we assume that these contributions are small. This is indeed true in the 
chirally symmetric regime where these contributions are proportional to the gauge-fixing parameter and therefore vanish
at least in the Landau gauge~\cite{Gies:2003dp}.

As we have not specified the running of the vector coupling in our toy model study of essential scaling in Sect.~\ref{sec:ESNJL},
we have not further specified the details of the gauge sector in Eq.~\eqref{eq:miransky_ansatz}. In fact, let us ignore the running of the gauge coupling in this
section, and consider the gauge coupling as a {\it scale-independent} ``external" parameter. The RG flow of the gauge coupling is then trivially
governed by
\be
\partial_t g^2 \equiv 0\,.\label{eq:mcoup}
\ee
This might be an acceptable approximation in the vicinity of an IR fixed point $g^2_{\ast}$. Nonetheless, the value of $g^2_{\ast}$ 
may still depend on other control parameters such as $\Nf$ or $\Nc$, see our discussion of QCD with many flavors in Sect.~\ref{sec:SUNYM}. 
\begin{figure}[!t]
\begin{center}
\includegraphics[scale=0.65]{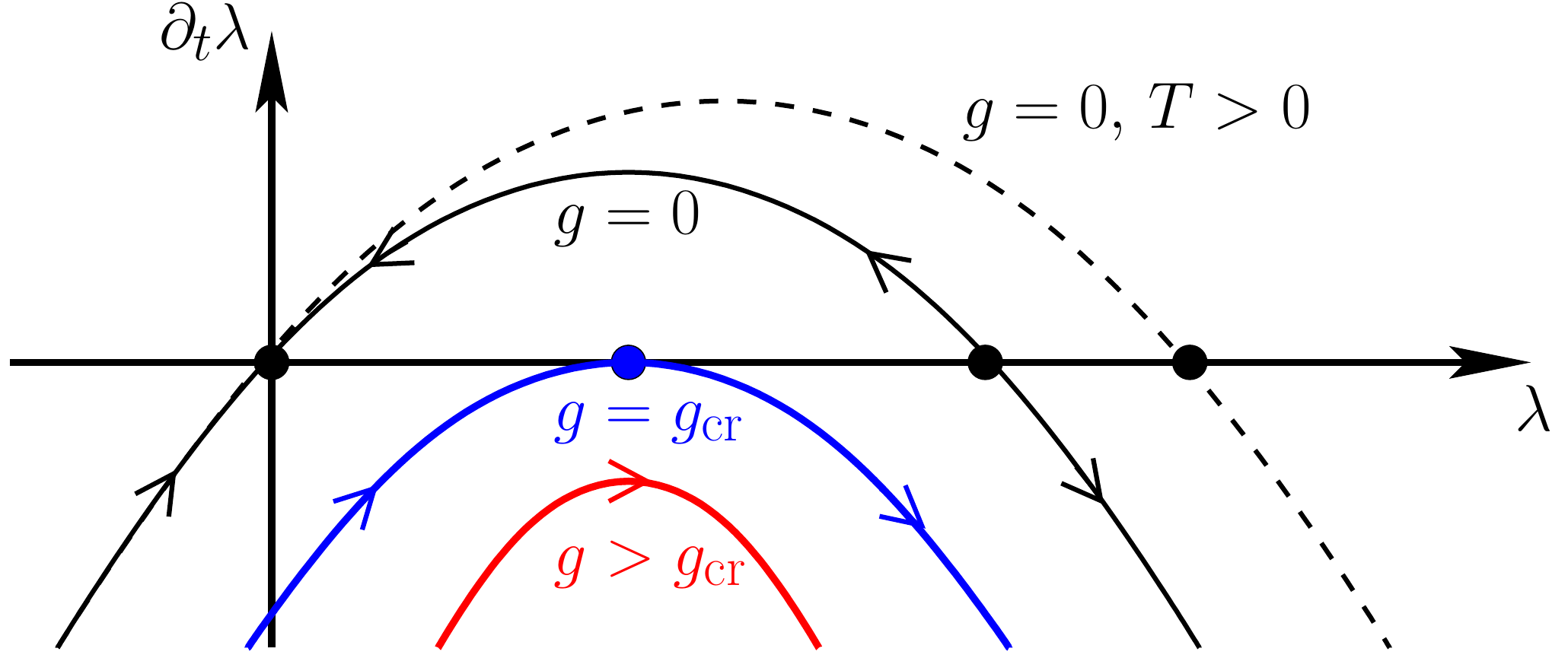}
\end{center}
\bigskip
\caption{Sketch of a typical $\beta$ function for the fermionic self-interactions $\lambda$, see Refs.~\cite{Gies:2005as,Braun:2010qs} and also \cite{Braun:2006jd} for
  the generalization to finite temperature): at vanishing gauge coupling, $g^2=0$, the Gau\ss ian fixed point $\lambda=0$ is IR
  attractive. For $g^2 = g^2_{\rm cr}$, the fixed-points merge due to a shift  of the parabola induced 
  by the gauge-field fluctuations~$\sim g^4$. For gauge couplings larger than the critical coupling $g^2>g^2_{\text{cr}}$, 
  no fixed points remain and the strength of the self-interactions increases rapidly, signaling the onset of 
  chiral symmetry breaking. The arrows indicate the direction of the flow towards the infrared. For increasing
  temperature, the parabolas become broader and higher. This is indicated by the dashed
  line.
} 
\label{fig:parab}
\end{figure}

In Fig.~\ref{fig:parab} we show a sketch for the $\beta_{\lambda}$ function, implicitly assuming that $a>0$, $b>0$ and $c>0$ in Eq.~\eqref{eq:4psiflow}. 
For a vanishing gauge coupling $g^2$ we find two fixed points, an IR attractive Gaussian fixed point at $\lambda=0$ and an IR repulsive fixed point at
$\lambda>0$. For increasing $g^2$ these fixed points approach each other and eventually merge for a {\it critical} value $g^2_{\rm cr}$, 
\be
g_{\rm cr}^2=\frac{d-2}{b + 2 \sqrt{ac}}\,.  
\ee
For $g^2 >g^2_{\rm cr}$ we strictly have $\partial_t \lambda<0$ and the four-fermion coupling then becomes a relevant operator and increases 
rapidly towards the IR, indicating the onset of (chiral) symmetry breaking. Thus, the four-fermion coupling $\lambda$ necessarily
diverges for $g^2 > g^2_{\rm cr}$ at a finite RG scale $\ksb=\ksb(g^2)$. 
Here, we assume that the initial conditions at the UV scale $k=\Lambda$ for the four-fermion coupling $\lambda$ are chosen such 
that $\lambda^{\rm UV}$ is smaller than the value of the IR repulsive fixed point, see Fig.~\ref{fig:parab}.  In beyond-standard model applications
$\lambda^{\rm UV}$ is sometimes considered to be a finite parameter, see e.~g. Ref.~\cite{Fukano:2010yv}. We therefore add that the exponential scaling
behavior discussed below can only be observed when $\lambda^{\rm UV}$ is chosen to be smaller than the value of its repulsive fixed point for a given~$g^2$.  
Otherwise, we expect a power-law-like scaling behavior as discussed in Sect.~\ref{sec:NJLPL}.

This picture of the emergence of chiral symmetry in gauge theories is not new but has been put 
forward in~\cite{Gies:2005as,Braun:2005uj,Braun:2006jd,Braun:2009ns} and
successfully employed for an analysis of the phase structure of QCD with various numbers of flavors and colors at zero and finite
temperature~\cite{Gies:2005as,Braun:2005uj,Braun:2006jd,Braun:2009ns}. Moreover, this picture has also been employed to study conformal scaling in
quantum field theories, see e.~g. Ref.~\cite{Kaplan:2009kr}.

Even though the symmetry breaking scale $k_{\text{SB}}$ is not a direct observable, it sets the scale for 
(chiral) observables~${\mathcal O}$ such as condensates, decay constants,
critical temperatures, etc.:
\be
{\mathcal O}=f_{\mathcal O}\,\ksb^{d_{\mathcal O}}\,,\label{eq:ObsScaling}
\ee
where $d_{\mathcal O}$ is the canonical mass dimension of the observable $\mathcal O$ and $f_{\mathcal O}$ is a function which does not depend on
$g^2_{\rm cr}$ but may depend on $g^2$ and other external parameters, e.~g., $\Nf$ and/or $\Nc$. The function $f_{\mathcal O}$ can be computed
systematically within certain approximations schemes such as large-$\Nc$ expansions or chiral perturbation theory, see Sect.~\ref{sec:ScalLEM} and
Refs.~\cite{Braun:2009ns,Jarvinen:2010ks}.

Let us now briefly discuss the scaling behavior of the symmetry-breaking scale $\ksb$ when~$g^2$ is varied by hand as a constant ``external" parameter. 
To this end, we have to solve the RG flow equation~\eqref{eq:4psiflow}. In close analogy to our study in Sect.~\ref{sec:ESNJL}, we obtain
the following result for the scale~$\ksb$:\footnote{We have chosen the initial conditions such 
that $\lambda^{\rm UV}=\lambda_{\rm max}$, where~$\lambda_{\rm max}$ denotes the position of the maximum of the
$\beta_{\lambda}$ function, i.e., the peak of the parabola in~Fig.~\ref{fig:parab}.}
\be
\ksb \propto \Lambda \theta(g^2-g^2_{\rm cr}) 
\exp\left( {-\frac{\pi}{2\epsilon\sqrt{g^2 - g^2_{\rm cr}}  }}
\right)\,.
\label{eq:miransky}
\ee
Here, $\epsilon$ is a numerical factor,
\be
\epsilon = \sqrt{\frac{(d-2)(2ac+b\sqrt{ac})}{b+2\sqrt{ac}}}\,,
\label{eq:defeps}
\ee
which in general depends on the details of the theory under consideration, e.~g. the number of colors and flavors in QCD.
In any case, we find an exponential scaling behavior of $\ksb$ for~$g^2$ close to $g^2_{\rm cr}$.  
Due to Eq.~\eqref{eq:ObsScaling} it is reasonable to expect that physical chiral observables~$\mathcal O$ inherit this scaling behavior from 
the symmetry breaking scale~$\ksb$.

Let us now discuss the consequences of the scaling law~\eqref{eq:miransky} when we apply our considerations 
to strongly-flavored gauge theories, such as QED${}_3$ or QCD with many fermion flavors. In these cases we may choose the IR fixed-point of 
the gauge coupling as an external parameter, i.~e. $g^2=g^2_{\ast}(N_f)$ in Eq.~\eqref{eq:miransky}. Depending on the $\Nf$ 
dependence of the coefficients $a$, $b$ and $c$ in the $\beta_{\lambda}$ function, the critical
value for the gauge coupling may depend on the number of flavors as well, $g^2_{\rm cr}=g^2_{\rm cr}(\Nf)$.
The critical number of fermion flavors $\Nfcr$ can then be obtained from the {\it criticality condition}
\be
g^2_{\rm cr}(\Nfcr)=g^2_{\ast}(\Nfcr).\label{eq:critcond}
\ee
This corresponds to the coupling value for which the two fixed points of the four-fermion coupling $\lambda$ merge and then annihilate 
each other for $g^2> g_{\text{cr}}^2 $. Expanding $g^2_{\ast}(\Nf)-g^2_{\rm cr}(\Nfcr)$ around $\Nfcr$,
\be
 g^2_{\ast}(\Nf)-g^2_{\rm cr}(\Nfcr)
= \alpha_1(\Nf\!-\!\Nfcr)+\alpha_2(\Nf\!-\!\Nfcr)^2+\dots,\label{eq:g2g2cr}
\ee
and plugging Eq.~\eqref{eq:g2g2cr} into Eq.~\eqref{eq:miransky}, we find the exponential $\Nf$-scaling of $\ksb$:
\be
\ksb\propto \Lambda\theta(\Nfcr-\Nf)\exp\left(
  -\frac{\pi(1-\frac{\alpha_2}{|\alpha_1|}|\Nfcr - \Nf|+\dots)}{2\epsilon\sqrt{|\alpha_1||\Nfcr 
        -  \Nf|} 
}\right).\label{eq:mcorrections}
\ee
We observe that the size of the regime for exponential scaling depends on the ratio $|\alpha_2/\alpha_1|$ which in turn depends on the theory 
under consideration. Thus, the size of the scaling regime may presumably be different in, e.~g., QCD and QED${}_3$. In Sect.~\ref{sec:SUNYM}
we compare these analytic findings with results from a numerical analysis of QCD with many quark flavors.

\subsubsection{Power-law Scaling}\label{sec:powerlaw} 
In this section we discuss how the running of the gauge coupling affects the RG flow of four-fermion couplings. In particular, we argue that (chiral)
symmetry breaking in strongly-flavored gauge theories is a multi-scale problem, in contrast to the scenario associated with Miransky scaling.  In
other words, the (chiral) symmetry breaking scale $\ksb$ and its scaling with the control parameters, e. g. the number of flavors $\Nf$,
depends on the scale fixing and its potential flavor dependence.
\begin{figure*}[t]
\centering\includegraphics[scale=0.65,clip=true]{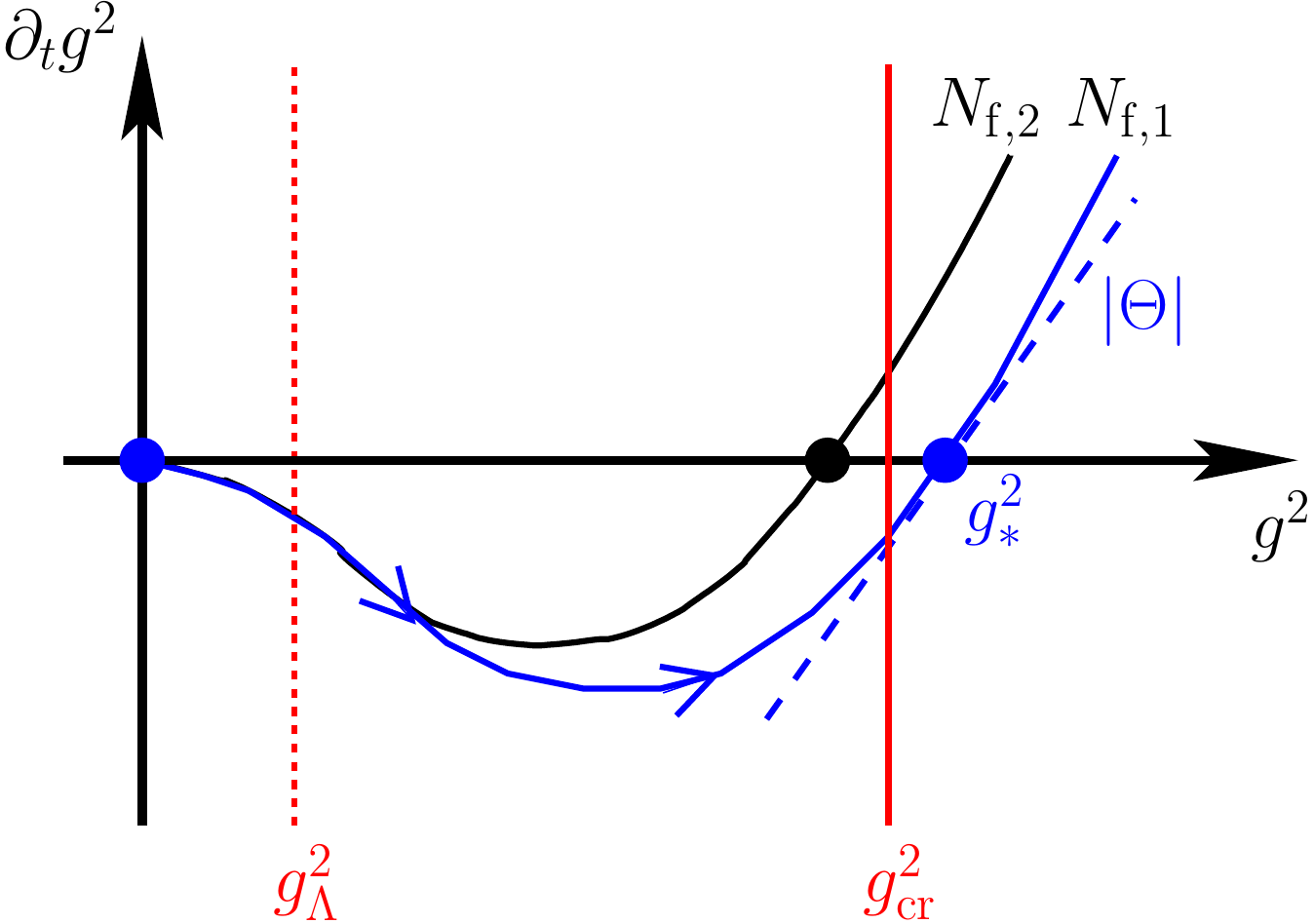}
\caption{Sketch of the $\beta_{g^2}$-function 
 for~$N_{{\rm f},1} <\Nfcr$ and~$N_{{\rm f},2} > \Nfcr$. 
  The slope of the $\beta_{g^2}$-function at the IR fixed-point
  corresponds to minus the critical exponent $\Theta$, see Eq.~\eqref{eq:FPR} and Eq.~\eqref{eq:genthetadef} for a general discussion.
  The vertical line to the right gives the value of $g^2_{\rm cr}$. The dotted vertical line to the left gives  the value of the gauge coupling at the UV scale~$\Lambda$ 
  which we keep fixed for all $\Nf$. However, the value of $g^2_{\rm cr}$ may depend on $\Nf$. The critical number of fermion
  flavors~$\Nfcr$  is determined by $g^2_{\rm cr}(\Nfcr)=g^2_{\ast}(\Nfcr)$, see Eq.~\eqref{eq:critcond}.
  The arrows indicate the direction of the flow towards the infrared.}
\label{fig:sketch2}
\end{figure*}

In the following, we include the running of the gauge coupling which goes beyond standard rainbow-ladder approaches employed in the context of strongly-flavored gauge
theories, see e.~g. Ref.~\cite{Appelquist:1998rb}. 

As we have argued in detail in Sect.~\ref{sec:fewflavor}, fixing the scale of theories with, say, different flavor numbers $\Nf$ by keeping the running
coupling at some scale $\Lambda$ (e.~g. $\tau$ mass) fixed to a certain value, seems to be a well accessible prescription for many non-perturbative
methods. In general, it is important to take care that this scale-fixing procedure is not (or as little as possible) spoilt by scheme dependences. The
latter constraint then essentially rules out $\LQCD$ as a proper scale in QCD to be kept fixed in theories with different flavor numbers.  
For what follows, we shall choose a mid-momentum scale for the scale fixing, lying in between the high-scale perturbative running and the more interesting
non-perturbative dynamics.  Thus, we fix the theories at any $\Nf$ by keeping the running coupling at some intermediate scale $\Lambda$ fixed to a
certain value, say $g^2_{\Lambda}\equiv g^2(\Lambda)$.

For a monotonically increasing coupling flow, the value of the non-trivial IR fixed point $g_{\ast}^2$ of the gauge coupling corresponds to the largest
possible coupling strength of the system in the conformal window, i.~e. for $\Nfcr<\Nf<\Nf^{\rm a.f.}$.  As both $g_\ast^2$ and $g_{\text{cr}}^2$ depend
on the number of flavors, the criticality condition $g_\ast^2(\Nfcr)=g_{\text{cr}}^2(\Nfcr)$ defines the lower end of the conformal window and thus the 
critical flavor number, see Sect.~\ref{sec:MS} and Fig.~\ref{fig:sketch1} for an illustration.

For $g_\ast^2>g_{\text{cr}}^2$, our model~\eqref{eq:miransky_ansatz} is below the conformal window and runs into the broken phase. Slightly below the 
conformal window, the running coupling $g^2$ exceeds the critical value while it is in the attractive domain
of the IR fixed point $g_\ast^2$. The flow in this fixed-point regime can approximately be described by the $\beta$-function expanded around the 
fixed point $g_\ast^2$:
\begin{equation}
\beta_{g^2}\equiv \pat g^2 =-\Theta\, (g^2\!-\! g_\ast^2) + {\mathcal O}((g^2  - g_\ast^2)^2)\,. \label{eq:FPR}
\end{equation}
The universal ``critical exponent" $\Theta$ denotes (minus) the first expansion coefficient and depends on~$\Nf$. We know
that~$\Theta <0$ for $\Nf\gtrsim \Nfcr$, since the fixed point is IR
attractive, see Fig.~\ref{fig:sketch2}. 
The flow equation~\eqref{eq:FPR} for the running coupling can then be solved analytically:
\begin{equation}
g^2(k)=g_\ast^2-\left(\frac{k}{k_0}\right)^{-\Theta}\,. \label{eq:FPsol}
\end{equation}
The scale $k_0$ corresponds to a scale where the system is already in the fixed-point regime. For the present fixed-point considerations, $k_0$
provides for all dimensionful scales. However, from the knowledge of the full RG trajectory, $k_0$ can be related to the initial scale $\Lambda$, say the $\tau$ mass
scale in QCD, by RG evolution. In the following we keep the scale~$k_0$ fixed, as we keep the UV scale~$\Lambda$ fixed.

As already discussed, a necessary condition for (chiral) symmetry breaking is that $g^2_\ast>g^2_{\text{cr}}$.  This implies that $g^2(k)$ 
exceeds $g_{\text{cr}}^2$ at some scale $k_{\text{cr}}$ which is implicitly defined by the criticality condition, $g_\ast^2(\Nfcr)=g_{\text{cr}}^2(\Nfcr)$, 
and therefore we have
\be
\kcr \geq \ksb\,,
\ee
where $\ksb$ is the scale at which the four-fermion coupling $\lambda$ diverges. Thus, $\kcr$ is an upper bound for the symmetry breaking
scale $\ksb$. From Eq.~\eqref{eq:FPsol} and the criticality condition~$g^2(k_{\text{cr}})= g^2_{\text{cr}}$, we derive an estimate for $\kcr$ valid in the
fixed-point regime
\be
\kcr\simeq k_0\, (g_\ast^2 -g_{\text{cr}}^2)^{-\frac{1}{\Theta}}. \label{eq:kcrest}
\ee
The scale $\kcr$ is dynamically generated. Note that $k_{\text{cr}}/k_0\to 0$ for $g^2_\ast\to g^2_{\text{cr}}$ from above. Due to
our scale-fixing procedure, this scale depends on $\Nf$ and $\Nfcr$ in a non-trivial way. Note that it is, in principle, possible to adjust the
initial value of the coupling at the initial scale~$\Lambda$ such that the scale~$\kcr$ is independent of $\Nf$ and $\Nfcr$. This procedure would be similar to keeping
the $\LQCD$ fixed for different values of~$\Nf$. As indicated in Sect.~\ref{sec:fewflavor}, 
we expect that such a scale-fixing procedure would, however, be strongly affected by  scheme dependences, at least in our truncation. Using
Eq.~\eqref{eq:g2g2cr} and a Taylor expansion of the critical exponent near the quantum phase transition,
\be
\Theta(\Nf)=\Theta_0 + \Theta_1 (\Nf-\Nfcr) +{\mathcal O}((\Nf-\Nfcr)^2)\,,
\label{eq:taylor}
\ee
we find the following $\Nf$ dependence of $\kcr$ for $\Nf\leq \Nfcr$:
\be
\kcr \simeq k_0|\Nfcr \!-\!\Nf|^{-\frac{1}{\Theta_0}} \label{eq:kcr} 
\left(\! 1 
\!-\! \frac{ |\Nfcr\! -\!
\Nf|}{\Theta_0}\left( \frac{\alpha_2}{|\alpha_1|} 
-\frac{\Theta_1}{\Theta_0}\ln (|\alpha_1| |\Nfcr\! -\! \Nf|)\right)\right) 
\!+\! \dots\,, 
\ee
where $\Theta_0:=\Theta(\Nfcr)$. Since $\kcr$ defines the scale at which the fixed-points in the $\beta$ function of the four-fermion coupling merge, the existence of a
finite $\kcr$ can be considered as a necessary condition for (chiral) symmetry breaking. Thus, we expect that the scale for a given IR 
observables~${\mathcal  O}$ for $\Nf\leq\Nfcr$ is set by~$\kcr$:
\be
{\mathcal O}= f_{\mathcal O} \kcr^{d_{\mathcal O}}\,, \label{eq:kcrO}
\ee
where $d_{\mathcal O}$ is again the canonical mass dimension and $f_{\mathcal O}$
dependes on $\Nf$ but not on $\Nfcr$, see also Eq.~\eqref{eq:ObsScaling}. However, we stress that $\kcr$ does
not include the full dependence of $\ksb$ on $(\Nf-\Nfcr)$, i. e.  $\kcr/\ksb$ is still a function of the control parameter, as we shall
discuss in the subsequent section.

Finally we would like to point out that the power-law scaling behavior discussed in this section is different from the power-law scaling
behavior discussed in purely fermionic models, such as the Gross-Neveu model. In the latter, the scaling behavior at the quantum phase transition
is governed by the critical exponent of the (relevant) four-fermion coupling. The scaling behavior in gauge theories, on the other hand, is governed
by the critical exponent of the gauge coupling which drives the fermions to criticality. The four-fermion couplings at the UV scale
are not considered to be free parameters in the present setup, as we have set them to zero at the UV scale.\footnote{Strictly speaking, we only require that
the initial values of the four-fermion couplings are smaller than the values of their IR repulsive fixed points.}

\subsubsection{Beyond Miransky Scaling}\label{sec:beyondmiransky}
Let us now discuss how the symmetry breaking scale $\ksb\leq \kcr$ depends on $(\Nf-\Nfcr)$. We consider again an action of the
form~\eqref{eq:miransky_ansatz}, and assume that $\Nf\lesssim\Nfcr$. The crucial new ingredient compared to the derivation of Miransky scaling is the
RG flow of the coupling.  We also assume that the system has already evolved from the initial UV scale ${\Lambda}$ to the scale $\kcr$ at which the fixed
points of the $\beta$ function of the four-fermion coupling have merged. Sufficiently close to $\Nfcr$, the flow of the gauge coupling is
governed by the fixed point regime for $g^2 > g^2_{\rm cr}$.  The running of the gauge coupling is then given by
\be
g^2(k) &=& g^2_{\ast} - (g^2_{\ast}-g^2_{\rm
  cr})\left(\frac{k}{\kcr}\right)^{-\Theta} 
= g^2_{\ast} - (\Delta g^2)\left(\frac{k}{\kcr}\right)^{-\Theta}\,,\label{eq:g2fp}
\ee
where $\Delta g^2=g^2_\ast-g^2_{\text{cr}}$, see Eq.~\eqref{eq:FPsol}. 
Recall that $g^2_{\ast}\sim \Nf$ and $\Delta g^2\sim |\Nfcr-\Nf|$. Plugging Eq.~\eqref{eq:g2fp} into
Eq.~\eqref{eq:4psiflow}, we find
\be
\beta_{\lambda}&\equiv&\partial_t\lambda=\beta_{\lambda}\Big|_{g^2_{\ast}}
-\frac{\partial \beta_{\lambda}}{\partial g^2}\Big|_{g^2_{\ast}}(\Delta g^2)\left(\frac{k}{\kcr}\right)^{-\Theta}
+\dots\label{eq:betacorr}\\
&=&\! (d\!-\!2)\lambda - a \lambda^2 -b\lambda g^2_{\ast} -c g^4_{\ast}
-\frac{\partial \beta_{\lambda}}{\partial g^2}\Big|_{g^2_{\ast}}\left(\!\frac{k}{k_0}\!\right)^{-\Theta}
\!\!\!\!\!\! +\dots, \nonumber
\ee
where we have used Eq.~\eqref{eq:kcrest}. Recall that $k\leq \kcr \ll k_0$ and $\Theta < 0$.  We observe that the zeroth order in $\Delta g^2$ coincides with
the $\beta_{\lambda}$ function for which we have found an (implicit) analytic solution for constant $g^2$ in Sect.~\ref{sec:MS}, yielding Miransky
scaling. We refer to this analytic solution as $\lambda_{g^2_{\ast}}$. The solution of the $\beta$-function~\eqref{eq:betacorr} can then be found by an
expansion around the solution $\lambda_{g^2_{\ast}}$:
\be
\lambda &=& \lambda_{g^2_{\ast}} + (\Delta g^2)\delta\lambda + \dots 
=\lambda_{g^2_{\ast}} + \left(\frac{\kcr}{k_0}\right)^{-{\Theta}}\delta\lambda + \dots
\,,
\ee
where~$\delta\lambda = -(\partial\lambda/\partial g^2)|_{g^2_{\ast}}$.
This expression allows us to systematically compute the scaling behavior for $\Nf\lesssim\Nfcr$. Since we are interested
in the (chiral) symmetry breaking scale $\ksb$ we have to solve $1/\lambda(\ksb)=0$ for $\ksb$. In zeroth order,
the scale $\ksb$ can be computed along the lines of our analysis in Sect.~\ref{sec:MS}. We find
\be
\ksb &\propto& \kcr \theta(\Nfcr-\Nf) \exp\left( {-\frac{\pi}{2\epsilon\sqrt{|\alpha_1||\Nfcr\! -\! \Nf|} }
}\right)\nn\\
&\simeq& k_0 \theta(\Nfcr-\Nf)|\Nfcr -\Nf|^{-\frac{1}{\Theta_0}} 
\exp\left( {-\frac{\pi}{2\epsilon\sqrt{|\alpha_1||\Nfcr\! -\! \Nf|} }}\right)\,,
\label{eq:ksbscaling}
\ee 
where we have used Eq.~\eqref{eq:kcr} in leading order. Higher order corrections to Eq.~\eqref{eq:ksbscaling} can be computed systematically
as outlined above and in the previous sections. Thus, we have found a {\it universal} correction to the exponential scaling 
behavior which is uniquely determined by the universal ``critical" exponent $\Theta$. A similar result 
has been suggested by Jarvinen and Sannino using a standard rainbow-ladder approach with a 
constant gauge coupling but a properly adjusted scale~\cite{Jarvinen:2010ks}. The presented RG analysis demonstrates
in a simple and systematic way that such a rainbow-ladder approach is indeed justified and yields the correct 
leading-order scaling behavior. In the context of the RG the scaling law~\eqref{eq:ksbscaling} 
has been first derived in Ref.~\cite{Braun:2010qs}.

Let us now turn to the scaling behavior of physical observables.  The scale of all (chiral) low-energy observables is set by $\ksb$. In other words, $\ksb$ represents
the UV cutoff of an effective theory at low energies, such as chiral perturbation theory and NJL-type models in case of QCD. At zero
temperature we therefore expect that a given IR observable $\mathcal O$ with
mass dimension $d_{\mathcal O}$ scales according to
\be
{\mathcal O}=f_{\mathcal O}(\Nf)\,\ksb^{d_{\mathcal O}}\,,\label{eq:slawcorr}
\ee
where $f_{\mathcal O}(\Nf)$ is a function which depends on~$\Nf$ but not on $\Nfcr$; in principle, it can 
be computed systematically in QCD using, e.~g., chiral perturbation theory or a large-$\Nc$ expansion, see Sect.~\ref{sec:ScalLEM}.

The scaling law~\eqref{eq:slawcorr} can be used as an ansatz to fit, e.~g., data from lattice simulations. This scaling law is remarkable for a number of reasons: first,
it relates two universal quantities with each other: quantitative values of observables and the IR critical exponent~$\Theta$. Second, it establishes a
quantitative connection between the (chiral) phase structure and the IR gauge dynamics which is encoded in~$\Theta$. Third, it is a parameter-free prediction following
essentially from scaling arguments. Last but not least, it shows that Miransky scaling and power-law scaling are simply two limits of the very same set of RG flows:
in the limit $|\Theta|\to\infty$ we find pure Miransky-scaling behavior, while we have pure power-law scaling in the limit $\Theta\to 0$.

At this point, we would like to emphasize again that the scaling behavior of any IR observable near $\Nfcr$ depends crucially on the
scale-fixing procedure applied in the first place. Still, the universal scaling will always show up at one or the other place and thus
cannot be removed, as stressed in Ref.~\cite{Braun:2009ns}. Our choice to fix the scale at, e.~g., the $\tau$-mass scale which is large enough not to be affected by 
chiral symmetry breaking is certainly not unique. In principle, the point where to fix the scale can be chosen as a free function of~$\Nf$. 
In Eq.~\eqref{eq:FPsol}, this would correspond to the choice of an arbitrary function $k_0=k_0(\Nf)$ for the global scale, which then appears also in 
the scaling relations~\eqref{eq:kcr} and~\eqref{eq:ksbscaling}. Indeed, an extreme choice would be given by measuring all dimensionful scales in units of a 
scale induced by chiral symmetry breaking (such as $\Tc$ or $f_{\pi}$). In this case, all chiral observables would jump non-analytically across
 $\Nf=\Nf^{\text{cr}}$. The scaling relations would then translate into scaling relations for other external scales. For example,
  the scale~$k_{g^2}$ at which the running coupling acquires a specific value would diverge with $\Nf\to\Nf^{\text{cr}}$ according to
  $k_{g^2}\sim |\Nf -\Nf^{\text{cr}}|^{-\frac{1}{|\Theta_0|}}\exp(c_{\rm M}/\sqrt{|\Nfcr-\Nf|})$, where~$c_{\rm M}=\pi/(2\epsilon\sqrt{|\alpha_1|})$. 
  This point of view establishes a different way of verifying the above scaling relations on the
  lattice.

Let us conclude this section with a discussion of the importance of the corrections to the exponential scaling
behavior due to the running of the gauge coupling. To this end, it is convenient to consider the logarithm of the (chiral) symmetry breaking scale~$\ksb$,
\begin{equation}
\ln \ksb = {\rm const.} - \frac{1}{\Theta_0}\ln |\Nfcr -\Nf| -\frac{\pi}{2\epsilon\sqrt{|\alpha_1||\Nfcr\! -\! \Nf|}}\,.
\end{equation}
This expression can be used to estimate the regime in which the corrections to the exponential scaling become subdominant.
For this, we compute the minimum of the function
\be
\frac{1}{|\Theta_0|}\ln |\Nfcr -\Nf| + \frac{\pi}{2\epsilon\sqrt{|\alpha_1||\Nfcr\! -\! \Nf|}}\label{eq:243t0}
\ee
with respect to $|\Nfcr -\Nf|$. In accordance with Eq.~\eqref{eq:taylor}, we assume $|\Nfcr -\Nf|<1$ here. From Eq.~\eqref{eq:243t0}, we can then 
estimate that corrections to the exponential scaling behavior are subdominant as long as
\be
|\Nf -\Nfcr|\lesssim \frac{\pi^2 |\Theta_0|^2}{16 \epsilon^2 |\alpha_1|}\,,\label{eq:size}
\ee
with $\epsilon$ being defined in Eq.~\eqref{eq:defeps}.  
Thus, corrections to Miransky scaling due to the running of the gauge coupling are
small when $|\Theta_0|\gg 1$ and large when $|\Theta_0|\ll 1$.  
In Sect.~\ref{sec:PLYM} we apply Eq.~\eqref{eq:size} to QCD to estimate the size of the regime in which
the exponential scaling behavior dominates. We will see that the exponential scaling behavior is dominantly 
visible only very close to~$\Nfcr$, provided that~$\Nfcr\approx 12$. This implies that
the $\Theta$-dependent universal corrections are more significant in QCD.

The role of $|\Theta|$ for the scaling behavior close to $\Nfcr$ can also be understood by simply looking at the $\beta _{g^2}$ function of 
the gauge coupling, see Fig.~\ref{fig:sketch2}. For $|\Theta|\gg 1$, the gauge coupling runs very fast into its IR fixed point once it has 
passed $g^2_{\rm cr}$. Thus, the situation for $g^2 > g^2_{\rm cr}$ is as close as possible to the situation studied in Sect.~\ref{sec:MS} where
the coupling has been simply approximated by a constant. For $|\Theta| \ll 1$ the gauge coupling runs very slowly (``walks") into its IR fixed 
point once it has passed $g^2_{\rm cr}$. This walking behavior for $g^2 \gtrsim g^2_{\rm cr}$ then gives rise to sizable corrections to the 
exponential scaling behavior.

\subsection{Scaling in Low-energy Models}\label{sec:ScalLEM}

In the previous sections we have stated that the dimensionless function~$f_{\mathcal O}$ in the scaling law~\eqref{eq:slawcorr}
can be computed explicitly with the aid of effective low-energy models.
In the following we use the example of low-energy models of QCD to demonstrate that this is indeed the case.
However, the subsequent analysis is by no means restricted to QCD. It can also be applied straightforwardly
to other gauge theories, such as QED${}_3$.

Let us now be explicit and compute the function~$f_{\mathcal O}$ for 
the pion decay constant~$f_{\pi}$. To this end, we employ a straightforward generalization of the simple ansatz~\eqref{Eq:HSTActionQM} 
to QCD with $\Nf$~flavors. For the sake of the argument, it suffices to consider the large-$\Nc$ limit.
Along the lines of our study in 
Sect.~\ref{sec:TQPTQCD}, it is then possible to derive a gap equation for the vacuum expectation value of the order parameter~$\langle \sigma\rangle\equiv f_{\pi}$,
see Ref.~\cite{Braun:2009ns}. Here, we skip the details and only state that the resulting expression for~$\langle \sigma\rangle$ is proportional to the square
of the Yukawa coupling times a purely fermionic
loop which yields a factor of~$\Nf\Nc$, see also Eq.~\eqref{eq:QMgapEq}.
This loop integral is UV divergent and needs to be regularized at an effective (regulator) scale~$\Lambda_{\rm H}$.

For momentum scales $p\lesssim \Lambda_{\rm H}$, we expect
a description in terms of such a hadronic low-energy model to be reasonable. 
We choose $\Lambda_{\rm H}=c_{\rm reg.}\ksb$, where $c_{\rm reg}\gtrsim 1$ is
a numerical factor independent of $\Nf$ and~$\Nc$. The gap equation for~$\langle \sigma\rangle$ 
can then be solved straightforwardly and yields\footnote{This can
be most easily seen from a rescaling of the Yukawa coupling by a factor~$\sqrt{\Nf\Nc}$.}
\be
\langle \sigma \rangle 
\propto \sqrt{\Nf\Nc}
k_0 \theta(\Nfcr-\Nf)|\Nfcr -\Nf|^{-\frac{1}{\Theta_0}} \exp\left( {-\frac{\pi}{2\epsilon\sqrt{|\alpha_1||\Nfcr\! -\! \Nf|} }}\right)\,,
\ee
where the last step holds near the conformal window, using the relation \eqref{eq:ksbscaling}.  Since $f_{\pi}\equiv\langle \sigma \rangle$, we have 
\be
f_{f_{\pi}}(\Nf)=\sqrt{\Nf}\,.
\ee

In the large-$\Nc$ approximation the $\Nf$-scaling behavior of $f_{\pi}$ and of the constituent quark mass~$m_{\psi}$ is identical.
Following our discussion in Sect.~\ref{sec:TQPTQCD}, we also expect that $m_{\psi}$ has a $\Nf$-scaling behavior near $\Nfcr$ which is
identical to that of the critical temperature~\mbox{$\Tc\sim m_{\psi}$}. The scaling behavior of other observables can be computed along these lines.

We would like to stress that the prefactor $\sqrt{\Nf\Nc}$ in the present example is an outcome of our large-$\Nc$
approximation of the low-energy sector. In general, we expect that any observable~$\mathcal O$ comes along with a complicated 
prefactor function $f_{\mathcal O}$ depending on the number of flavors~$\Nf$ and~$\Nc$. The 
determination of this function, e.~g. for the constituent mass, may become complicated, depending on the truncations made in the low-energy
sector. However, we emphasize that the prefactor function is independent of~$\Nfcr$ and therefore the $\Nf$ dependence coming from this
function does not modify the $|\Nf -\Nfcr|$-scaling. 

Finally we would like to add that the current quark mass is expected to modify the scaling relations away from the chiral limit. 
In order to study these modifications, a generalized Gell-Mann-Oakes-Renner relation 
based on the fixed-point scenario in many-flavor QCD has been advocated in Ref.~\cite{Sannino:2008pz}. Note that the scaling behavior of observables 
with the current quark mass in the (quasi-)conformal\footnote{Of couse, conformal invariance is broken explicitly when we allow for a finite current quark mass.
In this case, the theory never reaches the IR fixed point of the gauge coupling. 
However, the theory can still get close to the IR fixed-point of the gauge coupling for small quark masses and remain in its vicinity for a long RG time.}
phase of strongly-flavored gauge theories is of 
particular interest for lattice simulations and currently under investigation, see Refs.~\cite{DeGrand:2009mt,DelDebbio:2010ze,DelDebbio:2010jy}.

\subsection{Chiral ${\rm SU}(\Nc)$ Gauge Theories}\label{sec:SUNYM}

In this section we review numerical RG studies of strongly-flavored ${\rm SU}(\Nc)$ gauge theories from first principles. 
In Ref.~\cite{Gies:2005as} the zero-temperature phase diagram in the $(\Nf,\Nc)$-plane
has been computed using the functional RG.  The phase diagram in the plane spanned by the temperature and~$\Nf$
has first been computed in Refs.~\cite{Braun:2005uj,Braun:2006jd}. In Sect.~\ref{sec:RGYM} we briefly review the RG setup. The 
various phase diagrams are then discussed in Sect.~\ref{sec:PSFYM}. In Sects.~\ref{sec:MSYM} and~\ref{sec:PLYM} we present a quantitative 
study of scaling close to the quantum critical point~$\Nfcr$. This comprehensive 
analysis of scaling has first been performed in Ref.~\cite{Braun:2010qs}.

Before we begin with our discussion of strongly-flavored gauge theories, we would like to add a few words on the application of
the Wetterich equation to gauge theories and, in particular, to QCD.
The Wetterich equation has been employed for first-principles studies of QCD since the mid 1990s, where it started out with 
non-perturbative studies of the running of the gauge coupling~\cite{Reuter:1993kw}, gluon condensation~\cite{Reuter:1994zn,Reuter:1997gx}
and the momentum dependence of Yang-Mills propagators~\cite{Ellwanger:1995qf,Ellwanger:1996wy}.
Since then these studies have been refined and further developed from a technical point of view
(see Refs.~\cite{Litim:1998nf,Pawlowski:2005xe,Gies:2006wv,Rosten:2010vm} for reviews) but also for the
application to QCD phenomenology. Let us name a few examples. The running of the 
strong coupling has been computed on all scales at zero~\cite{Gies:2002af,Pawlowski:2003hq,Fischer:2004uk} and
at finite temperature~\cite{Braun:2005uj,Braun:2006jd}. The approach to chiral symmetry breaking, which we mainly review here,
has been studied from first principles in Refs.~\cite{Gies:2002hq,Gies:2005as,Braun:2005uj,Braun:2006jd}. 
Confinement has been investigated at zero~\cite{Pawlowski:2003hq,Fischer:2004uk}
as well as at finite temperature~\cite{Braun:2007bx,Marhauser:2008fz,Braun:2009gm,Braun:2010cy}. In particular, the results
for the deconfinement phase transition in Yang-Mills theories are in very good agreement with lattice simulations.
Moreover, the interrelation of quark confinement and chiral symmetry breaking has been analyzed in Refs.~\cite{Braun:2009gm,Braun:2011fw},
and the question of gluon condensation has been recently revisited in Ref.~\cite{Eichhorn:2010zc}. 
Last but not least, the emergence of hadronic states in the IR limit has been studied with so-called re-bosonization techniques in 
Refs.~\cite{Gies:2002hq,Braun:2008pi}. These studies include a detailed discussion of how to bridge the gap between the fundamental
degrees of freedom, namely quarks and gluons, and hadronic degrees of freedom as, e.~g., described by low-energy QCD models. 
These studies have recently motivated further studies
in this direction for QCD with two colors~\cite{Kondo:2010ts}.
\subsubsection{Renormalization Group Approach to Gauge Theories}\label{sec:RGYM}
In Refs.~\cite{Gies:2005as,Braun:2005uj,Braun:2006jd} the RG flow of QCD starting from 
quarks and gluons has been studied employing a covariant derivative expansion. A crucial ingredient for chiral symmetry breaking are the
scale-dependent gluon-induced quark self-interactions of the type included in Eq.~\eqref{eq:miransky_ansatz}.
We note that dynamical quarks influence the RG flow of QCD by qualitatively different mechanisms. First, quark fluctuations directly 
modify the running of the gauge coupling due to the screening nature of these fluctuations. On the other hand, gluon exchange between 
quarks induces quark self-interactions which can become relevant operators in the IR, as we have already discussed in the previous sections. 
These two mechanisms strongly influence each other as well. As we have seen, however, it is possible to disentangle the system once we 
accept that these fluctuations can be associated with different scales in the problem.

In the following we shall restrict ourselves to $d=4$ Euclidean space-time dimensions at vanishing temperature and work solely in the Landau gauge 
which is known to be a fixed point of the RG flow~\cite{Ellwanger:1995qf,Litim:1998qi}. Our strategy to study phases of strongly-interacting
gauge theories is now the same as applied before in the context of fermionic models: we consider the point-like limit and 
restrict our discussion to the RG flow in the chirally symmetric regime. This does not provide us with a direct access to 
the hadronic mass spectrum at low energies. However, it already allows us to map various phase diagrams in a clean and very controlled
way. In fact, it has been explicitly shown in Ref.~\cite{Gies:2005as} that the point-like limit is a reasonable approximation in the chirally symmetric regime,
where the regularization-scheme independence of universal quantities has been found to hold remarkably well in this limit.

For our study, we employ the following ansatz for the effective action which represents
the lowest nontrivial order in a consistent and systematic operator
expansion, see Ref.~\cite{Gies:2003dp} and also Refs.~\cite{Gies:2005as,Braun:2005uj,Braun:2006jd}:
\begin{eqnarray}
\Gamma_k&=&
\int d^4x \Bigg\{ \frac{1}{4} F_{\mu\nu}^{a} F_{\mu\nu}^{a}+ \bar{\psi}({\rm i}\fslash{\partial} 
+\bar{g}\fslash{A} 
)\psi 
+\frac{1}{2} \Big[
  \bar\lambda_-(\text{V--A}) +\bar\lambda_+ (\text{V+A})
  +\bar\lambda_\sigma (\text{S--P}) \nn\\
  && \qquad\qquad\qquad\qquad\quad +\bar\lambda_{\text{VA}}
  [2(\text{V--A})^{\text{adj}}\!+({1}/{\Nc})(\text{V--A})] \Big]\Bigg\} + \Gamma_{k}^{\text{gauge}}\,.
\label{equ::truncation}
\end{eqnarray}
We do not further specify $\Gamma_{k}^{\text{gauge}}$ since it is of no relevance of what follows. We only state that $\Gamma_{k}^{\text{gauge}}$
contains the gauge-fixing term and the ghost terms as well as it may also contain higher gluonic operators, e.~g., of the type~$\sim (F_{\mu\nu}^{a} F_{\mu\nu}^{a})^{n}$,
see e.~g. Refs.~\cite{Reuter:1994zn,Reuter:1997gx,Gies:2002af,Braun:2005uj,Braun:2006jd}. For details and reviews on gauge theories we refer the reader
to Refs.~\cite{Litim:1998nf,Pawlowski:2005xe,Gies:2006wv,Rosten:2010vm}.

The ansatz~\eqref{equ::truncation} for the effective action represents a straightforward generalization of the Fierz-complete ansatz~\eqref{equ::truncationQM1}
for the matter sector, see Sect.~\ref{sec:LowQCDF} for a detailed discussion. The definition of the 
four-fermion interaction channels can be found in Eqs.~\eqref{eq:VmAdef}-\eqref{eq::colorflavor}. This ansatz falls into
the QCD universality class when we set the various four-fermion interactions to zero at the initial RG scale, e.~g. at the Z-boson mass 
scale.

In our analysis, we neglect U${}_{\text{A}}(1)$-violating interactions induced by topologically non-trivial gauge configurations 
since we expect them to become relevant only inside the \xsb\ regime or for small $\Nf$. In addition, the lowest-order U${}_{\text{A}}$(1)-violating term 
schematically is $\sim (\yb\psi)^{\Nf}$, see e.~g. Refs.~\cite{tHooft:1976fv,Shifman:1979uw,Shuryak:1981ff,Schafer:1996wv,Pawlowski:1996ch}.
Thus, larger $\Nf$ correspond to larger RG ``irrelevance" by naive power-counting. Moreover, interactions of the type 
\mbox{$\sim(\yb\psi)^{\Nf}$} for $\Nf>3$ do not contribute directly to the flow of the four-fermion interactions due to the one-loop structure 
of the underlying RG equation for
the effective action, as discussed in Sect.~\ref{sec:example}.
 
 Using the truncated effective action \eqref{equ::truncation}, we obtain the following $\beta$ functions for the dimensionless couplings 
$\lambda_i=\bar{\lambda_i}/k^2$, see Refs.~\cite{Gies:2003dp,Gies:2005as}:
\begin{eqnarray}
\!\pat\lm
&=& 2\lm\!
    -4v_4 \lFBo\left[ \frac{3}{\Nc}g^2\lm
            -3g^2 \lva \right]
    \label{eq:lm}
-\frac{1}{8}v_{4}\lFB\left[\frac{12+9\Nc^2}{\Nc^2}g^{4} \right]
\\\nonumber
&&\; -8 v_4\lFna \Big\{-\Nf\Nc(\lm^2\!+\!\lp^2) + \lm^2
\!\!-2(\Nc\!+\!\Nf)\lm\lva
       +\Nf\lp\lsf + 2\lva^2 \Big\}, \label{eq:lmflow}
\\
\!\pat\lp &=&2 \lp\! -4v_4\lFBo \left[
-\frac{3}{\Nc}g^2\lp\right]
    \label{eq:lp}
-\frac{1}{8}v_{4}\lFB\left[
    -\frac{12\!+\! 3\Nc^2}{\Nc^2} g^{4} \right]
\nn\\
&&
\qquad\qquad 
-8 v_4  \lFna \Big\{\! - 3\lp^2 
- 2\Nc\Nf\lm\lp - 2\lp(\lm+(\Nc+\Nf)\lva)
\nn\\
&& \qquad\qquad\qquad\qquad\qquad\qquad\qquad\qquad\quad
        + \Nf\lm\lsf
+ \lva\lsf
        +\frac{1}{4}\lsf{}^2 \Big\},
\\
\!\pat\lsf
&=&
    2\lsf\! -4v_4 \lFBo \left[6\Cas\, g^2\lsf
    -6g^2\lp \right]\label{eq:lsf}
-\frac{1}{4} v_4 \lFB \Big[ -\frac{24
    -9\Nc^2}{\Nc}\, g^4  \Big] \nonumber\\
&&\qquad\qquad
 -8 v_4 \lFna
     \! \Big\{\! 2\Nc \lsf^2\!  -\! 2\lm\lsf\! - 2\Nf\lsf\lva\!\!
- \!6\lp\lsf\! \Big\}, \label{eq:lsffull}
\\
\!\pat\lva
&=& 2 \lva\!-4v_4 \lFBo \left[
\frac{3}{\Nc}g^2\lva -3g^2\lm \right]
    \label{eq:lva} 
-\frac{1}{8} v_4 \lFB \left[ -\frac{24 - 3\Nc^2}{\Nc}
g^4 \right]
\nn
    \\
&&\qquad\qquad -8 v_4 \lFna
  \Big\{\! - \!(\Nc+\Nf)\lva^2\! + 4\lm\lva\! \label{eq:lvaflow}
        - \frac{1}{4} \Nf \lsf^2\Big\}\,.
\end{eqnarray}
Here, $\Cas=(\Nc^2-1)/(2\Nc)$ is a Casimir operator of the gauge group,
and $v_4=1/(32\pi^2)$. The definition of the threshold functions $\lFna=\lF$, $\lFB=\lFBwa$ and $\lFBo=\lFBowa$
can be found in App.~\ref{app:regthres}. Recall that~$\eta_{\psi}$ vanishes in the point-like limit in Landau gauge.
In the numerical analysis of these flow equations, which we present below, we have dropped the contributions from the anomalous dimensions of 
the gauge coupling $\eta_{\rm A}=\beta_{g^2}/g^2$. This is justified since it has been found in Ref.~\cite{Braun:2010qs} 
that the contributions $\propto \eta_{\rm A}$ in the threshold functions do not strongly affect the results for $\Nfcr$. 
This can be also understood from an analytic point of view: we have $\eta_{\rm A}\to 0$ for $\Nf\to N_{\rm f}^{\rm a.f.}$ and $g^2< g^2_{\ast}$. 
Moreover, we can estimate $\eta_{\rm A}$ with the aid of the $\beta_{g^2}$ function in the $\MSbar$ scheme.
We find for $g^2< g^2_{\ast}$ that $|\eta_{\rm A}^{\rm 2-loop}| \lesssim 1$ for $\Nf \gtrsim 11$ and  $|\eta_{\rm A}^{\rm 4-loop}| \lesssim 0.5$ 
for $\Nf \gtrsim 8$. For our purposes, we therefore expect that these contributions may lead to quantitative corrections at most on the percent level.
However, these terms may become relevant for small~$\Nf$ and in the regime with broken chiral symmetry 
where we have $\eta_{\rm A} \sim {\mathcal O}(1)$.

Let us now further discuss the running of the gauge coupling. As mentioned above, the running coupling has been computed within the functional RG
approach~\cite{Reuter:1993kw,Gies:2002af,Pawlowski:2003hq,Braun:2005uj,Braun:2006jd}. However, we will closely follow the analysis
in Ref.~\cite{Braun:2010qs} and employ for simplicity the two- and four-loop result obtained in the $\MSbar$ scheme \cite{vanRitbergen:1997va,Czakon:2004bu}.
This is justified since our results for the phase boundary show a satisfactory convergence in the large-$\Nf$ regime. In the following
we will often restrict ourselves to the two-loop $\beta_{g^2}$ function, see Eq.~\eqref{eq:beta2loop}, as it already shows all qualitative features
and can be dealt with analytically. 

At this point a critical comment on the scheme dependence is in order: The chosen regularization scheme in the matter sector 
and the $\MSbar$ scheme do not coincide. This inconsistency results in an error for the estimate for the critical number of quark flavors,
i.~e. for the location of the quantum critical point. Since we are interested in the scaling behavior which is related to the
universal critical exponent $\Theta$ rather than in a high-precision determination of~$\Nfcr$, our results are only influenced
indirectly by this approximation.\footnote{Of course, the actual value of
  $\Theta_0=\Theta(\Nfcr)$ depends on the actual value of $\Nfcr$ which itself, as a universal quantity, depends on the difference of the
  scheme-dependent quantities $g^2_{\rm cr}$ and $g^2_{\ast}$.}
  Therefore the results using the four-loop running may not necessarily be considered as a more precise calculation. Instead, the
  difference between two-loop and four-loop $\MSbar$ results should be viewed as an estimate of the dependence of our results on the 
  quantitative details of the running gauge sector.
  
We close this section with a comment on gauge symmetry. Here, a subtlety becomes important. Naively, one may expect that
the running of the gauge coupling is modified due to the presence of finite four-fermion couplings. In fact, it is possible to construct
a 1PI diagram $\sim g\lambda_i$ with one external gluon line and two external fermion lines which potentially contributes to the running
of the quark-gluon vertex $\sim g\bar{\psi}\fslash{A}\psi$. Now a detailed analysis shows that the question of gauge invariance 
is intimately linked to the existence of such contributions~$\sim g\lambda_i$ to the running of the gauge  coupling: 
To render the RG flow gauge invariant we have to take into account regulator-dependent Ward-Takahashi identities~\cite{Ellwanger:1994iz,Reuter:1993kw}. 
In the present case, these symmetry constraints yield contributions to the running of the gauge coupling which depend on the quark self-interactions:
\be
\partial_{t}g^2&=& \beta_{g^2}-4 v_4
  \lFna\, \frac{g^2}{1-2v_4\lFna \sum c_i \lambda_i} \, \sum c_i\,
 \pat \lambda_i\,, \label{betaeq} 
\ee
where~$i\in\{\sigma,+,-,\text{VA}\}$ and $\beta_{g^2}$ denotes the standard $\beta_{g^2}$ function, e.~g. in the two-loop approximation. The dimensionless
factors~$c_i$ are given by
\be
c_{\sigma}=1+\Nf\,,\quad c_{+}=0\,,\quad c_{-}=-2\,,\quad c_{\textrm{VA}}=-2\Nf\,.
\ee
Apparently, the additional contributions on the right-hand side of Eq.~\eqref{betaeq} are proportional to the $\beta$-functions of the four-fermion couplings and
therefore vanish as long as the four-fermion couplings are at their fixed points, i.~e. as long as $g^2 \leq g^2_{\rm cr}$;
this has first been pointed out in Refs.~\cite{Gies:2003dp,Gies:2005as}. We conclude that 
these contributions $\propto \lambda_i$ do not  alter the scaling law~\eqref{eq:ksbscaling} in leading order.\footnote{In addition to 
the discussed next-to-leading order corrections to Eq.~\eqref{eq:ksbscaling},
these symmetry constraints may shift the fixed-point value $g^2_{\ast}$ of the gauge coupling and therefore cause additional 
higher-order corrections to the scaling law~\eqref{eq:ksbscaling}.}  In particular, the power-law behavior is unaffected by these corrections 
arising due to symmetry constraints. In the following we ignore these corrections in our numerical analysis.

In the regime with broken chiral symmetry in the ground state, the $\lambda_i$-dependent contributions may alter the
running of the gauge coupling. However, the fermions acquire a finite mass and diagrams with at least one internal
fermion line are expected to be suppressed compared to diagrams with no internal fermion lines: $\lFna (\epsilon_{\psi};\eta_{\psi})\to 0$
in the limit of a large dimensionless fermion mass~$\epsilon_{\psi}$. In any case, we are not
aiming at a study of the properties of QCD inside the broken regime but rather intend to map the phases of strongly-interacting gauge
theories by determining the parameter sets~$(\Nf,\Nc)$ for which the system remains in the chirally symmetric regime.

\subsubsection{Phases of Strongly-flavored ${\rm SU}(\Nc)$ Gauge Theories}\label{sec:PSFYM}
Now we  discuss phases of strongly-flavored gauge theories at zero and finite temperature. To 
simply determine the size of the conformal window at zero temperature, it suffices to consider the approximation of
a constant gauge coupling:
\be
\partial_t g^2=0\,.\nn
\ee
As discussed in Sect.~\ref{sec:MS}, the gauge coupling can then be considered as an ``external" $\Nf$-dependent 
parameter of the theory. To estimate the error of our truncation in the gauge sector, we choose the fixed-point value of the gauge-coupling 
at two-loop and four-loop level as the external parameter. This value corresponds to the largest possible IR value of the coupling inside of the conformal window.
For illustration, we give the fixed-point value at two-loop level which assumes a simple form:
\be
g^2_{\ast,{\rm 2-loop}}(\Nf)=\frac{16 (11 \Nc^2 - 2 \Nc \Nf) \pi^2}{  13 \Nc^2 \Nf - 34 \Nc^3 - 3 \Nf}
\,. \label{eq:2loopg2}
\ee

In the matter sector we employ two different truncations to which we refer as {\it one-channel} and {\it all-channels} approximation. 
The latter one is Fierz complete, i.~e.
we take into account the full set of flow equations~\eqref{eq:lm}-\eqref{eq:lva}.
In the one-channel approximation, on the other hand, we only take into account the RG flow of the scalar-pseudoscalar channel $\lambda_{\sigma}$
and set all other four-fermion couplings to zero:
\be
\!\pat\lsf
= 2\lsf 
- \frac{\Nc}{4\pi^2}\lsf^2 - \frac{3}{4\pi^2}\Cas\lsf g^2 - \frac{3}{256\pi^2}\left(\frac{9\Nc^2 -24}{\Nc}\right) g^4\,.
\label{eq:lsf1channel}
\ee

While the all-channels approximation can only be dealt with numerically, the one-channel approximation together with the two-loop
gauge-coupling fixed-point allows for analytic estimates for~$\Nfcr$ and the scaling behavior close to the (quantum)
phase transition. It is worth mentioning that the RG flow of the~$\lambda_{\sigma}$-coupling decouples from the other channels in the 
large-$\Nc$ limit, see Sect.~\ref{sec:LowQCDF}.
Recall that the associated (S--P)-channel is Fierz-equivalent to the interaction channel included in widely used QCD low-energy models, e.~g.
the quark-meson model.

For illustration, we first compute the critical value of the gauge coupling in the one-channel approximation using
the value of the IR fixed point of the two-loop gauge-coupling. We find
\begin{equation}
g^2_{\rm cr,one}= \frac{32\pi^2\left(2\Nc^3 -2\Nc-\sqrt{3\Nc^6 - 8\Nc^4} \right)}
{3(4+\Nc^4)}\!\stackrel{(\Nc=3)}{\approx}10.86\,,
\end{equation}
which does not depend on $\Nf$. In the all-channels approximation the critical value has to be computed numerically. As found in 
Ref.~\cite{Gies:2005as}, the resulting critical value $g^2_{\rm cr,all}$ of the gauge coupling then depends on $\Nf$: for a given number of colors,
$g^2_{\rm cr,all}$ decreases slightly with increasing~$\Nf$. 
\begin{figure}[!t]
\centering\includegraphics[scale=0.85]{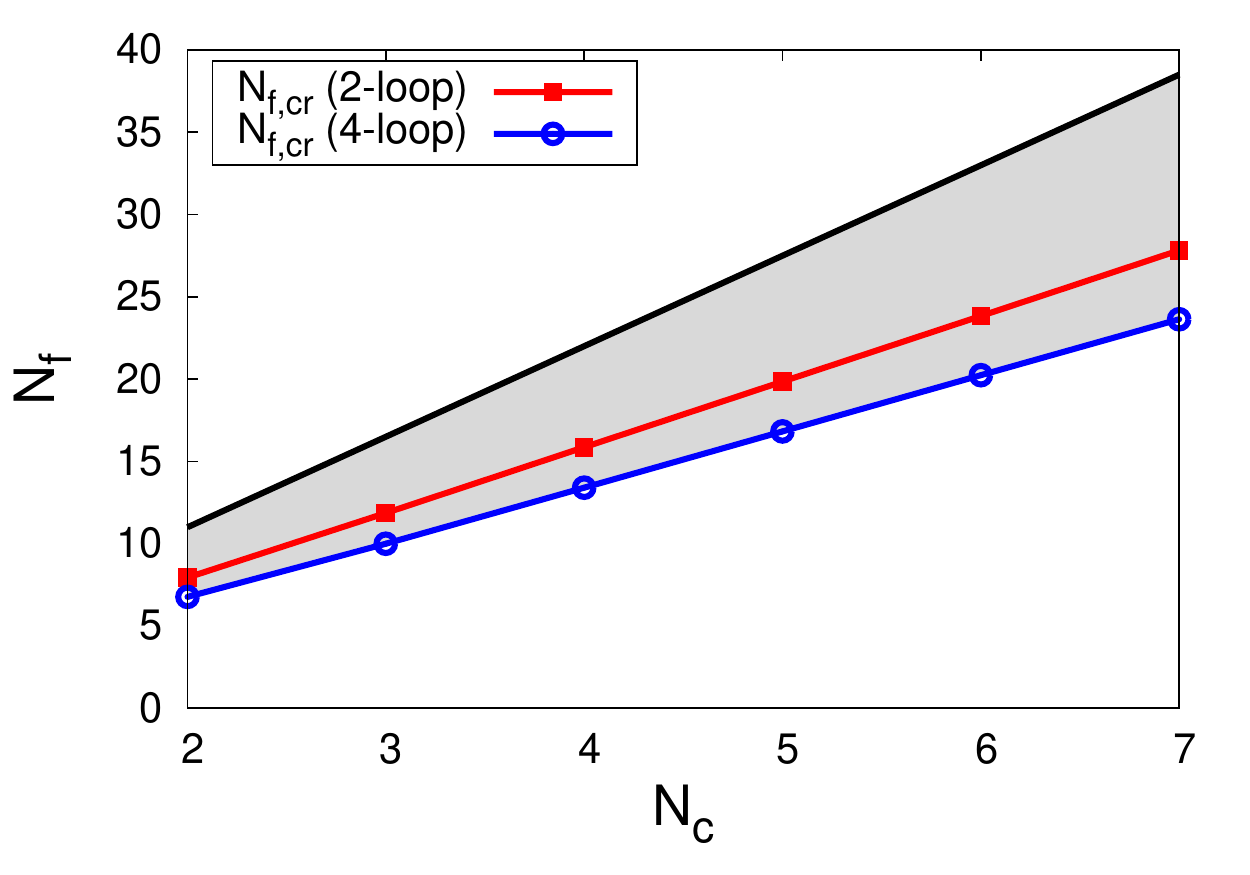} 
\caption{Phase diagram of strongly-flavored SU($\Nc$) gauge theories in the $(\Nf,\Nc)$-plane, see also Ref.~\cite{Gies:2005as}.
The upper solid (black) line gives the value of~$\Nf$ at which
asymptotic freedom is lost, $\Nf^{\text{a.f.}}=\frac{11}{2}\Nc$. The shaded area depicts the conformal window in the approximation
with a running gauge coupling at four-loop level. The result for~$\Nfcr$ 
as obtained from a running coupling at two-loop level is given by the red line. 
We observe that the conformal window is increased when we employ the gauge coupling
in the four-loop approximation instead of the two-loop approximation.
} 
\label{fig:NfNcPD}
\end{figure}

The fixed-point value $g^2_{\ast,{\rm 2-loop}}$ together with the critical value of the gauge coupling can be used to estimate the critical
number of quark flavors above which there is no chiral symmetry breaking in the IR. In accordance with the numerical results given in 
Ref.~\cite{Gies:2005as}, we find~\cite{Braun:2010qs}
\be
\Nfcr^{\rm one}\!=\! \frac{169 \Nc^6 \! - \! 136 \Nc^4 \! +\! 132 \Nc^2 \! - \! 68
   \sqrt{\Nc^4 \left(3 \Nc^2\! -\! 8\right)} \Nc^3}{58
 \Nc^5\!-\! 64 \Nc^3\! -\! 26 \sqrt{\Nc^4 \left(3
  \Nc^2\! -\! 8\right)} \Nc^2\! +\! 6 \sqrt{\Nc^4 \left(3
  \Nc^2\! -\! 8\right)}\! +\! 36 \Nc}
\!\stackrel{(\Nc =3)}{\approx}\! 11.7\,\label{eq:Nfcr1channel}
\ee
for the one-channel approximation. In the all-channels approximation we obtain
\be
\Nfcr^{\rm all}\approx 11.9\,
\ee
for~$\Nc=3$. We may use our analytic estimate for $\Nfcr$ from the one-channel approximation to estimate $\Nfcr$ in the limit $\Nc\to\infty$:
\be
\frac{\Nfcr^{\rm one}}{\Nc}=\frac{68 \sqrt{3}-169}{2 \left(13 \sqrt{3}-29\right)}\approx 4.0\,.\label{eq:Nfcr4Nc}
\ee
For the all-channels approximation with a running coupling in the two-loop and four-loop approximation, we 
find numerically that~$\Nfcr^{\rm all}/\Nc\approx 4.0$ and~$\Nfcr/\Nc\approx 3.4$ for large values of~$\Nc$, respectively, see Tab.~\ref{tab:PDYMresults}.
These results for $\Nfcr$ are in accordance with the results from Dyson-Schwinger equations
in the rainbow-ladder approximation, see e.~g. Refs.~\cite{Appelquist:1998rb,Dietrich:2006cm,Fukano:2010yv},
as well as with those from current lattice simulations~\cite{Kogut:1982fn,Gavai:1985wi,Fukugita:1987mb,Brown:1992fz,Damgaard:1997ut,
  Iwasaki:2003de,Catterall:2007yx,Appelquist:2007hu,Deuzeman:2008sc,Deuzeman:2009mh,Appelquist:2009ty,
  Fodor:2009wk,Fodor:2009ff,Pallante:2009hu}.
\begin{table*}[t]\center
\begin{tabular}{p{50pt}||p{40pt}|p{40pt}|p{40pt}|p{40pt}|p{40pt}|p{40pt}p{0pt}}
\centering $\Nc$ & \centering $2$ & \centering $3$ & \centering $4$ & \centering $5$ & \centering $6$ &\centering $7$ &\\ \hline\hline 
\centering $\Nfcr^{\rm one}$ & \centering $7.6$ & \centering $11.7$ & \centering $15.7$ & \centering $19.7$ & \centering $23.6$ &  \centering $27.6$ & \\
\centering $\Nfcr^{\rm all}$ & \centering $7.9$ & \centering $11.9$ & \centering $15.9$ & \centering $19.9$ & \centering $23.8$ & \centering $27.8$ &\\ 
\centering $\Nfcr^{\rm 4-loop}$ & \centering $6.8$ & \centering $10.0$ & \centering $13.4$ & \centering $16.8$ &  \centering $20.2$ &  \centering $23.6$ &\\ \hline 
\centering $\Delta\Nf^{\rm 2-loop}$ & \centering $0.36$ & \centering $0.27$ & \centering $0.31$ & \centering $0.36$ &\centering $0.42$ &\centering $0.48$ &
\end{tabular}
\caption{Critical number of flavors~$\Nfcr$ for various values of~$\Nc$ as obtained from different approximations: 
one-channel approximation with two-loop running gauge coupling ($\Nfcr^{\rm one}$), 
all-channels approximation with two-loop running gauge coupling ($\Nfcr^{\rm all}$), and
all-channels approximation with four-loop running gauge coupling ($\Nfcr^{\rm 4-loop}$).
The difference between $\Nfcr^{\rm all}$ and $\Nfcr^{\rm 4-loop}$
can be considered as an error estimate for the uncertainty arising due to the 
truncated gauge sector in our study.
In the bottom row, we give estimates for the size of the regime in which the exponential scaling behavior dominates.
These estimates have been obtained from Eq.~\eqref{eq:size} by using the one-channel approximation together with a
two-loop running gauge coupling. We observe that~$\Delta\Nf^{\rm 2-loop}$ increases only slightly with~$\Nc$, i.~e. $(\Delta \Nf^{\rm 2-loop}) /\Nc\ll 1$.
}
\label{tab:PDYMresults}
\end{table*}

In Fig.~\ref{fig:NfNcPD} we show the zero-temperature phase diagram of strongly-flavored SU($\Nc$) gauge theories in
the $(\Nf,\Nc)$-plane, see also Tab.~\ref{tab:PDYMresults}. Within the RG framework this phase diagram has first been computed in Ref.~\cite{Gies:2005as}.
The upper solid (black) line represents the boundary at which asymptotic freedom is lost. The shaded area
depicts the conformal window. We observe that the absolute size of the conformal window increases with~$\Nc$. However,
the relative size $(\Nf^{\text{a.f.}} - \Nfcr)/\Nf^{\text{a.f.}}\approx 3/11$ is approximately independent of~$\Nc$. In addition, we 
find that the size of the conformal window is increased when we employ a running coupling in the four-loop approximation 
instead of the two-loop approximation. To be specific, we obtain
\be
\Nfcr^{\rm 4-loop}\approx 10.0
\ee
for~$\Nc=3$ in the all-channels approximation and $\Nfcr^{\rm 4-loop}\approx 9.8$ in the one-channel approximation, in agreement 
with Ref.~\cite{Gies:2005as}. The difference between the two-loop and four-loop result 
can be viewed as an error estimate for the uncertainty arising due to the 
truncated gauge sector in our study. In addition to a test of the uncertainty in the gauge sector, the regularization scheme in the matter sector has been varied
in Ref.~\cite{Gies:2005as}. Such a variation leaves its imprint in the estimate for the critical value~$g^2_{\rm cr}$.
It turns out that the present truncation is remarkably stable under a variation of the scheme which instills further confidence 
in our present approach. To be specific, one finds $\Nf^{\text{cr}}=10.0\genfrac{}{}{0pt}{}{+1.6}{-0.7}$ for~$\Nc=3$ from 
a variation of the $\beta$-function of the gauge coupling and the regularization scheme.\footnote{Note that a variation of the $\beta$ function of the 
gauge coupling can be effectively considered 
as a variation of the truncation in the gauge sector.} This is in accordance with an another RG study of the conformal
phase transition in which the running gauge coupling has not been treated as an ``external" input but 
computed within the exactly same scheme as the matter sector~\cite{Braun:2005uj,Braun:2006jd}.

We would like to add that the phase diagram shown in Fig.~\ref{fig:NfNcPD} is also in accordance with the phase diagram found
by Dietrich and Sannino with Dyson-Schwinger equations in the rainbow-ladder approximation~\cite{Dietrich:2006cm}. On
the present level of accuracy of lattice simulations, the results for this phase diagram 
go well together with those from functional RG approaches~\cite{Gies:2005as,Braun:2005uj,Braun:2006jd} 
and Dyson-Schwinger approaches, see e.~g. Refs.~\cite{Dietrich:2006cm,Fukano:2010yv}.

\begin{figure}[!t]
\centering\includegraphics[scale=0.85]{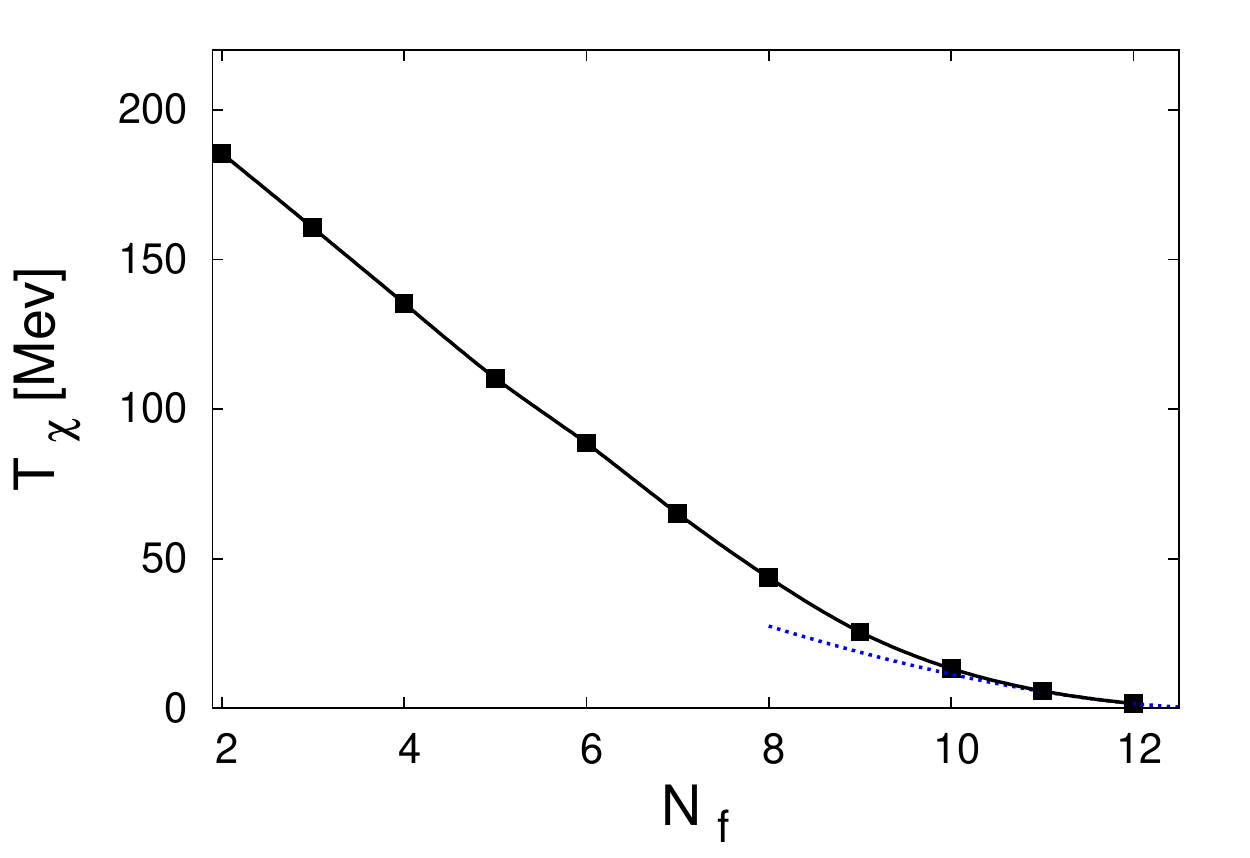} 
\caption{Chiral phase transition temperature~$\Tc$ as a function of the number of massless quark flavors $\Nf$ 
for $\Nf\geq2$, as obtained in Ref.~\cite{Braun:2006jd}. Strictly speaking, the shown results for~$\Tc$ only represent
an upper bound for the (exact) chiral phase transition temperature due to the presence of potentially strong
fluctuations of the Nambu-Goldstone modes in the (deep) IR sector of the theory, see main text for a detailed discussion. 
The flattening at $\Nf\gtrsim10$ is a consequence of the IR fixed-point
structure. The dotted line depicts the power-law scaling behavior near~$\Nfcr$, see Eq.~\eqref{eq:tctheta}. 
} 
\label{fig:TcNf}
\end{figure}

Finally, we would like to discuss the finite-temperature many-flavor phase boundary in QCD. In Fig.~\ref{fig:TcNf}
we show the chiral phase transition temperature~$\Tc$ as a function of the number of massless quark flavors~$\Nf$. 
This phase diagram has been computed for the first time in Refs.~\cite{Braun:2005uj,Braun:2006jd}. 
The results are in accordance with those obtained more recently with the aid of Dyson-Schwinger equations
in the rainbow-ladder approximation~\cite{Jarvinen:2010ks}.
We do not discuss the details
related to these computations here but only name the basic ingredients which have entered the RG study. First of all, the truncation in the matter sector
is identical to the one discussed in Sect.~\ref{sec:RGYM}. The resulting flow equations \eqref{eq:lmflow}-\eqref{eq:lvaflow} 
have been straightforwardly generalized to finite temperature. Further four-fermion operators, which may arise at finite temperature 
due to the broken Poincare invariance, have been neglected for the sake of simplicity. This appears to be a reasonable approximation
for large $\Nf$ where $\Tc$ becomes small and eventually approaches zero at the quantum critical point~$\Nfcr$. Moreover,
the chiral symmetry breaking scale was found to be larger than the phase transition temperature for small~$\Nf$.
Therefore these additional four-fermion operators are expected to be parametrically suppressed in the chirally symmetric 
regime.\footnote{Note that $T/k\gtrsim 1$ corresponds to large temperatures in the RG flow, whereas $T/k < 1$ corresponds
to low temperatures.}
The second important ingredient is the running of the gauge coupling at finite temperature which has been computed self-consistently 
within the functional RG framework in Refs.~\cite{Braun:2005uj,Braun:2006jd}. 
It was found that the running coupling (in Landau-DeWitt gauge) remains finite on all scales at finite temperature
and approaches the fixed point of the underlying 3$d$ Yang-Mills theory in the IR limit.
We stress that a naive generalization of the {\it perturbative} zero-temperature running of the gauge coupling is bound to
fail since the quarks decouple in the IR limit due their antiperiodic boundary conditions in Euclidean time direction. Thus, we have
effectively $\Nf\to 0$ at finite temperature for $k\to 0$ and we are left with the pure gauge coupling, 
even in the (quasi) conformal phase.\footnote{At finite temperature conformality is broken.} 
 
Let us now discuss the phase diagram in Fig.~\ref{fig:TcNf} from a physical point of view. In order to compute this diagram, 
the scale has been fixed at the $\tau$-mass scale to the same value for all~$\Nf$, see discussion in Sect.~\ref{sec:fewflavor}.
Due to the finite temperature of the system, the critical coupling $g_{\text{cr}}^2$ inherits a $T$-dependence from the quark modes which acquire a
thermal (Matsubara) mass. This leads to a quark decoupling, requiring stronger interactions for critical quark dynamics. In  
Fig.~\ref{fig:parab} this is indicated by the $\lambda_i$-parabolas becoming broader with a higher maximum. Hence, the annihilation 
of the Gau\ss ian fixed point by pushing the parabola below the $\lambda_i$ axis requires a larger gauge coupling. 
It follows that $g_{\text{cr}}^2 (T/k) \geq g_{\text{cr}}^2 (0)$.

At zero temperature and for small $\Nf$, the IR fixed point $g_\ast^2$ is far larger than~$g^2_{\text{cr}}$. Hence 
QCD is in the phase with broken chiral symmetry. 
For increasing~$T$, the temperature dependence of the coupling and that of $g^2_{\text{cr}}$ compete with each other.  
In accordance with our analytic estimate in Sect.~\ref{sec:fewflavor} we observe an almost linear decrease
of the critical temperature for small but increasing $\Nf$ with a slope of $\Delta \Tc =\Tc(\Nf)-\Tc(\Nf+1)\approx 25\,\mathrm{MeV}$ at small $\Nf$.  
The predicted relative difference for $\Tc$ for $\Nf=2$ and $3$ flavors of $2\Delta\Tc/(\Tc(\Nf\!=\!2)\!+\!\Tc(\Nf\!=\!3))\simeq 0.146$ is in good agreement with 
lattice studies \cite{Karsch:2000kv}. We conclude that the shape of the phase boundary for small $\Nf$ is basically
dominated by fermionic screening.

For the case of large flavor numbers, which is of particular interest here, the critical temperature decreases further and the phase
transition line terminates at the zero-temperature quantum phase transition at $\Nfcr$, representing the lower end 
of the conformal window. Within the approximations in Refs.~\cite{Braun:2005uj,Braun:2006jd} one finds that~$\Nfcr\approx 12.9$ 
and~$|\Theta_0| \approx 0.71$ for $\Nf\!=\!\Nfcr$; the discrepancy to the above given zero-temperature study 
can be essentially traced back to the differences in the running of the gauge coupling.

In Refs.~\cite{Braun:2005uj,Braun:2006jd} it was found that the scaling of the phase boundary for large $\Nf$ is consistent with the pure 
power-law scaling behavior~\eqref{eq:kcr}, as depicted by the dotted line in Fig.~\ref{fig:TcNf}:
\be
\Tc  \sim k_0 |\Nfcr -\Nf|^{\frac{1}{|\Theta_0|}}.\label{eq:tctheta}
\ee
From our discussion of scaling behavior this result is understandable, since the exponential scaling behavior sets in only 
very close to $\Nfcr$ and thus remains invisible in numerical fits over a wider $\Nf$-range, see Eq.~\eqref{eq:size}.
Of course, our analytic estimate for the scaling behavior of~$\Tc$ still remains an upper bound, even if
we took into account the exponential factor in Eq.~\eqref{eq:slawcorr}. This is due to the fact that strong fluctuations of Nambu-Goldstone modes 
in the IR may yield further corrections and lower the phase transition temperature, see e.~g. Ref.~\cite{Braun:2009si}.  
Whether these corrections at finite temperature yield additional corrections to the scaling behavior cannot be answered within
the scaling analysis presented here.\footnote{We would naively expect that corrections to Eq.~\eqref{eq:tctheta} 
can be only resolved in lattice simulations with very small masses for the pseudo Nambu-Goldstone modes and on very large lattice sizes.}
However, it may very well be that such corrections depend only on $\Nf$ but not on $\Nfcr$. 
\begin{figure}[!t]
\centering\includegraphics[scale=0.55]{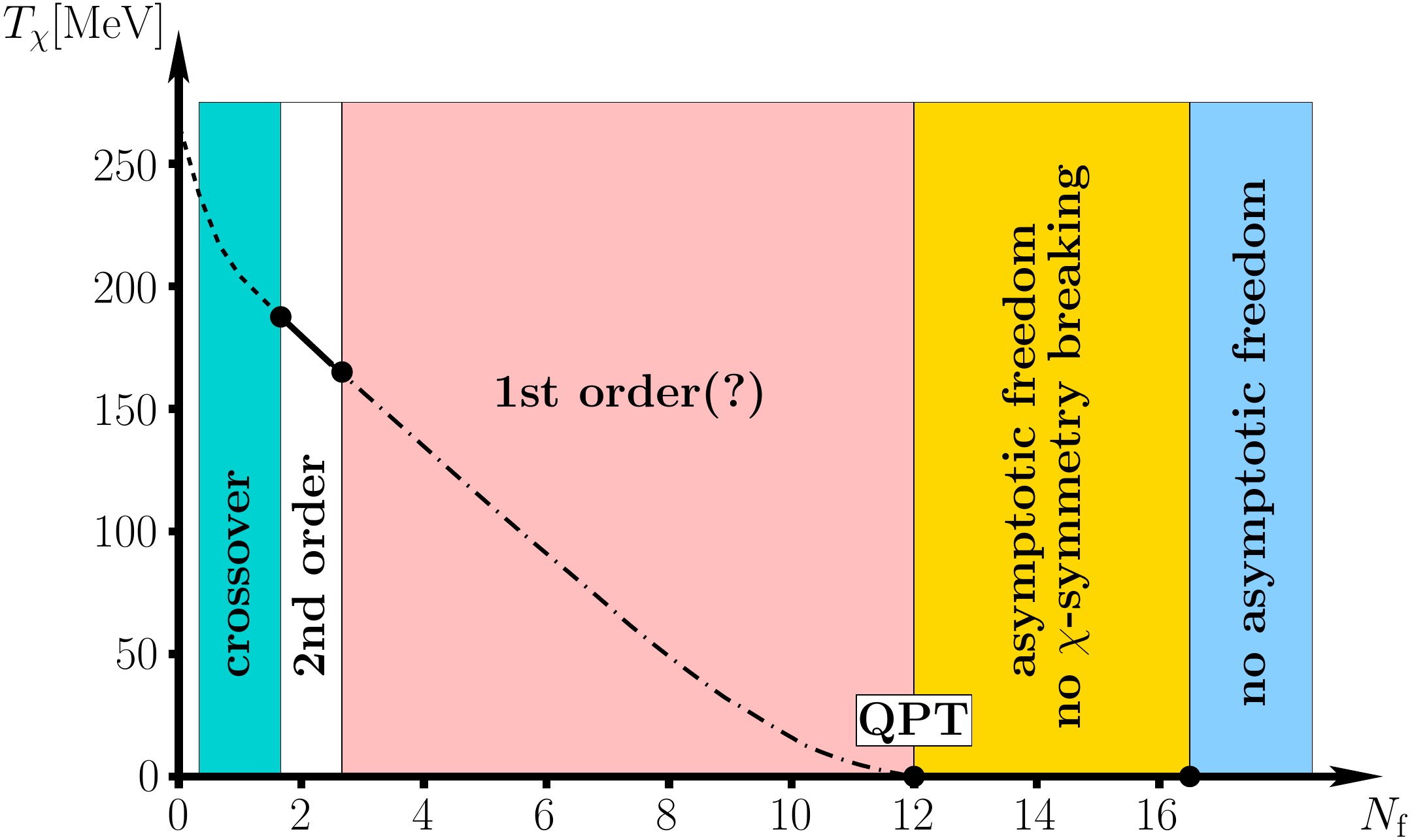} 
\caption{Sketch of the many-flavor QCD phase diagram (for massless quark flavors). 
For $\Nf<\Nfcr (\sim 12)$, we have different regimes which are distinguished by the nature of the
chiral phase transition. At $\Nf=\Nfcr$, the system undergoes a quantum phase transition (QPT). 
For $\Nfcr\leq\Nf\leq(11/2)\Nc=16.5$ (conformal window), there is no chiral symmetry breaking in the IR limit but the theory is still asymptotically free.
} 
\label{fig:TcNfsketch}
\end{figure}

We emphasize that further investigations of the finite-temperature scaling behavior close to the {\it quantum critical point}, $\Nf=\Nfcr$, is worthwhile. 
In particular, a study of the order of the nature of the finite-temperature phase transition seems to be rewarding since it may provide us with 
deep insights into the underlying chiral dynamics. 
An analysis in this direction based on RG arguments has been performed by Wilczek and Pisarski~\cite{Pisarski:1983ms}.
In Fig.~\ref{fig:TcNfsketch} we show a sketch of the many-flavor phase diagram which is inspired by our RG results depicted in Fig.~\ref{fig:TcNf}. 
For our discussion we shall assume that the number of massless
quark flavors can be indeed considered as a continuous control parameter. 
In the limit~$\Nf\to0$ we are left with a pure~SU($\Nc$) gauge theory. In this regime 
we do not have any chiral quark dynamics but only a deconfinement phase transition.
For~$0<\Nf\lesssim 2$, we then expect a 
crossover rather than an actual chiral phase transition.\footnote{We add that a generalization of 't~Hooft vertices
to non-integer values of~$\Nf$ may not be unique.}
This can be understood from a consideration of 
the lowest-order U${}_{\text{A}}$(1)-violating term(s) 
which are schematically given by $\sim (\yb\psi)^{\Nf}$, see e.~g. Refs.~\cite{tHooft:1976fv,Shifman:1979uw,Shuryak:1981ff,Schafer:1996wv}. For $\Nf=1$, 
such terms associated with topologically non-trivial gauge transformations act as a mass term for the quark fields which explicitly breaks the chiral
symmetry. The associated crossover line (dashed line in Fig.~\ref{fig:TcNfsketch}) is expected to end at some point below~$\Nf=2$. 
For~$\Nf=2$, QCD is assumed to fall in the O($4$)~universality class~\cite{Pisarski:1983ms}.
Recent lattice QCD data seem to be compatible with a second-order chiral phase transition for two massless quark flavors and~O$(4)$ scaling 
behavior at the phase boundary~\cite{Aoki:2006we,Aoki:2006br,Cheng:2006qk,Cheng:2009zi}.
The second-order phase transition line then most likely terminates in a ``critical endpoint" close to~$\Nf\lesssim 3$.
For~$\Nf=3$, evidence for a chiral first-order phase transition has been found in lattice QCD simulations, see e.~g. 
Refs.~\cite{Fucito:1984gt,Gavai:1987dk,Brown:1990ev,Karsch:2001nf}.
For larger values of~$\Nf$, only little is known about the nature of the phase transition.
As discussed above, the role of~U${}_{\text{A}}$(1)-violating terms~$\sim(\bar{\psi}\psi)^{\Nf}$ 
is probably subleading for~$\Nf\gtrsim 3$. From a simple ``entropy" argument, it seems reasonable to expect that the chiral phase transition is also of first order:
The order of the phase transition is (strongly) sensitive to the mismatch in the number of dynamical degrees of freedom below and above the 
chiral phase transition. Once the number of massless flavors~$\Nf$ exceeds some ``critical" value for a given~$\Nc$, this mismatch
potentially triggers a discontinuous behavior in the associated order parameter. This type of argument is similar to
arguments which seem to hold
in studies of, e.~g., the nature of the deconfinement transition in pure gauge theories~\cite{PepeWiese1,PepeWiese2,Braun:2010cy}. 
In these studies it has been found that the order of the deconfinement
phase transition changes from second to first order when the dimension of the gauge group is increased.
In any case, the line of first-order phase transitions hits the 
quantum critical point at~$T=0$ for~$\Nf\to\Nfcr$ ($\Nfcr\approx 12$ in Fig.~\ref{fig:TcNfsketch}). We stress that the phase transition in~$\Nf$-direction 
at~$T=0$ is expected to be continuous, whereas the phase transition in temperature
direction for~$\Nf\lesssim\Nfcr$ is presumably of first order. For~$\Nfcr\leq\Nf\leq (11/2)\Nc=16.5$, we are then in the conformal phase (regime) 
in which the theories are asymptotically free but we do not have chiral symmetry breaking in the IR limit.

At vanishing temperature, the analysis of the scaling behavior of IR observables is simplified compared to a scaling analysis at finite temperature
since dimensional reduction does not set in in the deep IR enhancing the Nambu-Goldstone modes. In the next two sections we shall therefore 
restrict ourselves to a quantitative analysis of scaling at vanishing temperature.
\begin{figure}[t]
\centering\includegraphics[scale=0.8]{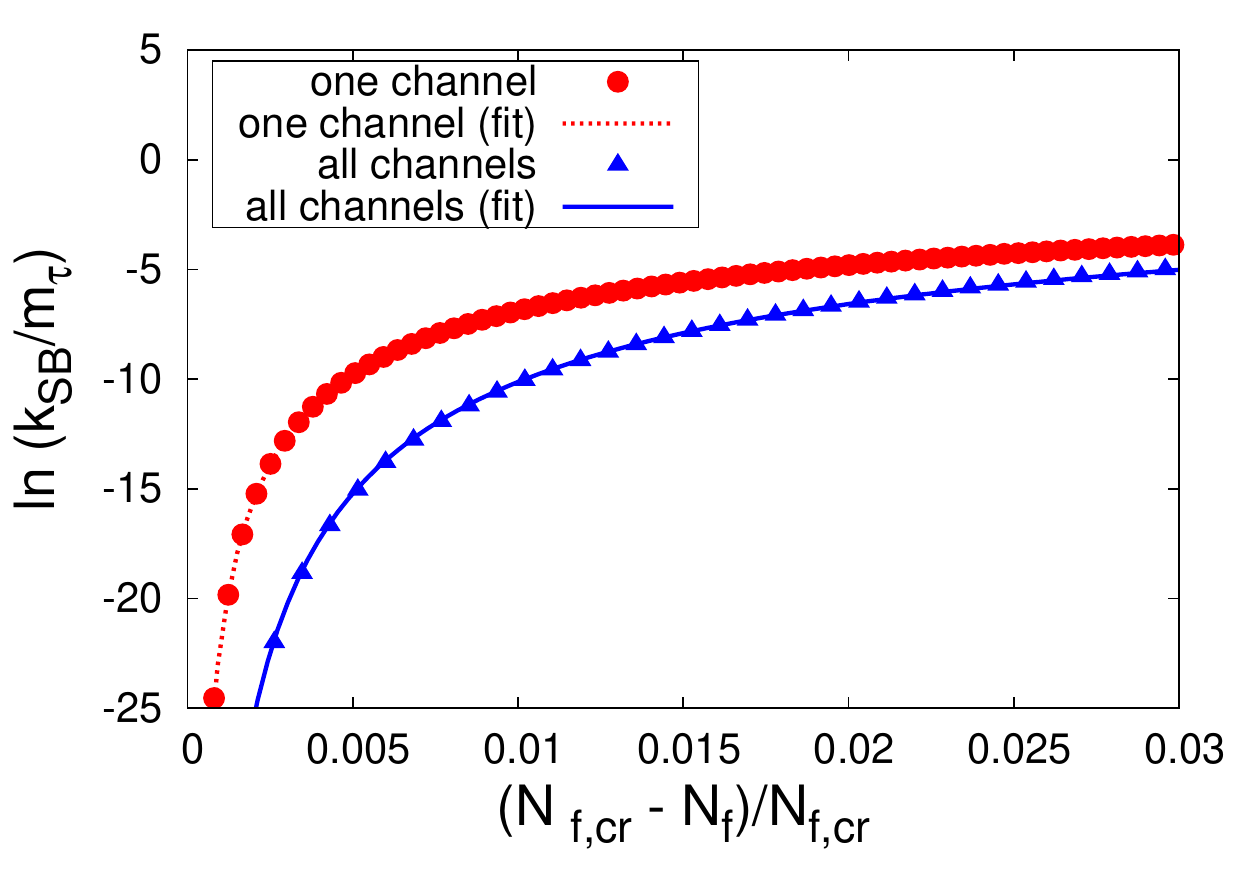}
\caption{
Logarithm of the (chiral) symmetry-breaking scale $\ln (\ksb/m_{\tau})$ as a function of the relative distance $(\Nfcr-\Nf)/\Nfcr$ 
from the quantum critical point for an $\Nf$-dependent but scale-independent, i.e. constant gauge
coupling. The corresponding fits are given in Eq.~\eqref{eq:Mirfits}. The figure has been taken from Ref.~\cite{Braun:2010qs}.
}
\label{fig:miransky}
\end{figure}
\subsubsection{Miransky Scaling}\label{sec:MSYM}
In the previous section we have computed the critical number of quark flavors~$\Nfcr$ as 
a function of~$\Nc$. To this end, it was not necessary to know the details of the running of the strong 
coupling~$g^2$. In the following we are interested in a quantitative study of the $\Nf$-scaling behavior of the symmetry breaking scale $\ksb$ 
close to the quantum critical point as encountered when the running of the gauge coupling is ignored,
\be
\partial_t g^2=0\,.\nn
\ee
To be precise, we shall consider a scenario in which the gauge coupling has assumed its IR fixed point value~$g^2_{\ast}$ 
for some $\Nf$ with $\Nf<\Nf^{\text{a.f.}}$ but $\Nf \gtrsim \Nfcr$. The fixed-point coupling then plays the role of an ``external"
parameter of the theory which can be changed by varying~$\Nf$ and/or~$\Nc$.
This allows us to increase the fixed-point coupling above the critical value~$g^2_{\rm cr}$ required for chiral symmetry breaking.
In Sect.~\ref{sec:MS} we have analyzed analytically the scaling behavior of physical observables for such a setup.
For our quantitative
study below, we employ the fixed-point value of the gauge-coupling at two-loop level.
Moreover, we restrict our quantitative analysis of exponential scaling behavior to the specific case $\Nc =3$.

In Fig.~\ref{fig:miransky} we show the results for $\ln(\ksb/\Lambda)$ as function of $(\Nfcr-\Nf)/\Nf$ as obtained from the 
one-channel (dots) and from the all-channels (triangles) approximation using $g^2_{\ast,{\rm 2-loop}}$ as a fixed input 
parameter.\footnote{Recall that the all-channels approximation is Fierz complete, while we only take into account the RG flow 
of the scalar-pseudoscalar channel $\lambda_{\sigma}$ in the one-channel approximation and set all other four-fermion couplings to 
zero.}
As initial conditions for the $\lambda_i$'s for a given $g^2_{\ast,{\rm 2-loop}}(\Nf)$ we have used the solution of the coupled set of linear 
equations
\be
\frac{\partial ( \partial_t\lambda_i)}{\partial \lambda_i}=0\,,
\ee
where $i\in\{+,-,\sigma,{\text{VA}}\}$. This corresponds to starting the flow at the maxima (extrema) of the parabolas.

We observe that for a given value of $\Nf$ the symmetry breaking scale $\ksb$ is smaller in the all-channels approximation compared to 
the one-channel approximation. The fits to the data points are also shown in Fig.~\ref{fig:miransky}. In agreement with the analytic 
results presented in Sect.~\ref{sec:MS} we find:
\be
\ln\ksb^{\rm one}\approx {\rm const.} - \frac{2.481}{|\Nfcr-\Nf|^{0.494}}\,,
\qquad
\ln\ksb^{\rm all}\approx {\rm const.} - \frac{3.932}{|\Nfcr-\Nf|^{0.516}}\,.\label{eq:Mirfits}
\ee
Thus, we clearly observe the expected exponential scaling behavior in the one-cannel and in the all-channels 
approximation for $\Nf\to\Nfcr$.  

The result from the one-channel approximation is in reasonable agreement with the analytic leading-order~(LO)
result found in Sect.~\ref{sec:MS}:
\be
\ln\ksb^{\rm LO}={\rm const.} - \frac{\pi}{2\epsilon\sqrt{|\alpha_1|
    |\Nfcr-\Nf|}} 
\approx {\rm const.} - \frac{2.386}{\sqrt{|\Nfcr-\Nf|}}\,. 
\ee
Note that $|\alpha_2/\alpha_1|\approx 0.273$. Differences to the analytic results are due to numerical errors of the fit and 
higher-order corrections which we have derived in Sect.~\ref{sec:MS}. 
\begin{figure}[t!]
\includegraphics[scale=0.6]{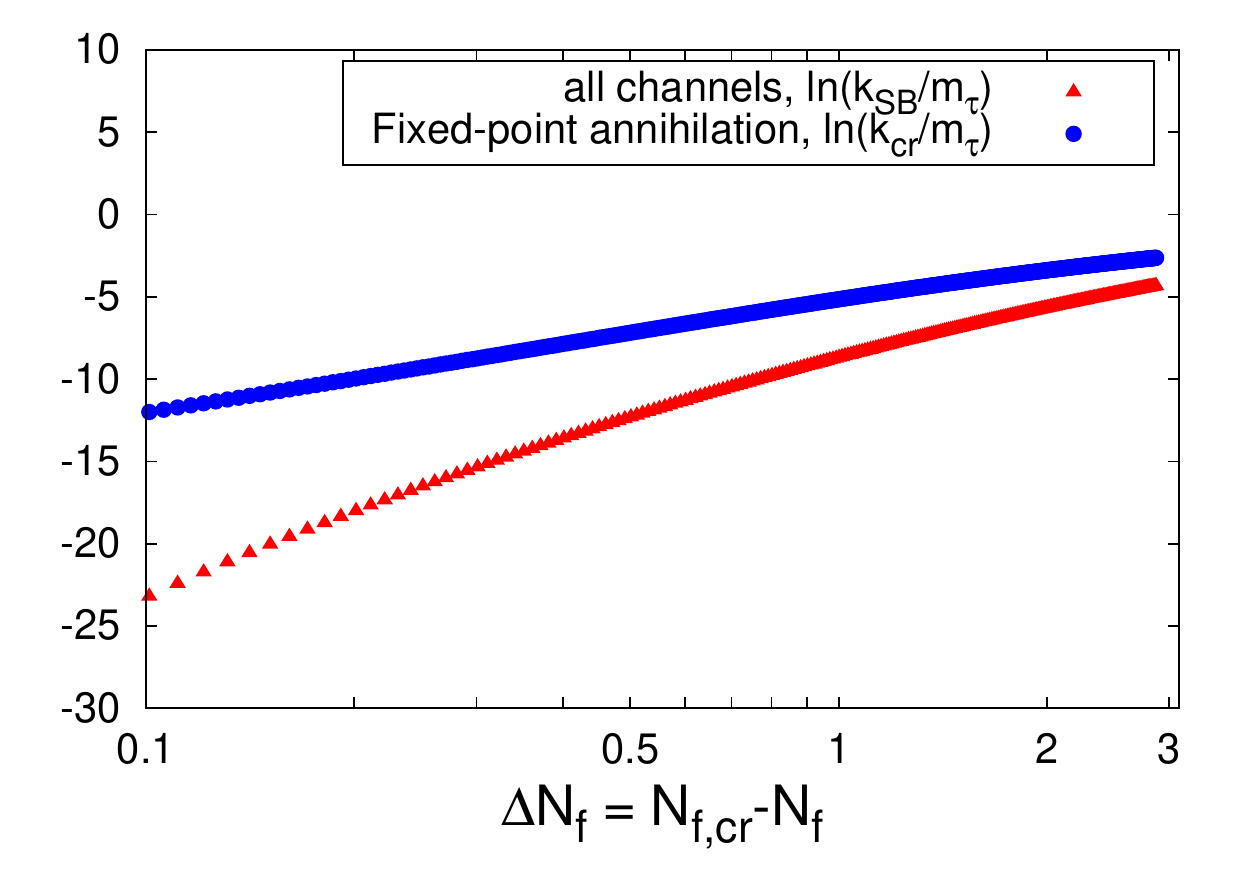}
\hfill
\includegraphics[scale=0.6]{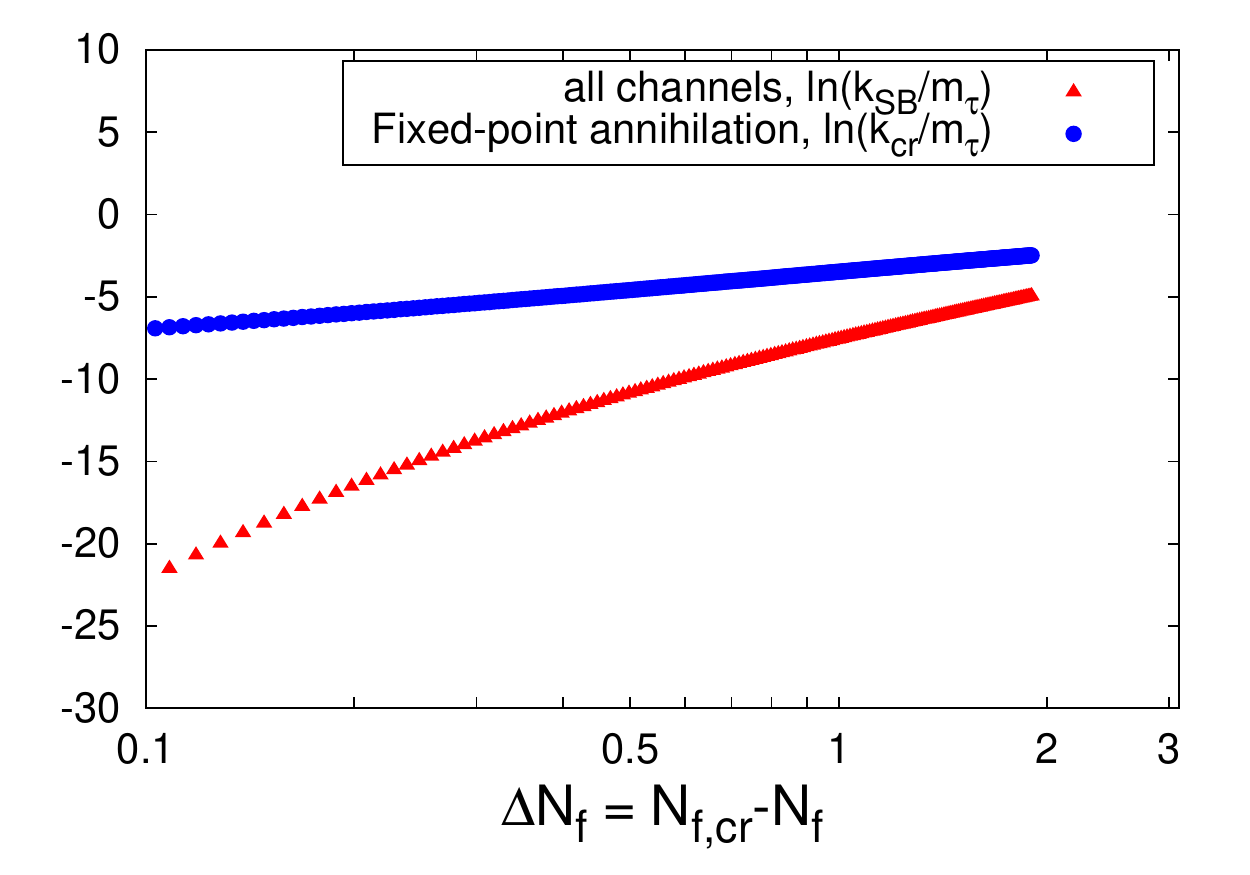}
\caption{Left panel: Double-logarithmic plot of $\Nf$ dependence of $\kcr$ and $\ksb$ as obtained from a study with a
  running coupling in the two-loop approximation. The criticality scale
    $\kcr$ (blue circles) is dominated by power-law scaling (straight line
    with slope $\sim|\Theta_0|^{-1}$ in
    this double-log plot), and clearly serves as an upper bound for the
    symmetry breaking scale $\ksb$ (red triangles), being a superposition of
    power-law and Miransky scaling. If the theories are probed at integer
    $\Nf$, i.e., $\Delta \Nf\gtrsim \mathcal{O}(1)$, the contribution due to
    Miransky scaling may not be visibile. 
    Right panel: Double-logarithmic plot of $\Nf$ dependence of $\kcr$ and $\ksb$ as obtained from a study with a
  running gauge coupling in the four-loop approximation. The 
    contributions due to Miransky scaling, roughly parameterized by the
    difference between $\kcr$ (blue circles) and $\ksb$ (red triangles),
    extend to larger values of $\Delta \Nf=\Nfcr -\Nf$, as the estimate for
    the critical exponent $\Theta_0=\Theta(\Nfcr)$ at four-loop level is larger than
  at two-loop level. In this perturbative estimate for the running coupling, the
  curves cannot be extended to larger values of $\Delta\Nf$, see main text.
  The figures have been taken from Ref.~\cite{Braun:2010qs}.
}
\label{fig:ksb24loop}
\end{figure}
\subsubsection{Power-law Scaling and Beyond}\label{sec:PLYM}
We now study scaling in a setup in which we take into account the (momentum) scale-dependence of the running gauge coupling. 
In order to compare the theories with different flavor numbers we fix the scales by keeping the running 
coupling at the $\tau$-mass scale $\Lambda=m_{\tau}$ fixed to $g^2(m_{\tau})/(4\pi)\approx 0.322$.
Since we apply the truncation~\eqref{equ::truncation} to QCD, we do not consider the four-fermion 
couplings $\lambda$ as independent external parameters as, e.g., in NJL-type low-energy QCD models, 
see Sect.~\ref{sec:LowQCDF}. More precisely, we impose the boundary condition $\lambda_i\to 0$ for $k\to \infty$ which 
guarantees that the $\lambda _i$'s at finite~$k$ are solely generated by quark-gluon dynamics, 
e.g., by 1PI ``box'' diagrams with 2-gluon exchange, see Fig.~\ref{fig:feynmanYM}(c). 
\begin{figure}[!t]
\begin{center}
\includegraphics[scale=0.58]{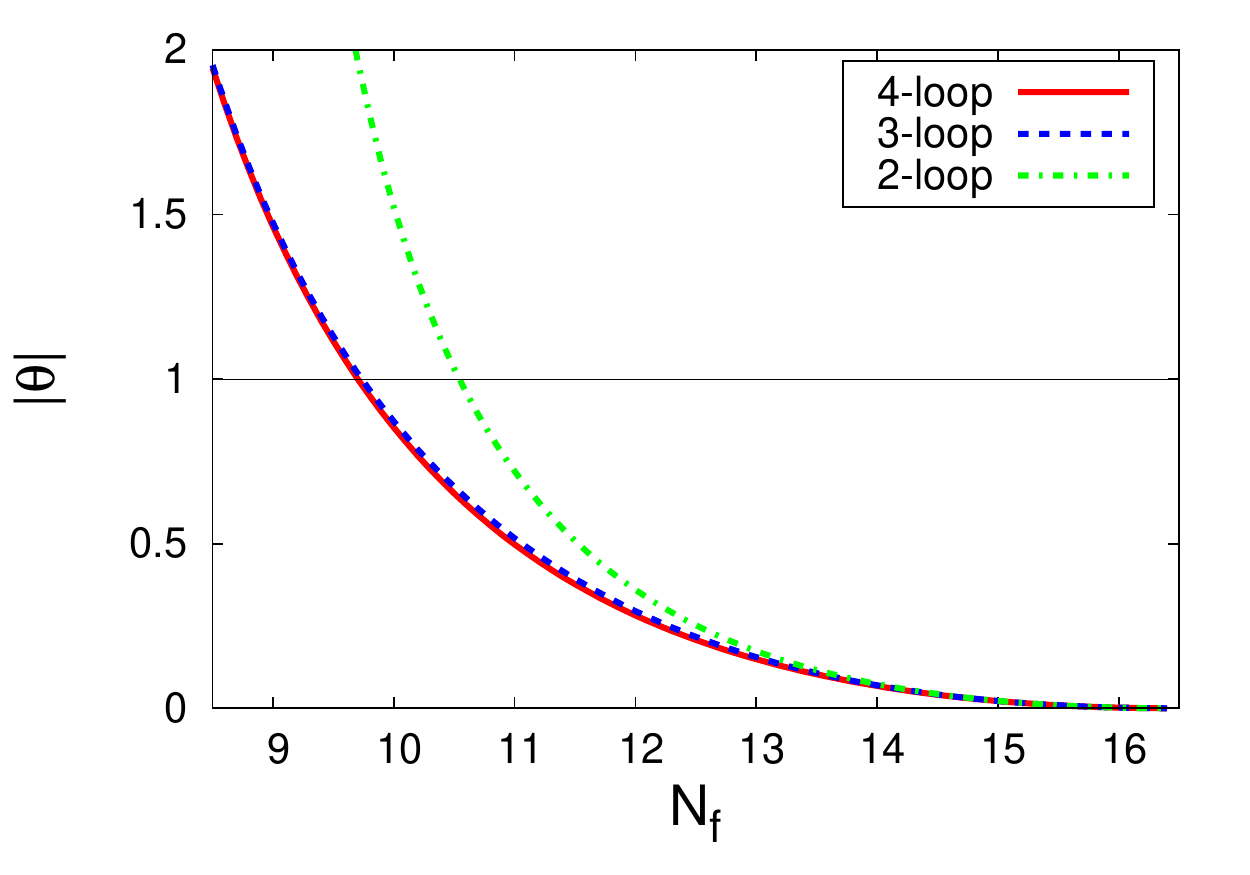} 
\hfill\includegraphics[scale=0.58]{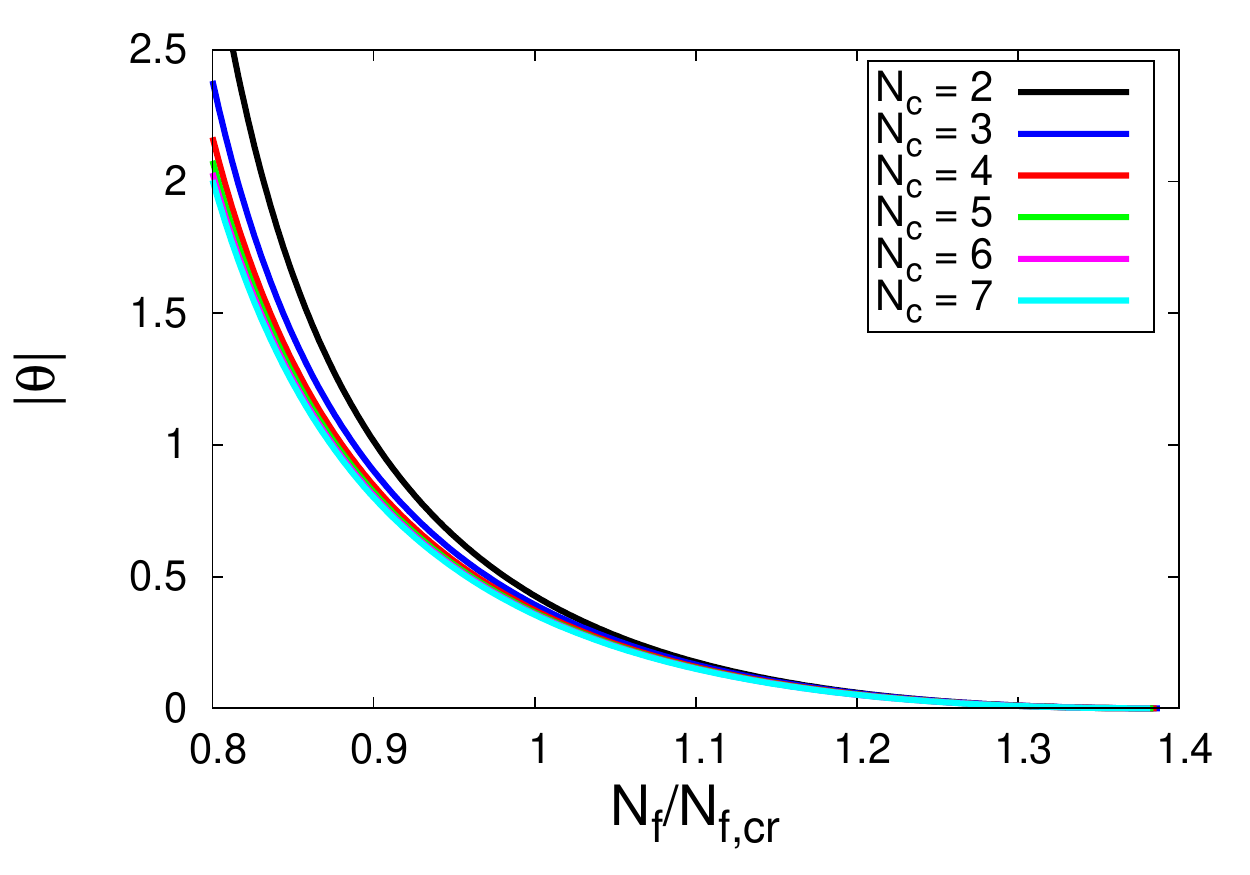} 
\end{center}
\caption{Left panel: Critical exponent $\Theta$ of the running gauge coupling at the CBZ fixed point for~$\Nc=3$
as a function of the number of flavors $\Nf$ as obtained from two-, three- and four-loop perturbation theory
in the $\MSbar$ scheme. Right panel: Critical exponent $\Theta$ of the two-loop running gauge coupling at the CBZ fixed point
as a function of~$\Nf/\Nfcr$ for $\Nc=2,3,\dots,7$ (from top to bottom). Recall that~$\Nfcr$ depends
 on~$\Nc$, see Sect.~\ref{sec:SUNYM}. 
} 
\label{fig:Theta}
\end{figure}

In Fig.~\ref{fig:ksb24loop} we show the results for the $\Nf$ dependence of the scales $\kcr$ and $\ksb$ for~$\Nc=3$ 
as obtained from a study
with a running gauge coupling in the two-loop and the four-loop approximation, respectively. 
The data points can be fitted to the analytic results for the scaling behavior of $\ksb$ and $\kcr$. For the 
all-channels approximation, we find
\be
\ln \kcr^{\rm 2-loop} &\approx& {\rm const.} + 2.566\, \ln |\Nf-\Nfcr|\,,\\
\ln \ksb^{\rm 2-loop} &\approx& {\rm const.} -
\frac{3.401}{|\Nf-\Nfcr|^{0.54}} 
+ 2.540 \ln |\Nf-\Nfcr|\,,\label{eq:2loopfit}
\ee
and
\be
\ln \kcr^{\rm 4-loop} &\approx& {\rm const.} + 1.180\, \ln |\Nf-\Nfcr|\,,\\
\ln \ksb^{\rm 4-loop} &\approx& {\rm const.} -
\frac{5.196}{|\Nf-\Nfcr|^{0.52}} 
+ 1.171 \ln |\Nf-\Nfcr|\,.\label{eq:4loopfit}
\ee
Thus, the fits are in reasonable agreement with our analytic predictions. For the multi-parameter fits~\eqref{eq:2loopfit} and~\eqref{eq:4loopfit}, we have
fixed the coefficient of the $\ln$-term which is the inverse critical exponent $\Theta_0=\Theta(\Nfcr)$. We emphasize that the predicted values for
the critical exponent $\Theta(\Nfcr)$ are substantially different for the running coupling in the two- and four-loop approximation, see left panel of
Fig.~\ref{fig:Theta}: 
\be
\frac{1}{|\Theta(\Nfcr)|}\approx 2.540\quad \text{(two-loop)}\,, 
\qquad
\frac{1}{|\Theta(\Nfcr)|}\approx 1.171\quad \text{(four-loop)}\,.
\ee
In Fig.~\ref{fig:ksb24loop} we observe that the critical exponent $\Theta$ clearly influences the scaling behavior close to the quantum critical 
point $\Nfcr$. In agreement with the analytic findings presented in Sect.~\ref{sec:QCGT}, the size of the regime with exponential scaling increases with 
increasing~$|\Theta|$. Using Eq.~\eqref{eq:size} we can give a quantitative estimate for the size of the regime in which the
exponential scaling behavior dominates. For the one-channel approximation, see Eq.~\eqref{eq:Nfcr1channel}, we find
\be
\Delta N_{\rm f}^{\rm 2-loop} := |\Nf-\Nfcr| \lesssim 0.27\,, 
\ee
where we have used the running coupling at two-loop level, see also Tab.~\ref{tab:PDYMresults}. 
Using a running coupling in the four-loop approximation ($\Nfcr^{\rm 4-loop}\approx 10.0$), the size of
this Miransky scaling regime can be estimated to be larger than one flavor. This is in agreement with the numerical results shown in
Fig.~\ref{fig:ksb24loop}. Since our studies rely on a perturbative estimate for the running coupling, the curves in Fig.~\ref{fig:ksb24loop}
cannot be extended to larger values of $\Delta\Nf= \Nfcr-\Nf$.  For instance, in the four-loop case, we have $\Nfcr\simeq9.8$. 
However, the CBZ fixed-point vanishes for~$\Nf\lesssim 8$. Our RG arguments based on expansions about an IR fixed
point thus only extend to $\Delta N_{{\rm f}}^{{\text{max}}} \simeq 1.8$, see right panel of Fig.~\ref{fig:ksb24loop}.
In non-perturbative functional studies in the Landau gauge, an IR fixed point appears to exist already in the pure gauge sector and thus also 
for $\Nf < \Nf^{\text{CBZ}}$, see Refs.~\cite{vonSmekal:1997is,vonSmekal:1997vx,Lerche:2002ep,Gies:2002af,Alkofer:2004it,Fischer:2004uk, 
Braun:2005uj,Braun:2006jd,Fischer:2008uz,Fischer:2009tn,Fischer:2006vf,Aguilar:2009nf}. In this case, no restriction on $\Delta \Nf$ arises.
\begin{figure*}[t]
\centering\includegraphics[scale=0.8]{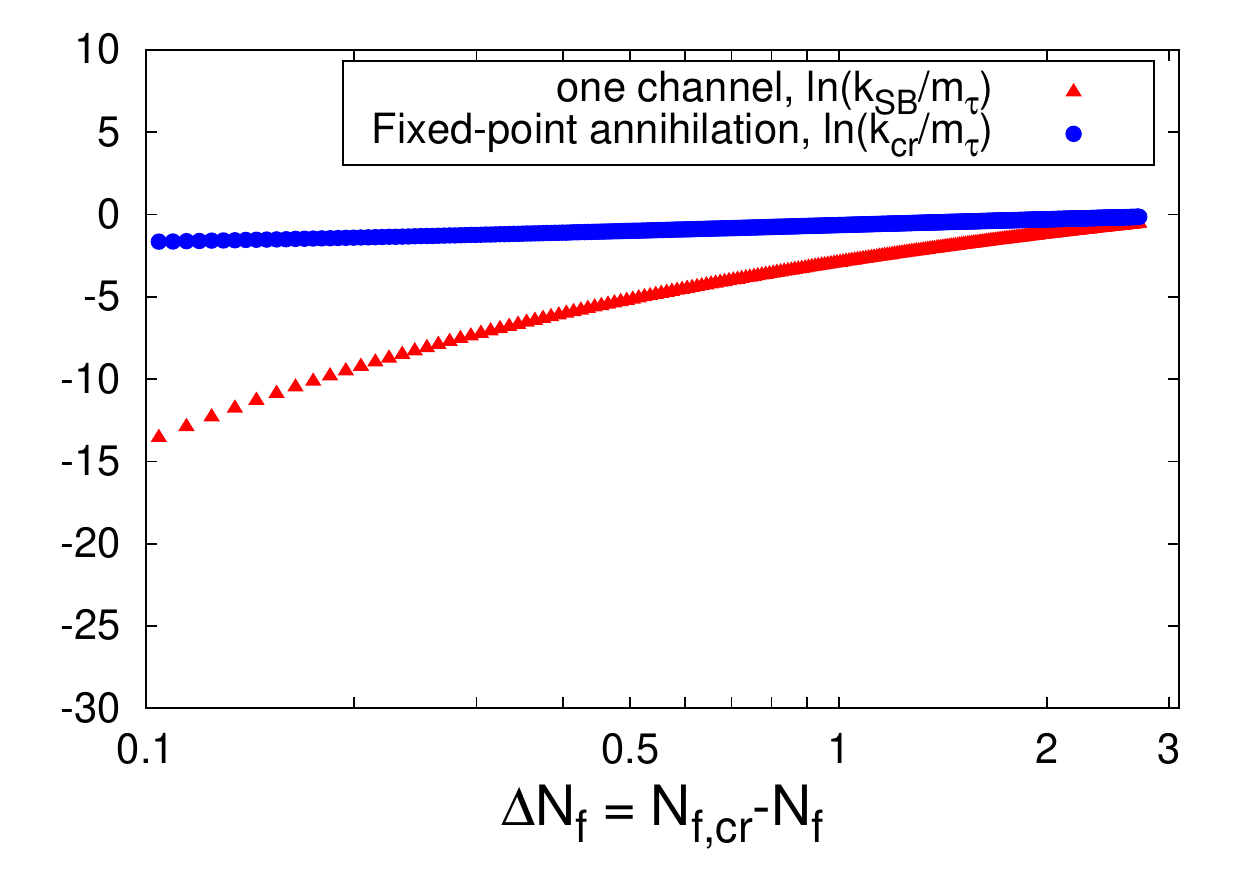}
\caption{$\Nf$ dependence of $\kcr$ (blue circles) and $\ksb$
  (red triangles) as obtained from a study with a specifically designed running gauge coupling for~$\Nc=3$, cf. Eq.~\eqref{eq:art2loop}. The associated 
  $\beta$ function allows us to vary the critical exponent $\Theta$ by hand.  Here, we show the results for
  $\Theta_0=|\Theta(\Nfcr)|\approx 4.3$ which can be compared to the results from the ``real" two-loop running coupling
  in the left panel of Fig.~\ref{fig:ksb24loop}. Contributions due to Miransky scaling are visible as deviations from a straight-line behavior (power law) 
  in this double-logarithmic plot. These results confirm our estimate that the Miransky-scaling window is
    larger for larger $|\Theta_0|$, whereas power-law scaling dominates for small $|\Theta_0|$, see left panel of Fig.~\ref{fig:ksb24loop}.
    The figure has been taken from Ref.~\cite{Braun:2010qs}.
    }
\label{fig:thetadep}
\end{figure*}

Let us now discuss our results in the light of lattice simulations. Since it is found that $\Nfcr\gtrsim 9$ in current lattice
simulations~\cite{Kogut:1982fn,Gavai:1985wi,Fukugita:1987mb,Brown:1992fz,Damgaard:1997ut,
  Iwasaki:2003de,Catterall:2007yx,Appelquist:2007hu,Deuzeman:2008sc,Deuzeman:2009mh,Appelquist:2009ty,
  Fodor:2009wk,Fodor:2009ff,Pallante:2009hu}, we expect that the pure exponential scaling behavior is difficult to resolve 
  and the corrections due to the running of the gauge coupling might be more relevant for a scaling analysis in 
  lattice simulations. From the viewpoint of such simulations, one might be interested in keeping the power of the ``Miransky" 
  term fixed to $1/2$ and use the scaling law~\eqref{eq:slawcorr}  
  to fit $\Nfcr$ and the critical exponent $\Theta_0=\Theta(\Nfcr)$. Recall that all chiral low-energy observables, such as the pion decay constant, 
  are expected to scale according to Eq.~\eqref{eq:slawcorr}.

In QCD, it appears to be a general feature that $\Theta_0\equiv \Theta(\Nfcr)$ decreases with $\Nfcr$. Estimates of $\Theta(\Nf)$ within two- and 
higher-loop approximations in the $\MSbar$ scheme are shown in the left panel of Fig.~\ref{fig:Theta}. Therefore, power-law
scaling is more prominent for larger $\Nfcr$. In particular, power-law scaling should be visible if theories are probed only for integer 
values of $\Nf$ as, e.g., on the lattice. 

Up to this point, we have restricted our scaling 
analysis to the case~$\Nc=3$. One may wonder whether the size of the Miransky scaling regime 
changes significantly when we vary the number of colors~$\Nc$. As discussed above,
this question can be essentially answered by looking at the value of the critical exponent~$\Theta$. In the right panel of Fig.~\ref{fig:Theta} we show
$\Theta$ as a function of~$\Nf/\Nfcr$ for~$\Nc=2,3,\dots,7$, where $\Nfcr=\Nfcr(\Nc)$ and $\Theta$ has been extracted from
the running coupling in the two-loop approximation. Interestingly, we observe that $\Theta$ at $\Nf=\Nfcr$ 
is approximately independent of~$\Nc$. This observation is in accordance with the result from a $1/\Nc$ expansion of~$\Theta_0\equiv\Theta(\Nfcr)$, which yields
\be
\Theta_0\equiv \Theta (\Nfcr) \approx -\frac{1}{3}-\frac{2}{9\Nc^2} + {\mathcal O}(1/\Nc^4)\,.\label{eq:Theta0LNcapp}
\ee
In order to obtain this simple form, we have used the estimate $\Nfcr = 4\Nc$, see Eq.~\eqref{eq:Nfcr4Nc}.
The large-$\Nc$ expansion of~$\Theta_0$ suggests that the exponential scaling behavior is only visible close to~$\Nfcr$, even for~$\Nc >3$.
Using Eq.~\eqref{eq:size} we can even give a 
direct estimate of the size of the regime in which the exponential scaling behavior dominates. The results
are listed in Tab.~\ref{tab:PDYMresults}. 
From this analysis we conclude that pure exponential scaling behavior seems to be difficult to resolve in, e.~g., lattice simulations, 
even for larger values of~$\Nc$.
Therefore the corrections due to the running of the gauge coupling (power-law scaling) might be more relevant
from a practical point of view in~QCD.

To further illustrate the important influence of the critical exponent $\Theta$ it is instructive to compute 
the scaling behavior of the scales $\kcr$ and $\ksb$ using a
specifically designed running gauge coupling. This coupling is inspired by the two-loop approximation modified by an artificial higher-order
term. The latter is constructed such that the critical exponent $\Theta$ can be changed by hand but the two-loop fixed point remains unchanged:
\be
\partial_t g^2 \equiv\beta_{g^2} = \beta_{g^2}^{\rm 2-loop} + \phi\, g^6(g^2 - g^2_{\ast,\rm 2-loop})\,,
\label{eq:art2loop}
\ee
where the parameter $\phi$ allows us to change $\Theta$ without changing $\Nfcr$. 
In Fig.~\ref{fig:thetadep} we present our results for $\ksb$ and $\kcr$ for $\phi=0.003$ (i.~e. $|\Theta(\Nfcr)|\approx 4.3$) as obtained from
the one-channel approximation for~$\Nc=3$. The comparison of these results with the results from the "real" two-loop running coupling 
with~$\Theta_0\approx 0.39$ 
(left panel of Fig.~\ref{fig:ksb24loop}) clearly confirms that the size of the exponential-scaling regime depends strongly on~$\Theta$.

In our scaling analysis we have mainly studied the behavior of the symmetry breaking scale~$\ksb$
which sets the scale for low-energy observables, see e.~g. Eq.~\eqref{eq:slawcorr} and Sect.~\ref{sec:ScalLEM}.
Although we expect that chiral observables scale according to the behavior of~$\ksb$, an explicit computation with the aid of functional 
methods and lattice simulations is still appealing. In Ref.~\cite{Gromenko:2008tn}, the quark mass, the chiral condensate and the pion decay 
constant have been computed within a truncated set of Dyson-Schwinger equations for many-flavor QCD. Signatures of the quantum critical point have been
identified and the critical exponents have been extracted from a pure power-law fit to the numerical data available for
$\Nf<\Nf^{\text{cr}}$. In the light of our scaling relations Eqs.~\eqref{eq:kcrO} and~\eqref{eq:slawcorr},  
the results of Ref.~\cite{Gromenko:2008tn} unfortunately remain somewhat inconclusive for two reasons: First, the exponential factor 
in Eq.~\eqref{eq:slawcorr} has not been taken into account in the fit. Second, $\Nfcr$ has been fitted for each chiral IR observable 
separately yielding slightly different values. While neglecting the exponential scaling factor might be reasonable due to the fact that the
Miransky scaling window is expected to be small in QCD, the uncertainty in $\Nfcr$ arising from the fitting procedure 
is likely to spoil the fit for the critical exponent. We expect that a more careful analysis in the vicinity of the quantum critical point 
can easily put our scaling relation to test. 

\subsection{Excursion: Confinement and Chiral Symmetry Breaking}\label{sec:CSBCONF}
Up to this point we have restricted our discussion to chiral symmetry breaking in gauge theories, such as QCD. We have totally left aside
effects arising from the confining dynamics in QCD. However, we were mainly interested in phase transitions in the limit of many massless quark flavors.
In this case, there is no good order parameter for confinement available anyway and it seems reasonable to expect that nonanalyticities in the correlation functions 
are mainly dominated by the chiral degrees of freedom. Now we shall discuss QCD with a small number of quark flavors. There, we may expect
that the confining dynamics significantly affects the chiral dynamics at the finite-temperature phase transition.

In order to gain some insight into the interrelation of quark confinement and chiral symmetry breaking we analyze how the 
order parameter for confinement influences the chiral fixed-point structure of the theory. The relation of both has been studied in detail
in Ref.~\cite{Braun:2011fw}. Here, we only review the main arguments and restrict ourselves to the large-$\Nc$ limit.
Before we start with our analysis, however, we would like to discuss some issues arising in studies of the relation of quark confinement
and chiral symmetry breaking.

The deconfinement phase transition has been studied in pure SU($\Nc$) gauge theories ($\Nf\to 0$) and
QCD with both lattice simulations, see e.~g. Refs.~\cite{Cheng:2006qk,Aoki:2006br,Aoki:2006we,Aoki:2009sc,Panero:2009tv,%
Cheng:2009zi,Datta:2010sq,Borsanyi:2010zi,Bazavov:2010pg,Kanaya:2010qd}, 
and functional continuum 
methods~\cite{Braun:2007bx,Marhauser:2008fz,Fischer:2009wc,Fischer:2009gk,Braun:2009gm,Fischer:2010fx,Braun:2010cy,Pawlowski:2010ht}.
Concerning chiral symmetry breaking, we have seen that the (chiral) condensate~$\langle \bar{\psi}\psi\rangle$ 
serves as an order parameter. A finite chiral condensate implies that the chiral ${\rm SU}(\Nf)_{\rm L}\otimes {\rm SU}(\Nf)_{\rm R}$ flavor
symmetry of QCD is broken. Concerning the confinement phase transition, on the other hand, an order parameter can be constructed
from the so-called Polyakov-loop variable:
\be\label{eq:Polloop}
L[A_0]=\frac{1}{\Nc}\, {\mathcal P}\,\exp\left({\rm i}\bar{g}\int_0^{\beta}
dx_0\,A_0(x_0,\vec{x})
\right)\,,
\ee
where $\beta=1/T$ is the inverse temperature and $\bar{g}$ denotes the bare gauge coupling; $\mathcal P$ stands for path ordering.
In QCD with $\Nc$ colors and infinitely heavy quarks this quantity is related to the
operator that generates a static quark, i.~e. an infinitely heavy quark~\cite{Polyakov:1978vu}. 
Loosely speaking, the logarithm of the expectation value $\langle \tr _{\rm F} L[A_0] \rangle$ can be related to the 
free energy~$F_q$ of a static quark. To be more specific, we can interpret it as
half of the free energy $F_{q\bar q}$ of a static quark--anti-quark pair at infinite distance. Here, the trace~$\tr_{\rm F}$ is evaluated in the fundamental
representation. Moreover, the expectation value 
$\langle \tr _{\rm F} L\rangle$ is an order parameter for center symmetry breaking of the 
underlying gauge group~\cite{Greensite:2003bk}. To see this, we consider
gauge transformations $U_z(x_0,x)$ with $U_z^{-1}(0,\vec x) U_z(\beta,\vec x) =z$, where $z\in \CZ$ is an element of the center $\CZ$ of the
gauge group. Under such a transformation the Polyakov loop is multiplied with a center element~$z$, $\langle \tr _{\rm F} L\rangle \to z\, \langle \tr _{\rm F} L\rangle$. Thus, 
a center-symmetric confining disordered ground state with $F_q \to\infty$ is ensured by $\langle \tr _{\rm F} L\rangle=0$. In turn,
deconfinement with $F_q <\infty$ is signaled by $\langle \tr _{\rm F} L\rangle\neq 0$. This  consideration 
implies center-symmetry breaking in the ordered phase.

The relation of quark confinement and chiral symmetry breaking in QCD is indeed not yet fully understood. 
As the chiral and the deconfinement phase transition are related to different symmetries of the theory, 
it is difficult to establish a simple (analytic) relation between both. Even worse, the deconfinement phase transition turns into a crossover
in the presence of dynamical quarks since the latter break explicitly the underlying center symmetry.\footnote{The explicit symmetry
breaking becomes stronger the smaller the current quark masses are.} As there is no unique way to define the critical temperature 
associated with a crossover, a proof of an exact coincidence of the two transitions seems to be impossible in any case.
On the other hand, both phase transitions are driven by the gauge degrees of freedom in QCD. This is readily apparent for the deconfinement phase transition.
For the chiral phase transition the relation to the gauge degrees is more indirect.\footnote{As a matter fact, the chiral phase transition in QCD
can be investigated with NJL-type models (and in Polyakov-loop extended versions thereof). In these models the 
quark interactions are considered as parameters which are tuned by hand to fit low-energy observables, see our discussion in
Sect.~\ref{sec:NJLLQCD}.} However, we have seen in the previous sections that the quark self-interactions are dynamically generated and driven to criticality 
by the gauge degrees of freedom: once the gauge coupling exceeds a critical value, the quark sector is driven to criticality without 
requiring any fine-tuning. This observation may suggest that there might be a deeper relation between the chiral dynamics in the 
matter sector and the confining dynamics in the gauge sector and serves as a motivation for the subsequent analysis.

In Polyakov-loop extended low-energy models a background field~$\langle A_0\rangle$ is introduced
to study some aspects of quark confinement and the associated phase transition~\cite{Meisinger:1995ih,Pisarski:2000eq,Mocsy:2003qw,%
Fukushima:2003fw,Megias:2004hj,Ratti:2005jh,Sasaki:2006ww,Schaefer:2007pw,Hell:2008cc,Mizher:2010zb,Skokov:2010wb,Herbst:2010rf,Skokov:2010uh,Hell:2011ic}.
This background field can be related to the Polyakov variable $L[A_0]$. In fact, it has been shown that $\tr_{\rm F}\,L[\bfe]$ serves as an order parameter 
for quark confinement in Polyakov-Landau-DeWitt gauge~\cite{Braun:2007bx,Marhauser:2008fz}, where $\bfe$ is an element of the 
Cartan subalgebra and denotes the ground state of the associated order-parameter potential in the adjoint gauge 
algebra.\footnote{Strictly speaking, we have to distinguish between the background temporal gauge field 
in Landau-DeWitt gauge and its expectation value associated with the order parameter for confinement, 
see Refs.~\cite{Braun:2007bx,Braun:2010cy}. We skip this subtlety here since it is of  no importance for our present analysis and refer to $\bfe$ 
as the position of the ground-state of the order-parameter potential.} This potential can be computed, e.~g., from the knowledge of gauge 
correlation functions, as first demonstrated in a first-principles RG study~\cite{Braun:2007bx,Braun:2010cy}. In any case,
the order parameter $\tr_{\rm F}\,L[\bfe]$ is related to the standard Polyakov loop $\langle \tr _{\rm F} L[A_0]\rangle$ via the Jensen inequality,
\be
\tr_{\rm F}\,L[\bfe]\geq \langle \tr_{\rm F}\, L[A_0]\rangle\,.
\ee
We would like to add that one of the underlying approximations in (Polyakov-loop extended) low-energy model studies is 
to set $\tr_{\rm F}\,L[\bfe] = \langle \tr_{\rm F}\,L[A_0]\rangle$. This opens up the possibility to incorporate results for the Polyakov 
loop $\langle \trf L[A_0]\rangle$ as obtained from lattice simulations in these studies. It is then found that the chiral and the deconfinement phase 
transition lie indeed close to each other at small values of the quark chemical potential, as it is found 
to be the case in lattice QCD simulations~\cite{Cheng:2006qk,Aoki:2006br,Aoki:2006we,Aoki:2009sc,Cheng:2009zi,Borsanyi:2010zi,Bazavov:2010pg,Kanaya:2010qd}. 

Recently, so-called dual observables arising from a variation of the boundary conditions of the fermions in time-like direction have been 
introduced~\cite{Gattringer:2006ci} and employed for a study of the relation of quark confinement and chiral symmetry 
breaking at finite temperature~\cite{Synatschke:2007bz,Bilgici:2008qy,Kashiwa:2008bq,Sakai:2008py,Bilgici:2009tx,%
Braun:2009gm,Fischer:2009wc,Fischer:2009gk,Fischer:2010fx,Zhang:2010ui,Mukherjee:2010cp,Gatto:2010qs}. 
These dual observables relate the spectrum of the Dirac operator to the order parameter for confinement, namely the Polyakov loop. 
The introduction of these observables constitutes an important formal advance which allows us to gain a deeper insight into the underlying 
dynamics at the QCD phase boundary. However, they do not allow us to fully resolve the question regarding the relation of 
quark confinement and chiral symmetry breaking. 

In the following we aim to shed more light on the question under which circumstances the chiral and the deconfinement transition lie close to each other.
To this end, we analyze the deformation of the RG fixed-point structure of chiral four-fermion interactions due to confining gauge dynamics.
Technically speaking, this means that we couple the order parameter for confinement to the RG flow 
of four-fermion interactions. As discussed in detail in Sect.~\ref{subsec:bos}, the latter
can be related to the order parameter for chiral symmetry breaking by means of partial bosonization. As we shall
see, this yields an intimate relation between the chiral and the deconfinement order parameter which suggests the existence of a dynamical locking 
mechanism for the chiral phase transition. To keep our discussion of the general mechanisms as simple as possible,
we restrict ourselves to $\Nf=2$ massless quark flavors with~$\Nc$ colors and employ the following ansatz for the effective action:
\be
\Gamma_k [\bar{\psi},\psi,\bfe]=\int d^4 x \Big\{ \bar{\psi}\left(
{\rm i}\fslash{\partial} + \bar{g}\gamma_0 \langle A_0\rangle\right)
\psi 
 + \frac{1}{2}\bar{\lambda}_{\sigma}\left[ (\bar{\psi}\psi)^2 
\!-\! (\bar{\psi}\vec{\tau} \gamma_5\psi)^2\right]\Big\}\,,
\label{eq:fermionic_actionCONF}
\ee
where the $\tau_i$ represent the Pauli matrices and couple the spinors in flavor space. 

The action~\eqref{eq:fermionic_actionCONF} can be considered as an
ansatz for a QCD low-energy model. In fact, we have discussed this action for~$\bfe\equiv 0$ in 
the context of QCD low-energy models in Sect.~\ref{sec:NJLLQCD}, where we have shown that the RG flow 
of the $\bar{\lambda}_{\sigma}$-interaction decouples from the RG flows of other allowed four-fermion interaction channels
in the large-$\Nc$ limit. Of course, fermionic self-interactions are fluctuation-induced in full QCD, e.~g. by two-gluon exchange, 
and are therefore not fundamental, see our discussion in the previous sections. However, we are here rather interested in studying
how the fixed-point structure of four-fermion interaction is deformed under the influence of confining dynamics. For such a 
general discussion, we expect the ansatz~\eqref{eq:fermionic_actionCONF} to be sufficient.

Based on the action~\eqref{eq:fermionic_actionCONF} we have discussed quantum and thermal phase transitions in QCD low-energy 
models in Sect.~\ref{sec:TQPTQCD} for $\bfe\equiv 0$. Let us now turn to a discussion of the fixed-point structure for finite~$\bfe$.
The value of the background field $\bfe$ is determined by the ground state of the 
associated order-parameter potential, see Refs.~\cite{Braun:2007bx,Braun:2010cy}. As discussed above, this
ground-state value~$\bfe$ is directly related to an order parameter for confinement, namely~$ \trf L[\bfe]$. In the following we do not need to know 
the exact values of~$\bfe$ and~$ \trf L[\bfe]$. We only need to know about
some general properties of the confinement order parameter.

For temperatures much larger than the deconfinement phase-transition temperature~$\Td$ we have $\bfe=0$, i.~e. $\trf L[\bfe] =1$. 
On the other hand, the position $\bfe$ of the ground state in the confined phase of pure ${\rm SU}(\Nc)$ Yang-Mills theory is uniquely 
determined up to center transformations by~\cite{Braun:2007bx,Braun:2010cy}
\be
\trf (L[\bfe])^n = 0 
\label{eq:polcoord}
\ee
with $(n\mod\Nc)=1,\dots,\Nc-1$. These conditions determine the $\Nc -1$ coordinates~$\{\phi^{(a)}\}$ of $\bfe$:
\be
\beta \bar{g} \bfe &=& 2\pi\sum_{T^a \in {\rm Cartan}} T^a \phi^{(a)} 
= 2\pi\sum_{T^a \in {\rm Cartan}} T^a v^{(a)} |\phi|\,,\quad v^2=1\,,\label{eq:A0span}
\ee
where the $T^a$'s denote the generators of the underlying ${\rm SU}(\Nc)$ gauge group in the fundamental representation.\footnote{The dimension 
of the Cartan subalgebra is $N_c-1$.} Concerning the parameterization of~$\bfe$, it turns out that it is convenient to introduce the eigenvalues $\nu _l$ 
of the hermitian matrix in Eq.~\eqref{eq:A0span}:
\be
\nu_l = {\rm spec}\left\{ (T^{a}v^{a})_{ij}\,\; |\; v^2=1 \right\}\,.
\ee
Finally, we have
\be
\frac{1}{\Nc}\left|\trf (L[\bfe])^n\right| 
 \leq \frac{1}{\Nc^n}\,
\ee
for $n\in {\mathbb N}$. Note that the ground-state value~$\bfe$ is shifted in QCD with dynamical quarks and yields a small but finite order 
parameter in the confined phase. 

We now have set the stage for a discussion of the fixed-point structure of the four-fermion coupling~$\lambda_{\sigma}$.
The flow equation of the latter can be computed along the lines of Sect.~\ref{subsec:simpleex}. In the point-like limit we 
find~\cite{Braun:2011fw}:
\be
\beta_{\lambda_{\sigma}}\equiv\partial_t \lambda_{\sigma} = 2\lambda_{\sigma}
- 16\Big(2 + \frac{1}{\Nc}\Big) v_3
\sum_{l=1}^{N_c} l_{1}^{\rm (F),(4)}(\tau,0,\nu_l |\phi|)\,\lambda_{\sigma}^2\,,
\label{eq:lpsi_flowCONF}
\ee
where $v_3=1/(8\pi^2)$ and~$\lambda_{\sigma}=k^2 \bar{\lambda}_{\sigma}$. Since we work in the point-like limit, we 
have~\mbox{$\eta_{\psi}=0$}. To derive this equation we have employed a 3$d$ regulator function; the
definition of the background-field dependent threshold function can be found in App.~\ref{app:regthres}. Moreover, we have exploited the 
fact that the fermion propagator $(\Gamma ^{(2)})^{-1}$ (inverse two-point function) can be spanned by the generators of the 
Cartan subalgebra as follows:
\be
\left(\Gamma ^{(2)}[\{\nu_l |\phi|\}]\right)^{-1}_{ij}=\frac{1}{N_c} \left(\Gamma ^{(2)}_{0}[\{\nu_l |\phi|\}]\right)^{-1} \mathbbm{1}_{ij} 
+ \sum_{T^a \in {\rm Cartan}} \left(\Gamma ^{(2)}_{a}[\{\nu_l |\phi|\}]\right)^{-1} T^a_{ij}\,.
\label{eq:prop_expCONF}
\ee 
Here, the $T^{a}_{ij}$'s denote the generators in the fundamental (color) representation.
The expansion coefficients on the right-hand side can be computed straightforwardly by using ${\rm tr}_{\rm F}T^{a}T{^b}=\frac{1}{2}\delta_{ab}$
and ${\rm \tr}_{\rm F}T^a =0$.

For vanishing temperature as well as temperatures much larger than the deconfinement phase-transition temperature~$\Td$, 
the fixed-point structure is identical to the one discussed in Sect.~\ref{sec:TQPTQCD} 
since $\bfe$ tends to zero for $T\gg \Td$ and~$\bfe\equiv 0$ for~$T=0$. For finite~$\bfe$, 
the pseudo fixed-point $\lambda_{\sigma}^{\ast}$ depends on $\bfe$ and the dimensionless temperature~$\tau=T/k$. Within the
present approximation, the value of the pseudo-fixed point~$\lambda_{\sigma}^{\ast}$ can be given in closed form:
\be
\lambda_{\sigma}^{\ast}(\tau,\bfe) &=&\left( \frac{1}{ \pi^2}
\left(2\! +\! \frac{1}{\Nc}\right)\sum_{l=1}^{\Nc} l_1^{\rm (F),(4)}(\tau,0,\nu_l|\phi|)  
\right)^{-1}
\nn\\
&=&\left( \frac{1}{\lambda_{\sigma}^{\ast}(0,0)}
+ \frac{1}{6\pi^2}\left(2\!+\!\frac{1}{\Nc}\right) \sum_{n=1}^{\infty}(-\Nc)^{n}\Big[\trf (L[\bfe])^n \right. \nn\\
&& \qquad\qquad\qquad \qquad\qquad\qquad
 \left.
 +\, \trf (L^{\dagger}[\bfe])^n\Big]
\left(
1+\frac{n}{\tau}
\right){\rm e}^{-\frac{n}{\tau}}
\right)^{-1}\,.
\label{eq:lfpa0CONF}
\ee
The specific form in the second line has been obtained with the regulator function~\eqref{eq:fermreg}.
However, we stress that the general form of the asymptotic series~\eqref{eq:lfpa0CONF} holds for {\it any}
regulator function, as can be shown by means of Poisson resummation 
techniques. Note that the series~\eqref{eq:lfpa0CONF} effectively represents a low-temperature expansion and that
the sum over the $\tau$-dependent terms in the second line of Eq.~\eqref{eq:lfpa0CONF}
is closely related to the geometric series.
\begin{figure}[t]
\centering\includegraphics[scale=0.85]{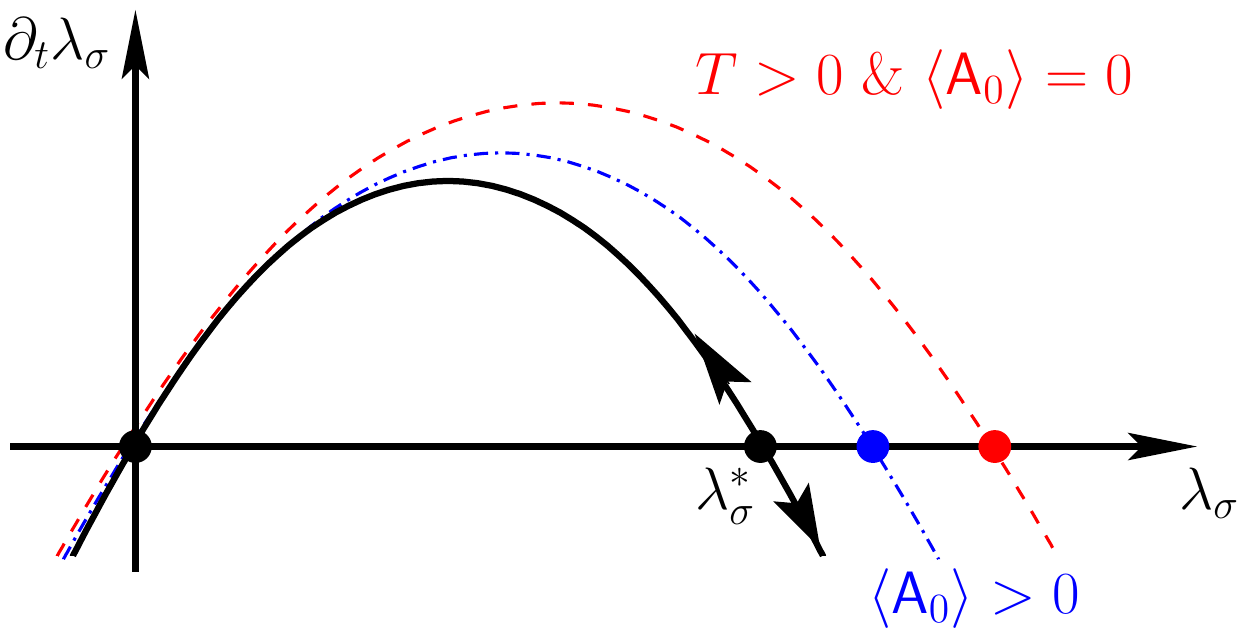}
\caption{Sketch of the $\beta_{\lambda_{\sigma}}$ function of the four-fermion interaction for 
$T=0$ (black/solid line), a given finite value of the temperature $T$ and $\bfe=0$ 
(red/dashed line) and the same temperature $T$ but $\bfe >0$ (blue/dashed-dotted line), see Ref.~\cite{Braun:2011fw}.
The arrows indicate the direction of the RG flow towards the infrared.
}
\label{fig:parabolaCONF}
\end{figure}

Using Eq.~\eqref{eq:polcoord} it follows that all finite-temperature corrections to the (pseudo) fixed-point value vanish identically in the confined 
phase for $\Nc\to\infty$, provided that the ground-state value $\bfe$ is identical in SU($\Nc$) Yang-Mills theory and
QCD with dynamical fermions. Of course, this assumption is not exactly fulfilled but for physical quark masses it is reasonable 
to assume 
\be
\trf L[\bfe]\ll 1
\ee
for $T\lesssim \Td$. Thus, we have found that 
\be
\lambda_{\sigma}^{\ast}(0,0)\equiv\lambda_{\sigma}^{\ast}(\tau,\bfe)
\label{eq:fpequivCONF}
\ee
in the limit $\Nc\to\infty$, independent of the temperature for $T\lesssim \Td$. 
On the other hand, we have 
\be
\lambda_{\sigma}^{\ast}(\tau,\bfe)\to \lambda_{\sigma}^{\ast}(\tau,0)\quad\text{for}\quad\bfe\to 0\,.
\ee
With the same reasoning it also follows that
\be
\beta_{\lambda_{\sigma}}(0,0)\equiv\beta_{\lambda_{\sigma}}(\tau,\bfe)\label{eq:betaequal}
\ee
for $T\lesssim \Td$ and $\Nc\to\infty$, see also Fig.~\ref{fig:parabolaCONF} for illustration.
This means that for $T<\Td$ the question of whether chiral symmetry is spontaneously broken or not is in fact {\it independent} 
of the temperature, but depends only on the choice of the initial condition~$\lambda_{\sigma}^{\rm UV}$ relative to its fixed-point 
value~$\lambda_{\sigma}^{\ast}$ at \mbox{$T=0$}.  
We add that Eqs.~\eqref{eq:fpequivCONF}-\eqref{eq:betaequal} are regularization-scheme 
independent statements and that $\lambda_{\sigma}^{\ast}(\tau,\bfe)$
interpolates continuously for a given finite value of $\tau$ between $\lambda_{\sigma}^{\ast}(0,0)$ 
and $\lambda_{\sigma}^{\ast}(\tau,0)$.
Note that we have not specified the precise value of~$\Td$ since it simply does not enter our
analysis. 

Provided that we choose an initial value $\lambda_{\sigma}^{\rm UV}> \lambda_{\sigma}^{\ast}(0,0)$, it follows immediately from 
Eq.~\eqref{eq:betaequal} that 
\be
\Tc \geq \Td\,\label{eq:TcTd}
\ee
for $\Nc\to\infty$, see Ref.~\cite{Braun:2011fw}. 
This means that the chiral phase transition is locked in due to the confining dynamics in the gauge sector.
Loosely speaking, thermal fluctuations of the quark fields, which tend to restore the chiral symmetry, are 
suppressed since they are directly linked to the deconfinement order parameter.
Thus, we have found that the restoration of chiral symmetry is intimately connected 
to the confining dynamics in the gauge sector. These findings emerging from a
non-perturbative analysis of the fermionic fixed-point structure confirm the results of a {\it mean-field} study 
by Meisinger and Ogilvie~\cite{Meisinger:1995ih}. 
\begin{figure*}[t]
\centering\includegraphics[scale=0.75,clip=true]{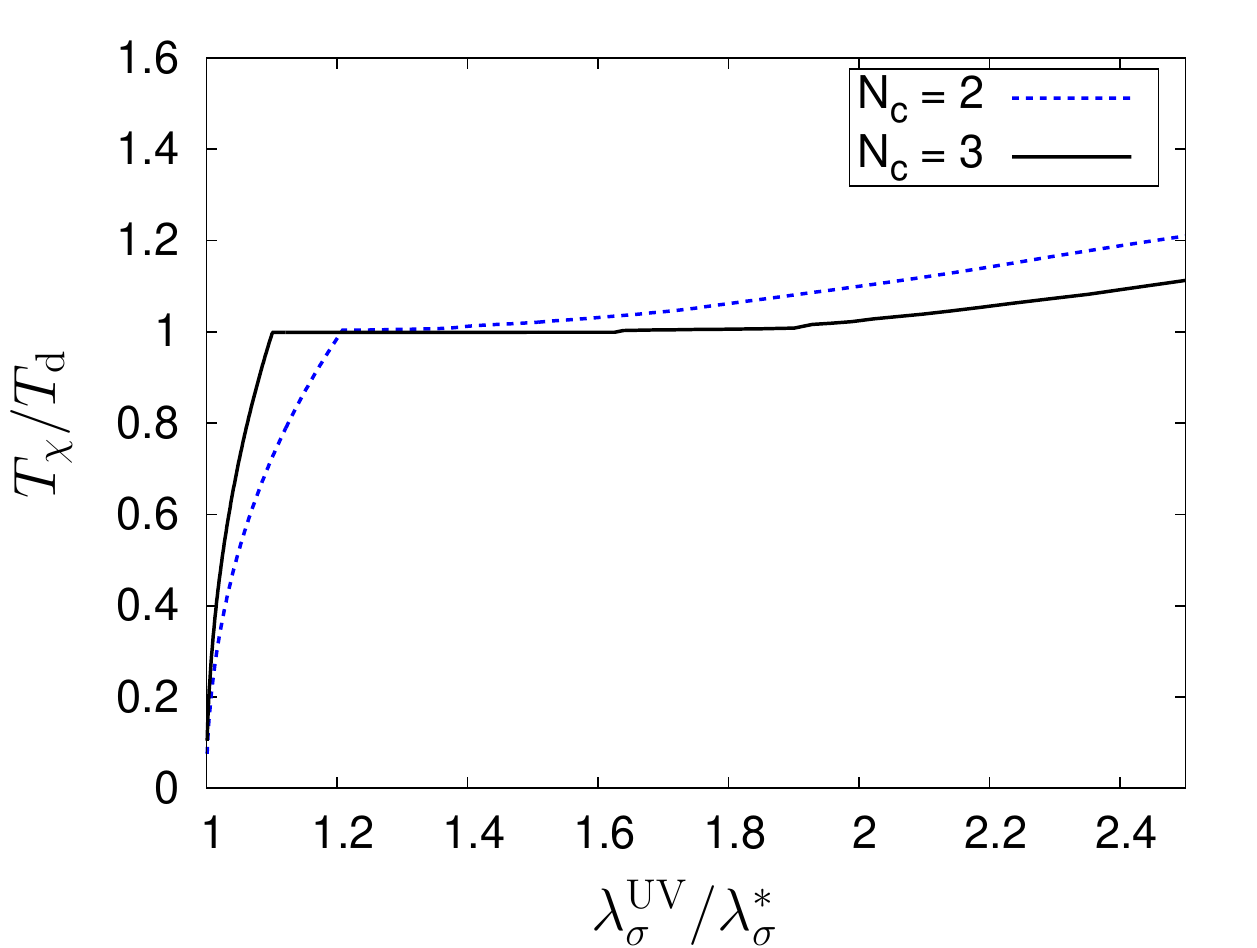}
\caption{Ratio~$T_{\chi}/T_{\rm d}$ of the chiral and the deconfinement phase transtion temperature as 
a function of~$\lambda^{\rm UV}_{\sigma}/\lambda_{\sigma}^{\ast}$ for~$N_{\rm c}=2,3$. In the large-$\Nc$ limit,
the lower end of the locking window~(i.~e. the regime with $T_{\chi}/T_{\rm d}=1$) is given by~$\lambda^{\rm UV}_{\sigma}/\lambda_{\sigma}^{\ast}=1$.
The figure has been taken from Ref.~\cite{Braun:2011fw}. 
}
\label{fig:TcTdlambda}
\end{figure*}

Let us now discuss how our (simplified) analysis relates to (full) QCD. In QCD, we only 
have a single input parameter, e.~g. the value of the strong
coupling~$g^2$ at a given scale which then determines~$\LQCD$. Thus, we have
\be
\Tc \sim \LQCD\,,\nn
\ee
see also Sect.~\ref{sec:fewflavor}. In the present analysis the chiral transition temperature~$\Tc$  depends on two 
parameters, namely the value of the background field $\bfe$ and the initial 
condition~$\lambda_{\sigma}^{\rm UV}$. Nevertheless, Eq.~\eqref{eq:TcTd} is a parameter-free statement which 
simply follows from an analysis of the effect of gauge dynamics on the fixed-point structure in the matter sector. In particular, we 
have only made use of general properties of the deconfinement order parameter and the fact that
$\lambda^{\rm UV}_{\sigma}>\lambda_{\sigma}^{\ast}(0,0)$ is a {\it necessary} condition for chiral symmetry breaking
at $T=0$ and $\bfe=0$.  Of course, the initial condition $\lambda_{\sigma}^{\rm UV}$ is not a free parameter in QCD but originally 
generated by quark-gluon interactions at high (momentum) scales. In a given
regularization scheme the value of $\lambda_{\sigma}^{\rm UV}$ can therefore in principle be related to the value of the 
strong coupling $g^2$ at, e.~g., the $\tau$ mass scale, see our discussion of chiral symmetry breaking in the previous
sections. We would like to point out that neither the value of~$g^2$ at some scale nor the value 
of~$\lambda_{\sigma}^{\rm UV}$ on a given RG trajectory is a physical observable. However, their values can be related to 
physical low-energy observables. Recall that the value of~$\lambda_{\sigma}^{\rm UV}$
determines the critical scale $\ksb$ which sets the scale for IR observables, see e.~g. Eq.~\eqref{eq:lambdacrOQM}.

Since our general statements do not depend on the actual value of the deconfinement temperature~$\Td$, one may wonder
which role this quantity plays at all in our analysis. We have argued that $\Tc$ depends on both~$\lambda_{\sigma}^{\rm UV}$ and~$\bfe$ in our study.
The presence of the background field~$\bfe$ implies the existence of the transition temperature~$\Td$ which defines a scale in the theory.
For~$\bfe=0$, on the other hand, we have argued that  $\Tc\sim \ksb$. In this case, the scale $\ksb$ is eventually determined by our choice
for $\lambda_{\sigma}^{\rm UV}$.
If we now take into account the background field~$\bfe$, then the chiral phase transition temperature~$\Tc$ is locked in and
we necessarily have $\Tc\geq \Td$ in the large-$\Nc$ limit, see Eq.~\eqref{eq:TcTd}. Thus, 
the chiral phase transition temperature for all theories which would allow for $\Tc \leq \Td$ for $\bfe=0$ is 
shifted such that $\Tc\simeq\Td$. 
This observation allows us to define a ``locking window" for the parameter~$\lambda_{\sigma}^{\rm UV}$
in which the chiral phase transition~$\Tc$ and the deconfinement phase transition~$\Td$ lie close to each other~\cite{Braun:2011fw}.
The upper end of this window can 
be estimated by the smallest value for $\lambda_{\sigma}^{\rm UV}$ for which $\Tc$ for $\bfe=0$ is still
larger than~$\Td$. The lower end of this window is given by~$\lambda_{\sigma}^{\ast}(0,0)$ in the large-$\Nc$ limit.
Whereas $\lambda_{\sigma}^{\ast}$ and $\lambda_{\sigma}^{\rm UV}$ are scheme-dependent quantities, the mere 
existence of such a window in parameter space can be viewed as a universal statement. 
Since~$\lambda_{\sigma}^{\rm UV}$ sets the scale for physical low-energy
observables, the existence of a ``locking window" for~$\lambda_{\sigma}^{\rm UV}$
suggests the existence of a corresponding window for the values of low-energy observables, such as the pion
decay constant~$f_{\pi}$. In Ref.~\cite{Braun:2011fw} 
the existence of the latter has indeed been confirmed. It was also found that
that the physical value for the pion decay constant~($f_{\pi}\approx 90\,\text{MeV}$) is compatible
with the almost coinciding phase transition temperatures observed in lattice QCD 
simulations~\cite{Cheng:2006qk,Aoki:2006br,Aoki:2006we,Aoki:2009sc,Cheng:2009zi,Borsanyi:2010zi,Bazavov:2010pg,Kanaya:2010qd} 
and in functional first-principles studies~\cite{Braun:2009gm,Fischer:2011mz}.

In our analysis we have concentrated on the limit $\Nc\to\infty$. One may suspect that finite-$\Nc$ corrections alter our conclusions. In fact,
we observe that terms with
\be
n\mod\Nc=0
\ee
contribute to the right-hand side of Eq.~\eqref{eq:lfpa0CONF} and to the RG flow of $\lambda_{\sigma}$, when we go beyond the
large-$\Nc$ limit. Strictly speaking, Eq.~\eqref{eq:fpequivCONF} then holds only for $\tau\ll 1$ but not 
for arbitrary values of~$\tau=T/k$. However, this does not necessarily imply that we do not have a finite range of values for the
initial condition~$\lambda_{\sigma}^{\rm UV}$ anymore in which the chiral and the deconfinement phase transition are tightly linked.
It only implies that the lower end of the window for $\lambda_{\sigma}^{\rm UV}$ is shifted to larger values
compared to the large-$\Nc$ limit where the lower end is given by~$\lambda_{\sigma}^{\ast}(0,0)$. 
In Ref.~\cite{Braun:2011fw}, finite-$\Nc$ corrections have been explicitly
taken into account. It was found that the locking mechanism for the chiral phase transition is still present for finite~$\Nc$, 
in particular for~$\Nc=2$ and~$\Nc=3$. In Fig.~\ref{fig:TcTdlambda} we show~$T_{\chi}/T_{\rm d}$ as 
a function of~$\lambda^{\rm UV}_{\sigma}/\lambda_{\sigma}^{\ast}$ for~$N_{\rm c}=2,3$. 
For the computation of~$T_{\chi}/T_{\rm d}$, the results from Ref.~\cite{Braun:2007bx}
for~$\langle A_0\rangle$ for the corresponding SU($\Nc$) Yang-Mills theory have been used to
solve the flow equation for~$\lambda_{\sigma}$, see Eq.~\eqref{eq:lpsi_flowCONF}.

We would like to close our discussion of confinement and chiral symmetry breaking 
with a few (critical) comments concerning the approximations underlying our present analysis.
First, it is clear that the confinement order parameter in full QCD receives contributions from Feynman diagrams
with at least one internal fermion line. These contributions tend to lower the deconfinement phase transition temperature.\footnote{Recall that
we only have a deconfinement crossover in the presence of dynamical quarks.}
We anticipate that our analytic findings are not (strongly) affected by this approximation since they rely on very general properties 
of the confinement order parameter. Therefore we still expect that a window in parameter space exists in which the chiral and the deconfinement 
phase transition lie close to each other. However, our estimate for the dependence of~$\Tc$ on~$\lambda_{\sigma}^{\rm UV}$
and the size of the ``locking window" will change quantitatively when we take into account
the corrections to the confinement oder parameter due to quark fluctuations. Second, our ansatz~\eqref{eq:fermionic_actionCONF} 
for the effective action is not complete with respect to Fierz transformations; for example, we have dropped the vector-channel 
interaction $\sim (\bar{\psi}\gamma_{\mu}\psi)^2$.
Such interactions would also contribute to the RG flow of the four-fermion interaction $\lambda_{\sigma}$. 
At finite temperature the minimal set of point-like four-fermion interactions is larger than at 
vanishing temperature, since the Poincare invariance is broken by the heat bath. If we allow for a finite $\bfe$, the minimal set 
is even larger than in the case of a vanishing background field~$\bfe$. This is due to the
fact that a finite background field $\bfe$ distinguishes a direction in color space. For example, 
our expansion~\eqref{eq:prop_expCONF} of the fermion propagator suggests that a finite background
field $\bfe$ gives rise to additional point-like interactions of the type~$\sim (\bar{\psi}T^{(3)}\psi)^2$
and~$\sim (\bar{\psi}T^{(8)}\psi)^2$ for $\Nc=3$. However, the additional diagrams are of the same
topology as the one shown in Fig.~\ref{fig:feynmanYM}~(a), if we drop gluon-induced terms. We therefore expect that the inclusion of additional 
four-fermion interactions associated with a Fierz-complete basis is important
for a quantitative computation of the chiral phase transition temperature beyond the large-$\Nc$ limit.
Such an analysis is beyond the scope of this review. 
A study of the impact of the confinement order-parameter on contributions
to the RG flow arising from $1$PI diagrams with one- or two internal gluon lines, see Fig.~\ref{fig:feynmanYM}~(b) and~(c), certainly constitutes a further
important improvement.
As discussed in Sect.~\ref{sec:PSFYM}, the inclusion of such terms in the RG flow of four-fermion interactions opens up the possibility 
to remove the parameter $\lambda_{\sigma}^{\rm UV}$, 
so that we are left with a single input parameter for the gauge and the matter sector, namely $g^2$ at a given UV scale.
In any case, the locking mechanism for the chiral and deconfinement phase transition discussed here already
provides a simple explanation for the observed almost-coincidence of both phase transitions.

\subsection{Fermions in Higher Representations and QED-like Theories}\label{sec:outlookYM}

In this section we briefly discuss the applicability of the present approach to non-abelian gauge theories with fermions in higher representations, 
e.~g. the adjoint representation,
and to QED-like theories in $d=2\!+\!1$ space-time dimensions. The latter class of theories is of great interest since the associated models may
serve as effective theories for graphene, see e.~g. Refs.~\cite{PhysRevLett.95.146801,PhysRevLett.96.256802,PhysRevB.79.165425}.
The general discussion of scaling behavior in gauge theories in Sect.~\ref{sec:QCGT} indeed also holds for QED-like theories in~$2<d<4$
space-time dimensions, as it essentially relies on only two assumptions (minimal requirements): 
\begin{itemize}
\item[(i)] The existence of a non-trivial IR fixed-point of the gauge coupling for a large number
of fermions flavors, i.~e. in the chirally symmetric regime.
\item[(ii)] The existence of an IR-attractive Gau\ss ian as well as an IR-repulsive non-trivial fixed-point in the RG flows 
of the four-fermion couplings in the limit of vanishing gauge 
coupling.\footnote{This assumption might be violated in~$d=2$, e.~g., if the anomalous dimension of the fermions is 
zero in the limit of vanishing gauge coupling.}
\end{itemize}
These assumptions are fulfilled for QED in $2<d<4$ space-time dimensions. 

The existence of a critical number $\Nfcr$ of fermion
flavors in~QED${}_3$ has been confirmed in several studies, see e.~g. Refs.~\cite{Pisarski:1984dj,Appelquist:1986fd,Appelquist:1988sr,Atkinson:1989fp,%
  Pennington:1990bx,Curtis:1992gm,Burden:1990mg,Maris:1995ns,Gusynin:1995bb,Maris:1996zg,Fischer:2004nq}. 
  However, the scaling behavior of physical observables as a function of~$\Nf$ close to the quantum critical point~$\Nfcr$ has not yet been analyzed
in great detail. Only little is known about the (precise) size of the regime in which Miransky (exponential) scaling is dominant. It might well be
that the critical exponent~$\Theta$ associated with the gauge coupling is large at the phase transition~(i.~e. $|\Theta(\Nfcr)|\gtrsim 1$). As a 
consequence, the scaling
of physical observables close to~$\Nfcr$ would be governed mainly by an exponential behavior, and the {\it universal} power-law 
corrections  associated with~$\Theta$ might be parametrically suppressed. An analysis of the presently available data for
QED${}_3$ in this  direction as well as  an independent RG study therefore seem to be worthwhile. Although~$\Nf$ does not correspond to
an experimentally accessible parameter of the theory, an analysis of the $\Nf$-scaling behavior might provide us with important insights into
the dynamics underlying chiral symmetry breaking in graphene.

A further fruitful extension of the discussed RG approach is represented by the study of gauge theories with fermions in 
higher representations, e.~g. fermions in the adjoint representation. To this end, the use of computer algebra systems might be
advisable~\cite{Huber:2011qr}.
Comprehension of these classes of gauge theories underlies (walking) technicolor-like scenarios for the Higgs sector 
and it is therefore important for our understanding of physics beyond the standard model.
The quantum phase transition, which occurs in such  
theories for large~$\Nf$, has been studied using both Dyson-Schwinger equations as well as lattice simulations, see e.~g. 
Refs.~\cite{Dietrich:2006cm,Catterall:2008qk,Sannino:2008pz,DelDebbio:2008zf,DelDebbio:2009fd,Sannino:2009aw,Patella:2010dj,Fukano:2010yv,Mojaza:2010cm}.
An RG study in this direction can be used to benchmark presently available results for the critical number of fermion flavors in these theories 
and therefore contribute to a better understanding of dynamical symmetry breaking in gauge theories.
Since $\Nfcr$ for a given number of colors is smaller in QCD with adjoint fermions than in QCD with fermions in the fundamental representation,
it is tempting to speculate whether the dynamics close to the quantum phase transition is 
strongly affected by the confining dynamics in the gauge sector. In this respect, an analysis of the interrelation of confining and chiral dynamics
along the lines of Sect.~\ref{sec:CSBCONF} could be rewarding.



%
\section{Summary}\label{sec:summary}

We have reviewed RG approaches to various strongly interacting fermionic theories, ranging from non-relativistic many-body problems to relativistic 
gauge theories. 
We have shown that an analysis of the fixed-point structure of such theories allows us to study {\it universal} long-range behavior associated
with quantum and thermal phase transitions in a clean and controlled way. 
Due to the intimate relation between phase transitions in a given theory and its fixed-point structure, evidence for the existence of 
fixed points can be verified in experiments. For example, we have shown in Sect.~\ref{sec:coldgases} that the existence of a
non-trivial (IR repulsive) fixed-point in the four-fermion coupling is tightly linked to the 
observed universal behavior in experiments with ultracold atomic Femi gases in the limit of a broad {\it Feshbach} resonance. 

The analogue of the fixed-point associated with universal behavior in ultracold Fermi gases also exists 
in the Gross-Neveu model in $2<d<4$ space-time dimensions. For example, Gross-Neveu-type models
play an important role in the context of (relativistic) superconductors. 
In any case, the existence of a non-trivial IR
repulsive fixed-point in the Gross-Neveu model in $2<d<4$ space-time dimensions
is associated with a second-order quantum phase transition, see Sect.~\ref{sec:GNmodel}. 
We have argued that theories
with such a quantum phase transition are guaranteed to be asymptotically safe, i.~e. 
nonperturbatively renormalizable. It is still an open question whether the converse is also true.
An answer to this question might provide us with further insights into the
asymptotic-safety scenario which underlies the quantization of gravity within a conventional path-integral approach.

In QCD low-energy models, we have seen that a non-trivial IR-repulsive fixed-point for the four-fermion couplings exists as well, see Sect.~\ref{sec:NJLLQCD}.
In these models, the fermions play the role of (constituent) quarks. The mass of these fermions can be tuned by varying the four-fermion couplings.
We have analyzed Fierz ambiguities in these models and shown that the dynamics close to the finite-temperature phase 
boundary can be easily understood in terms of the fixed-point structure of the theory.

In QCD-like gauge theories, the fermionic interactions do not represent free parameters. On the contrary, 
the running gauge coupling can drive the fermion sector to criticality, resulting in chiral symmetry breaking without
any fine-tuning of the fermionic couplings, see Sect.~\ref{sec:gaugetheories}. 
This is generically true for 
asymptotically free (chiral) gauge theories with $N_{\rm f}$ (massless) 
fermion flavors, such as QCD or even effective theories for graphene. 
A detailed analysis of the fixed-point structure in this class of theories provides a quantitative 
determination of the quantum phase transition that occurs for large values of~$\Nf$.

In Sects.~\ref{sec:QCGT} and~\ref{sec:SUNYM} we have reviewed our understanding of phases of strongly-flavored
gauge theories and the scaling behavior of their mass spectrum close to a quantum phase transition. We have discussed that
essentially three different types of scaling behavior can occur close to such a phase transition: 
power-law behavior, exponential behavior, or a combination thereof. In the first and the third case, 
the scaling behavior in leading order is governed by a {\it universal} critical exponent which is determined by the 
symmetries and the dimensionality of the theory. Interestingly, this critical exponent also determines the infrared dynamics
of the gauge sector. Therefore an analysis of the $\Nf$-scaling behavior of observables in asymptotically free gauge theories
allows us to probe the infrared gauge dynamics. 

In summary, our review of universal aspects of various strongly interacting theories shows that 
a transfer of knowledge between studies of strongly-interacting hadronic matter and (non-relativistic) many-body problems is important and 
inspiring in order to gain a deeper insight into the dynamics which underlie the equation of state of 
hadronic matter as well as the generation of bound states in many-body problems.


\begin{acknowledgement}
The author is deeply indebted to H.~Gies, B.~Klein and J.~M.~Pawlowski 
for uncountable discussions, encouragement and collaboration throughout the years.  
Moreover, the author is very grateful to~C.~S.~Fischer, H.~Gies, A.~Janot, J.~M.~Pawlowski, J.~Polonyi, D.~D.~Scherer 
and~A.~Schwenk for very enjoyable and fruitful collaborations on the topics presented here. 
It is also a great pleasure for the author to thank~S.~Diehl,  D.~D.~Dietrich, 
J.~E.~Drut, A.~Eichhorn, R.~J.~Furnstahl, T.~K.~Herbst, L.~Janssen, M.~Ku, M.~P.~Lombardo, H.-J.~Pirner,
F.~Sannino, B.-J.~Schaefer, M.~M.~Scherer and A.~Wipf for various discussions which have left their traces in this manuscript.
For critical comments on the manuscript, the author is very grateful to H.~Gies and B.~Klein.
The author acknowledges support by DFG grant BR~4005/2-1 and the DFG research training group GRK~1523/1.
\end{acknowledgement}

\appendix


%
\section{Conventions}\label{sec:conv}
\subsection{Units}\label{app:units}

In our studies of {\it relativistic} quantum field theories set $\hbar=c=k_{\rm B}=1$. As a consequence of this convention, 
the SI units for length (meter, m) and temperature (Kelvin, K) are related to the energy unit~$\text{MeV}$ as follows
\be
1\;\text{m}=10^{15}\;\text{fm}\approx 5.1\times 10^{12}\;\frac{1}{\text{MeV}}\qquad\text{and}\qquad
1\;\text{K}\approx 8.6\times 10^{-11}\;\text{MeV}\,.\nn
\ee

If not indicated otherwise, we set $\hbar=k_{\rm B}=2m=1$ in our studies of {\it non-relativistic} quantum field theories, 
where $m$ is the mass (parameter) of the fermions. As for relativistic quantum field theories, length and inverse momenta
then have the same dimension, i.~e.
\be
\rm [length]=[momentum]^{-1}\,.\nn
\ee
Moreover, temperature and energy have the same dimension. From our choice $2m=1$, it follows that the dimensions of energy 
and squared momenta are identical. Thus, we have
\be
\rm [temperature]=[energy]=2\times [momentum]\,.\nn
\ee
Finally is worth mentioning that in our conventions relativistic and non-relativistic spinors have the same
mass dimension:
\be
[\psi]_{\rm rel.}=[\psi]_{\rm non-rel.}=\frac{d}{2}\times {\rm [momentum]}\,,\nn
\ee
where $d$ denotes the number of space dimensions.
Note that this is not true for the scalar fields~$\phi$. In this case, we have
\be
[\phi]_{\rm rel.}=\frac{d-1}{2}\times {\rm [momentum]}\,,\quad\text{and}\quad 
[\phi]_{\rm non-rel.}=[\psi]_{\rm non-rel.}=\frac{d}{2}\times {\rm [momentum]}\,.\nn
\ee
\subsection{Minkowski- and Euclidean Space-Time}
The coordinates in $d$-dimensional Euclidean space-time and Minkowski (M) space-time are related by
\be
&& \qquad\qquad\qquad\quad x_{{\rm M},0}= -\I x_{0}\,,\quad x_{{\rm M},i}=x_i\,, \nn \\
&& g ^{\mu\nu}_{\rm M} x_{{\rm M},\mu}x_{{\rm M},\nu}=
x_{{\rm M},0}^{2}-\vec{x}^{\,2}_{{\rm M}}  = - x_{0}^{2} - \vec{x}^{\,2} =  - g^{\mu\nu} x_{\mu}x_{\nu}=-x^{2}\,, \nn
\ee
where $\mu,\nu=0,\dots,d\!-\! 1$ and correspondingly for the momenta. The metric tensor in Euclidean space-time is given by the Kronecker-Delta, 
$g^{\mu\nu}=\delta ^{\mu\nu}$, whereas we have the metric tensor $g^{\mu\nu}_{\rm M}=\text{diag}(+,-,-,\dots,-)$ in 
Minkowski space-time. 

\subsection{Fourier Transformation}

Our conventions for Fourier transformations in $d$-dimensional Euclidean space are summarized. For fermion fields we employ
\be
\psi(x)&=& \int \frac{d^d p}{(2\pi)^d}\psi(p)\, \E^{\I p_{\mu}x_{\mu}}\,,\\
\bar{\psi}(x)&=& \int \frac{d^d p}{(2\pi)^d}\bar{\psi}(p)\, \E^{-\I p_{\mu}x_{\mu}}\,.
\ee
For bosonic fields we use
\be
\phi(x)&=& \int \frac{d^d p}{(2\pi)^d}\phi(p)\, \E^{\I p_{\mu}x_{\mu}}\,.
\ee
Our conventions for the Fourier transformation of the fields imply that
\be
\int d^d x\, \E^{-i p_{\mu}x_{\mu}} =(2\pi)^d \delta^{(d)}(p)\,.
\ee

\section{Dirac Algebra}\label{sec:dirac}
\subsection{Clifford Algebra in $d=4$ (Euclidean) Space-Time Dimensions}\label{subsec:CA}

We work exclusively in Euclidean space-time in this work and restrict our quantitative discussions 
to~$d=3$ and~$d=4$ space-time dimensions. The Dirac algebra is then defined through
\be
\label{eq:diracalgebra}
\left\{\gamma_{\mu},\gamma_{\nu}\right\}&=&\gamma_{\mu}\gamma_{\nu}+\gamma_{\nu}\gamma_{\mu}
=2\delta_{\mu\nu}\mathbbm{1}\,, \label{eqapp:gammaantcom}
\\
\left(\gamma_{\mu}\right)^{\dagger}&=&\gamma_{\mu}\,,
\\
\gamma_{5}&=&\gamma_{1}\gamma_{2}\gamma_{3}\gamma_{0}\,,
\\
\sigma_{\mu\nu}&=&\frac{i}{2}[\gamma_{\mu},\gamma_{\nu}]
=\frac{i}{2}(\gamma_{\mu}\gamma_{\nu}-\gamma_{\nu}\gamma_{\mu})\,.
\ee

\subsection{Clifford Algebra in $d=3$ (Euclidean) Space-Time Dimensions}\label{subsec:CAd3}
For our studies of quantum field theories in $d=3$ Euclidean space-time dimensions, we employ a four-component
representation for the $\gamma$-matrices. The explicit representation of our choice for the $4\times
4$ representation of the Dirac algebra can be written as
\be
\gamma_0=\tau_3\otimes\tau_3\,,\quad
\gamma_1=\tau_3\otimes\tau_1\,,\quad\gamma_2=\tau_3\otimes\tau_2\,.  
\ee
Here, the $\tau_i$ denote the Pauli matrices which satisfy $\tau_i
\tau_j = \delta_{ij}\tau_0 + \mathrm{i}\epsilon_{ijk}\tau_k$, with
$i,j,k=1,2,3$ and $\tau_0=\mathbbm{1}_2$ is a $2\times2$ unit matrix. The
$\gamma$-matrices satisfy the anticommutation relation given in Eq.~\eqref{eqapp:gammaantcom}
Moreover, we have two additional $4\times4$ matrices which anticommute with all
$\gamma_{\mu}$ and with each other: 
\be
\gamma_3=-\tau_{1}\otimes\tau_{0}\,,\quad\gamma_5=\tau_2\otimes\tau_0\,,\quad\gamma_{3}^{\,2}=\gamma_{5}^{\,2}=\mathbbm{1}\,.
\ee
On the other hand, the matrix $\gamma_{35}\equiv\mathrm{i}\gamma_{3}\gamma_{5}$ commutes with $\gamma_{\mu}$ 
and anticommutes with $\gamma_{3}$ and $\gamma_{5}$.

\subsection{Fierz Transformations}
The Clifford Algebra defined in Sect.~\ref{subsec:CA} is spanned by $16$ basis elements~$\gamma^{(A)}$:
\be
\{\gamma^{(A)}\}
:=\{\mathbbm{1}, \gamma_0,\gamma_1,\gamma_2,\gamma_3,\sigma_{03},\sigma_{13},\sigma_{23},\sigma_{01},\sigma_{12},\sigma_{20},
\I\gamma_0\gamma_5,\I\gamma_1\gamma_5,\I\gamma_2\gamma_5,\I\gamma_3\gamma_5,\gamma_5
\}\,,
\ee
which obey
\be
\tr\left\{ \gamma ^{(A)} \gamma^{(B)}\right\}=4\,\delta_{AB}\,.
\ee
This basis is complete:
\be
\frac{1}{4}\sum_A \gamma_{ad}^{(A)}\gamma^{(A)}_{ef}=\delta_{af}\delta_{ed}\,. 
\ee
From the completeness relation it is straightforward to expand two matrices, e.~g. $M^{(1)}_{ab} M^{(2)}_{cd}$,
in terms of the basis elements $\gamma^{(A)}$: 
\be
M^{(1)}_{ab} M^{(2)}_{cd} = \frac{1}{4} 
\sum_{A}\gamma^{(A)}_{ad}\, \sum_{e,f} (M^{(2)}_{ce} \gamma^{(A)}_{ef} M^{(1)}_{fb} )\,.
\ee
This expression corresponds to the general expression given in Eq.~\eqref{eq:genFierz}. Defining 
\be
O_{\text{S}}=\mathbbm{1} \,,\quad O_{\text{V}}=\gamma _{\mu}\,,\quad O_{\text{T}}=\frac{1}{\sqrt{2}}\sigma _{\mu\nu}\,,\quad 
O_{\text{A}}=\gamma _{\mu}\gamma _{5}\,\quad\text{and}\quad O_{\text{P}}=\gamma _5\,,
\ee
we then obtain the following Fierz identities:
\be
\label{eq:fierzIDapp}
(\bar{\psi}_{a}O_{\text{X}}\psi_{b})(\bar{\psi}_{c}O_{\text{X}}\psi_{d})
=\sum_{\text{Y}}C_{\text{XY}}
(\bar{\psi}_{a}O_{\text{Y}}\psi_{d})(\bar{\psi}_{c}O_{\text{Y}}\psi_{b}),
\ee
where $\text{X,Y = S,V,T,A,P}$ and
\be
\label{eqfierzMAT}
C_{\text{XY}}=\frac{1}{4}\left(
\begin{array}{ccccc}
-1  & -1 & -1 & 1 & -1 \\
-4   & 2 & 0 & 2  & 4  \\
-6   & 0 & 2 & 0 & -6 \\
4  &  2 & 0 & 2 & -4 \\
-1   & 1  & -1 & -1 & -1 \\
\end{array}
\right).
\ee
In Sect.~\ref{sec:example}, we study a NJL~model with one fermion species at zero and at finite temperature.
In this special case, the combination 
\be
(\bar{\psi}O_{\text{V}}\psi)^2+(\bar{\psi}O_{\text{A}}\psi)^2\nn
\ee
is invariant under Fierz transformations. Due to the relation
\be
(\bar{\psi}O_{\text{V}}\psi)^2-(\bar{\psi}O_{\text{A}}\psi)^2
+2[(\bar{\psi}O_{\text{S}}\psi)^2-(\bar{\psi}O_{\text{P}}\psi)]=0\,,
\ee
we can transform the combination
\be
(\bar{\psi}O_{\text{V}}\psi)^2-(\bar{\psi}O_{\text{A}}\psi)^2\nn
\ee
completely into scalar and pseudo-scalar channels.

\section{SU($N$) Algebra}
In this appendix we give our conventions for the generators of the SU($N$) Lie-groups. The 
group~SU($N$) of unitary matrices $U$ of rank~$N$ with determinant $\det U =1$ 
has $N^2-1$ generators $T^{a}$ which obey the commutation relations
\be
\left[T^{a},T^{b}\right]=\I f^{abc} T^{c}\,,
\ee
where $f^{abc}$ are the (anti-symmetric) structure constants of the group, and $a,b,c$ take 
the values $1,\dots,N^2 -1$. The normalization of the generators is given by
 \be
\Tr\left\{T^a T^b\right\}=\frac{1}{2}\delta^{ab}\,.
\ee
Moreover, the generators fulfill the following (Fierz) identities:
\be
\sum _{a} (T^a)_{\alpha\beta}(T^a)_{\gamma\delta}=\frac{1}{2}\delta _{\alpha\delta} \delta_{\beta\gamma} 
-\frac{1}{2N}\delta _{\alpha\beta} \delta_{\gamma\delta} 
\ee
and
\be
\sum _{a} \left\{(T^a)_{\alpha\beta}(T^a)_{\gamma\delta}+\frac{1}{N} (T^a)_{\alpha\delta}(T^a)_{\beta\gamma}\right\}
=\frac{N^2 -1}{2 N^2}\delta _{\alpha\delta} \delta_{\beta\gamma}\,.
\ee
For SU($2$), the generators are related to the Pauli matrices $\tau ^a$ via $T^a =\frac{1}{2}\tau ^a$ 
and the structure constants $f^{abc}$ are given by the (standard) totally antisymmetric tensor~$\epsilon ^{abc}$.
The generators for the group $SU(3)$ can be expressed in terms of the Gell-Mann matrices $\lambda ^a$ 
via~$T^a =\frac{1}{2}\lambda ^a$.

\section{Regulator Functions and Threshold Functions}\label{app:regthres}
\subsection{Regulator Functions}
In the computation of the RG flow equations a regulator function needs to be specified which determines the regularization scheme.
For explicit calculations we employ {\it optimized} regulator functions at zero and at finite 
temperature~\cite{Litim:2000ci,Litim:2001fd,Litim:2001up,Litim:2006ag,Blaizot:2006rj}. 

If not indicated otherwise, we employ the following so-called spatial regulator functions whenever we study a {\it relativistic} 
theory at zero {\it and} at finite temperature. To be specific, we choose
\be
R_{\rm B}(\vec{p}^{\,2})=\vec{p}^{\,2}\left(\frac{k^2}{\vec{p}^{\,2}}\!-\!1\right)\theta(k^2\!-\!\vec{p}^{\,2})
\equiv \vec{p}^{\,2} r_{\rm B} \left({ \frac{\vec{p}^{\,2}}{k^2}}\right)\,
\label{eq:bosreg}
\ee
for the bosonic degrees of freedom, whereas we choose
\be
R_{\psi}(\vec{p})=-\fslash{\vec{p}}\left(\sqrt{\frac{k^2}{\vec{p}^{\,2}}}\!-\!1\right)\theta(k^2\!-\!\vec{p}^{\,2})
\equiv -\fslash{\vec{p}}\, r_{\psi} \left({ \frac{\vec{p}^{\,2}}{k^2}}\right)
\label{eq:fermreg}
\ee
for the fermionic degrees of freedom. In many cases these regulator functions open up the possibility 
to perform analytically the Matsubara sums as well as the momentum integrals appearing in the 1PI diagrams.

For cases in which we consider a theory only at vanishing temperature, we
use the following so-called covariant regulator functions for the bosons and fermions, respectively:
\be
R_{\rm B}(p^{\,2})&=& p^{\,2}\left(\frac{k^2}{p^{\,2}}\!-\!1\right)\theta(k^2\!-\! p^{\,2})
\equiv p^{\,2} r_{\rm B} \left({ \frac{p^{\,2}}{k^2}}\right)\,,\label{eq:bosreg2}\\
R_{\psi}(p)&=& -\fslash{p}\left(\sqrt{\frac{k^2}{p^{\,2}}}\!-\!1\right)\theta(k^2\!-\! p^{\,2})
\equiv \fslash{p}\, r_{\psi} \left({ \frac{p^{\,2}}{k^2}}\right)\,.
\label{eq:fermreg2}
\ee

For our studies of {\it non-relativistic} fermionic many-body problems we use~\cite{Diehl:2009ma} 
\be
R_k ^{\psi}(\vec{p}^2)=k^2 \, r_{\psi}({\mathcal Z})\,\quad\text{with}\quad {\mathcal Z}=(\vec{p}^{\,2}-\mu)/k^2\,,\label{eq:nr_cutoff}
\ee
where
\be
r_{\psi}({\mathcal Z})=(\text{sign}({\mathcal Z})-{\mathcal Z})\theta(1-|{\mathcal Z}|)\,\label{eq:nr_cutoff_shape}
\ee
and $\vec{p}^{\,2}$ denotes the square of the spatial momentum. 

In the next two sections we list the threshold functions which appear in the RG flow equations. These functions represent the $1$PI diagrams 
contributing to the RG flow of the studied couplings. Note that we have adopted the conventions introduced in 
Refs.~\cite{Jungnickel:1995fp}. The list of threshold functions in the two subsequent sections is not exhaustive. We only list
those functions which have been employed explicitly in this review.

\subsection{Threshold Functions for Covariant Regulators}\label{app:covregfcts}
In this section we give the general definition of the threshold functions as obtained for covariant regulators. 
For (explicit) evaluation of the momentum integrations we have used the (optimized) regulator
functions defined in Eqs.~\eqref{eq:bosreg2} and~\eqref{eq:fermreg2}. In order to define the threshold functions, it is convenient to define 
dimensionless propagators for the bosons~(B) and the fermions~($\psi$), respectively:
\be
\tilde{G} _{\rm B} (\omega)=\frac{1}{  x(1+r_{\rm B}) + \omega}
\ee 
and
\be
\tilde{G} _{\psi} (\omega)=\frac{1}{  x(1+r_{\psi})^2 + \omega}\,,
\ee
where~$x=p^2/k^2$.

The threshold functions representing purely bosonic 1PI diagrams in the flow equations of 
bosonic self-interactions are given by
\be
l_0 ^{\rm (d)} (\omega;\eta_{\rm B})&=& \frac{1}{2}\int _0 ^{\infty} dx\, x^{\frac{d}{2}}
(\partial _t r_{\rm B} - \eta_{\rm B} r_{\rm B})\,\tilde{G} _{\rm B} (\omega)\nn\\
&=&\frac{2}{d}\left(1-\frac{\eta_{\rm B}}{d+2} \right)\frac{1}{1+\omega}\,,
\label{eq:l0_BosLoopDefSR}
\ee
where $\eta_{\rm B}\equiv -\partial_t \ln Z_{\rm B}$. Bosonic threshold functions of order $n$ can then be 
obtained from Eq.~\eqref{eq:l0_BosLoopDefSR} by taking
derivatives with respect to the dimensionless mass parameter~$\omega$:
\be
\frac{\partial}{\partial \omega} l_n^{\rm (d)} (\omega;\eta_B) 
= -(n + \delta_{n,0})\, l_{n+1}^{\rm (d)}  (\omega;\eta_B)\,.
\ee

The threshold functions representing purely fermionic 1PI diagrams in the flow equations of 
bosonic self-interactions as well as fermionic self-interactions are given by
\be
l_0 ^{(\rm F),(d)} (\omega;\eta_{\psi})&=& \int _0 ^{\infty} dx\, x^{\frac{d}{2}} 
(\partial _t r_{\psi} - \eta_{\psi}r_{\psi})(1 + r_{\psi})
\tilde{G} _{\psi} (\omega)\nn\\
&=& \frac{2}{d}\left( 1 - \frac{\eta_{\psi}}{d+1}\right)\frac{1}{1 +\omega}\,,
\label{eq:l0_FerLoopDefSR}
\ee
where~$\eta_{\psi} = -\partial_t \ln Z_{\psi}$\,. Again, higher-order fermionic threshold functions can be found by taking derivatives 
with respect to the dimensionless mass parameter~$\omega$:
\be
\frac{\partial}{\partial \omega} l_n ^{\rm (F),(d)} (\omega;\eta_{\psi}) 
= -(n + \delta_{n,0})\, l_{n+1}^{\rm (F),(d)} (\omega;\eta_{\psi})\,.\label{eq:l0_FerLoopN}
\ee

Let us now turn to the threshold functions representing mixed boson-fermion diagrams. For example, these functions
enter the flow equations of Yukawa couplings and flow equations of four-fermion couplings. We have
\be
l_{1,1} ^{\rm (FB),(d)} (\omega_{\psi},\omega_{\rm B};\eta_{\psi},\eta_{\rm B})=-\frac{1}{2}
\int _0 ^{\infty} dx\, x^{\frac{d-2}{2}}\tilde{\partial}_t \,\left[\tilde{G} _{\psi} (\omega_{\psi})\tilde{G} _{\rm B} (\omega_{\rm B})\right]\,.
\ee
In order to evaluate the integral over~$x$, we use\footnote{Here, we only give the explicit expressions for the formal derivatives for the 
(optimized) regulator functions given in Eqs.~\eqref{eq:bosreg2} and~\eqref{eq:fermreg2}.} 
\be
\tilde{\partial}_t \Big|_{\psi}&=&\left( \frac{1}{x^{1/2}} - \eta_{\psi}\left(\frac{1}{x^{1/2}} - 1\right)\right)\theta(1-x)\frac{\partial}{\partial r_{\psi}}\,,\\
\tilde{\partial}_t \Big|_{\rm B}&=&\left( \frac{2}{x} - \eta_{\rm B}\left(\frac{1}{x} - 1\right)\right)\theta(1-x)\frac{\partial}{\partial r_{\rm B}}\,,
\ee
where the first and the second line states how the formal derivative $\tilde{\partial} _t$ acts on the fermion and boson propagator, respectively.
This yields
\be
&&l_{1,1} ^{\rm (FB),(d)} (\omega_{\psi},\omega_{\rm B};\eta_{\psi},\eta_{\rm B})\nn\\
&& \quad\; =\frac{2}{d}\frac{1}{(1\!+\!\omega_{\psi})(1\!+\!\omega_{\rm B})} 
\left\{\left(1\!-\!\frac{\eta_{\psi}}{d\!+\! 1} \right)\frac{1}{1\!+\!\omega_{\psi}} + \left(1\!-\!\frac{\eta_{\rm B}}{d\!+\!2} \right)\frac{1}{1\!+\!\omega_{\rm B}}
\right\}.
\ee
The threshold function~$l_{1,2} ^{\rm (FB),(d)}$ is defined as follows:
\be
l_{1,2} ^{\rm (FB),(d)} (\omega_{\psi},\omega_{\rm B};\eta_{\psi},\eta_{\rm B})&=& -\frac{1}{2}
\int _0 ^{\infty} dx\, x^{\frac{d-2}{2}}\tilde{\partial}_t \,\left[\tilde{G} _{\psi} (\omega_{\psi})\left(\tilde{G} _{\rm B} (\omega_{\rm B})\right)^2\right]\nn\\
&=& \frac{2}{d}\frac{1}{(1\!+\!\omega_{\psi})(1\!+\!\omega_{\rm B})^{2}}
\left\{\frac{1}{1\!+\!\omega_{\psi}}
\left(1\!-\!\frac{\eta_\psi}{d\!+\! 1}\right)\right. \nn\\
&& \qquad \qquad\qquad\qquad\quad\qquad\;\quad \left.
+ \frac{2}{1\!+\!\omega_{\rm B}}
\left(1\!-\!\frac{\eta_{\rm B}}{d\!+\! 2}\right)\right\}.
\ee

The threshold functions entering the RG flow equations of the wave-function renormalizations read
\be
m_{1,2} ^{{\rm (FB),(d)}} (\omega_{\psi},\omega_{\rm B};\eta_{\psi},\eta_{\rm B}) 
&=&\frac{1}{2}
\int _0 ^{\infty} \!dx  x^{\frac{d}{2}}\tilde{\partial}_t \,
\Big[(1+r_{\psi})\tilde{G} _{\psi} (\omega_{\psi})\frac{d}{dx}\tilde{G} _{\rm B} (\omega_{\rm B})\Big] \nn\\
&=& \left(1-\frac{\eta_{\rm B}}{d+1}\right)\frac{1}{(1+\omega_{\psi})(1+\omega_{\rm B})^2}\,
\ee
and
\be
m_{4} ^{\rm (F),(d)}(\omega;\eta_{\psi})
&=& -\frac{1}{2} \int _0 ^{\infty} \!dx x^{\frac{d+2}{2}}\tilde{\partial}_t
\left[ \frac{d}{dx} (1+r_{\psi})\tilde{G} _{\psi} (\omega)\right]^2 \nn\\
&=&\frac{1}{(1\!+\!\omega)^4}+\frac{1\!-\!\eta_{\psi}}{d\!-\! 2}\frac{1}{(1\!+\!\omega)^3}
\!-\!\left(\frac{1\!-\!\eta_{\psi}}{2d\!-\! 4} +\frac{1}{4}\right)\frac{1}{(1\!+\!\omega)^2}\,.
\ee

Finally, we define the threshold functions which appear in the flow equations of fermionic theories with an explicitly 
broken chiral symmetry:
\be
\tilde{l}_1^{\rm (F),(d)}(\omega;\eta_{\psi})&=& -\frac{\omega}{2} \int _0 ^{\infty} \!dx x^{\frac{d-2}{2}}\tilde{\partial}_t
\left(\tilde{G} _{\psi} (\omega)\right)^2
\nn\\
&\stackrel{(\eta_{\psi}=0)}{=}&
\frac{4}{d}\frac{\omega}{(1+\omega)^3}
\,
\ee
and
\be
\hat{l}_1^{\rm (F),(d)}(\omega;\eta_{\psi})&=& -\frac{1}{2} \int _0 ^{\infty} \!dx x^{\frac{d}{2}}\tilde{\partial}_t
\left[
(1+r_{\psi})^2\left(\tilde{G} _{\psi} (\omega)\right)^2
\right]\nn\\
&\stackrel{(\eta_{\psi}=0)}{=}&
 \frac{2}{d}\frac{1-\omega}{(1+\omega)^3}\,.
\ee
Note that
\be
{l}_1^{\rm (F),(d)}(\omega;\eta_{\psi})&=& \hat{l}_1^{\rm (F),(d)}(\omega;\eta_{\psi}) + \tilde{l}_1^{\rm (F),(d)}(\omega;\eta_{\psi})\,,
\ee
since
\be
\tilde{\partial}_t \Big|_{\psi}=(\partial_t r_{\psi}-\eta_{\psi}r_{\psi})\frac{\partial}{\partial r_{\psi}}\,.
\ee
Moreover, we have
\be
{b}_1^{\rm (F),(d)}(\omega;\eta_{\psi})&=& -\frac{\omega}{2}\int _0 ^{\infty} \!dx x^{\frac{d-2}{2}}\tilde{\partial}_t
\tilde{G} _{\psi} (\omega)
\nn\\
&\stackrel{(\eta_{\psi}=0)}{=}&
\frac{2}{d}\frac{\omega}{(1+\omega)^2}
\,.
\ee
\subsection{Threshold Functions for Dimensionally Reduced Regulators}
Spatial (or thermal) regulator functions are mostly applied in the context of finite-temperature studies.
However, their application is not limited to these kind of investigations. Below we give the general definition of the resulting threshold 
functions employed in this review. For the (explicit) evaluation of the momentum integrations we have used the regulator 
functions given in Eqs.~\eqref{eq:bosreg} and~\eqref{eq:fermreg}.

In order to define the threshold functions, it is convenient to define dimensionless propagators for the bosons~(B) and 
the fermions~($\psi$), respectively:
\be
\tilde{G} _{\rm B} (x_0,\omega)=\frac{1}{ \hat{z}_{\rm B} x_0 + x(1+r_{\rm B}) + \omega}
\ee 
and
\be
\tilde{G} _{\psi} (x_0,\omega)=\frac{1}{ \hat{z}_{\phi}^2 x_0 + x(1+r_{\psi})^2 + \omega}\,,
\ee
where~$x=\vec{p}^{\,2}/k^2$. Here, we have
dressed the ratio of the wave-function renormalizations longitudinal and transversal to the heat-bath,~$\hat{z}_{\rm B} =Z_{\rm B}^{\|}/Z_{\rm B}^{\perp}$
and~$\hat{z}_{\psi} =Z_{\psi}^{\|}/Z_{\psi}^{\perp}$.

First, we define the threshold functions which appear in the RG flow equations for the
bosonic self-interactions. For the purely bosonic loops, we find
\be
l_0 ^{\rm (d)} (\tau,\omega;\eta_{\rm B},\hat{z}_{\rm B})&=&\frac{\tau}{2}\sum_{n=-\infty}^{\infty}\int _0 ^{\infty} dx\, x^{\frac{d-1}{2}} 
(\partial _t r_{\rm B} - \eta_{\rm B} r_{\rm B})\,\tilde{G} _{\rm B} (\tilde{\omega}_n^2,\omega)\nn\\
&=&\frac{2}{d-1}\frac{1}{\sqrt{1+\omega}}\left(1-\frac{\eta_{\rm B}}{d+1} \right)
\left(\frac{1}{2} + \bar{n}_{\rm B}(\tau,\omega) \right)\,,\label{eq:l0_BosLoopDef}
\ee
where $\eta_{\rm B}\equiv -\partial_t \ln Z_{\rm B}^{\perp}$, $\tau=T/k$ denotes the dimensionless 
temperature and $\tilde{\omega}=2\pi n\tau$ denotes the dimensionless bosonic Matsubara frequencies. 
The function $n_{\rm B}$ represents the Bose-Einstein distribution function
\be
\bar{n}_{\rm B}(\tau,\omega)=\frac{1}{\E ^{\sqrt{1+\omega}/\tau} -1}\,.
\ee
Bosonic threshold functions of order $n$ can then be obtained from Eq.~\eqref{eq:l0_BosLoopDef} by taking
derivatives with respect to the dimensionless mass parameter $\omega$:
\be
\frac{\partial}{\partial \omega} l_n^{\rm (d)} (\tau,\omega;\eta_B,\hat{z}_{\rm B}) 
= -(n + \delta_{n,0})\, l_{n+1}^{\rm (d)}  (\tau,\omega;\eta_B,\hat{z}_{\rm B})\,.
\ee

For the purely fermionic loops contributing to the flow equations of the bosonic self-interactions but also
to the RG flow of the four-fermion coupling, we find
\be
l_0 ^{(\rm F),(d)} (\tau,\omega,\mu;\eta_{\psi},\hat{z}_{\psi})&=&\tau\sum_{n=-\infty}^{\infty}\int _0 ^{\infty} dx\, x^{\frac{d-1}{2}} 
(\partial _t r_{\psi} \!-\!\eta_{\psi}r_{\psi})(1\!+\! r_{\psi})
\tilde{G} _{\psi} ((\tilde{\nu}_n+ 2\pi\tau\mu)^2,\omega)\nn\\
&=&\!\frac{1}{d\!-\! 1}\frac{1}{\sqrt{1\!+\!\omega}}\left( 1 \!-\!\frac{\eta_{\psi}}{d}\right)\!
\left(1 \!-\! \bar{n}_{\psi} (\tau,{\rm i}\mu,\omega) \!-\! \bar{n}_{\psi} (\tau,\!-{\rm i}\mu,\omega) \right).
\label{eq:l0_FerLoopDef}
\ee
Here, we have introduced the dimensionless fermionic Matsubara frequencies $\tilde{\nu}_n=(2n+1)\pi\tau$
and~$\eta_{\psi}\equiv -\partial_t \ln Z_{\psi}^{\perp}$.
The function $\bar{n}_{\psi}$ denotes the Fermi-Dirac distribution function:
\be
\bar{n}_{\psi}(\tau,\mu,\omega)=\frac{1}{\E ^{(\sqrt{1+\omega}/\tau) + 2\pi\mu} +1}\,.
\ee
Higher-order fermionic threshold functions can again be found by taking derivatives with respect to the dimensionless
mass parameter~$\omega$:
\be
\frac{\partial}{\partial \omega} l_n ^{\rm (F),(d)} (\tau,\omega,\mu;\eta_{\psi},\hat{z}_{\psi}) 
= -(n + \delta_{n,0})\, l_{n+1}^{\rm (F),(d)} (\tau,\omega,\mu;\eta_{\psi},\hat{z}_{\psi})\,.\label{eq:l0_FerLoopN3d}
\ee

Finally, we give the definition of the threshold function which appears in the RG flow equations of
the Yukawa coupling. We have
\be
&& l_{1,1}^{\rm (FB),(d)}(\tau,\omega_{\psi},\omega_{\rm B};\eta_{\psi},\eta_{\rm B},\hat{z}_{\psi},\hat{z}_{\rm B})\nn\\
&&\qquad\qquad\qquad =-\frac{\tau}{2}\,\sum_{n=-\infty}^{\infty}
\int _0 ^{\infty} dx\, x^{\frac{d-3}{2}}\tilde{\partial}_t \,\left[\tilde{G} _{\psi} (\tilde{\nu}_n^2,\omega_{\psi})
\tilde{G} _{\rm B} (\tilde{\nu}_n^2,\omega_{\rm B})\right]\,.
\ee
To evaluate the integral over $x$ (spatial momenta), we have to take derivatives with respect to the regulator 
function. For the regulator functions~\eqref{eq:bosreg} and~\eqref{eq:fermreg} these derivatives are given by 
\be
\tilde{\partial}_t \Big|_{\psi}&=& \left(\frac{1}{x^{1/2}} - \eta_{\psi}\left(\frac{1}{x^{1/2}} - 1\right)\right)\theta(1-x)\frac{\partial}{\partial r_{\psi}}\,,\\
\tilde{\partial}_t \Big|_{\rm B}&=&\left(\frac{2}{x} - \eta_{\rm B}\left(\frac{1}{x} - 1\right)\right)\theta(1-x)\frac{\partial}{\partial r_{\rm B}}\,,
\ee
where the first and the second line defines how the formal derivative $\tilde{\partial} _t$ acts on 
the fermion propagator and the boson propagator, respectively. These expressions are only valid for the spatial regulator functions given in 
Eqs.~\eqref{eq:bosreg2} and~\eqref{eq:fermreg2}. Recall that~$x$ refers to the squared {\it spatial} (loop) momentum.
In this review, we have only used general properties of the threshold function~$l_{1,1}^{\rm (FB),(d)}$ (for $\mu=0$), but
we have not employed it in the numerical evaluations of the flow equations. Therefore we only give the general definition of this function here.


\bibliographystyle{adp}
\bibliography{bibliography}

\providecommand{\WileyBibTextsc}{}
\let\textsc\WileyBibTextsc
\providecommand{\othercit}{}
\providecommand{\jr}[1]{#1}
\providecommand{\etal}{~et~al.}


\begin{thebibliography}{[100]}

\bibitem{Braun-Munzinger:2001ip}
 \textsc{P.~Braun-Munzinger},  \textsc{D.~Magestro},  \textsc{K.~Redlich},  and
   \textsc{J.~Stachel},
 \jr{Phys. Lett.} \textbf{B518}, 41--46 (2001).


\bibitem{Feshbach}
 \textsc{H.~Feshbach},
 \jr{Annals of Physics} \textbf{5}(4), 357 -- 390 (1958).


\bibitem{Zwierlein}
 \textsc{M.\,W. Zwierlein},  \textsc{C.\,H. Schunck},  \textsc{A.~Schirotzek},
  and  \textsc{W.~Ketterle},
 \jr{Nature} \textbf{442}, 54--58 (2006).


\bibitem{Partridge}
 \textsc{W.~PartridgeGuthrie B. amd~Li.},  \textsc{R.\,I. Kamar},
  \textsc{Y.\,a. Liao},  and  \textsc{R.\,G. Hulet},
 \jr{Science} \textbf{311}, 503--505 (2006).


\bibitem{Partridge:2006zz}
 \textsc{G.\,B. Partridge},  \textsc{W.~Li},  \textsc{Y.\,A. Liao},
  \textsc{R.\,G. Hulet},  \textsc{M.~Haque},  and  \textsc{H.\,T.\,C.
  Stoof},
 \jr{Phys. Rev. Lett.} \textbf{97}(19), 190407 (2006).


\bibitem{Regal}
 \textsc{Q.~Chen},  \textsc{C.\,A. Regal},  \textsc{M.~Greiner},
  \textsc{D.\,S. Jin},  and  \textsc{K.~Levin},
 \jr{Phys. Rev. A} \textbf{73}(4), 041601 (2006).


\bibitem{Nascimbene}
 \textsc{S.~Nascimb{\`e}ne},  \textsc{N.~Navon},  \textsc{K.\,J. Jiang},
  \textsc{F.~Chevy},  and  \textsc{C.~Salomon},
 \jr{Nature} \textbf{463}(7284), 1057--1060 (2010).


\bibitem{Carlson:2003zz}
 \textsc{J.~Carlson},  \textsc{S.\,Y. Chang},  \textsc{V.\,R. Pandharipande},
  and  \textsc{K.\,E. Schmidt},
 \jr{Phys. Rev. Lett.} \textbf{91}(5), 050401 (2003).


\bibitem{Astrakharchik}
 \textsc{G.\,E. Astrakharchik},  \textsc{J.~Boronat},  \textsc{J.~Casulleras},
  \textsc{Giorgini},  and  \textsc{S.},
 \jr{Phys. Rev. Lett.} \textbf{93}(20), 200404 (2004).


\bibitem{Bulgac:2005pj}
 \textsc{A.~Bulgac},  \textsc{J.\,E. Drut},  and  \textsc{P.~Magierski},
 \jr{Phys. Rev. Lett.} \textbf{96}, 090404 (2006).


\bibitem{Bulgac:2007ah}
 \textsc{A.~Bulgac},  \textsc{J.\,E. Drut},  and  \textsc{P.~Magierski},
 \jr{Phys. Rev. Lett.} \textbf{99}, 120401 (2007).


\bibitem{Burovski}
 \textsc{E.~Burovski},  \textsc{N.~Prokof'ev},  \textsc{B.~Svistunov},  and
  \textsc{M.~Troyer},
 \jr{Phys. Rev. Lett.} \textbf{96}(16), 160402 (2006).


\bibitem{Ceperley}
 \textsc{V.\,K. Akkineni},  \textsc{D.\,M. Ceperley},  and
  \textsc{N.~Trivedi},
 \jr{Phys. Rev. B} \textbf{76}(16), 165116 (2007).


\bibitem{Bulgac:2008zz}
 \textsc{A.~Bulgac},  \textsc{J.\,E. Drut},  and  \textsc{P.~Magierski},
 \jr{Phys. Rev.} \textbf{A78}, 023625 (2008).


\bibitem{CombMora}
 \textsc{R.~Combescot} and  \textsc{C.~Mora},
 \jr{EPL (Europhysics Letters)} \textbf{68}(1), 79 (2004).


\bibitem{Chevy2}
 \textsc{F.~Chevy},
 \jr{Phys. Rev. Lett.} \textbf{96}(13), 130401 (2006).


\bibitem{Chevy}
 \textsc{F.~Chevy},
 \jr{Phys. Rev. A} \textbf{74}(6), 063628 (2006).


\bibitem{Stoof}
 \textsc{K.\,B. Gubbels},  \textsc{M.\,W.\,J. Romans},  and  \textsc{H.\,T.\,C.
  Stoof},
 \jr{Phys. Rev. Lett.} \textbf{97}(21), 210402 (2006).


\bibitem{Duan}
 \textsc{W.~Yi} and  \textsc{L.\,M. Duan},
 \jr{Phys. Rev. A} \textbf{73}(3), 031604 (2006).


\bibitem{Forbes}
 \textsc{A.~Bulgac} and  \textsc{M.\,M. Forbes},
 \jr{Phys. Rev. A} \textbf{75}(3), 031605 (2007).


\bibitem{Diehl:2007ri}
 \textsc{S.~Diehl},  \textsc{H.~Gies},  \textsc{J.\,M. Pawlowski},  and
  \textsc{C.~Wetterich},
 \jr{Phys. Rev.} \textbf{A76}, 053627 (2007).


\bibitem{Pilati}
 \textsc{S.~Pilati} and  \textsc{S.~Giorgini},
 \jr{Phys. Rev. Lett.} \textbf{100}(3), 030401 (2008).


\bibitem{Stoof2}
 \textsc{K.\,B. Gubbels} and  \textsc{H.\,T.\,C. Stoof},
 \jr{Phys. Rev. Lett.} \textbf{100}(14), 140407 (2008).


\bibitem{Recati}
 \textsc{A.~Recati},  \textsc{C.~Lobo},  and  \textsc{S.~Stringari},
 \jr{Phys. Rev. A} \textbf{78}(2), 023633 (2008).


\bibitem{Bloch}
 \textsc{I.~Bloch},  \textsc{J.~Dalibard},  and  \textsc{W.~Zwerger},
 \jr{Rev. Mod. Phys.} \textbf{80}(3), 885--964 (2008).


\bibitem{Haussmann}
 \textsc{R.~Haussmann} and  \textsc{W.~Zwerger},
 \jr{Phys. Rev. A} \textbf{78}(6), 063602 (2008).


\bibitem{Diehl:2009ma}
 \textsc{S.~Diehl},  \textsc{S.~Floerchinger},  \textsc{H.~Gies},
  \textsc{J.\,M. Pawlowski},  and  \textsc{C.~Wetterich},
 \jr{Annalen Phys.} \textbf{522}, 615--656 (2010).


\bibitem{Gies:2002hq}
 \textsc{H.~Gies} and  \textsc{C.~Wetterich},
 \jr{Phys. Rev.} \textbf{D69}, 025001 (2004).


\bibitem{Gies:2005as}
 \textsc{H.~Gies} and  \textsc{J.~Jaeckel},
 \jr{Eur. Phys. J.} \textbf{C46}, 433--438 (2006).


\bibitem{Braun:2005uj}
 \textsc{J.~Braun} and  \textsc{H.~Gies},
 \jr{Phys. Lett.} \textbf{B645}, 53--58 (2007).


\bibitem{Braun:2006jd}
 \textsc{J.~Braun} and  \textsc{H.~Gies},
 \jr{JHEP} \textbf{06}, 024 (2006).


\bibitem{Braun:2008pi}
 \textsc{J.~Braun},
 \jr{Eur. Phys. J.} \textbf{C64}, 459--482 (2009).


\bibitem{Braun:2009si}
 \textsc{J.~Braun},
 \jr{Phys. Rev.} \textbf{D81}, 016008 (2010).


\bibitem{Bardeen:1957mv}
 \textsc{J.~Bardeen},  \textsc{L.\,N. Cooper},  and  \textsc{J.\,R.
  Schrieffer},
 \jr{Phys. Rev.} \textbf{108}, 1175--1204 (1957).


\bibitem{Melik}
 \textsc{L.\,P. Gorkov} and  \textsc{T.\,K. Melik-Barkhudarov},
 \jr{Sov. Phys. JETP} \textbf{13}, 1018 (1961).


\bibitem{Heiselberg:2000ya}
 \textsc{H.~Heiselberg},  \textsc{C.\,J. Pethick},  \textsc{H.~Smith},  and
  \textsc{L.~Viverit},
 \jr{Phys. Rev. Lett.} \textbf{85}, 2418--2421 (2000).


\bibitem{Arnold:2001mu}
 \textsc{P.\,B. Arnold} and  \textsc{G.\,D. Moore},
 \jr{Phys. Rev. Lett.} \textbf{87}, 120401 (2001).


\bibitem{Kashurnikov:2001zz}
 \textsc{V.\,A. Kashurnikov},  \textsc{N.\,V. Prokof'ev},  and  \textsc{B.\,V.
  Svistunov},
 \jr{Phys. Rev. Lett.} \textbf{87}, 120402 (2001).


\bibitem{Blaizot:2004qa}
 \textsc{J.\,P. Blaizot},  \textsc{R.~Mendez~Galain},  and
  \textsc{N.~Wschebor},
 \jr{Europhys. Lett.} \textbf{72}, 705--711 (2005).


\bibitem{Schmidt:2011zu}
 \textsc{R.~Schmidt} and  \textsc{T.~Enss},
 \jr{Phys. Rev.} \textbf{A83}, 063620 (2011).


\bibitem{Ku:2008vk}
 \textsc{M.~Ku},  \textsc{J.~Braun},  and  \textsc{A.~Schwenk},
 \jr{Phys. Rev. Lett.} \textbf{102}, 255301 (2009).


\bibitem{Thies:2003kk}
 \textsc{M.~Thies} and  \textsc{K.~Urlichs},
 \jr{Phys. Rev.} \textbf{D67}, 125015 (2003).


\bibitem{deForcrand:2006zz}
 \textsc{P.~de~Forcrand} and  \textsc{U.~Wenger},
 \jr{PoS} \textbf{LAT2006}, 152 (2006).


\bibitem{Nickel:2009wj}
 \textsc{D.~Nickel},
 \jr{Phys. Rev.} \textbf{D80}, 074025 (2009).


\bibitem{Kojo:2009ha}
 \textsc{T.~Kojo},  \textsc{Y.~Hidaka},  \textsc{L.~McLerran},  and
  \textsc{R.\,D. Pisarski},
 \jr{Nucl. Phys.} \textbf{A843}, 37--58 (2010).


\bibitem{Colangelo:2002hy}
 \textsc{G.~Colangelo},  \textsc{S.~D{\"u}rr},  and  \textsc{R.~Sommer},
 \jr{Nucl. Phys. Proc. Suppl.} \textbf{119}, 254--256 (2003).


\bibitem{AliKhan:2003cu}
 \textsc{A.~Ali~Khan} \etal{},
 \jr{Nucl. Phys.} \textbf{B689}, 175--194 (2004).


\bibitem{Colangelo:2003hf}
 \textsc{G.~Colangelo} and  \textsc{S.~D{\"u}rr},
 \jr{Eur. Phys. J.} \textbf{C33}, 543--553 (2004).


\bibitem{Guagnelli:2004ww}
 \textsc{M.~Guagnelli} \etal{},
 \jr{Phys. Lett.} \textbf{B597}, 216--221 (2004).


\bibitem{Colangelo:2004xr}
 \textsc{G.~Colangelo} and  \textsc{C.~Haefeli},
 \jr{Phys. Lett.} \textbf{B590}, 258--264 (2004).


\bibitem{Braun:2004yk}
 \textsc{J.~Braun},  \textsc{B.~Klein},  and  \textsc{H.\,J. Pirner},
 \jr{Phys. Rev.} \textbf{D71}, 014032 (2005).


\bibitem{Colangelo:2005gd}
 \textsc{G.~Colangelo},  \textsc{S.~Durr},  and  \textsc{C.~Haefeli},
 \jr{Nucl. Phys.} \textbf{B721}, 136--174 (2005).


\bibitem{Braun:2005fj}
 \textsc{J.~Braun},  \textsc{B.~Klein},  \textsc{H.\,J. Pirner},  and
  \textsc{A.\,H. Rezaeian},
 \jr{Phys. Rev.} \textbf{D73}, 074010 (2006).


\bibitem{Braun:2005gy}
 \textsc{J.~Braun},  \textsc{B.~Klein},  and  \textsc{H.\,J. Pirner},
 \jr{Phys. Rev.} \textbf{D72}, 034017 (2005).


\bibitem{Braun:2010vd}
 \textsc{J.~Braun},  \textsc{B.~Klein},  and  \textsc{P.~Piasecki},
 \jr{Eur. Phys. J.} \textbf{C71}, 1576 (2011).


\bibitem{Klein:2010tk}
 \textsc{B.~Klein},  \textsc{J.~Braun},  and  \textsc{B.\,J. Schaefer},
 \jr{PoS} \textbf{LATTICE2010}, 193 (2010).


\bibitem{Braun:2010pp}
 \textsc{J.~Braun},  \textsc{B.~Klein},  and  \textsc{B.\,J. Schaefer},
 \jr{in preparation} (2010).


\bibitem{Kohn:1965zzb}
 \textsc{W.~Kohn} and  \textsc{L.\,J. Sham},
 \jr{Phys. Rev.} \textbf{140}(4A), A1133--A1138 (1965).


\bibitem{Hubbard:1959ub}
 \textsc{J.~Hubbard},
 \jr{Phys. Rev. Lett.} \textbf{3}, 77--80 (1959).


\bibitem{Stratonovich}
 \textsc{R.~Stratonovich},
 \jr{Dokl. Akad. Nauk.} \textbf{115}, 1097 (1957).


\bibitem{Mueller1}
 \textsc{T.\,N. De~Silva} and  \textsc{E.\,J. Mueller},
 \jr{Phys. Rev. A} \textbf{73}(5), 051602 (2006).


\bibitem{Sensarma}
 \textsc{R.~{Sensarma}},  \textsc{W.~{Schneider}},  \textsc{R.\,B. {Diener}},
  and  \textsc{M.~{Randeria}},
 \jr{arXiv:0706.1741}.


\bibitem{Dobaczewski:2001ed}
 \textsc{J.~Dobaczewski},  \textsc{W.~Nazarewicz},  and  \textsc{P.\,G.
  Reinhard},
 \jr{Nucl. Phys.} \textbf{A693}, 361--373 (2001).


\bibitem{Stoitsov:2003pd}
 \textsc{M.\,V. Stoitsov},  \textsc{J.~Dobaczewski},  \textsc{W.~Nazarewicz},
  \textsc{S.~Pittel},  and  \textsc{D.\,J. Dean},
 \jr{Phys. Rev.} \textbf{C68}, 054312 (2003).


\bibitem{Bender:2003jk}
 \textsc{M.~Bender},  \textsc{P.\,H. Heenen},  and  \textsc{P.\,G.
  Reinhard},
 \jr{Rev. Mod. PHys.} \textbf{75}, 121--180 (2003).


\othercit
\bibitem{Anderson}
 \textsc{P.\,W. Anderson},
The Theory of Superconductivity in the HighTc Cuprate Superconductors
  (Princeton University Press, Princeton, 1997).


\othercit
\bibitem{FuldeBook}
 \textsc{P.~Fulde},
Electron Correlations in Molecules and Solids (Springer, Heidelberg, 1991).


\bibitem{Honerkamp1}
 \textsc{{C. Honerkamp}},
 \jr{Eur. Phys. J. B} \textbf{21}(1), 81--91 (2001).


\bibitem{Honerkamp2}
 \textsc{{C. Honerkamp}},  \textsc{{M. Salmhofer}},  and  \textsc{{T.M.
  Rice}},
 \jr{Eur. Phys. J. B} \textbf{27}(1), 127--134 (2002).


\bibitem{Honerkamp3}
 \textsc{{R. Gersch}},  \textsc{{C. Honerkamp}},  \textsc{{D. Rohe}},  and
  \textsc{{W. Metzner}},
 \jr{Eur. Phys. J. B} \textbf{48}(3), 349--358 (2005).


\bibitem{Honerkamp4}
 \textsc{M.~{Salmhofer}},  \textsc{C.~{Honerkamp}},  \textsc{W.~{Metzner}},
  and  \textsc{{Oliver}},
 \jr{Progress of Theoretical Physics} \textbf{112}(December), 943--970 (2004).


\bibitem{PSSB:PSSB2221030242}
 \textsc{J.~Mertsching} and  \textsc{H.\,J. Fischbeck},
 \jr{physica status solidi (b)} \textbf{103}(2), 783--795 (1981).


\bibitem{Schnetz:2004vr}
 \textsc{O.~Schnetz},  \textsc{M.~Thies},  and  \textsc{K.~Urlichs},
 \jr{Ann. Phys.} \textbf{314}, 425--447 (2004).


\bibitem{Drut:2007zx}
 \textsc{J.\,E. {Drut}} and  \textsc{D.\,T. {Son}},
 \jr{Phys. Rev.} \textbf{B77}(7), 075115 (2008).


\bibitem{Gies:2010st}
 \textsc{H.~Gies} and  \textsc{L.~Janssen},
 \jr{Phys. Rev.} \textbf{D82}, 085018 (2010).


\bibitem{Gies:2009da}
 \textsc{H.~Gies},  \textsc{L.~Janssen},  \textsc{S.~Rechenberger},  and
  \textsc{M.\,M. Scherer},
 \jr{Phys. Rev.} \textbf{D81}, 025009 (2010).


\othercit
\bibitem{Jaeckel:2003uz}
 \textsc{J.~Jaeckel},
Effective actions for strongly interacting fermionic systems,
PhD thesis, University of Heidelberg, 2003, hep-ph/0309090.


\bibitem{Braun:2009ns}
 \textsc{J.~Braun} and  \textsc{H.~Gies},
 \jr{JHEP} \textbf{05}, 060 (2010).


\bibitem{Braun:2010qs}
 \textsc{J.~Braun},  \textsc{C.\,S. Fischer},  and  \textsc{H.~Gies},
 \jr{(to appear in Phys. Rev. D), arXiv:1012.4279}.


\bibitem{Weinberg:1979bn}
 \textsc{S.~Weinberg},
 \jr{Phys. Rev.} \textbf{D19}, 1277--1280 (1979).


\bibitem{Holdom:1981rm}
 \textsc{B.~Holdom},
 \jr{Phys. Rev.} \textbf{D24}, 1441 (1981).


\bibitem{Hong:2004td}
 \textsc{D.\,K. Hong},  \textsc{S.\,D.\,H. Hsu},  and
  \textsc{F.~Sannino},
 \jr{Phys. Lett.} \textbf{B597}, 89--93 (2004).


\bibitem{Sannino:2004qp}
 \textsc{F.~Sannino} and  \textsc{K.~Tuominen},
 \jr{Phys. Rev.} \textbf{D71}, 051901 (2005).


\bibitem{Dietrich:2005jn}
 \textsc{D.\,D. Dietrich},  \textsc{F.~Sannino},  and
  \textsc{K.~Tuominen},
 \jr{Phys. Rev.} \textbf{D72}, 055001 (2005).


\bibitem{Dietrich:2006cm}
 \textsc{D.\,D. Dietrich} and  \textsc{F.~Sannino},
 \jr{Phys. Rev.} \textbf{D75}, 085018 (2007).


\bibitem{Ryttov:2007sr}
 \textsc{T.\,A. Ryttov} and  \textsc{F.~Sannino},
 \jr{Phys. Rev.} \textbf{D76}, 105004 (2007).


\bibitem{Antipin:2009wr}
 \textsc{O.~Antipin} and  \textsc{K.~Tuominen},
 \jr{Phys. Rev.} \textbf{D81}, 076011 (2010).


\bibitem{Sannino:2009za}
 \textsc{F.~Sannino},
 \jr{Acta Phys. Polon.} \textbf{B40}, 3533--3743 (2009).


\bibitem{Wilson:1971bg}
 \textsc{K.\,G. Wilson},
 \jr{Phys. Rev.} \textbf{B4}, 3174--3183 (1971).


\bibitem{Wilson:1971dh}
 \textsc{K.\,G. Wilson},
 \jr{Phys. Rev.} \textbf{B4}, 3184--3205 (1971).


\bibitem{Wilson:1973jj}
 \textsc{K.\,G. Wilson} and  \textsc{J.\,B. Kogut},
 \jr{Phys. Rept.} \textbf{12}, 75--200 (1974).


\bibitem{Wegner:1972ih}
 \textsc{F.\,J. Wegner} and  \textsc{A.~Houghton},
 \jr{Phys. Rev.} \textbf{A8}, 401--412 (1973).


\bibitem{Nicoll:1977hi}
 \textsc{J.\,F. Nicoll} and  \textsc{T.\,S. Chang},
 \jr{Phys. Lett.} \textbf{A62}, 287--289 (1977).


\bibitem{Polchinski:1983gv}
 \textsc{J.~Polchinski},
 \jr{Nucl. Phys.} \textbf{B231}, 269--295 (1984).


\bibitem{Wetterich:1992yh}
 \textsc{C.~Wetterich},
 \jr{Phys. Lett.} \textbf{B301}, 90--94 (1993).


\bibitem{Shankar:1993pf}
 \textsc{R.~Shankar},
 \jr{Rev. Mod. Phys.} \textbf{66}, 129--192 (1994).


\bibitem{Litim:1998nf}
 \textsc{D.\,F. Litim} and  \textsc{J.\,M. Pawlowski},
 \jr{in {\it The Exact Renormalization Group}, Eds.~Krasnitz et al., World
  Scientific,} pp.\,168 (1999), hep--th/9901063.


\othercit
\bibitem{SalmhoferBook}
 \textsc{M.~Salmhofer},
Renormalization: An Introduction (Springer, Heidelberg, 1999).


\bibitem{Bagnuls:2000ae}
 \textsc{C.~Bagnuls} and  \textsc{C.~Bervillier},
 \jr{Phys. Rept.} \textbf{348}, 91 (2001).


\bibitem{Berges:2000ew}
 \textsc{J.~Berges},  \textsc{N.~Tetradis},  and  \textsc{C.~Wetterich},
 \jr{Phys. Rept.} \textbf{363}, 223--386 (2002).


\bibitem{Polonyi:2001se}
 \textsc{J.~Polonyi},
 \jr{Central Eur. J. Phys.} \textbf{1}, 1--71 (2003).


\bibitem{Bogner:2003wn}
 \textsc{S.\,K. Bogner},  \textsc{T.\,T.\,S. Kuo},  and
  \textsc{A.~Schwenk},
 \jr{Phys. Rept.} \textbf{386}, 1--27 (2003).


\bibitem{Delamotte:2003dw}
 \textsc{B.~Delamotte},  \textsc{D.~Mouhanna},  and
  \textsc{M.~Tissier},
 \jr{Phys. Rev.} \textbf{B69}, 134413 (2004).


\bibitem{Pawlowski:2005xe}
 \textsc{J.\,M. Pawlowski},
 \jr{Annals Phys.} \textbf{322}, 2831--2915 (2007).


\bibitem{Gies:2006wv}
 \textsc{H.~Gies},
 \jr{hep-ph/0611146}.


\bibitem{Delamotte:2007pf}
 \textsc{B.~Delamotte},
 \jr{cond-mat/0702365}.


\bibitem{Bogner:2009bt}
 \textsc{S.\,K. Bogner},  \textsc{R.\,J. Furnstahl},  and
  \textsc{A.~Schwenk},
 \jr{Prog. Part. Nucl. Phys.} \textbf{65}, 94--147 (2010).


\bibitem{Rosten:2010vm}
 \textsc{O.\,J. Rosten},
 \jr{arXiv:1003.1366}.


\bibitem{Kopietz:2010zz}
 \textsc{P.~Kopietz},  \textsc{L.~Bartosch},  and  \textsc{F.~Schutz},
 \jr{Lect. Notes Phys.} \textbf{798}, 1--380 (2010).


\bibitem{Honerkamp5}
 \textsc{W.~Metzner},  \textsc{M.~Salmhofer},  \textsc{C.~Honerkamp},
  \textsc{V.~Meden},  and  \textsc{K.~Schonhammer},
 \jr{arXiv:1105.5289}.


\bibitem{Stevenson:1981vj}
 \textsc{P.\,M. Stevenson},
 \jr{Phys. Rev.} \textbf{D23}, 2916 (1981).


\othercit
\bibitem{Pokorski:1987ed}
 \textsc{S.~Pokorski},
Gauge field theories (Cambridge, UK: Univ. Press, 1987).


\bibitem{Jungnickel:1995fp}
 \textsc{D.\,U. Jungnickel} and  \textsc{C.~Wetterich},
 \jr{Phys. Rev.} \textbf{D53}, 5142--5175 (1996).


\bibitem{Bonini:1994kp}
 \textsc{M.~Bonini},  \textsc{M.~D'Attanasio},  and
  \textsc{G.~Marchesini},
 \jr{Nucl. Phys.} \textbf{B437}, 163--186 (1995).


\bibitem{Ellwanger:1995qf}
 \textsc{U.~Ellwanger},  \textsc{M.~Hirsch},  and  \textsc{A.~Weber},
 \jr{Z. Phys.} \textbf{C69}, 687--698 (1996).


\bibitem{Ellwanger:1996wy}
 \textsc{U.~Ellwanger},  \textsc{M.~Hirsch},  and  \textsc{A.~Weber},
 \jr{Eur. Phys. J.} \textbf{C1}, 563--578 (1998).


\bibitem{Reuter:1996ub}
 \textsc{M.~Reuter},
 \jr{hep-th/9602012}.


\bibitem{Reuter:1997gx}
 \textsc{M.~Reuter} and  \textsc{C.~Wetterich},
 \jr{Phys. Rev.} \textbf{D56}, 7893--7916 (1997).


\bibitem{Freire:2000bq}
 \textsc{F.~Freire},  \textsc{D.\,F. Litim},  and  \textsc{J.\,M.
  Pawlowski},
 \jr{Phys. Lett.} \textbf{B495}, 256--262 (2000).


\bibitem{Morris:2000fs}
 \textsc{T.\,R. Morris},
 \jr{JHEP} \textbf{12}, 012 (2000).


\bibitem{Arnone:2005fb}
 \textsc{S.~Arnone},  \textsc{T.\,R. Morris},  and  \textsc{O.\,J.
  Rosten},
 \jr{Eur. Phys. J.} \textbf{C50}, 467--504 (2007).


\bibitem{Abbott:1980hw}
 \textsc{L.\,F. Abbott},
 \jr{Nucl. Phys.} \textbf{B185}, 189 (1981).


\bibitem{Abbott:1981ke}
 \textsc{L.\,F. Abbott},
 \jr{Acta Phys. Polon.} \textbf{B13}, 33 (1982).


\bibitem{Gies:2001nw}
 \textsc{H.~Gies} and  \textsc{C.~Wetterich},
 \jr{Phys. Rev.} \textbf{D65}, 065001 (2002).


\bibitem{Gies:2002kd}
 \textsc{H.~Gies} and  \textsc{C.~Wetterich},
 \jr{Acta Phys. Slov.} \textbf{52}, 215--220 (2002).


\bibitem{Floerchinger:2009uf}
 \textsc{S.~Floerchinger} and  \textsc{C.~Wetterich},
 \jr{Phys. Lett.} \textbf{B680}, 371--376 (2009).


\bibitem{Floerchinger:2010da}
 \textsc{S.~Floerchinger},
 \jr{Eur. Phys. J.} \textbf{C69}, 119--132 (2010).


\bibitem{Litim:2002xm}
 \textsc{D.\,F. Litim} and  \textsc{J.\,M. Pawlowski},
 \jr{Phys. Rev.} \textbf{D66}, 025030 (2002).


\bibitem{Litim:2001fd}
 \textsc{D.\,F. Litim},
 \jr{Int. J. Mod. Phys.} \textbf{A16}, 2081--2088 (2001).


\bibitem{Litim:2000ci}
 \textsc{D.\,F. Litim},
 \jr{Phys. Lett.} \textbf{B486}, 92--99 (2000).


\bibitem{Litim:2001up}
 \textsc{D.\,F. Litim},
 \jr{Phys. Rev.} \textbf{D64}, 105007 (2001).


\bibitem{Nambu:1961tp}
 \textsc{Y.~Nambu} and  \textsc{G.~Jona-Lasinio},
 \jr{Phys. Rev.} \textbf{122}, 345--358 (1961).


\bibitem{Nambu:1961fr}
 \textsc{Y.~Nambu} and  \textsc{G.~Jona-Lasinio},
 \jr{Phys. Rev.} \textbf{124}, 246--254 (1961).


\bibitem{Klevansky:1992qe}
 \textsc{S.\,P. Klevansky},
 \jr{Rev. Mod. Phys.} \textbf{64}, 649--708 (1992).


\bibitem{Berges:1997eu}
 \textsc{J.~Berges},  \textsc{D.\,U. Jungnickel},  and
  \textsc{C.~Wetterich},
 \jr{Phys. Rev.} \textbf{D59}, 034010 (1999).


\bibitem{Schaefer:1999em}
 \textsc{B.\,J. Schaefer} and  \textsc{H.\,J. Pirner},
 \jr{Nucl. Phys.} \textbf{A660}, 439--474 (1999).


\bibitem{Braun:2003ii}
 \textsc{J.~Braun},  \textsc{K.~Schwenzer},  and  \textsc{H.\,J.
  Pirner},
 \jr{Phys. Rev.} \textbf{D70}, 085016 (2004).


\bibitem{Schaefer:2004en}
 \textsc{B.\,J. Schaefer} and  \textsc{J.~Wambach},
 \jr{Nucl. Phys.} \textbf{A757}, 479--492 (2005).


\bibitem{Nakano:2009ps}
 \textsc{E.~Nakano},  \textsc{B.\,J. Schaefer},  \textsc{B.~Stokic},
  \textsc{B.~Friman},  and  \textsc{K.~Redlich},
 \jr{Phys. Lett.} \textbf{B682}, 401--407 (2010).


\bibitem{Jaeckel:2002rm}
 \textsc{J.~Jaeckel} and  \textsc{C.~Wetterich},
 \jr{Phys. Rev.} \textbf{D68}, 025020 (2003).


\bibitem{Walecka:1995mi}
 \textsc{J.\,D. Walecka},
 \jr{Oxford Stud. Nucl. Phys.} \textbf{16}, 1--610 (1995).


\bibitem{Aoki:1999dw}
 \textsc{K.\,I. Aoki},  \textsc{K.~Morikawa},  \textsc{J.\,I. Sumi},
  \textsc{H.~Terao},  and  \textsc{M.~Tomoyose},
 \jr{Phys. Rev.} \textbf{D61}, 045008 (2000).


\bibitem{Tetradis:1993ts}
 \textsc{N.~Tetradis} and  \textsc{C.~Wetterich},
 \jr{Nucl. Phys.} \textbf{B422}, 541--592 (1994).


\bibitem{Litim:2002cf}
 \textsc{D.\,F. Litim},
 \jr{Nucl. Phys.} \textbf{B631}, 128--158 (2002).


\bibitem{Bervillier:2007rc}
 \textsc{C.~Bervillier},  \textsc{A.~Juttner},  and  \textsc{D.\,F.
  Litim},
 \jr{Nucl. Phys.} \textbf{B783}, 213--226 (2007).


\bibitem{Benitez:2009xg}
 \textsc{F.~Benitez} \etal{},
 \jr{Phys. Rev.} \textbf{E80}, 030103 (2009).


\bibitem{Goldstone:1961eq}
 \textsc{J.~Goldstone},
 \jr{Nuovo Cim.} \textbf{19}, 154--164 (1961).


\bibitem{Goldstone:1962es}
 \textsc{J.~Goldstone},  \textsc{A.~Salam},  and  \textsc{S.~Weinberg},
 \jr{Phys. Rev.} \textbf{127}, 965--970 (1962).


\bibitem{Gasser:1987ah}
 \textsc{J.~Gasser} and  \textsc{H.~Leutwyler},
 \jr{Phys. Lett.} \textbf{B188}, 477 (1987).


\bibitem{Braun:2010tt}
 \textsc{J.~Braun},  \textsc{H.~Gies},  and  \textsc{D.\,D. Scherer},
 \jr{Phys. Rev.} \textbf{D83}, 085012 (2011).


\bibitem{Braun:2009gm}
 \textsc{J.~Braun},  \textsc{L.\,M. Haas},  \textsc{F.~Marhauser},  and
  \textsc{J.\,M. Pawlowski},
 \jr{Phys. Rev. Lett.} \textbf{106}, 022002 (2011).


\bibitem{Gies:2003dp}
 \textsc{H.~Gies},  \textsc{J.~Jaeckel},  and  \textsc{C.~Wetterich},
 \jr{Phys. Rev.} \textbf{D69}, 105008 (2004).


\bibitem{Eichhorn:2011pc}
 \textsc{A.~Eichhorn} and  \textsc{H.~Gies},
 \jr{arXiv:1104.5366}.


\bibitem{Braun:2007td}
 \textsc{J.~Braun} and  \textsc{B.~Klein},
 \jr{Phys. Rev.} \textbf{D77}, 096008 (2008).


\bibitem{Braun:2008sg}
 \textsc{J.~Braun} and  \textsc{B.~Klein},
 \jr{Eur. Phys. J.} \textbf{C63}, 443--460 (2009).


\bibitem{Ejiri:2009ac}
 \textsc{S.~Ejiri} \etal{},
 \jr{Phys. Rev.} \textbf{D80}, 094505 (2009).


\bibitem{Engels:2011km}
 \textsc{J.~Engels} and  \textsc{F.~Karsch},
 \jr{arXiv:1105.0584}.


\bibitem{Miransky:1988gk}
 \textsc{V.\,A. Miransky} and  \textsc{K.~Yamawaki},
 \jr{Mod. Phys. Lett.} \textbf{A4}, 129--135 (1989).


\bibitem{Miransky:1996pd}
 \textsc{V.\,A. Miransky} and  \textsc{K.~Yamawaki},
 \jr{Phys. Rev.} \textbf{D55}, 5051--5066 (1997).


\bibitem{Berezinskii}
 \textsc{V.\,L. Berezinskii},
 \jr{Sov. Phys. JETP} \textbf{32}, 493 (1971).


\bibitem{Berezinskii2}
 \textsc{V.\,L. Berezinskii},
 \jr{Sov. Phys. JETP} \textbf{34}, 610 (1972).


\bibitem{Kosterlitz:1973xp}
 \textsc{J.\,M. Kosterlitz} and  \textsc{D.\,J. Thouless},
 \jr{J. Phys.} \textbf{C6}, 1181--1203 (1973).


\bibitem{Kaplan:2009kr}
 \textsc{D.\,B. Kaplan},  \textsc{J.\,W. Lee},  \textsc{D.\,T. Son},  and
  \textsc{M.\,A. Stephanov},
 \jr{Phys. Rev.} \textbf{D80}, 125005 (2009).


\othercit
\bibitem{HGPD}
 \textsc{H.~Gies},
(private communication).


\bibitem{Berges:2000ur}
 \textsc{J.~Berges} and  \textsc{J.~Cox},
 \jr{Phys. Lett.} \textbf{B517}, 369--374 (2001).


\bibitem{Aarts:2001qa}
 \textsc{G.~Aarts} and  \textsc{J.~Berges},
 \jr{Phys. Rev.} \textbf{D64}, 105010 (2001).


\bibitem{Berges:2004yj}
 \textsc{J.~Berges},
 \jr{AIP Conf. Proc.} \textbf{739}, 3--62 (2005).


\bibitem{Gasenzer:2008zz}
 \textsc{T.~Gasenzer} and  \textsc{J.\,M. Pawlowski},
 \jr{Phys. Lett.} \textbf{B670}, 135--140 (2008).


\othercit
\bibitem{Alkofer:2009zz}
 \textsc{R.~Alkofer},  \textsc{H.~Gies},  and  \textsc{B.\,J. Schaefer} (eds.),
Aspects of non-equilibrium quantum field theory: From cosmology to table-top
  experiments. Proceedings, 46. Internationale Universitaetswochen fuer
  Theoretische Physik, Winter School, IUTP 46, Schladming, Austria, 2009.


\bibitem{BraunFT}
 \textsc{J.~Braun},
 \jr{(in preparation)}.


\bibitem{Litim:2006ag}
 \textsc{D.\,F. Litim} and  \textsc{J.\,M. Pawlowski},
 \jr{JHEP} \textbf{11}, 026 (2006).


\bibitem{Blaizot:2006rj}
 \textsc{J.\,P. Blaizot},  \textsc{A.~Ipp},  \textsc{R.~Mendez-Galain},  and
  \textsc{N.~Wschebor},
 \jr{Nucl. Phys.} \textbf{A784}, 376--406 (2007).


\othercit
\bibitem{Fisher:1971ks}
 \textsc{M.\,E. Fisher},
The theory of critical point singularities,
 in: Varenna 1970, Proceedings, Critical Phenomena, edited by M.\,S. Green,
  (Academic Press, New York, 1971),  pp.\,1--99.


\bibitem{Fisher:1972zza}
 \textsc{M.\,E. Fisher} and  \textsc{M.\,N. Barber},
 \jr{Phys. Rev. Lett.} \textbf{28}, 1516--1519 (1972).


\othercit
\bibitem{Goldenfeld:1992qy}
 \textsc{N.~Goldenfeld},
Lectures on phase transitions and the renormalization group, Frontiers in
  physics (Addison-Wesley, Reading, USA, 1992).


\bibitem{Zinn-Justin:2002ru}
 \textsc{J.~Zinn-Justin},
 \jr{Int. Ser. Monogr. Phys.} \textbf{113}, 1--1054 (2002).


\bibitem{Karsch:2010ya}
 \textsc{F.~Karsch},
 \jr{Prog. Theor. Phys. Suppl.} \textbf{186}, 479--484 (2010).


\bibitem{Dietrich:2010yw}
 \textsc{D.\,D. Dietrich},
 \jr{Phys. Rev.} \textbf{D82}, 065007 (2010).


\bibitem{DelDebbio:2010jy}
 \textsc{L.~Del~Debbio} and  \textsc{R.~Zwicky},
 \jr{Phys. Lett.} \textbf{B700}, 217--220 (2011).


\bibitem{DelDebbio:2010ze}
 \textsc{L.~Del~Debbio} and  \textsc{R.~Zwicky},
 \jr{Phys. Rev.} \textbf{D82}, 014502 (2010).


\bibitem{PTP.112.943}
 \textsc{M.~Salmhofer},  \textsc{C.~Honerkamp},  \textsc{W.~Metzner},  and
  \textsc{O.~Lauscher},
 \jr{Progress of Theoretical Physics} \textbf{112}(6), 943--970 (2004).


\bibitem{Hagen:2007ew}
 \textsc{G.~Hagen} \etal{},
 \jr{Phys. Rev.} \textbf{C76}, 034302 (2007).


\bibitem{Otsuka:2009cs}
 \textsc{T.~Otsuka},  \textsc{T.~Suzuki},  \textsc{J.\,D. Holt},
  \textsc{A.~Schwenk},  and  \textsc{Y.~Akaishi},
 \jr{Phys. Rev. Lett.} \textbf{105}, 032501 (2010).


\bibitem{Friman:2011vm}
 \textsc{B.~Friman} and  \textsc{A.~Schwenk},
 \jr{arXiv:1101.4858}.


\bibitem{Anderson14071995}
 \textsc{M.\,H. Anderson},  \textsc{J.\,R. Ensher},  \textsc{M.\,R. Matthews},
  \textsc{C.\,E. Wieman},  and  \textsc{E.\,A. Cornell},
 \jr{Science} \textbf{269}(5221), 198--201 (1995).


\bibitem{PhysRevLett.75.1687}
 \textsc{C.\,C. Bradley},  \textsc{C.\,A. Sackett},  \textsc{J.\,J. Tollett},
  and  \textsc{R.\,G. Hulet},
 \jr{Phys. Rev. Lett.} \textbf{75}(9), 1687--1690 (1995).


\bibitem{PhysRevLett.75.3969}
 \textsc{K.\,B. Davis},  \textsc{M.\,O. Mewes},  \textsc{M.\,R. Andrews},
  \textsc{N.\,J. van Druten},  \textsc{D.\,S. Durfee},  \textsc{D.\,M. Kurn},
  and  \textsc{W.~Ketterle},
 \jr{Phys. Rev. Lett.} \textbf{75}(22), 3969--3973 (1995).


\bibitem{DeMarco10091999}
 \textsc{B.~DeMarco} and  \textsc{D.\,S. Jin},
 \jr{Science} \textbf{285}(5434), 1703--1706 (1999).


\othercit
\bibitem{Sakurai:2011zz}
 \textsc{J.\,J. Sakurai} and  \textsc{J.~Napolitano},
{Modern quantum physics} (Addison-Wesley (Boston, USA), 2011).


\bibitem{Diehl:2007th}
 \textsc{S.~Diehl},  \textsc{H.~Gies},  \textsc{J.\,M. Pawlowski},  and
  \textsc{C.~Wetterich},
 \jr{Phys. Rev.} \textbf{A76}, 021602 (2007).


\bibitem{Bartosch2009b}
 \textsc{L.~Bartosch},  \textsc{P.~Kopietz},  and  \textsc{A.~Ferraz},
 \jr{Physical Review B} \textbf{80}, 104514 (2009).


\bibitem{Floerchinger:2008qc}
 \textsc{S.~Floerchinger},  \textsc{M.~Scherer},  \textsc{S.~Diehl},  and
  \textsc{C.~Wetterich},
 \jr{Phys. Rev.} \textbf{B78}, 174528 (2008).


\bibitem{Floerchinger:2009pg}
 \textsc{S.~Floerchinger},  \textsc{M.\,M. Scherer},  and
  \textsc{C.~Wetterich},
 \jr{Phys. Rev.} \textbf{A81}, 063619 (2010).


\bibitem{Scherer:2010sv}
 \textsc{M.\,M. Scherer},  \textsc{S.~Floerchinger},  and
  \textsc{H.~Gies},
 \jr{arXiv:1010.2890}.


\bibitem{Blaizot:2003vz}
 \textsc{J.\,P. Blaizot},  \textsc{R.~Mendez-Galain},  and
  \textsc{N.~Wschebor},
 \jr{cond-mat/0311460}.


\bibitem{Blaizot:2008xx}
 \textsc{J.\,P. Blaizot},
 \jr{arXiv:0801.0009}.


\bibitem{Blaizot:2005xy}
 \textsc{J.\,P. Blaizot},  \textsc{R.~Mendez~Galain},  and
  \textsc{N.~Wschebor},
 \jr{Phys. Lett.} \textbf{B632}, 571--578 (2006).


\bibitem{Blaizot:2005wd}
 \textsc{J.\,P. Blaizot},  \textsc{R.~Mendez-Galain},  and
  \textsc{N.~Wschebor},
 \jr{Phys. Rev.} \textbf{E74}, 051116 (2006).


\bibitem{Blaizot:2006vr}
 \textsc{J.\,P. Blaizot},  \textsc{R.~Mendez-Galain},  and
  \textsc{N.~Wschebor},
 \jr{Phys. Rev.} \textbf{E74}, 051117 (2006).


\bibitem{Diehl:2005an}
 \textsc{S.~Diehl} and  \textsc{C.~Wetterich},
 \jr{Phys. Rev.} \textbf{A73}, 033615 (2006).


\bibitem{Lee:2010qp}
 \textsc{J.\,W. Lee},  \textsc{M.\,G. Endres},  \textsc{D.\,B. Kaplan},  and
  \textsc{A.\,N. Nicholson},
 \jr{PoS} \textbf{LATTICE2010}, 197 (2010).


\bibitem{Endres:2011er}
 \textsc{M.\,G. Endres},  \textsc{D.\,B. Kaplan},  \textsc{J.\,W. Lee},  and
  \textsc{A.\,N. Nicholson},
 \jr{arXiv:1106.5725}.


\bibitem{BraunDSMMS}
 \textsc{J.~Braun},  \textsc{S.~Diehl},  and  \textsc{M.\,M. Scherer},
 \jr{(in preparation)}.


\bibitem{Fulde}
 \textsc{P.~Fulde} and  \textsc{R.\,A. Ferrell},
 \jr{Phys. Rev.} \textbf{135}(3A), A550--A563 (1964).


\bibitem{Hohenberg:1964zz}
 \textsc{P.~Hohenberg} and  \textsc{W.~Kohn},
 \jr{Phys. Rev.} \textbf{136}(3B), B864--B871 (1964).


\bibitem{Furnstahl:2007xm}
 \textsc{R.\,J. Furnstahl},
 \jr{nucl-th/0702040}.


\bibitem{Drut:2009ce}
 \textsc{J.\,E. {Drut}},  \textsc{R.\,J. {Furnstahl}},  and
  \textsc{L.~{Platter}},
 \jr{Progress in Particle and Nuclear Physics} \textbf{64}(January), 120--168
  (2010).


\bibitem{Engel}
 \textsc{J.~Engel},
 \jr{Phys. Rev. C} \textbf{75}(1), 014306 (2007).


\bibitem{Jennings}
 \textsc{B.\,G. Giraud},  \textsc{B.\,K. Jennings},  and  \textsc{B.\,R.
  Barrett},
 \jr{Phys. Rev. A} \textbf{78}(3), 032507 (2008).


\bibitem{Leeuwen}
 \textsc{R.~van Leeuwen},
 \jr{Adv. Quant. Chem.} \textbf{43}, 25 (2003).


\bibitem{Polonyi:2001uc}
 \textsc{J.~Polonyi} and  \textsc{K.~Sailer},
 \jr{cond-mat/0108179}.


\bibitem{Puglia2003145}
 \textsc{S.\,J. Puglia},  \textsc{A.~Bhattacharyya},  and  \textsc{R.\,J.
  Furnstahl},
 \jr{Nuclear Physics A} \textbf{723}(1-2), 145 -- 180 (2003).


\bibitem{PTP.92.833}
 \textsc{R.~Fukuda},  \textsc{T.~Kotani},  \textsc{Y.~Suzuki},  and
  \textsc{S.~Yokojima},
 \jr{Progress of Theoretical Physics} \textbf{92}(4), 833--862 (1994).


\bibitem{1997cond.mat..2247V}
 \textsc{M.~{Valiev}} and  \textsc{G.\,W. {Fernando}},
 \jr{cond-mat/9702247}.


\bibitem{PhysRevB.66.155113}
 \textsc{J.~Polonyi} and  \textsc{K.~Sailer},
 \jr{Phys. Rev. B} \textbf{66}(15), 155113 (2002).


\bibitem{Schwenk:2004hm}
 \textsc{A.~Schwenk} and  \textsc{J.~Polonyi},
 \jr{nucl-th/0403011}.


\bibitem{BSP}
 \textsc{J.~Braun},  \textsc{J.~Polonyi},  and  \textsc{A.~Schwenk},
 \jr{(in preparation)}.


\bibitem{Alexandrou:1988jg}
 \textsc{C.~Alexandrou},  \textsc{J.~Myczkowski},  and  \textsc{J.\,W.
  Negele},
 \jr{Phys. Rev.} \textbf{C39}, 1076--1087 (1989).


\othercit
\bibitem{BraunRBRG}
 \textsc{J.~Braun},
Functional renormalization group and density functional theory,
(talk given at the ReisensbuRG workshop of DFG FOR 723), 2009.


\othercit
\bibitem{SchwenkDFTRG}
 \textsc{A.~Schwenk},
{R}{G} for nuclear forces and nuclear structure,
(talk given at the INT workshop on ``New applications of the renormalization
  group method in nuclear, particle and condensed matter physics"), 2010.


\bibitem{Hands:1995jq}
 \textsc{S.~Hands},  \textsc{S.~Kim},  and  \textsc{J.\,B. Kogut},
 \jr{Nucl. Phys.} \textbf{B442}, 364--390 (1995).


\bibitem{Castorina:2003kq}
 \textsc{P.~Castorina},  \textsc{M.~Mazza},  and  \textsc{D.~Zappala},
 \jr{Phys. Lett.} \textbf{B567}, 31--38 (2003).


\bibitem{Hofling:2002hj}
 \textsc{F.~Hofling},  \textsc{C.~Nowak},  and  \textsc{C.~Wetterich},
 \jr{Phys. Rev.} \textbf{B66}, 205111 (2002).


\bibitem{Gies:2009hq}
 \textsc{H.~Gies} and  \textsc{M.\,M. Scherer},
 \jr{Eur. Phys. J.} \textbf{C66}, 387--402 (2010).


\bibitem{ZinnJustin:2002ru}
 \textsc{J.~Zinn-Justin},
 \jr{Int. Ser. Monogr. Phys.} \textbf{113}, 1--1054 (2002).


\bibitem{Pisarski:1983ms}
 \textsc{R.\,D. Pisarski} and  \textsc{F.~Wilczek},
 \jr{Phys. Rev.} \textbf{D29}, 338--341 (1984).


\bibitem{Reuter:1996cp}
 \textsc{M.~Reuter},
 \jr{Phys. Rev.} \textbf{D57}, 971--985 (1998).


\bibitem{Litim:2001hk}
 \textsc{D.\,F. Litim} and  \textsc{J.\,M. Pawlowski},
 \jr{Phys. Lett.} \textbf{B516}, 197--207 (2001).


\bibitem{Litim:2010tt}
 \textsc{D.\,F. Litim} and  \textsc{D.~Zappala},
 \jr{Phys. Rev.} \textbf{D83}, 085009 (2011).


\bibitem{Karkkainen:1993ef}
 \textsc{L.~Karkkainen},  \textsc{R.~Lacaze},  \textsc{P.~Lacock},  and
  \textsc{B.~Petersson},
 \jr{Nucl. Phys.} \textbf{B415}, 781--796 (1994).


\bibitem{Gracey:1993kc}
 \textsc{J.\,A. Gracey},
 \jr{Int. J. Mod. Phys.} \textbf{A9}, 727--744 (1994).


\bibitem{Hands:1992be}
 \textsc{S.~Hands},  \textsc{A.~Kocic},  and  \textsc{J.\,B. Kogut},
 \jr{Ann. Phys.} \textbf{224}, 29--89 (1993).


\bibitem{Weinberg:1976xy}
 \textsc{S.~Weinberg},
 \jr{Lectures presented at Int. School of Subnuclear Physics, Ettore Majorana,
  Erice, Sicily, Jul 23 - Aug 8, 1976}.


\bibitem{Weinberg:1996kw}
 \textsc{S.~Weinberg},
 \jr{hep-th/9702027}.


\bibitem{Dou:1997fg}
 \textsc{D.~Dou} and  \textsc{R.~Percacci},
 \jr{Class. Quant. Grav.} \textbf{15}, 3449--3468 (1998).


\bibitem{Lauscher:2001ya}
 \textsc{O.~Lauscher} and  \textsc{M.~Reuter},
 \jr{Phys. Rev.} \textbf{D65}, 025013 (2002).


\bibitem{Litim:2003vp}
 \textsc{D.\,F. Litim},
 \jr{Phys. Rev. Lett.} \textbf{92}, 201301 (2004).


\bibitem{Codello:2006in}
 \textsc{A.~Codello} and  \textsc{R.~Percacci},
 \jr{Phys. Rev. Lett.} \textbf{97}, 221301 (2006).


\bibitem{Machado:2007ea}
 \textsc{P.\,F. Machado} and  \textsc{F.~Saueressig},
 \jr{Phys. Rev.} \textbf{D77}, 124045 (2008).


\bibitem{Codello:2007bd}
 \textsc{A.~Codello},  \textsc{R.~Percacci},  and  \textsc{C.~Rahmede},
 \jr{Int. J. Mod. Phys.} \textbf{A23}, 143--150 (2008).


\bibitem{Eichhorn:2009ah}
 \textsc{A.~Eichhorn},  \textsc{H.~Gies},  and  \textsc{M.\,M. Scherer},
 \jr{Phys. Rev.} \textbf{D80}, 104003 (2009).


\bibitem{Groh:2010ta}
 \textsc{K.~Groh} and  \textsc{F.~Saueressig},
 \jr{J. Phys.} \textbf{A43}, 365403 (2010).


\bibitem{Eichhorn:2010tb}
 \textsc{A.~Eichhorn} and  \textsc{H.~Gies},
 \jr{Phys. Rev.} \textbf{D81}, 104010 (2010).


\bibitem{Fischer:2006at}
 \textsc{P.~Fischer} and  \textsc{D.\,F. Litim},
 \jr{AIP Conf. Proc.} \textbf{861}, 336--343 (2006).


\bibitem{Gies:2003ic}
 \textsc{H.~Gies},
 \jr{Phys. Rev.} \textbf{D68}, 085015 (2003).


\bibitem{Gies:2009sv}
 \textsc{H.~Gies},  \textsc{S.~Rechenberger},  and  \textsc{M.\,M.
  Scherer},
 \jr{Eur. Phys. J.} \textbf{C66}, 403--418 (2010).


\bibitem{Percacci:2009fh}
 \textsc{R.~Percacci} and  \textsc{O.~Zanusso},
 \jr{Phys. Rev.} \textbf{D81}, 065012 (2010).


\bibitem{GellMann:1964nj}
 \textsc{M.~Gell-Mann},
 \jr{Phys. Lett.} \textbf{8}, 214--215 (1964).


\bibitem{Zweig:1981pd}
 \textsc{G.~Zweig},
 \jr{CERN-TH-401}.


\bibitem{Fritzsch:1973pi}
 \textsc{H.~Fritzsch},  \textsc{M.~Gell-Mann},  and
  \textsc{H.~Leutwyler},
 \jr{Phys. Lett.} \textbf{B47}, 365--368 (1973).


\bibitem{Nambu:1974zg}
 \textsc{Y.~Nambu},
 \jr{Phys. Rev.} \textbf{D10}, 4262 (1974).


\bibitem{Bardeen:1976tm}
 \textsc{W.\,A. Bardeen} and  \textsc{R.\,B. Pearson},
 \jr{Phys. Rev.} \textbf{D14}, 547 (1976).


\bibitem{Greenberg:1976ph}
 \textsc{O.\,W. Greenberg} and  \textsc{C.\,A. Nelson},
 \jr{Phys. Rept.} \textbf{32}, 69--121 (1977).


\othercit
\bibitem{Alkofer:1995mv}
 \textsc{R.~Alkofer} and  \textsc{H.~Reinhardt},
Chiral quark dynamics (Berlin, Germany: Springer, 1995).


\bibitem{Fukushima:2003fw}
 \textsc{K.~Fukushima},
 \jr{Phys. Lett.} \textbf{B591}, 277--284 (2004).


\bibitem{Ratti:2005jh}
 \textsc{C.~Ratti},  \textsc{M.\,A. Thaler},  and  \textsc{W.~Weise},
 \jr{Phys. Rev.} \textbf{D73}, 014019 (2006).


\bibitem{Sasaki:2006ww}
 \textsc{C.~Sasaki},  \textsc{B.~Friman},  and  \textsc{K.~Redlich},
 \jr{Phys. Rev.} \textbf{D75}, 074013 (2007).


\bibitem{Skokov:2010wb}
 \textsc{V.~Skokov},  \textsc{B.~Stokic},  \textsc{B.~Friman},  and
  \textsc{K.~Redlich},
 \jr{Phys. Rev.} \textbf{C82}, 015206 (2010).


\bibitem{Herbst:2010rf}
 \textsc{T.\,K. Herbst},  \textsc{J.\,M. Pawlowski},  and  \textsc{B.\,J.
  Schaefer},
 \jr{Phys. Lett.} \textbf{B696}, 58--67 (2011).


\bibitem{Skokov:2010uh}
 \textsc{V.~Skokov},  \textsc{B.~Friman},  and  \textsc{K.~Redlich},
 \jr{Phys. Rev.} \textbf{C83}, 054904 (2011).


\bibitem{Vogl:1989ea}
 \textsc{U.~Vogl},  \textsc{M.\,F.\,M. Lutz},  \textsc{S.~Klimt},  and
  \textsc{W.~Weise},
 \jr{Nucl. Phys.} \textbf{A516}, 469--495 (1990).


\bibitem{Klimt:1989pm}
 \textsc{S.~Klimt},  \textsc{M.\,F.\,M. Lutz},  \textsc{U.~Vogl},  and
  \textsc{W.~Weise},
 \jr{Nucl. Phys.} \textbf{A516}, 429--468 (1990).


\bibitem{Klimt:1990ws}
 \textsc{S.~Klimt},  \textsc{M.\,F.\,M. Lutz},  and  \textsc{W.~Weise},
 \jr{Phys. Lett.} \textbf{B249}, 386--390 (1990).


\bibitem{Schaefer:2008hk}
 \textsc{B.\,J. Schaefer} and  \textsc{M.~Wagner},
 \jr{Phys. Rev.} \textbf{D79}, 014018 (2009).


\bibitem{Fukushima:2008wg}
 \textsc{K.~Fukushima},
 \jr{Phys. Rev.} \textbf{D77}, 114028 (2008).


\bibitem{Hell:2009by}
 \textsc{T.~Hell},  \textsc{S.~Rossner},  \textsc{M.~Cristoforetti},  and
  \textsc{W.~Weise},
 \jr{Phys. Rev.} \textbf{D81}, 074034 (2010).


\bibitem{Jungnickel:1997yu}
 \textsc{D.\,U. Jungnickel} and  \textsc{C.~Wetterich},
 \jr{Eur. Phys. J.} \textbf{C2}, 557--567 (1998).


\bibitem{Jendges:2006yk}
 \textsc{L.~Jendges},  \textsc{B.~Klein},  \textsc{H.\,J. Pirner},  and
  \textsc{K.~Schwenzer},
 \jr{hep-ph/0608056}.


\bibitem{tHooft:1976fv}
 \textsc{G.~'t~Hooft},
 \jr{Phys. Rev.} \textbf{D14}, 3432--3450 (1976).


\bibitem{Shifman:1979uw}
 \textsc{M.\,A. Shifman},  \textsc{A.\,I. Vainshtein},  and  \textsc{V.\,I.
  Zakharov},
 \jr{Nucl. Phys.} \textbf{B163}, 46 (1980).


\bibitem{Shuryak:1981ff}
 \textsc{E.\,V. Shuryak},
 \jr{Nucl. Phys.} \textbf{B203}, 93 (1982).


\bibitem{Schafer:1996wv}
 \textsc{T.~Schafer} and  \textsc{E.\,V. Shuryak},
 \jr{Rev. Mod. Phys.} \textbf{70}, 323--426 (1998).


\bibitem{Pawlowski:1996ch}
 \textsc{J.\,M. Pawlowski},
 \jr{Phys. Rev.} \textbf{D58}, 045011 (1998).


\bibitem{Stephanov:2007fk}
 \textsc{M.\,A. Stephanov},
 \jr{PoS} \textbf{LAT2006}, 024 (2006).


\bibitem{Braun:2011fw}
 \textsc{J.~Braun} and  \textsc{A.~Janot},
 \jr{arXiv:1102.4841}.


\bibitem{Goldberger:1958tr}
 \textsc{M.\,L. Goldberger} and  \textsc{S.\,B. Treiman},
 \jr{Phys. Rev.} \textbf{110}, 1178--1184 (1958).


\bibitem{Meisinger:1995ih}
 \textsc{P.\,N. Meisinger} and  \textsc{M.\,C. Ogilvie},
 \jr{Phys. Lett.} \textbf{B379}, 163--168 (1996).


\bibitem{Pisarski:2000eq}
 \textsc{R.\,D. Pisarski},
 \jr{Phys. Rev.} \textbf{D62}, 111501 (2000).


\bibitem{Mocsy:2003qw}
 \textsc{A.~Mocsy},  \textsc{F.~Sannino},  and  \textsc{K.~Tuominen},
 \jr{Phys. Rev. Lett.} \textbf{92}, 182302 (2004).


\bibitem{Megias:2004hj}
 \textsc{E.~Megias},  \textsc{E.~Ruiz~Arriola},  and  \textsc{L.\,L.
  Salcedo},
 \jr{Phys. Rev.} \textbf{D74}, 065005 (2006).


\bibitem{Schaefer:2007pw}
 \textsc{B.\,J. Schaefer},  \textsc{J.\,M. Pawlowski},  and
  \textsc{J.~Wambach},
 \jr{Phys. Rev.} \textbf{D76}, 074023 (2007).


\bibitem{Hell:2008cc}
 \textsc{T.~Hell},  \textsc{S.~Roessner},  \textsc{M.~Cristoforetti},  and
  \textsc{W.~Weise},
 \jr{Phys. Rev.} \textbf{D79}, 014022 (2009).


\bibitem{Mizher:2010zb}
 \textsc{A.\,J. Mizher},  \textsc{M.\,N. Chernodub},  and  \textsc{E.\,S.
  Fraga},
 \jr{Phys. Rev.} \textbf{D82}, 105016 (2010).


\bibitem{Hell:2011ic}
 \textsc{T.~Hell},  \textsc{K.~Kashiwa},  and  \textsc{W.~Weise},
 \jr{Phys. Rev.} \textbf{D83}, 114008 (2011).


\bibitem{PhysRevLett.95.146801}
 \textsc{V.\,P. Gusynin} and  \textsc{S.\,G. Sharapov},
 \jr{Phys. Rev. Lett.} \textbf{95}(14), 146801 (2005).


\bibitem{PhysRevLett.96.256802}
 \textsc{V.\,P. Gusynin},  \textsc{S.\,G. Sharapov},  and  \textsc{J.\,P.
  Carbotte},
 \jr{Phys. Rev. Lett.} \textbf{96}(25), 256802 (2006).


\bibitem{PhysRevB.79.165425}
 \textsc{J.\,E. Drut} and  \textsc{T.\,A. L\"ahde},
 \jr{Phys. Rev. B} \textbf{79}(16), 165425 (2009).


\bibitem{Caswell:1974gg}
 \textsc{W.\,E. Caswell},
 \jr{Phys. Rev. Lett.} \textbf{33}, 244 (1974).


\bibitem{Banks:1981nn}
 \textsc{T.~Banks} and  \textsc{A.~Zaks},
 \jr{Nucl. Phys.} \textbf{B196}, 189 (1982).


\bibitem{Pisarski:1984dj}
 \textsc{R.\,D. Pisarski},
 \jr{Phys. Rev.} \textbf{D29}, 2423 (1984).


\bibitem{Appelquist:1986fd}
 \textsc{T.\,W. Appelquist},  \textsc{M.\,J. Bowick},  \textsc{D.~Karabali},
  and  \textsc{L.\,C.\,R. Wijewardhana},
 \jr{Phys. Rev.} \textbf{D33}, 3704 (1986).


\bibitem{Appelquist:1988sr}
 \textsc{T.~Appelquist},  \textsc{D.~Nash},  and  \textsc{L.\,C.\,R.
  Wijewardhana},
 \jr{Phys. Rev. Lett.} \textbf{60}, 2575 (1988).


\bibitem{Atkinson:1989fp}
 \textsc{D.~Atkinson},  \textsc{P.\,W. Johnson},  and
  \textsc{P.~Maris},
 \jr{Phys. Rev.} \textbf{D42}, 602--609 (1990).


\bibitem{Pennington:1990bx}
 \textsc{M.\,R. Pennington} and  \textsc{D.~Walsh},
 \jr{Phys. Lett.} \textbf{B253}, 246--251 (1991).


\bibitem{Curtis:1992gm}
 \textsc{D.\,C. Curtis},  \textsc{M.\,R. Pennington},  and
  \textsc{D.~Walsh},
 \jr{Phys. Lett.} \textbf{B295}, 313--319 (1992).


\bibitem{Burden:1990mg}
 \textsc{C.\,J. Burden} and  \textsc{C.\,D. Roberts},
 \jr{Phys. Rev.} \textbf{D44}, 540--550 (1991).


\bibitem{Maris:1995ns}
 \textsc{P.~Maris},
 \jr{Phys. Rev.} \textbf{D52}, 6087--6097 (1995).


\bibitem{Gusynin:1995bb}
 \textsc{V.\,P. Gusynin},  \textsc{A.\,H. Hams},  and
  \textsc{M.~Reenders},
 \jr{Phys. Rev.} \textbf{D53}, 2227--2235 (1996).


\bibitem{Maris:1996zg}
 \textsc{P.~Maris},
 \jr{Phys. Rev.} \textbf{D54}, 4049--4058 (1996).


\bibitem{Fischer:2004nq}
 \textsc{C.\,S. Fischer},  \textsc{R.~Alkofer},  \textsc{T.~Dahm},  and
  \textsc{P.~Maris},
 \jr{Phys. Rev.} \textbf{D70}, 073007 (2004).


\bibitem{Dagotto:1989td}
 \textsc{E.~Dagotto},  \textsc{A.~Kocic},  and  \textsc{J.\,B. Kogut},
 \jr{Nucl. Phys.} \textbf{B334}, 279 (1990).


\bibitem{Hands:1989mv}
 \textsc{S.~Hands} and  \textsc{J.\,B. Kogut},
 \jr{Nucl. Phys.} \textbf{B335}, 455 (1990).


\bibitem{Hands:2002dv}
 \textsc{S.\,J. Hands},  \textsc{J.\,B. Kogut},  and  \textsc{C.\,G.
  Strouthos},
 \jr{Nucl. Phys.} \textbf{B645}, 321--336 (2002).


\bibitem{Hands:2004bh}
 \textsc{S.\,J. Hands},  \textsc{J.\,B. Kogut},  \textsc{L.~Scorzato},  and
  \textsc{C.\,G. Strouthos},
 \jr{Phys. Rev.} \textbf{B70}, 104501 (2004).


\bibitem{Kondo:1991yk}
 \textsc{K.\,i. Kondo},  \textsc{S.~Shuto},  and  \textsc{K.~Yamawaki},
 \jr{Mod. Phys. Lett.} \textbf{A6}, 3385--3396 (1991).


\bibitem{Appelquist:1996dq}
 \textsc{T.~Appelquist},  \textsc{J.~Terning},  and  \textsc{L.\,C.\,R.
  Wijewardhana},
 \jr{Phys. Rev. Lett.} \textbf{77}, 1214--1217 (1996).


\bibitem{Appelquist:1997dc}
 \textsc{T.~Appelquist} and  \textsc{S.\,B. Selipsky},
 \jr{Phys. Lett.} \textbf{B400}, 364--369 (1997).


\bibitem{Velkovsky:1997fe}
 \textsc{M.~Velkovsky} and  \textsc{E.\,V. Shuryak},
 \jr{Phys. Lett.} \textbf{B437}, 398--402 (1998).


\bibitem{Appelquist:1998rb}
 \textsc{T.~Appelquist},  \textsc{A.~Ratnaweera},  \textsc{J.~Terning},  and
  \textsc{L.\,C.\,R. Wijewardhana},
 \jr{Phys. Rev.} \textbf{D58}, 105017 (1998).


\bibitem{Harada:2000kb}
 \textsc{M.~Harada} and  \textsc{K.~Yamawaki},
 \jr{Phys. Rev. Lett.} \textbf{86}, 757--760 (2001).


\bibitem{Sannino:1999qe}
 \textsc{F.~Sannino} and  \textsc{J.~Schechter},
 \jr{Phys. Rev.} \textbf{D60}, 056004 (1999).


\bibitem{Harada:2003dc}
 \textsc{M.~Harada},  \textsc{M.~Kurachi},  and  \textsc{K.~Yamawaki},
 \jr{Phys. Rev.} \textbf{D68}, 076001 (2003).


\bibitem{Terao:2007jm}
 \textsc{H.~Terao} and  \textsc{A.~Tsuchiya},
 \jr{arXiv:0704.3659}.


\bibitem{Poppitz:2009uq}
 \textsc{E.~Poppitz} and  \textsc{M.~Unsal},
 \jr{JHEP} \textbf{09}, 050 (2009).


\bibitem{Armoni:2009jn}
 \textsc{A.~Armoni},
 \jr{Nucl. Phys.} \textbf{B826}, 328--336 (2010).


\bibitem{Sannino:2009qc}
 \textsc{F.~Sannino},
 \jr{Phys. Rev.} \textbf{D80}, 065011 (2009).


\bibitem{Sannino:2009me}
 \textsc{F.~Sannino},
 \jr{Nucl. Phys.} \textbf{B830}, 179--194 (2010).


\bibitem{Kogut:1982fn}
 \textsc{J.\,B. Kogut} \etal{},
 \jr{Phys. Rev. Lett.} \textbf{48}, 1140 (1982).


\bibitem{Gavai:1985wi}
 \textsc{R.\,V. Gavai},
 \jr{Nucl. Phys.} \textbf{B269}, 530 (1986).


\bibitem{Fukugita:1987mb}
 \textsc{M.~Fukugita},  \textsc{S.~Ohta},  and  \textsc{A.~Ukawa},
 \jr{Phys. Rev. Lett.} \textbf{60}, 178 (1988).


\bibitem{Brown:1992fz}
 \textsc{F.\,R. Brown} \etal{},
 \jr{Phys. Rev.} \textbf{D46}, 5655--5670 (1992).


\bibitem{Damgaard:1997ut}
 \textsc{P.\,H. Damgaard},  \textsc{U.\,M. Heller},  \textsc{A.~Krasnitz},  and
   \textsc{P.~Olesen},
 \jr{Phys. Lett.} \textbf{B400}, 169--175 (1997).


\bibitem{Iwasaki:2003de}
 \textsc{Y.~Iwasaki},  \textsc{K.~Kanaya},  \textsc{S.~Kaya},
  \textsc{S.~Sakai},  and  \textsc{T.~Yoshie},
 \jr{Phys. Rev.} \textbf{D69}, 014507 (2004).


\bibitem{Catterall:2007yx}
 \textsc{S.~Catterall} and  \textsc{F.~Sannino},
 \jr{Phys. Rev.} \textbf{D76}, 034504 (2007).


\bibitem{Appelquist:2007hu}
 \textsc{T.~Appelquist},  \textsc{G.\,T. Fleming},  and  \textsc{E.\,T.
  Neil},
 \jr{Phys. Rev. Lett.} \textbf{100}, 171607 (2008).


\bibitem{Deuzeman:2008sc}
 \textsc{A.~Deuzeman},  \textsc{M.\,P. Lombardo},  and
  \textsc{E.~Pallante},
 \jr{Phys. Lett.} \textbf{B670}, 41--48 (2008).


\bibitem{Deuzeman:2009mh}
 \textsc{A.~Deuzeman},  \textsc{M.\,P. Lombardo},  and
  \textsc{E.~Pallante},
 \jr{Phys. Rev.} \textbf{D82}, 074503 (2010).


\bibitem{Appelquist:2009ty}
 \textsc{T.~Appelquist},  \textsc{G.\,T. Fleming},  and  \textsc{E.\,T.
  Neil},
 \jr{Phys. Rev.} \textbf{D79}, 076010 (2009).


\bibitem{Fodor:2009wk}
 \textsc{Z.~Fodor},  \textsc{K.~Holland},  \textsc{J.~Kuti},
  \textsc{D.~Nogradi},  and  \textsc{C.~Schroeder},
 \jr{Phys. Lett.} \textbf{B681}, 353--361 (2009).


\bibitem{Fodor:2009ff}
 \textsc{Z.~Fodor},  \textsc{K.~Holland},  \textsc{J.~Kuti},
  \textsc{D.~Nogradi},  and  \textsc{C.~Schroeder},
 \jr{PoS} \textbf{LAT2009}, 055 (2009).


\bibitem{Pallante:2009hu}
 \textsc{E.~Pallante},
 \jr{PoS} \textbf{LAT2009}, 015 (2009).


\bibitem{DeGrand:2010ba}
 \textsc{T.~DeGrand},
 \jr{arXiv:1010.4741}.


\bibitem{Kusafuka:2011fd}
 \textsc{Y.~Kusafuka} and  \textsc{H.~Terao},
 \jr{arXiv:1104.3606}.


\bibitem{Miransky:1984ef}
 \textsc{V.\,A. Miransky},
 \jr{Nuovo Cim.} \textbf{A90}, 149--170 (1985).


\bibitem{Miransky:1985aq}
 \textsc{V.\,A. Miransky} and  \textsc{P.\,I. Fomin},
 \jr{Sov. J. Part. Nucl.} \textbf{16}, 203 (1985).


\bibitem{Chivukula:1996kg}
 \textsc{R.\,S. Chivukula},
 \jr{Phys. Rev.} \textbf{D55}, 5238--5240 (1997).


\bibitem{Appelquist:1998xf}
 \textsc{T.~Appelquist} and  \textsc{F.~Sannino},
 \jr{Phys. Rev.} \textbf{D59}, 067702 (1999).


\bibitem{Bethke:2009jm}
 \textsc{S.~Bethke},
 \jr{Eur. Phys. J.} \textbf{C64}, 689--703 (2009).


\bibitem{Karsch:2000kv}
 \textsc{F.~Karsch},  \textsc{E.~Laermann},  and  \textsc{A.~Peikert},
 \jr{Nucl. Phys.} \textbf{B605}, 579--599 (2001).


\bibitem{Schaefer:2009ui}
 \textsc{B.\,J. Schaefer},  \textsc{M.~Wagner},  and
  \textsc{J.~Wambach},
 \jr{Phys. Rev.} \textbf{D81}, 074013 (2010).


\bibitem{Braun:2006wu}
 \textsc{J.~Braun},
 \jr{hep-ph/0611145}.


\bibitem{Fukano:2010yv}
 \textsc{H.\,S. Fukano} and  \textsc{F.~Sannino},
 \jr{Phys. Rev.} \textbf{D82}, 035021 (2010).


\bibitem{Jarvinen:2010ks}
 \textsc{M.~Jarvinen} and  \textsc{F.~Sannino},
 \jr{JHEP} \textbf{02}, 081 (2011).


\bibitem{Sannino:2008pz}
 \textsc{F.~Sannino},
 \jr{Phys. Rev.} \textbf{D80}, 017901 (2009).


\bibitem{DeGrand:2009mt}
 \textsc{T.~DeGrand} and  \textsc{A.~Hasenfratz},
 \jr{Phys. Rev.} \textbf{D80}, 034506 (2009).


\bibitem{Reuter:1993kw}
 \textsc{M.~Reuter} and  \textsc{C.~Wetterich},
 \jr{Nucl. Phys.} \textbf{B417}, 181--214 (1994).


\bibitem{Reuter:1994zn}
 \textsc{M.~Reuter} and  \textsc{C.~Wetterich},
 \jr{hep-th/9411227}.


\bibitem{Gies:2002af}
 \textsc{H.~Gies},
 \jr{Phys. Rev.} \textbf{D66}, 025006 (2002).


\bibitem{Pawlowski:2003hq}
 \textsc{J.\,M. Pawlowski},  \textsc{D.\,F. Litim},  \textsc{S.~Nedelko},  and
  \textsc{L.~von Smekal},
 \jr{Phys. Rev. Lett.} \textbf{93}, 152002 (2004).


\bibitem{Fischer:2004uk}
 \textsc{C.\,S. Fischer} and  \textsc{H.~Gies},
 \jr{JHEP} \textbf{10}, 048 (2004).


\bibitem{Braun:2007bx}
 \textsc{J.~Braun},  \textsc{H.~Gies},  and  \textsc{J.\,M. Pawlowski},
 \jr{Phys. Lett.} \textbf{B684}, 262--267 (2010).


\bibitem{Marhauser:2008fz}
 \textsc{F.~Marhauser} and  \textsc{J.\,M. Pawlowski},
 \jr{arXiv:0812.1144}.


\bibitem{Braun:2010cy}
 \textsc{J.~Braun},  \textsc{A.~Eichhorn},  \textsc{H.~Gies},  and
  \textsc{J.\,M. Pawlowski},
 \jr{Eur. Phys. J.} \textbf{C70}, 689--702 (2010).


\bibitem{Eichhorn:2010zc}
 \textsc{A.~Eichhorn},  \textsc{H.~Gies},  and  \textsc{J.\,M.
  Pawlowski},
 \jr{Phys. Rev.} \textbf{D83}, 045014 (2011).


\bibitem{Kondo:2010ts}
 \textsc{K.\,I. Kondo},
 \jr{Phys. Rev.} \textbf{D82}, 065024 (2010).


\bibitem{Litim:1998qi}
 \textsc{D.\,F. Litim} and  \textsc{J.\,M. Pawlowski},
 \jr{Phys. Lett.} \textbf{B435}, 181--188 (1998).


\bibitem{vanRitbergen:1997va}
 \textsc{T.~van Ritbergen},  \textsc{J.\,A.\,M. Vermaseren},  and
  \textsc{S.\,A. Larin},
 \jr{Phys. Lett.} \textbf{B400}, 379--384 (1997).


\bibitem{Czakon:2004bu}
 \textsc{M.~Czakon},
 \jr{Nucl. Phys.} \textbf{B710}, 485--498 (2005).


\bibitem{Ellwanger:1994iz}
 \textsc{U.~Ellwanger},
 \jr{Phys. Lett.} \textbf{B335}, 364--370 (1994).


\bibitem{Aoki:2006we}
 \textsc{Y.~Aoki},  \textsc{G.~Endrodi},  \textsc{Z.~Fodor},  \textsc{S.\,D.
  Katz},  and  \textsc{K.\,K. Szabo},
 \jr{Nature} \textbf{443}, 675--678 (2006).


\bibitem{Aoki:2006br}
 \textsc{Y.~Aoki},  \textsc{Z.~Fodor},  \textsc{S.\,D. Katz},  and
  \textsc{K.\,K. Szabo},
 \jr{Phys. Lett.} \textbf{B643}, 46--54 (2006).


\bibitem{Cheng:2006qk}
 \textsc{M.~Cheng} \etal{},
 \jr{Phys. Rev.} \textbf{D74}, 054507 (2006).


\bibitem{Cheng:2009zi}
 \textsc{M.~Cheng} \etal{},
 \jr{Phys. Rev.} \textbf{D81}, 054504 (2010).


\bibitem{Fucito:1984gt}
 \textsc{F.~Fucito},  \textsc{S.~Solomon},  and  \textsc{C.~Rebbi},
 \jr{Phys. Rev.} \textbf{D31}, 1460 (1985).


\bibitem{Gavai:1987dk}
 \textsc{R.\,V. Gavai},  \textsc{J.~Potvin},  and
  \textsc{S.~Sanielevici},
 \jr{Phys. Rev. Lett.} \textbf{58}, 2519 (1987).


\bibitem{Brown:1990ev}
 \textsc{F.\,R. Brown} \etal{},
 \jr{Phys. Rev. Lett.} \textbf{65}, 2491--2494 (1990).


\bibitem{Karsch:2001nf}
 \textsc{F.~Karsch},  \textsc{E.~Laermann},  and  \textsc{C.~Schmidt},
 \jr{Phys. Lett.} \textbf{B520}, 41--49 (2001).


\bibitem{PepeWiese1}
 \textsc{M.~{Pepe}} and  \textsc{U.\,J. {Wiese}},
 \jr{Nuclear Physics B} \textbf{768}(April), 21--37 (2007).


\bibitem{PepeWiese2}
 \textsc{K.~{Holland}},  \textsc{M.~{Pepe}},  and  \textsc{U.\,J.
  {Wiese}},
 \jr{Nuclear Physics B} \textbf{694}(August), 35--58 (2004).


\bibitem{vonSmekal:1997is}
 \textsc{L.~von Smekal},  \textsc{R.~Alkofer},  and  \textsc{A.~Hauck},
 \jr{Phys. Rev. Lett.} \textbf{79}, 3591--3594 (1997).


\bibitem{vonSmekal:1997vx}
 \textsc{L.~von Smekal},  \textsc{A.~Hauck},  and  \textsc{R.~Alkofer},
 \jr{Ann. Phys.} \textbf{267}, 1 (1998).


\bibitem{Lerche:2002ep}
 \textsc{C.~Lerche} and  \textsc{L.~von Smekal},
 \jr{Phys. Rev.} \textbf{D65}, 125006 (2002).


\bibitem{Alkofer:2004it}
 \textsc{R.~Alkofer},  \textsc{C.\,S. Fischer},  and  \textsc{F.\,J.
  Llanes-Estrada},
 \jr{Phys. Lett.} \textbf{B611}, 279--288 (2005).


\bibitem{Fischer:2008uz}
 \textsc{C.\,S. Fischer},  \textsc{A.~Maas},  and  \textsc{J.\,M.
  Pawlowski},
 \jr{Annals Phys.} \textbf{324}, 2408--2437 (2009).


\bibitem{Fischer:2009tn}
 \textsc{C.\,S. Fischer} and  \textsc{J.\,M. Pawlowski},
 \jr{Phys. Rev.} \textbf{D80}, 025023 (2009).


\bibitem{Fischer:2006vf}
 \textsc{C.\,S. Fischer} and  \textsc{J.\,M. Pawlowski},
 \jr{Phys. Rev.} \textbf{D75}, 025012 (2007).


\bibitem{Aguilar:2009nf}
 \textsc{A.\,C. Aguilar},  \textsc{D.~Binosi},  \textsc{J.~Papavassiliou},  and
   \textsc{J.~Rodriguez-Quintero},
 \jr{Phys. Rev.} \textbf{D80}, 085018 (2009).


\bibitem{Gromenko:2008tn}
 \textsc{O.~Gromenko},
 \jr{arXiv:0710.1591}.


\bibitem{Aoki:2009sc}
 \textsc{Y.~Aoki} \etal{},
 \jr{JHEP} \textbf{06}, 088 (2009).


\bibitem{Panero:2009tv}
 \textsc{M.~Panero},
 \jr{Phys. Rev. Lett.} \textbf{103}, 232001 (2009).


\bibitem{Datta:2010sq}
 \textsc{S.~Datta} and  \textsc{S.~Gupta},
 \jr{Phys. Rev.} \textbf{D82}, 114505 (2010).


\bibitem{Borsanyi:2010zi}
 \textsc{S.~Borsanyi} \etal{},
 \jr{arXiv:1011.4230}.


\bibitem{Bazavov:2010pg}
 \textsc{A.~Bazavov} and  \textsc{P.~Petreczky},
 \jr{PoS} \textbf{LATTICE2010}, 169 (2010).


\bibitem{Kanaya:2010qd}
 \textsc{K.~Kanaya},
 \jr{AIP Conf. Proc.} \textbf{1343}, 57--62 (2011).


\bibitem{Fischer:2009wc}
 \textsc{C.\,S. Fischer},
 \jr{Phys. Rev. Lett.} \textbf{103}, 052003 (2009).


\bibitem{Fischer:2009gk}
 \textsc{C.\,S. Fischer} and  \textsc{J.\,A. Mueller},
 \jr{Phys. Rev.} \textbf{D80}, 074029 (2009).


\bibitem{Fischer:2010fx}
 \textsc{C.\,S. Fischer},  \textsc{A.~Maas},  and  \textsc{J.\,A.
  Muller},
 \jr{Eur. Phys. J.} \textbf{C68}, 165--181 (2010).


\bibitem{Pawlowski:2010ht}
 \textsc{J.\,M. Pawlowski},
 \jr{arXiv:1012.5075}.


\bibitem{Polyakov:1978vu}
 \textsc{A.\,M. Polyakov},
 \jr{Phys. Lett.} \textbf{B72}, 477--480 (1978).


\bibitem{Greensite:2003bk}
 \textsc{J.~Greensite},
 \jr{Prog. Part. Nucl. Phys.} \textbf{51}, 1 (2003).


\bibitem{Gattringer:2006ci}
 \textsc{C.~Gattringer},
 \jr{Phys. Rev. Lett.} \textbf{97}, 032003 (2006).


\bibitem{Synatschke:2007bz}
 \textsc{F.~Synatschke},  \textsc{A.~Wipf},  and  \textsc{C.~Wozar},
 \jr{Phys. Rev.} \textbf{D75}, 114003 (2007).


\bibitem{Bilgici:2008qy}
 \textsc{E.~Bilgici},  \textsc{F.~Bruckmann},  \textsc{C.~Gattringer},  and
  \textsc{C.~Hagen},
 \jr{Phys. Rev.} \textbf{D77}, 094007 (2008).


\bibitem{Kashiwa:2008bq}
 \textsc{K.~Kashiwa},  \textsc{M.~Matsuzaki},  \textsc{H.~Kouno},
  \textsc{Y.~Sakai},  and  \textsc{M.~Yahiro},
 \jr{Phys. Rev.} \textbf{D79}, 076008 (2009).


\bibitem{Sakai:2008py}
 \textsc{Y.~Sakai},  \textsc{K.~Kashiwa},  \textsc{H.~Kouno},  and
  \textsc{M.~Yahiro},
 \jr{Phys. Rev.} \textbf{D77}, 051901 (2008).


\bibitem{Bilgici:2009tx}
 \textsc{E.~Bilgici} \etal{},
 \jr{Few Body Syst.} \textbf{47}, 125--135 (2010).


\bibitem{Zhang:2010ui}
 \textsc{B.~Zhang},  \textsc{F.~Bruckmann},  \textsc{C.~Gattringer},
  \textsc{Z.~Fodor},  and  \textsc{K.\,K. Szabo}(2010).


\bibitem{Mukherjee:2010cp}
 \textsc{T.\,K. Mukherjee},  \textsc{H.~Chen},  and  \textsc{M.~Huang},
 \jr{Phys. Rev.} \textbf{D82}, 034015 (2010).


\bibitem{Gatto:2010qs}
 \textsc{R.~Gatto} and  \textsc{M.~Ruggieri},
 \jr{Phys. Rev.} \textbf{D82}, 054027 (2010).


\bibitem{Fischer:2011mz}
 \textsc{C.\,S. Fischer},  \textsc{J.~Luecker},  and  \textsc{J.\,A.
  Mueller},
 \jr{Phys. Lett.} \textbf{B702}, 438--441 (2011).


\bibitem{Huber:2011qr}
 \textsc{M.\,Q. Huber} and  \textsc{J.~Braun},
 \jr{arXiv:1102.5307}.


\bibitem{Catterall:2008qk}
 \textsc{S.~Catterall},  \textsc{J.~Giedt},  \textsc{F.~Sannino},  and
  \textsc{J.~Schneible},
 \jr{JHEP} \textbf{11}, 009 (2008).


\bibitem{DelDebbio:2008zf}
 \textsc{L.~Del~Debbio},  \textsc{A.~Patella},  and  \textsc{C.~Pica},
 \jr{Phys. Rev.} \textbf{D81}, 094503 (2010).


\bibitem{DelDebbio:2009fd}
 \textsc{L.~Del~Debbio},  \textsc{B.~Lucini},  \textsc{A.~Patella},
  \textsc{C.~Pica},  and  \textsc{A.~Rago},
 \jr{Phys. Rev.} \textbf{D80}, 074507 (2009).


\bibitem{Sannino:2009aw}
 \textsc{F.~Sannino},
 \jr{Phys. Rev.} \textbf{D79}, 096007 (2009).


\bibitem{Patella:2010dj}
 \textsc{A.~Patella},  \textsc{L.~Del~Debbio},  \textsc{B.~Lucini},
  \textsc{C.~Pica},  and  \textsc{A.~Rago},
 \jr{PoS} \textbf{LATTICE2010}, 068 (2010).


\bibitem{Mojaza:2010cm}
 \textsc{M.~Mojaza},  \textsc{C.~Pica},  and  \textsc{F.~Sannino},
 \jr{Phys. Rev.} \textbf{D82}, 116009 (2010).


\end{thebibliography}

\end{document}